\documentclass[b5paper,10pt]{book}
\usepackage[T1]{fontenc} 
\usepackage{lmodern}
\usepackage{graphicx}
\usepackage{epsfig}
\usepackage{rotating}
\usepackage{amssymb}
\usepackage{dsfont}
\usepackage{psfrag}
\usepackage{amsmath,euscript,array,mathrsfs}
\usepackage{bbold}
\usepackage{epsf}
\usepackage{slashed}
\usepackage{cancel}
\usepackage{background}
\usepackage{cite}

\usepackage{caption,subcaption,wrapfig}
\usepackage[colorlinks=true,linktocpage=true,linkcolor=blue,citecolor=blue,urlcolor=blue]{hyperref}

\usepackage{tikz}
\usetikzlibrary{decorations.markings,decorations.pathmorphing}

\usepackage{titlesec}
\newcommand*{\mytitlefont}{\fontfamily{pag}\selectfont}

\titleformat{\chapter}[display]
{\mytitlefont\Large}
{\filright\MakeUppercase{\chaptertitlename} \Huge\thechapter}
{1ex}
{\titlerule\vspace{1ex}\filleft}
[\vspace{1ex}\titlerule]
\newcommand{\beq}{\begin{equation}}
\newcommand{\eeq}{\end{equation}}
\newcommand{\be}{\begin{equation}}
\newcommand{\ee}{\end{equation}}
\newcommand{\beqs}{\begin{eqnarray}}
\newcommand{\eeqs}{\end{eqnarray}}

\def\section#1{%
	\vspace{4ex}	
	\noindent
	\refstepcounter{section}%
	\addcontentsline{toc}{section}{$\quad$\thesection $\ $ #1}%
	{ 
		\mytitlefont	 \Large 
		\strut\S \thesection \quad
		#1}%
	\par\vspace{1ex}
}

\def\subsection#1{
	\vspace{2ex}	
	\noindent
	\refstepcounter{subsection}%
	\addcontentsline{toc}{subsection}{#1}%
	{ 
		\mytitlefont	 \strut \S \thesubsection \quad  #1}%
	\par\vspace{1ex}
}

\usepackage{fancyhdr}
\fancypagestyle{plain}{\fancyhf{}} 
\newcommand{\dd}{\mathrm{d}}
\newcommand{\sac}{\, , \qquad}

\newcommand{\AAA}{\mathcal{A}}
\newcommand{\FF}{\mathcal{F}}
\newcommand{\OO}{\mathcal{O}}
\newcommand{\GG}{\mathcal{G}}
\newcommand{\MM}{\mathcal{M}}
\newcommand{\NN}{\mathcal{N}}
\newcommand{\LL}{\mathcal{L}}
\newcommand{\BB}{\mathcal{B}}
\newcommand{\B}{\mathds{B}}
\newcommand{\Zinter}{\mathds{Z}}
\newcommand{\C}{\mathds{C}}
\newcommand{\R}{\mathds{R}}
\newcommand{\CP}{\mathds{C}\mathds{P}}
\newcommand{\parent}[1]{\left(#1\right)}
\newcommand{\parentsq}[1]{\left[#1\right]}
\newcommand{\rhot}{\overline{\rho}}
\newcommand{\Bconf}{\mathds{B}_8^{\rm{conf}}}
\newcommand{\Binf}{\mathds{B}_8^{\infty}}
\newcommand{\BOP}{\mathds{B}_8^{\rm{OP}}}
\newcommand{\Bp}{\mathds{B}_8^{+}}
\newcommand{\Bm}{\mathds{B}_8^{-}}
\newcommand{\Bpm}{\mathds{B}_8^{\pm}}
\newcommand{\ffunc}{\mathsf{F}}
\newcommand{\ls}{\ell_s}

\newcommand{\eqq}[1]{(\ref{#1})}
\newcommand{\Sec}[1]{section~\ref{#1}}
\def\cO{{\cal O}}

\def\d{\dd}

\def\pot{V}
\def\Z{{\cal Z}}
\def\W{{\cal W}}
\def\k{\kappa}

\def\lym{\lambda}

\newcommand{\mt}[1]{\textrm{\scriptsize #1}}

\definecolor{c1}{RGB}{189,64,8}
\definecolor{c2}{RGB}{247,128,38}
\definecolor{c3}{RGB}{253,190,17}
\definecolor{c4}{RGB}{68,199,33}
\definecolor{c5}{RGB}{28,173,133}
\definecolor{c6}{RGB}{39,146,182}
\definecolor{c7}{RGB}{2,83,148}
\definecolor{c8}{RGB}{29,53,133}
\definecolor{c9}{rgb}{.5,0,.5}
\definecolor{gray2}{rgb}{.42,.42,.42}

\begin{document}
	$\mbox{}$
	\backgroundsetup{
		scale=1,
		angle=0,
		opacity=1,  
		contents={\includegraphics[]{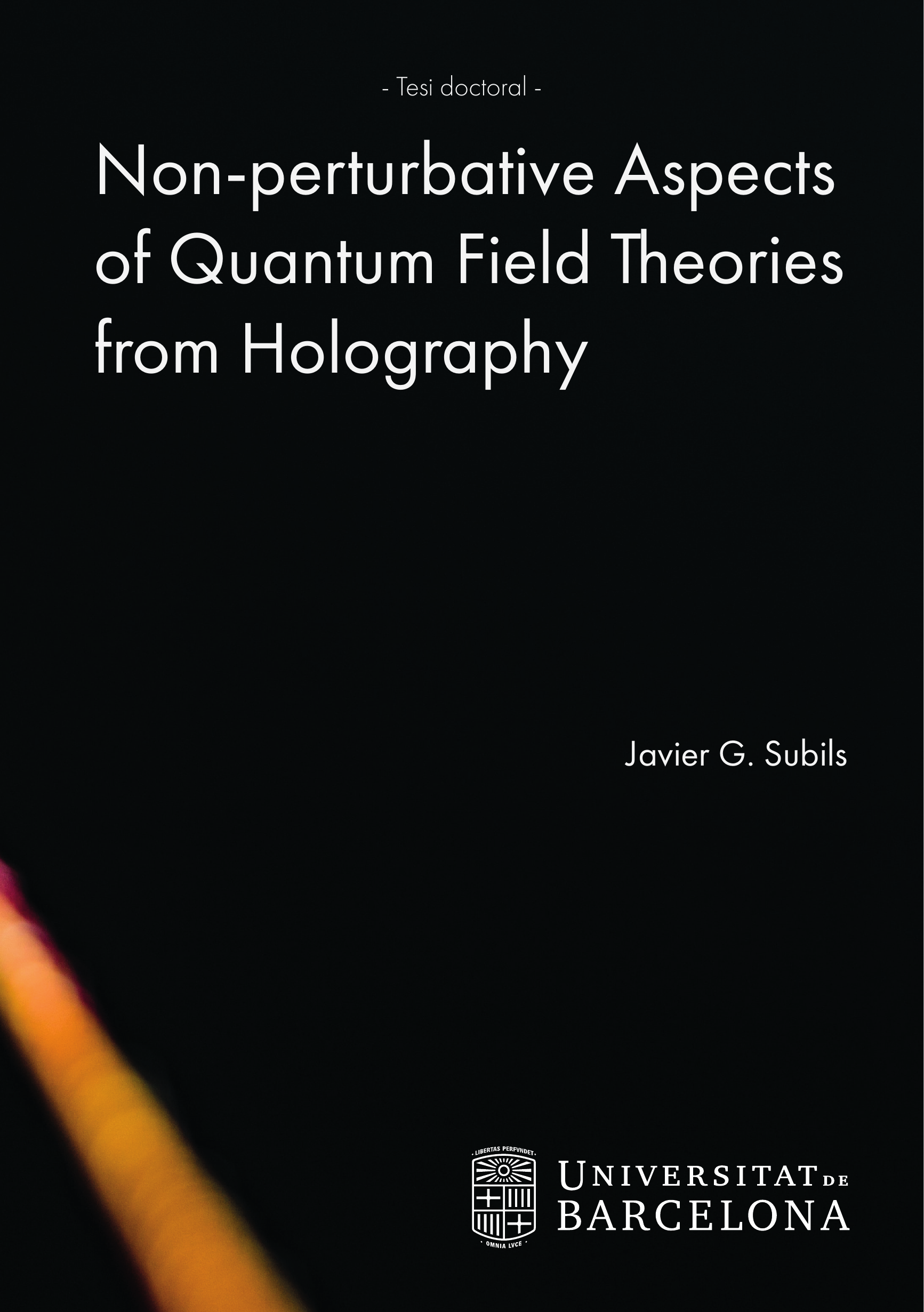}}
	}
	\newpage
	$\mbox{}$
	\thispagestyle{empty}
	\backgroundsetup{
		scale=1,
		angle=0,
		opacity=1,  
		contents={\includegraphics[]{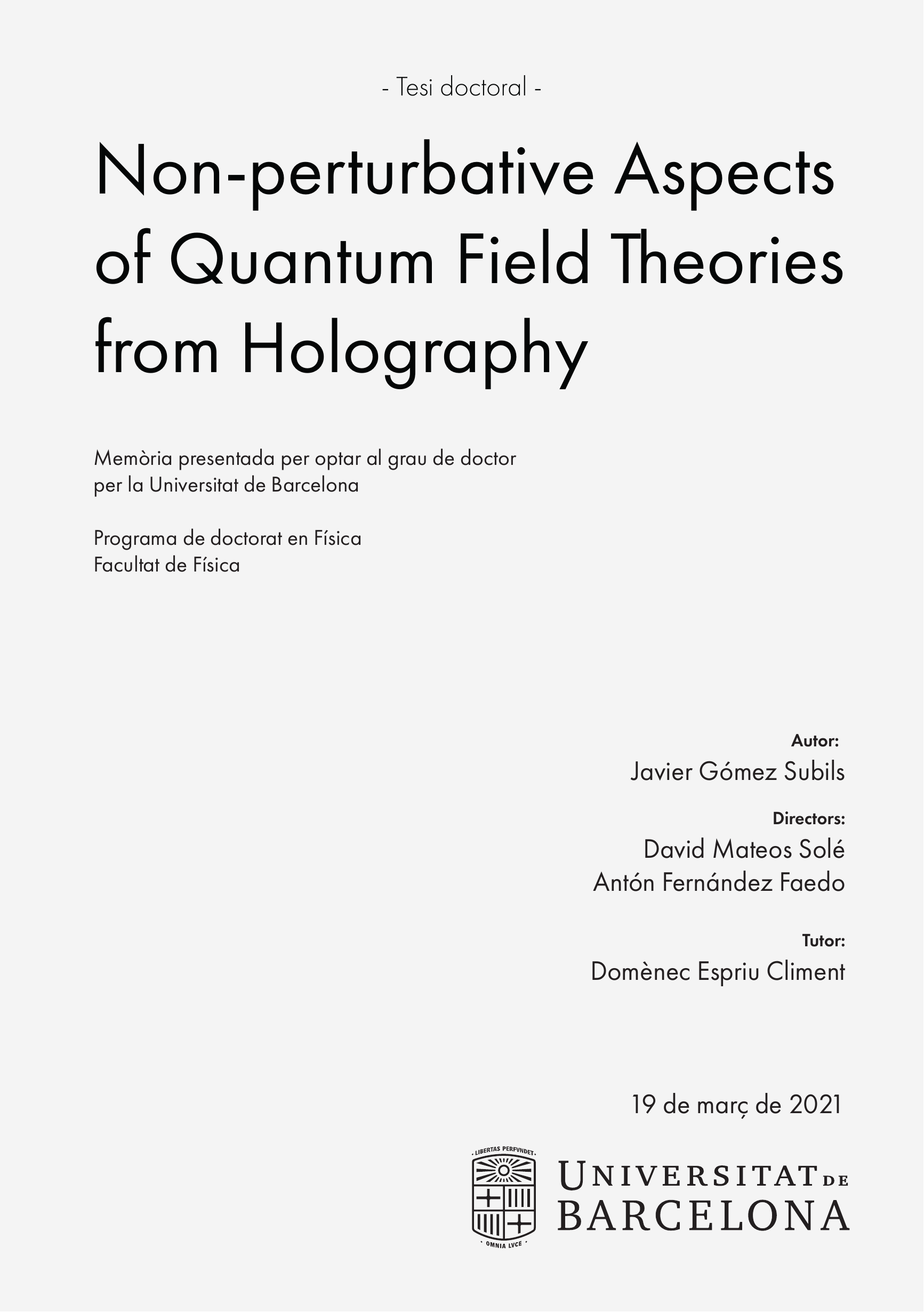}}
	}
	\frontmatter
	\backgroundsetup{
		scale=1,
		angle=0,
		opacity=1,  
		contents={}
	}	
	\voffset=-.5cm
	\textheight = 540pt
	\setcounter{page}{5}

\newpage
	
\thispagestyle{empty}
	
	$\mbox{}$
	\vspace{5cm}
	\begin{flushright}
		\textit{a mis padres}
	\end{flushright}

\newpage

\thispagestyle{empty}
$\mbox{}$

 \newpage

\tableofcontents

\newpage

\thispagestyle{empty}

	$\mbox{}$
\vspace{5cm}
\begin{center}
	{\it "Throughout the universe we can find any number of constant \\ and secretly interwoven relationships."} \cite{Francis}
\end{center}
\newpage

\thispagestyle{empty}
$\mbox{}$

\newpage

\chapter{Preface}

This thesis gathers the main results I have obtained during my PhD studies, from 2016 to 2021. These studies have been performed at the Gravity and Strings group of the \textit{Universitat de Barcelona}, under the supervision of David Mateos and Ant\'on F. Faedo. We have been driven by the wish of understanding the strongly coupled regime of quantum field theories and, for that purpose, we relied on the gauge/gravity correspondence.

The thesis is organised as follows:

\begin{itemize}
	\item Chapter \ref{Chapter1_IntroductionThesis} is an introduction and contextualises its topic. It starts with a brief explanation of string theory and continues reviewing how the AdS/CFT correspondence was born. At the end, it revisits some holographic models which are relevant for our work.
	\item Chapter \ref{Chapter2_B8family} collects the family of supergravity solutions we studied in \cite{Faedo:2017fbv}, dual to super Yang--Mills like gauge theories.
	\item In Chapter \ref{Chapter3_observables}, based on \cite{Faedo:2017fbv,Elander:2018gte,Jokela:2020wgs}, we study different physical properties of these theories. Namely, we discover that generically they posses a mass gap but are not confining. We also show their spectrum and question what information we can extract from entanglement measures.
	\item In Chapter \ref{Chapter4_thermo} we study the phase diagram of the theories by constructing black brane solutions, which are the gravity duals of thermal states in the field theory side. We encounter a rich structure, with first and second order phase transitions and a triple point. This Chapter is based on  \cite{Elander:2020rgv}.
	\item In addition, in Chapter \ref{Chapter5_HoloCCFTs} we revisit the concept of complex conformal field theories and study their gravitational duals, as in \cite{Faedo:2019nxw}. This result was obtained from the intuition we gained studying the former theories and questioning about walking regimes in Renormalisation Group flows.
	\item Finally, in Chapter \ref{Chapter6_Transport} we discuss the application of gauge/gravity dualities from \cite{Hoyos:2020hmq}, concerning the physics of neutron stars.
\end{itemize}

\newpage
\thispagestyle{empty}

\chapter{Acknowledgements}

In the first place, I wish to thank my supervisor David Mateos. It has been
a great pleasure and honour to work with him these years. He has provided me with guidance while allowing me to pursue my own projects freely. Thus, leading me to become an independent researcher.
Many of the academic and scientific skills I have developed during this period
are thanks to him and his advice.

Secondly, I want to acknowledge that this project would not have seen its end without the help of Ant\'on. Starting as the postdoc I would ask for help when I got stuck, he ended as a supervisor of this work. I very much appreciate he fostering me going on with the projects when they seemed to fall apart. Not to mention he being among those who got me fond of bouldering. But that is of course a different topic.

In addition, I would like to express my gratitude to my collaborators Carlos, Daniel, David, Javier and Matti. Working with outstanding scientists like them has been a fruitful experience which has positively influenced me, not only in the academic aspect. I must also mention other people such as Arttu, Christiana, Diego and
Yago, from whom I have benefited many stimulating discussions.

Haluaisin kiitt\"a\"a my\"os Nikoa ja Aleksia mahdollisuudesta ty\"oskennell\"a kolme kuukautta heid\"an kanssaan Fysiikan tutkimuslaitoksessa ja heid\"an yst\"av\"allisest\"a vastaanotostaan. Muistankin ilolla muita laitoksen j\"aseni\"a, kuten Joonasta, Oscaria ja Jania. Heid\"an ansiostaan Suomi on ollut minulle odotettua l\"ampim\"ampi paikka. Lis\"aksi haluaisin kiitt\"a\"a Tavastt\"ahden v\"ake\"a, erityisesti Jyri\"a, hienoista hetkist\"a, joita vietettiin yhdess\"a asuessani siell\"a.\footnote{Also, I would like to thank Santi, not only for his friendship, but also for his valuable comments and advice concerning the design of the cover of this thesis.}

My office colleagues Alan, Jairo, Nikos, Albert, Isabel and Arnan have been irreplaceable partners in this journey. I cannot forget either about the other PhD students of the group: M\'ikel, Marija, David and Raimon; other friends from the faculty such as Adri\`a, Albert, Andreu, Chiranjib, Claudia, Gl\`oria, Iv\' an, the two Joseps and the two Marcs; and also Cristian, \'I\~nigo, Iv\'an, Miguel and \'Oscar.

Gracias a todos aquellos que hab\'eis estado a mi lado durante todo este tiempo. Hab\'eis demostrado ser amigos de verdad al interesaros sobre qu\'e hac\'ia exactamente en mi doctorado (prefiero no mencionar las veces que os cuestionabais si todo eso serv\'ia para algo). O me escuchabais cuando, con no mucha circunspecci\'on, me daba por hablar de la tesis. Lo hab\'eis demostrado en eso, y en tantos otros momentos en que hemos salido a correr, subido monta\~nas, jugado a juegos de mesa, arreglado el mundo, hablado de la vida, hackeado YouTube, organizado convivencias... Ser\'ia injusto que intentara nombraros, porque estoy seguro de que me dejar\'ia a muchos.

Finalmente, 
quiero agradecer a mi familia, especialmente a mis padres, el haberme apoyado siempre en todas las decisiones que he ido tomando. Doy gracias por haberles tenido siempre a mi lado. Sin ellos, nada de esto hubiera estado posible.

\vfill

{\small This thesis and the research stay at the University of Helsinki have been supported by the FPU program of the Spanish Government, fellowships  FPU15/02551 and EST18/00331.}

\vspace{2mm}

{\small \textbf{Cover image:} Photo by \href{https://unsplash.com/@amir_v_ali}{amirali mirhashemian} on \href{https://unsplash.com/}{Unsplash}.}
\mainmatter

\chapter{Introduction}\label{Chapter1_IntroductionThesis}

Building models to understand how Nature works is all Physics is about. Simplifying those models at the same time that they describe a wider variety of phenomena is all Theoretical Physics is about. Unfortunately, when constructing such models we often encounter regions of parameters where the mathematics we use break down. At that moment, we need to either improve our mathematical knowledge of the model or change it. When this happens, however, \textit{dualities} may come into rescue, linking \textit{a priori} unrelated schemes in mysterious and appealing ways, showing that we are not the ones who create neither Nature nor Mathematics, but just those who discover them. This very philosophical discussion, with which the reader may not agree, goes of course beyond the scope of this thesis. Nevertheless, it allows us to introduce the \textit{gauge/gravity duality}, also referred to as \textit{AdS/CFT correspondence} or simply \textit{holography}.\footnote{Strictly speaking, these concepts are not synonyms, but we will use them interchangeably in this thesis.}

Motivated by string theory, the gauge/gravity duality conjectures equivalences between theories. By doing so, it is able to relate observables of the strongly coupled regime of quantum field theories (QFTs) to observables of classical theories of gravity in higher dimensions. Therefore, it provides a powerful and geometrical way to tackle problems which the perturbative approach to QFTs cannot address. Despite lacking of a prove of the correspondence, it has passed many non-trivial tests. Captivated by its success, we plan to understand some non-perturbative aspects of QFTs by means of their gravity duals.

In this Introduction, we review the key ingredients of string theory that we use in the rest of the thesis, as well as the main holographic models that preceded and motivated our study. We will be following mainly \cite{Uranga, Tong:2009np,CasalderreySolana:2011us} (see also \cite{Polchinski} and \cite{PravosFernandez:2019hpe,Bea:2016ekp}).

\newpage
\section{(Super)string theory in a nutshell}

\subsection{String theory has strings}

String theory was born in the context of strong interactions, when a theory of a quantised string attempted to reproduce Veneziano's amplitude \cite{Veneziano:1968yb}, proposed to explain some characteristics of new experimental data. In the early seventies, the Nambu--Goto action \cite{Nambu:1970,Goto:1971ce} appeared for the bosonic string and soon after Ramond, Neveu an Schwarz \cite{Ramond:1971gb,Neveu:1971rx} constructed the fermionic string. However, as soon as Veneziano's amplitude ceased to reproduce the new experimental results, string theory was abandoned in favour of a new and very promising theory for the description of strong interactions: Quantum Chromodynamics (QCD). Instead of being thrown into the trash of useless and forgotten theories, it was reinterpreted at a much more fundamental level by Scherk and Schwarz. In this deeper sense is how we think of it today: as a quantum theory of gravity \cite{Scherk:1974ca}.

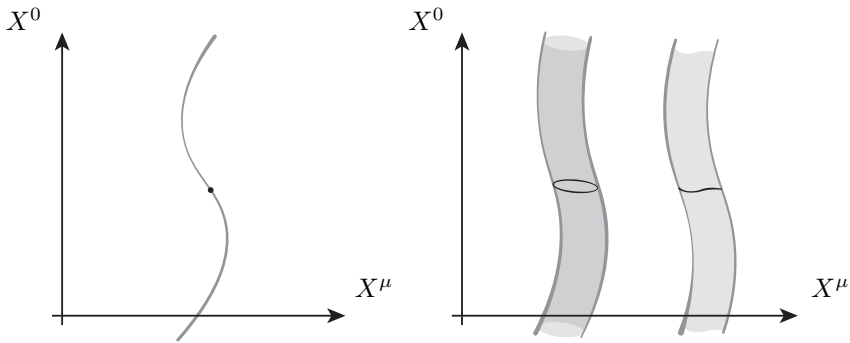
\begin{figure}[t]
	\begin{center}\centering
		\includegraphics[width=\textwidth]{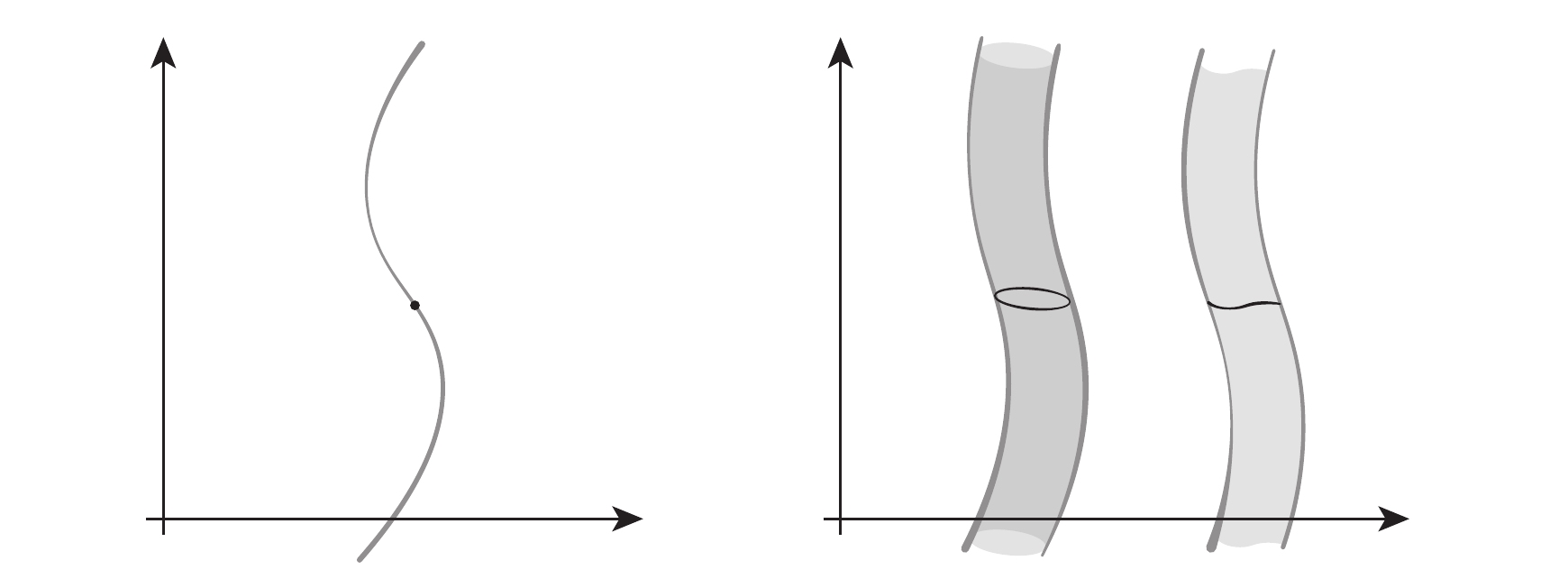} 
		\put(-180,120){$X^0$}
		\put(-30,20){$X^\mu$}
		\put(-330,120){$X^0$}
		\put(-200,20){$X^\mu$}
		\caption{\small (Left) Worldline of a point-like particle propagating in spacetime. (Right) Worldshhets of closed and open string-like particles propagating in spacetime. In both pictures, the vertical and horizontal axes represent the flow of time and the spacial directions respectively.}
		\label{fig:worldsheet}
	\end{center}
\end{figure}

At this fundamental level, the proposal of string theory is that the smallest constituents of the universe are one-dimensional objects called \textit{strings}. They play the role that point-like particles play in QFTs. The length of such string, denoted by $\ell_s$, is the unique parameter of the theory and (its inverse) sets a fundamental energy scale. The state of a single string will be specified by its embedding in the ($D+1$)-dimensional spacetime where it lives. The position at proper string time $\sigma^0\in[\tau_1,\tau_2]$ of a point on the string $\sigma ^1\in[0,2\pi]$  is given by the coordinate functions $X^M(\sigma^\alpha)$ with $M=0\,,\dots\,,D$. The two-dimensional surface obtained in this way is referred to as the string \textit{worldsheet}, which is nothing but the string analogue of point-particles' \textit{worldline} (see Figure \ref{fig:worldsheet}). Note that strings can be both open or closed, depending on whether $X^M(\tau,0) = X^M(\tau,2\pi)$ or not.  The action of the string is given by its area, which is the Nambu--Goto action we mentioned earlier
\begin{equation}\label{eq:NambuGoto}
S_{\text{NG}}(X^M) = \frac{1}{2\pi\ell_s^2}\int_\Sigma \dd^2\sigma\ \sqrt{-\gamma}\,.
\end{equation}
Here $\Sigma$ is the string worldsheet and $\gamma$ is the determinant of the metric induced on the worldsheet by the flat spacetime metric $\eta_{MN}$,
\begin{eqnarray}
\gamma_{ab} = \frac{\partial X^M}{\partial \sigma^a}\frac{\partial X^N}{\partial \sigma^b} \eta_{MN}\,.
\end{eqnarray}
It is convenient to consider an equivalent action known as the Polyakov action
\begin{equation}\label{eq:Polyakov}
S_{\text{Pol}}(X^M,\gamma_{ab}) = \frac{1}{4\pi\ell_s^2}\int_\Sigma \dd^2\sigma \sqrt{-\gamma} \, \gamma^{ab}\, \partial_aX^M\partial_bX^N\, \eta_{MN}\,.
\end{equation}
Note $\gamma_{ab}$ has been promoted to a dynamical degree of freedom. Now that we have the action of the system that we want to study, we follow the standard quantisation procedure. First, we solve the equations of motion and decompose them in Fourier modes. Then, we promote the Fourier coefficients to creation and annihilation operators, in such a way that we obtain the quantised string spectrum. It is important to highlight that this spectrum consists of the different vibrational modes of the string, each of which has a given mass and spin from the spacetime point of view. As a result, a tower of states whose masses are separated by $\sim \ell_s^{-1}$ emerges. At low energies, we can just keep the massless sector and ignore the rest of the states. These massless states decompose into a traceless symmetric component $G_{MN}$, an anti-symmetric part $B_{MN}$ and a trace $\Phi$. 

Note that the state associated to the $G_{MN}$ field is a massless spin-2 particle. This should make us think of gravity, and we will call this particle the \textit{graviton}. In fact, there are general arguments 
\cite{Feynman:1996kb} that any theory of interacting massless spin-2 particles must be equivalent to general relativity. Interestingly, even though we have started with a flat metric, we see that spacetime becomes dynamical. The anti-symmetric $B_{MN}$ tensor, which from the mathematical point of view is a 2-form, is referred to as the \textit{Kalb--Ramond field}. Finally, the scalar field corresponding to the trace is called the \textit{dilaton}. Should we look a bit closer to the spectrum, we realise that Lorentz invariance fixes the number of dimensions to $D=26$ and that it has a tachyon.

Everything we have said so far is for a single string. Interactions between strings are introduced perturbatively and geometrically We postulate that one string can split into two and that two strings can join at a vertex of strength ${\bf g}_s$, which is not a free parameter but given by the expectation value of the dilaton, ${\bf g}_s = e^\Phi$. Consequently, it will in principle depend on the position in spacetime.  We may still speak of \textit{the string coupling constant} $g_s$, meaning the value of the dilaton at infinity, $g_s = e^{\Phi(\infty)}$.

\begin{figure}[t]
	\begin{center}\centering
\includegraphics[width=.9\textwidth]{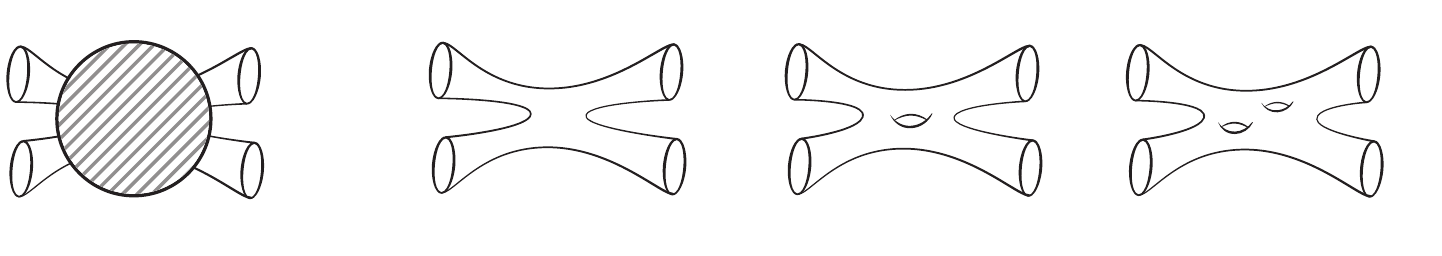} 
\put(-10,25){$+\cdots$}
\put(-42,0){${\bf g}_s^6$}
\put(-81,25){$+$}
\put(-117,0){${\bf g}_s^4$}
\put(-157,25){$+$}
\put(-192,0){${\bf g}_s^2$}
\put(-240,25){$=$}
		\caption{\small Genus expansion of the two-to-two closed strings scattering amplitude.}
		\label{fig:genusExpansion}
	\end{center}
\end{figure}
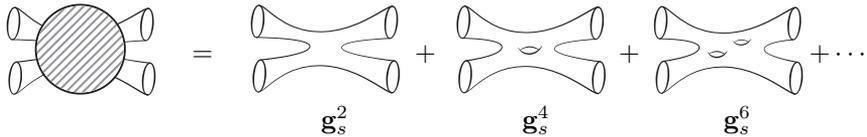

For illustration, imagine we want to study the scattering of two-to-two strings. Following the logic of Feynman diagrams, it will be given by the sum over all possible topologies that link the initial and the final state, as shown in Figure \ref{fig:genusExpansion}. Note that a genus expansion arises, in which the contribution of a surface with $h$ holes turns out to be weighted by a factor ${\bf g}_s^{2h-2}$. The form of the final amplitude is
\begin{equation}
\mathscr{A} = \sum_{g_0}^{\infty}\ {\bf g}_s^{2h-2} \, f_g(\ell_s)\,,
\end{equation}
where $f_s(\ell_s)$ depends on the details of each topology.

Up to this point, what we have been describing is the bosonic string theory. Since our universe contains fermions such as electrons, it is desirable to include fermionic degrees of freedom in string theory, which happen to be fermionic modes on the string worldsheet. The resulting theory is supersymmetric and named after this property as superstring theory. Together with the graviton, the dilaton and the Kalb--Ramond field (which in the context of superstring theory is referred to as the \textit{Neveu--Schwarz (NS) 2-form}), there will be additional fields. If fact, rather than ending with a unique theory, this leads to a set of them whose characteristics we mention next. Their low energy ($E\ll \ell_s^{-1}$) effective actions give rise to different \textit{supergravity theories}.

\renewcommand{\labelitemi}{$\blacksquare$}
\begin{itemize}
	\item \textbf{Type IIA.} In this case the theory contains exclusively closed oriented strings. By oriented we mean that, when interactions are considered, only orientable manifolds are included in the genus expansion. Together with the graviton, the NS 2-form and the dilaton, it contains a 1-form $(C_1)_M$ and a 3-form $(C_3)_{MNP}$. These $p$-form fields are gauge potentials and the action is invariant under the gauge transformation $C_p\mapsto C_p+\dd\Lambda_{p-1}$. Out of them we construct the field strengths $F_{n+1} = \dd C_{n}$, which contain the gauge invariant information and are referred to as the \textit{Ramond--Ramond (RR) fields}. Note also that the degrees of freedom encoded by the $C_p$ form could alternatively be encoded by its dual $(8-p)$-form $C_{8-p}$ via Hodge duality acting on their field strengths:
	\begin{equation}
	F_{p+1} = *F_{9-p}\,,
	\end{equation}
	with $*$ the Hodge star in 10 dimensions.
	The low energy effective action of type IIA string theory is then that of non-chiral $\NN = 2$ and $D=10$ supergravity.
	
	\item\textbf{Type IIB} is also a theory of closed oriented strings. Apart from $G_{MN}$, $B_{MN}$ and $\Phi$, it is endowed with an additional 2-form $(C_2)_{MN}$ and a \mbox{4-form}  $(C_4)_{MNPQ}$ RR fields which are gauge potentials. The field strength of the latter is self-dual. In this case, the low-energy effective theory is chiral $\NN=2$ and $D=10$ supergravity.
	
	\item\textbf{Heterotic SO($32$)}. This string theory is again a theory of closed oriented strings as well. Together with the graviton, the NS 2-form and the dilaton, it is supplemented with 496 gauge bosons $A^a_M$ that correspond to the generators of SO($32$). The low energy effective action turns out to be that of $\NN=1$ and $D=10$ supergravity coupled to SO($32$) gauge vector multiplets.
	
	\item \textbf{Heterotic $E_8\times E_8$} is the same as the previous but replacing the group SO($32$) by $E_8\times E_8$.
	
	\item\textbf{Type I.} 
	This theory contains closed and open unoriented strings, so non-orientable manifolds (such as the M\"obius strip) are included in the genus expansion. Together with the graviton and the dilaton fields, it has a $C_2$ RR 2-form field and 496 gauge bosons that correspond to the generators of SO($32$). Its low-energy effective action is that of $\NN=1$ and $D=10$ supergravity, coupled to SO($32$) gauge vector multiplets.
\end{itemize}

The present status of string theory, which we have just summarised, might seem unsatisfactory and incomplete. And it is for two reasons. First, going beyond the asymptotic expansion in the string coupling is still pending work. Second, we have seen that string theory is not unique but a set of theories. Indeed, these five theories are equally acceptable as perturbative theories of quantum gravity. Nevertheless, the different low energy limits we have explained can be related via dualities. In fact, note that two of the five superstring theories (type I and heterotic SO($32$)) have an equivalent low energy description ($\NN=1$ supergravity in $D=10$ plus Yang--Mills theory with SO($32$) gauge group). This fact leads to the belief that there must be a unique quantum theory,  which would contain the previous theories as different limits of it. This theory would be eleven-dimensional and its fundamental objects would be membranes rather that strings. That is why it is referred to as \textit{M-Theory}.

\subsection{String theory has Dp-branes}

Taking a step forward the perturbative perspective we have adopted so far, we want to consider non-perturbative states of string theory. Of course this discussion is going to be limited by the fact that we still lack of a complete non-perturbative formulation of string theory. However, we can make use of the low energy descriptions we have just mentioned to try to understand such states.

This non-perturbative objects we are interested are extended objects called \textit{branes}. Here we focus on Dp-branes, which appear with $p=1$, $3$, $5$, $7$ and $9$ in type IIB string theory or 
$p=0$, $2$, $4$, $6$ and $8$ in type IIA. These objects can be thought of as solitons: they are stable field configurations which solve the supergravity equations of motion. Typically, they have a localised core and asymptote to flat space. We can regard them as classical excitations over the vacuum of the theory and think of them as field configurations generated by some source sitting at the core of the solution. Because supergravity is an effective theory, it does not provide a microscopic description of these branes. Put differently, these states do not correspond to oscillation states of the string. 

The precise expression of the Dp-brane solution is\footnote{We follow the conventions from \cite{Faedo:2014ana}. In particular, note that (without loss of generality) we have normalised the dilaton in a $g_s$-independent way, so that the actual local string coupling would be $g_s e^{\Phi}$.}
\begin{eqnarray}
\label{eq:Dpbrane}
\dd s^2 &=& G_{MN}\dd x^M\dd x^N = h^{-\frac{1}{2}}\eta_{\mu\nu} \dd x^\mu\dd x^\nu+ h^{\frac{1}{2}}\,\delta_{nm}\,\dd x^n\dd x^m\, \nonumber\\[2mm]
e^{2\Phi} &=& h^{\frac{3-p}{2}}\nonumber \\ [2mm]
h(r) &=& 1+\frac{L^{7-p}}{r^{7-p}} \,,\nonumber \\ [2mm]
F_{8-p} &=& (7-p)\, L^{7-p}\, \omega_{8-p}\,,
\end{eqnarray}
where the indexes $\mu,\, \nu = 0,\dots,p$ correspond to the time and the coordinates along which the brane is extended whereas $m,\, n = p+1,\dots,9$ correspond to those transverse to the brane. Also, $r^2= (x^{p+1})^2,\dots,(x^9)^2$ and $\omega_{8-p}$ is the volume form of the $(8-p)$-sphere. Note that the NS form is set to zero in this case, $B_{MN}=0$. The dimensionful constant $L$, which is arbitrary in the supergravity approximation, is in the quantum theory fixed by the requirement that the Dp-brane charge is quantised,
\begin{equation}\label{eq:quant}
\int_{\text{S}^{8-p}} F_{8-p}\,=\, 2\kappa_{10}^2\, T_p\, N\,,
\end{equation}
where the ten-dimensional Newton's constant and the Dp-brane tension are given by
\begin{equation}\label{eq:newtons_tension}
\frac{1}{2\kappa_{10}^2}\,=\, \frac{2\pi}{(2\pi\ell_s)^8\, g_s^2}\,,\qquad T_p \,=\, \frac{1}{(2\pi\ell_s)^p\, g_s\, \ell_s}
\end{equation}
respectively. From \eqref{eq:quant} and \eqref{eq:newtons_tension} we get an expression for $L$, 
\begin{equation}\label{eq:generalLquant}
L^{7-p} \,=\, \frac{(2\pi\ell_s)^{7-p}}{(7-p)\, V_{8-p}}\, g_sN\,.
\end{equation}
Here $V_{8-p}= \int_{\text{S}^{8-p}} \omega_{8-p}$ is the volume of the $(8-p)$-sphere. Because of the supergravity approximations, these solutions are only going to be valid at the lowest order in $\ell_s$ (curvature lengths larger than the string length) and at leading order in $g_s$. 

These objects will play a crucial role in this thesis, since they are the starting point of the gauge/gravity duality, as we explain in the next Section.

\section{The AdS/CFT correspondence}

\subsection{A duality to rule them all}

In type IIB string theory, a stack of $N$ D3-branes can be regarded from two different points of view. On the one hand, it can be thought of as a defect in flat spacetime where open strings can end. If the low energy limit of this description is taken, open strings attached to the brane decouple from closed strings propagating in ten-dimensional spacetime, as well as closed strings from each other. In this limit, the massive modes can be integrated out. The massless spectrum of the open strings contains six scalar fields $\phi^i$, $i=1,\dots,6$, gauge fields $A_\mu$ and four Weyl fermions, all of which are in the adjoint representation of SU($N$).\footnote{More precisely, the group would be U($N$), but the a U($1$) factor inside U($N$) is free and can be decoupled.} The low-energy effective action for this matter content turns out to be that of the $\NN=4$ super Yang--Mills (SYM) theory with gauge group SU($N$) in ($3+1$) dimensions, which is a superconformal field theory (SCFT). For definiteness, let us give the bosonic part of the Lagrangian that is obtained in this limit:
\begin{equation}
\mathcal{L} = \frac{1}{g_{\text{YM}}^2}\text{Tr}\parent{\frac{1}{4}F^{\mu\nu}F_{\mu\nu} + \frac{1}{2}D_\mu \phi^iD^\mu \phi^i + [\phi^i, \phi^j]^2}
\end{equation}
where $g_{\text{YM}}^2 = 2\pi g_s$ is the Yang--Mills (YM) coupling constant and $F_{\mu\nu}$ and $D_\mu$ are the field strength and covariant derivative corresponding to $A_\mu$ respectively. Note that this description turns useful when $g_s N$, which controls the loop expansion of the theory, is small. In such scenario, we are dealing with weakly coupled open strings. In gauge theory language this translates to a small \mbox{'t~Hooft} coupling 
\begin{equation}\label{eq:Hooft}
\lambda = g_{\text{YM}}^2 N
\end{equation}
and physics can be addressed via perturbation theory.

On the other hand, D3-branes are massive objects, and hence they bend the fabric of spacetime according to the principles of general relativity. As we discussed, Dp-branes are solutions of supergravity. In particular, \eqref{eq:Dpbrane} with $p=3$ is a solution to the type IIB supergravity equations of motion. The spacetime metric in this case reads
\begin{equation}\label{D3brane}
\dd s^2 = h^{-\frac{1}{2}}\parent{-\dd t^2 + \dd x_1^2+ \dd x_2^2+ \dd x_3^2} + h^{\frac{1}{2}}\parent{\dd r^2 + r^2\dd \Omega_5^2}\, .
\end{equation}
Recall that the $x_i$ coordinates inside the first parenthesis correspond to the coordinates in the directions along which the brane extends. The second parenthesis describes the part of the metric transverse to the brane, which here is just six-dimensional flat space. We have chosen to write it as a \textit{cone} over the five-sphere, given by the solid angle $\dd\Omega_5^2$ in the five-sphere and the radial direction $r$.  From \eqref{eq:Dpbrane} and \eqref{eq:generalLquant}, and taking into account that the volume of the five-sphere is $V_5=\pi^3$, the warp factor $h$ is given by
\begin{equation} \label{eq:Randls}
h(r) = 1+\frac{L^4}{r^4}\,,\qquad \frac{L^4}{\ell_s^4} = 4\pi g_s N\,.
\end{equation}
Recall that $L$ is related to the number of branes because we imposed that their charge is quantised. The dilaton is constant in this case.

The metric \eqref{D3brane} has two interesting separated regions. The region near the origin in the transverse directions, which is the region $r\ll L$ (\textit{i.e.} near the D3-branes) is approximated to
\begin{equation}\label{eq:AdS5S5}
\dd s^2 \approx \frac{r^2}{L^2}\parent{-\dd t^2 + \dd x_1^2+ \dd x_2^2+ \dd x_3^2} + L^2\ \frac{\dd r^2}{r^2} + L^2\dd \Omega_5^2 \, =\,  \dd s^2_ {\text{AdS}_5}\,+\, L^2\,\dd \Omega_5^2\,.
\end{equation}
As stated, this is the metric of the product of five-dimensional Anti-de Sitter AdS$_5$ space times the round five-sphere S$^5$ of radius $L$. On the opposite limit $r\gg L$, that is to say, very far away from the branes, \eqref{D3brane} reduces to ten-dimensional flat space. In conclusion, the spacetime sourced by a stack of $N$ coincident D3-branes can be thought of as a distant flat region plus an AdS$_5$ throat near them.

What is the the low energy limit in this case? It consist of the excitations that have arbitrarily low energy when measured from the asymptotic region. Naively, low energy (and thus, non interacting) closed strings propagating far away from the brane must be considered. However, the whole tower of states corresponding to closed strings living in the throat needs to be included, since they get redshifted when measured from infinity. Note that the description at hand is useful when $L/\ell_s \gg1$ in such a way that we need not deal with sub-string-scale geometry and higher order curvature corrections can be neglected. Interestingly, from \eqref{eq:Randls} we conclude that this is the limit $g_sN\gg1$. If we further demand that $g_s$ is small, so that strings are not interacting, this limit is accomplished by taking the large $N$ limit.

In conclusion, we have found that a stack of $N$ coincident D3-branes can be described by
\begin{itemize}
	\item[--] a three-dimensional defect to which open strings can be attached, living in ten-dimensional flat spacetime where close strings can propagate. In the low energy limit, we obtain $\NN = 4$ SYM theory in $(3+1)$ dimensions decoupled from free closed strings propagating in flat ten-dimensional spacetime.
	\item[--] the curved spacetime \eqref{D3brane} where only closed strings can propagate. In the low energy limit, we obtain type IIB closed string theory in AdS$_5\times$S$^5$ decoupled from free closed strings propagating in flat ten-dimensional spacetime.
\end{itemize}
Therefore, it is natural to conjecture that $\NN = 4$ SYM theory in $(3+1)$ dimensions is dual to type IIB string theory in AdS$_5\times$S$^5$ \cite{Maldacena:1997re}.  Moreover, as we mentioned earlier, the regime in which these two descriptions are useful is not the same. As a consequence, perturbation theory results performed in the gauge theory side at small coupling are related to higher curvature corrections of the gravitational theory. Conversely, classical gravity in AdS$_5\times$S$^5$ provides a way to understand the strongly coupled regime of the CFT. 

As a consistency check of the conjecture, we can examine if the symmetries on both parts of the duality match. The $\NN=4$ SYM theory is invariant under Conf$(1,3)\times$SU($4$). The first factor corresponds to the conformal group, consisting of the Poincar\' e group, four special conformal transformations and the dilatation symmetry, generated by the dilatation operator
	\begin{equation}
	D:\quad x^\mu\ \to\ \Lambda \, x^\mu\,.
	\end{equation}
	The second factor is the R-symmetry of the theory, $\text{SU}(4)\simeq\text{SO}(6)$. As a result, the group is isomorphic to the isometry group of the AdS$_5\times$S$^5$ metric \eqref{eq:AdS5S5}, which is SO$(2,4)\times$SO$(6)$. The first factor corresponds to the isometries of AdS and is isomorphic to Conf$(1,3)$. The second is the isometry group of the five-sphere.
	
	Let us examine dilatations closer. The coordinates $x^\mu$ in \eqref{eq:AdS5S5} are those parallel to the brane, and hence should be identified with the gauge theory coordinates. In the gravity side, the isometry is accompanied with the rescaling of $r$
	\begin{equation}
	r\ \to\ \Lambda^{-1} \, r\,.
	\end{equation}
	Short-distance physics in the gauge theory are then associated to physics near the boundary of AdS (large $r$). The radial coordinate should be then identified with the energy scale of the Renormalisation Group (RG) flow in the gauge theory.

For the sake of illustration, let us apply the duality to study the strongly coupled regime of the gauge theory at finite temperature before finishing this Section. This is accomplished by considering a black brane solution in AdS$_5\times$S$^5$. Its metric
\begin{eqnarray}
\label{eq:blackbrane_D3}
\dd s^2 &=& \frac{r^2}{L^2}\parent{-f(r)\dd t^2 + \dd x_1^2+ \dd x_2^2+ \dd x_3^2} + \frac{L^2}{f(r)}\ \frac{\dd r^2}{r^2} + L^2\,  \dd \Omega_5^2\,, \nonumber \\[2mm]
f(r)&=&1-\frac{r_H^4}{r^4}\,,
\end{eqnarray}
gives the relation between the entropy density $S$, set by the area of the horizon; and the temperature $T$, fixed by demanding regularity at the horizon. The result is
\begin{equation}
S\ =\ \frac{\pi^2}{2} \ N^2\ T^3\,,
\end{equation}
which differs \cite{Gubser:1996de} from the weak coupling result by a factor of $3/4$. Note that the parametric dependence $S\propto N^2$ signals that the gauge theory is deconfined.

Unfortunately, the world is not described by $\NN=4$ SYM. At some point, we may be interested in studying non-perturbative aspects of QFTs that are not CFTs. Remarkably, the application of the gauge/gravity correspondence is not constrained to the example we have just reviewed. Rather, it happens to be instructive in many other scenarios. To see that, let us first discuss how considering other Dp-branes --and not just D3's-- allows us to describe theories which are not conformally invariant.

\subsection{Non-conformal holography}

Soon after this first realisation of holography, analogous connections between gauge theories and classical gravity were explored \cite{Itzhaki:1998dd} considering other types of Dp-branes. The generalisation states that maximally supersymmetric SYM in $d = p + 1$ dimensions with gauge group SU$(N)$ can be realised as the worldvolume theory of $N$ coincident Dp-branes in type II string theory in the the decoupling limit, \textit{i.e.} $\ell_s\to 0$, while the Yang--Mills coupling, which now reads
\begin{equation}\label{eq:YMcoupling}
g_{\text{YM}}^2 \,=\, 2\pi g_s\, (2\pi\ell_s)^{p-3}
\end{equation}
is kept fixed. Note that both the YM coupling $g_{\text{YM}}$ and the 't~Hooft coupling $\lambda$, given by \eqref{eq:YMcoupling} and \eqref{eq:Hooft} respectively, will be dimensionful as long as $p\neq3$. We review the case of D2-branes, since it is the one we are mostly interested in. From \eqref{eq:Dpbrane} we see that the D2-brane solution to type IIA supergravity equations of motion in the decoupling limit reads\footnote{Recall that, in contrast to \cite{Itzhaki:1998dd}, we work (without loss of generality) with a $g_s$-independent normalised dilaton. Because of that, $e^{2\Phi}|_{\text{there}}=g_s\, e^{2\Phi}|_{\text{here}}$.}

\begin{eqnarray}\label{eq:D2brane}
\dd s^2 &=& \Big(\frac{r}{L}\Big)^{\frac{5}{2}}\,\left(-\dd t^2 + \dd x_1^2 +\dd x_2^2\right) +\Big(\frac{L}{r}\Big)^{\frac{5}{2}}\parent{\dd r^2 + r^2\dd\Omega_6^2}, \nonumber\\[2mm]
e^{2\Phi} &=& \left(\frac{L}{r}\right)^{\frac{5}{2}}\nonumber \\ [2mm]
L^5 &=& 6\pi^2\ell_s^6g_{\text{YM}}^2\nonumber \\ [2mm]
F_{6} &=& 5\, L^{5}\, \omega_{6}\,,
\end{eqnarray}
Following \cite{Itzhaki:1998dd}, we will use $U = r/\ell_s^2$ to translate the radial coordinate in the bulk to the gauge theory energy scale. In this case, the conjecture we are left with is that $\NN=8$ SYM theory in $(2+1)$ dimensions is dual to type IIA closed string theory in the spacetime given by \eqref{eq:D2brane}. An appropriate way of thinking of the decoupling limit in the present case is 
\begin{equation}
U=\frac{r}{\ell_s^2} = \text{fixed}\,,\qquad g_{\text{YM}}^2 = \frac{g_s}{\ell_s} =\text{fixed}\,, \qquad \ell_s\to 0\,.
\end{equation}

Note that, as we mentioned earlier, in this case the YM coupling is dimensionful with dimensions of energy. Thus, the effective coupling of the gauge theory at a given scale is going to run as $g_{\text{eff}}^2\sim g_{\text{YM}}^2N/U$, which complicates a little bit the discussion on which description is reliable in each regime. In the ultraviolet (UV) region $U\gg g_{\text{YM}}^2 N$, where the effective coupling $g_{\text{eff}}$ is small compared to the energy scale,  we can use perturbative SYM theory. In this regime the supergravity description is not reliable because curvatures are large. At intermediate energies, however, the type IIA description is going to be the valid one as long as both the dilaton and the curvature in ten dimensions are small in string units. This constrains the energy to be within the range
\begin{equation}\label{eq:energyrange}
g_{\text{YM}}^2 N^{\frac{1}{5}}\ll U \ll g_{\text{YM}}^2 N\,.
\end{equation}
Notice that \eqref{eq:energyrange} already imposes that we are in the large $N$ limit. Finally, when $U<g_{\text{eff}}^2N^{\frac{1}{5}}$ the dilaton becomes large. This signals that type IIA supergravity ceases to be reliable, since the string theory is becoming strongly coupled. Nevertheless, one of the dualities we mentioned earlier comes into rescue in this case: type IIA can be uplifted to eleven dimensional supergravity. In order to do so, an extra circle has to be added to the metric
\begin{equation}\label{eq:uplift_to11D}
\dd s_{11}^2 = e^{\frac{4\phi}{3}}(\ell_p\dd \varphi + A^\mu \dd x_\mu)^2 + e^{-\frac{2\phi}{3}}\dd s_{10}^2\,.
\end{equation}
with $\varphi\in (0,2\pi)$. From the eleven dimensional perspective, we see that the dilaton diverging is just signalling that the local value of the radius of the M-theory circle becomes larger than the Planck length. If this happens, the eleven-dimensional supergravity description can be trusted as long as the curvature in 11D Planck units
\begin{equation}
\ell_p^2 R\sim \left( \frac{1}{N}\cdot \frac{g_{\text{YM}}^2}{U}\right)^{\frac{1}{3}}
\end{equation}
is small. For large $N$, this happens if $U > g_{\text{YM}}^2$. It was argued in \cite{Itzhaki:1998dd} that in the opposite case $g_{\text{YM}}^2 > U$ a different solution corresponding to M2-branes localised on the M-Theory circle should be considered, in which case the resulting 11D curvature would be $\sim N^{-\frac{1}{3}}$, which is small in the large $N$ limit. For very small energies the geometry becomes that of AdS$_4\times$S$^7$, which corresponds to the low energy superconformal field theory with SO($8$) R-symmetry from \cite{Maldacena:1997re}.

The holographic duality we have exposed in this Section is just an example that illustrates the fact that gauge/gravity duality can be applied to a variety of setups. The interested reader may want to find out more examples in \cite{Itzhaki:1998dd}. We have decided to review the D2-brane case because it is going to be one of the main characters in this thesis. For instance, it is going to be our starting point to construct gravitational duals which will be dual to a family of three-dimensional gauge theories with interesting non-perturbative phenomena, such as having a discrete spectrum with a mass gap. One of them will be a confining theory. Before going into that, we wish to end up this Introduction by discussing how holography can be used in order to approach more realistic physical systems. In particular, in Section \ref{sec_Holo_Conf}, we will see how confinement appears in holographic setups.

\subsection{Top-down versus bottom-up}

Up to this point, we have discovered gauge/gravity duals arising from different limits of string theories. We first discussed the pioneering relation between a classical theory of gravity in AdS$_5$ and a strongly coupled CFT in four dimensions starting with a stack of $N$ D3-branes. Later, we mentioned that this kind of correspondence is also found when other brane setups are considered. This approach is the so called \textit{top-down} holography. It precisely consist in starting with some concrete and fixed string model in 10 or 11 dimensional supergravity. More complicated setups like intersecting branes, smearing of them in the compact space or branes displaced from the rest can be considered, introducing interesting physics. In this procedure, the details of the dual gauge theory (\textit{i.e.} its Lagrangian) are typically known or at least understood. These field theories often keep some amount of supersymmetry and the hope is that some qualitative insight for more realistic theories can be gained out of studying them.

Strongly-coupled phenomena in QFTs can also be addressed by \textit{bottom-up} holography. In this approach holographic models are built following generic principles of gauge/gravity duality. In this case the starting point for describing some strongly-coupled gauge theory is some effective gravity model with which the gauge theory is tried to be mimicked. For instance, one could try to mimic QCD in the Veneziano limit \cite{Gubser:2008ny,Alho:2013hsa,Jarvinen:2011qe,Jokela:2018ers}, understand heavy ion collisions \cite{Chesler:2010bi,Casalderrey-Solana:2013aba,Casalderrey-Solana:2013sxa,Chesler:2015wra,Chesler:2015bba,Chesler:2015lsa,Chesler:2016ceu,Attems:2016tby,Attems:2017zam,Attems:2018gou}, study properties of neutron stars \cite{Hoyos:2016zke,Hoyos:2016cob,Ecker:2017fyh,Fadafa:2019euu,Ishii:2019gta,Hoyos:2020hmq} or even condensed matter systems \cite{Hartnoll:2016apf,Ammon:2015wua,Zaanen:2015oix,Baggioli:2019rrs}. In this approach there is a lot of freedom in the construction of the models, stemming from the absence of a concrete higher dimensional configuration which determines their details. Instead, the matter content, the form of the potentials for the fields and the parameters in them are fitted to experimental data or aimed to reproduce some physical phenomena.

There is an interesting interplay between top-down and bottom-up holography. The former teaches the imprints that its string theory origin leaves into the gravity side of the duality. Guided by this, the latter helps us to construct simpler models that pick only the relevant ingredients and has phenomenological interest.

For the most part of this thesis, we will be using the top-down approach to understand a family of gauge theories with interesting infrared (IR) dynamics. More specifically, we will investigate how confinement arises in the strongly coupled regime from the gravitational perspective, and question which observables signal that a theory is confining. Finally, in the last two Chapters, we will take the bottom-up approach. In Chapter~\ref{Chapter5_HoloCCFTs} we use simple bottom-up models in order to study the phenomena of walking dynamics, by which a theory can develop a hierarchy of scales. We will be considering there a simple model in Einstein-dilaton gravity in five dimensions. This approach will also be used in Chapter~\ref{Chapter6_Transport}, where our ultimate goal will be to understand transport in strongly coupled quark matter. 

\vfill
\newpage

\section{Holographic confinement}\label{sec_Holo_Conf}

Let us make clear in this Section what we mean by confining behaviour. In this thesis, we will say that a theory is confining if it exhibits linearly growing potential between quarks and anti-quarks at large separation. For completeness, let us review different notions of confinement since it is a rather ambiguous concept. 

Several notions of confinement can be found in the literature, and they are not equivalent. A well known fact is that in pure glue SU($N$) YM theory it is possible to consider the Polyakov loop as a well defined order parameter, which gets a non-zero expectation value in the confined phase. If we consider infinitely massive, non-dynamical quarks in the vacuum of this theory, they are attracted by a constant force at large distances, corresponding to the linear quark-antiquark potential we mentioned. This behaviour is sometimes referred to as ``fulfilling an \textit{area law}'', since it is equivalent to the statement that large Wilson loops scale with the area surrounded by them.

Another notion of confinement examines whether, in the large $N$ limit, the free energy is of order $\OO(N^0)$ -- reflecting the contributions of colour singlet hadrons in the confined phase-- or of order $\OO(N^2)$ -- reflecting the contribution of gluons in the deconfined one \cite{Thorn:1980iv}. This is related to the spectrum consisting of colour singlets.

The picture changes, however, when the quarks become dynamical. In that case, one cannot take two quarks arbitrarily far from one another. This would require more and more energy and, eventually, pair production from the vacuum would generate two separated mesons. A refined notion of confinement states then that there are only colour neutral particles in the spectrum of the theory. There is an issue with this definition though, since it also holds for gauge--Higgs theories deep in the Higgs regime. There, there are only Yukawa forces and neither linearly rising Regge trajectories nor colour electric flux tubes appear. This criterion was referred to as ``C confinement'' in \cite{Greensite:2017ajx}, as opposed to "S$_\text{c}$-confinement'', which is an extension of the area law criterion to gauge plus matter theories. It is important to keep in mind that in the presence of flavour degrees of freedom there is no local order parameter for confinement since these phases can be continuously connected with Coulomb and Higgs phases (see, however, recent work \cite{Cherman:2020hbe}).

Now that we have reviewed the main criteria for confinement, and stated which one we use, let us see how confinement appears in holographic constructions.

\subsection{The AdS soliton}\label{sec:AdSsoliton}

The first confining holographic setup ever considered is that of \cite{Witten:1998zw}. Consider $\NN=4$ SYM at a finite temperature $T$. In Euclidean signature, the system lives on $\R^3 \times \text{S}^1$, where the time direction is compactified in a circle of period $\beta=1/T$. At length scales larger than the size of the circle, this theory is effectively described by pure YM theory in three dimensions, which is obtained by performing a Kaluza--Klein reduction on the circle. By doing so, all fermionic modes acquire a tree-level mass of order $1/\beta$ after imposing antiperiodic boundary conditions in the circle, since this constraint projects out the zero-mode. Despite obeying periodic boundary conditions, the scalars also acquire a mass at the quantum level due to their coupling to the fermions. Therefore, in the IR the theory reduces to YM in three dimensions, which is confining and has a mass gap. The Lorentzian version of the theory is obtained by analytical continuation of one of the non-compact three directions we are left with, in such a way that the original finite temperature solution was just a ``theoretical device'' to end up with this confining theory.

The above procedure can be reproduced holographically. The gravity dual of $\NN= 4$ theory at finite temperature has already been given in \eqref{eq:blackbrane_D3}. Guided by the process just described  in the field theory side, we analytically continue the time coordinate $t$ there, $t\mapsto i x_0$. Recall that in that case, the period of the Euclidean time was fixed in such a way that the geometry ends smoothly at the horizon, where $f(r)$ vanishes. In fact, demanding regularity at the horizon of the Riemannian metric is what fixes the Hawking temperature $T$ of the solution. Now, we recover Lorentzian signature by analytically continuing one of the other $x_i$ directions to Euclidean time. Taking, for example, $x_3\mapsto i t$ we are left with the so called \textit{AdS soliton} metric \cite{Witten:1998zw,Horowitz:1998ha}
\begin{equation}\label{eq:AdSSoliton}
\dd s^2 = \frac{r^2}{L^2}\parent{-\dd t + \dd x_1^2+\dd x_2^2+f(r)\dd x_0^2}\ + \ \frac{L^2}{r^2}\parent{\frac{\dd r^2}{f(r)} + r^2\dd\Omega_5^2}
\end{equation}
Note that now it is the circumference winding the $x_0$ \textit{spatial} direction the one that shrinks to zero size at $r=r_0$. The smooth closing off of the metric at a particular value of the radial coordinate signals the presence of a mass gap, since the variation of the metric along the radial coordinate geometrically implements the RG flow of the gauge theory. Another way to see this is that the five-sphere is keeping its finite size at the bottom of the geometry, where space ends. Consequently, not all the fluctuations of the metric, related to the spectrum of the theory, are going to be allowed, but only those that satisfy the appropriate boundary conditions. Thus, we naively expect a discrete spectrum with a mass gap, and that is indeed the case \cite{Brower:2000rp}.
\begin{figure}[t]
	\begin{center}
		\includegraphics[width=\textwidth]{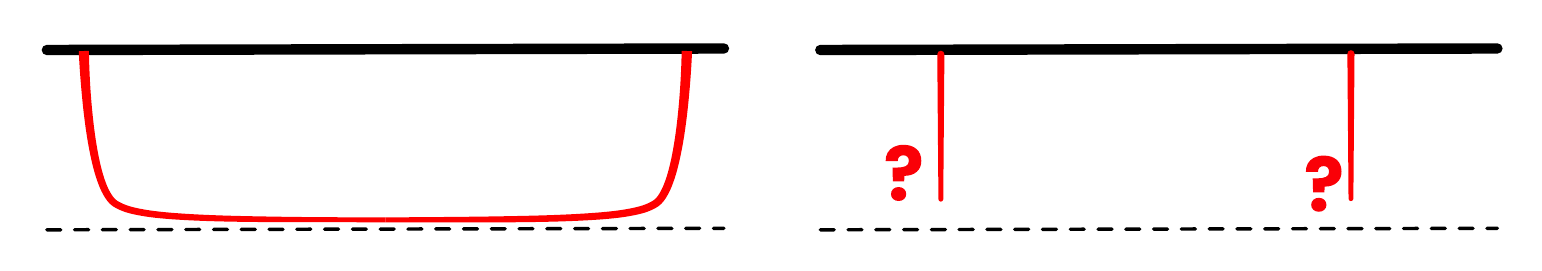} 
		\caption{\small \small (Left) Connected string configuration (thick red curve) in the calculation of the quark-antiquark potential in string theory. The top, continuous, black, horizontal line represents the boundary on which the gauge theory resides. The bottom, dashed, black, horizontal line is the place where the geometry ends smoothly. (Right) Disconnected configuration that is not allowed due to charge conservation, since the endpoints of the strings have no place to end.}	
		\label{fig:stringsqq}
	\end{center}
\end{figure}

The feature we are mostly interested in of this supergravity solution is that it provides a simple geometric picture for confinement, as illustrated in Figure~\ref{fig:stringsqq}. Indeed, we can compute the quark anti-quark potential of a gauge theory by means of its gravity dual, where the flux tube between them is described by a string whose ends are attached to the boundary of spacetime \cite{Maldacena:1998im,Rey:1998ik}. 
If the supergravity solution ends smoothly, for large separations it becomes advantageous for the string to place most of it length near the regular end of the geometry, where it attains a constant, minimum energy per unit length. This leads to a linear quark-antiquark potential at large separations and thus, confinement. Crucially, for this picture to hold the string cannot break apart due to charge conservation. The reason for this is that the end points of the strings carry charge and because of that they can only be attached to branes or fall into horizons. 

In conclusion, the spacetime \eqref{eq:AdSSoliton} provides the holographic picture of a theory with a mass gap and confinement. Notice that both phenomena are linked to the fact that a cycle in the geometry shrinks smoothly at some particular value of the radial coordinate, associated to an energy scale. This raises the expectation that string theory duals of gauge theories that posses a mass gap are also expected to exhibit confinement. The theories we will study in this thesis, however, will provide a counterexample to this expectation.

Now that we know that holography can explain confinement in a geometrical way, it has additional phenomenological interest. For instance, this very same model has been used to simulate real time evolution of plasma balls surrounded by a confining background recently \cite{Bantilan:2020pay}. Considering additional ingredients permits to incorporate more phenomena present in QCD, as we shall see next.

\subsection{The Witten--Sakai--Sugimoto model}

Now that we have obtained a holographic description of confinement, we are tempted to engineer some model based on the ideas from above in order to approach QCD phenomenology. Of course holography ---at least, in the way we know it today--- is not going to provide a perfect dual for QCD. On the one hand, holographic results are only reliable at strong coupling. For that reason we do not expect them to describe the high energy regime of QCD, where the coupling becomes small due to asymptotic freedom. In that regime perturbative QCD works spectacularly well. On the other hand, gauge/gravity duality is trustworthy at large number of colours and in QCD $N=3$. Having said so, we still think that holography can provide very valuable insights into the strongly coupled non-perturbative regime of QCD.

The Witten--Sakai--Sugimoto (WSS) model \cite{Sakai:2004cn}, which is nicely reviewed in \cite{Rebhan:2014rxa}, is a refinement of the AdS soliton. Note that \eqref{eq:AdSSoliton} describes an effectively three-dimensional theory at the IR, since low energy observers will not excite degrees of freedom corresponding to the compact spacial dimension. Nevertheless, the theory is still four-dimensional in the UV, as we will comment on later. If we want to describe an effective, infrared, four-dimensional theory, we shall start with D4-branes in type IIA string theory and proceed analogously. We consider the black brane solution in Euclidean signature and declare one of the coordinates corresponding to a spacial direction to become the time coordinate. 

This model also describes fundamental matter. 
In holography, it is quite straightforward to introducing $N_f$ flavours in the limit $N_f\ll N$ \cite{Aharony:1998xz,Karch:2000gx,Karch:2002sh}.
It corresponds to the addition of $N_f$ D-brane probes (\textit{i.e.} without considering their backreaction on the metric). For this reason, the WSS model is endowed with $N_f$ probe D8-$\overline{\text{D8}}$ brane pairs transverse to the compact S$^1$, which further mimic U($N_f$)$_L\times$U($N_f$)$_R$ chiral symmetry in QCD. In this way, many low-energy QCD phenomena are reproduced, such as confinement, chiral symmetry breaking, a spectrum of mesons and glueballs, baryons and effects from the axial anomaly. Recently, this model has also been used to study nucleation of bubbles and gravitational wave production from first-order phase transitions of strongly coupled gauge theories in the early universe \cite{Bigazzi:2020avc}.
Unfortunately, the fact that WSS model becomes five-dimensional in the UV is its major shortcoming. The reason is that the scale at which this happens cannot be made much larger that 1 GeV without leaving the supergravity approximation, once the observables are related to those of QCD. This is a direct consequence of the shrinking cycle (an S$^1$ in this case) belonging to the gauge theory directions. This drawback can be surpassed by constructing a geometry whose shrinking cycle belongs to the compact part of the ten dimensional metric, as we discuss next.

\subsection{The Klebanov--Strassler model}\label{sec:KlebanovStrassler}

From the two previous models we got the intuition that when a cycle in the geometry of the gravitational description is contracting smoothly, a confining behaviour and a mass gap in the dual gauge theory may arise. However, as we have just pointed out, the cycle being part of the gauge theory directions causes the theory to have an extra spacial dimension in the UV.

This issue can be circumvented by constructing a geometry whose shrinking cycle is internal. In this Section we want to present one of such geometries. The first step is to substitute the five-sphere in \eqref{D3brane} by a different five-dimensional Einstein manifold $X_5$. If in the D3-brane metric \eqref{D3brane} the space transverse to the branes was $\R^6$, which can be thought of as a cone over the five-sphere, now we replace it by a (Ricci flat, six-dimensional) cone over $X_5$. Think of this as placing the stack of $N$ D3-branes at the tip of the cone over $X_5$. Such construction and the intuition gained by the previous constructions, leads us to conjecture that type IIB string theory on AdS$_5\times$X$_5$ is dual to the low-energy limit of the world volume theory of the D3-branes sitting at the singularity.

The transverse six-dimensional space of the model we are interested in here is given by the following equation in $\C^4$:
	\begin{equation} \label{eq:conifold}
	\sum_{n=1}^4 z_n^2 = 0\,.
	\end{equation}
	This space is referred to as \textit{the conifold} and it is a cone over a compact manifold denoted by $X_5 = T^{1,1}$. This is a coset space $T^{1,1} = (\text{SU}(2)\times\text{SU}(2)/U(1))$ 
with topology S$^2\,\times\,$S$^3$. In the so called Klebanov--Witten (KW) model \cite{Klebanov:1998hh}, the D3-branes are place at the singularity in such a way that the resulting $\NN = 1$ supersymmetric theory is still a superconformal field theory and has gauge group SU($N$)$\times$SU($N$). Even though we have not obtained a confining theory yet, the fact that the compact space topology is a product of two spheres will allow us to collapse one of them while keeping the other finite. This will introduce confinement.

For that purpose, note that for many six-dimensional spaces like \eqref{eq:conifold}, there are also fractional D3-branes which can only live within the singularity \cite{Morrison:1998cs,Gimon:1996rq,Douglas:1996xg,Gubser:1998fp,Klebanov:1999rd}. We can think of these fractional D3-branes as D5-branes wrapped over 2-cycles that collapse at the singularity. If $M$ of those fractional branes are added to the singular point of the conifold, the gauge group is modified to SU($N+M$)$\times$SU($N$). Even though the theory is still supersymmetric, it is no longer conformal. In contrast, the relative coupling of the group factors runs logarithmically \cite{Klebanov:1999rd}, which spoils the UV description of the theory. In this context, the Klebanov--Tseylin solution \cite{Klebanov:2000nc} solves the supergravity equations to all orders in $M/N$. However, this construction runs into a singular metric at the IR, after the D3-brane charge eventually becomes negative. We cannot trust this solution at the IR and therefore we are not allowed to question whether the theory is confining.

The singularity is resolved by the Klebanov--Strassler (KS) solution \cite{Klebanov:2000hb}. This is done by considering a deformation of the conifold. Indeed, replace \eqref{eq:conifold} by the deformed conifold
\begin{equation}
\label{eq:deformed_conifold}
\sum_{n=1}^4 z_n^2 = \epsilon^2\,.
\end{equation}
By doing so, the ten-dimensional metric becomes regular at the IR. Interestingly, the S$^2$ factor of $T^{1,1}$ now smoothly  shrinks to zero as the S$^3$ remains finite at the IR. This is precisely what we were looking for, and this theory is indeed confining. 

Let us pause in how the IR of the theory is approached. Recall that the UV of the theory has gauge group SU($N+M$)$\,\times\,$SU($N$) and a logarithmically running coupling. It is possible to study the strong coupling regime by considering the Seiberg dual of the SU($N+M$) factor, which becomes SU($N-M$). Surprisingly, after doing so, the SU($N$)$\,\times\,$SU($N-M$) we arrive to is the same as the original one under the replacement $N\mapsto N-M$. 
In a sense, the RG flow is self-similar and can be though of as a cascade of Seiberg dualities. This RG flow cascade finishes when the gauge group becomes SU($M+p$)$\,\times\,$SU($p$) with $0\leq p < M$. The remaining $p$ colours should come from $p$ branes placed at various points in the background When $p = 0$, the IR theory is SU($M$) and therefore the theory is confining. This can be checked by computing the quark-antiquark potential in this background. When $p>0$, however, confinement is lost. From the supergravity point of view, quarks can simply be attached to one of the $p$ remaining D3-branes. This is the holographic picture of external sources being screened by dynamical quarks.

\section{Mass gap without confinement}

\begin{figure}[t]
	\begin{center}
		\includegraphics[width=\textwidth]{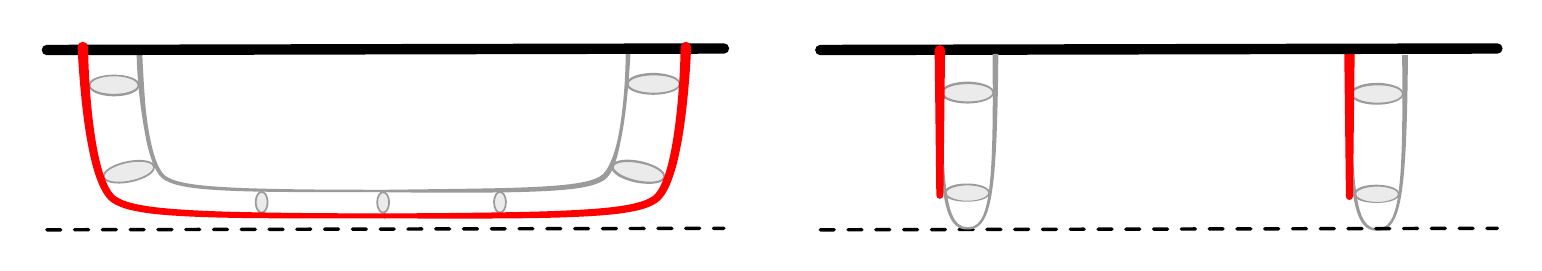} 
		\caption{\small (Left) Connected membrane configuration in the calculation of the quark-antiquark potential in M-theory. The projection of the membrane onto the non-compact directions is represented by the thick, red curve. The M-theory circle at each point is represented next to it by the grey circles. The top, continuous, black, horizontal line represents the boundary on which the gauge theory resides. The bottom, dashed, black, horizontal line is the place where the geometry ends smoothly. (Right) Disconnected membrane configuration  allowed by charge conservation, since the membrane closes off smoothly at the bottom of the geometry and hence it has a cigar-like topology with no boundary.}
		\label{fig:membreanesqq}
	\end{center}
\end{figure}

In all the examples explained earlier, the fact that a circle collapses has caused the dual field theory to be confining, supporting what we illustrated in Figure~\ref{fig:stringsqq}. Recall that this motivated the expectation that string theory duals of gauge theories that possess a mass gap will be confining theories.

The theories we study in this thesis provide a counterexample to this expectation. The key reason for that is that the regular description is given by eleven-dimensional supergravity instead of type IIA. The existence of a mass gap or the confining nature of the theory can only be reliably addressed in the former description. In eleven dimensions, the radial coordinate will indeed end at a finite value, at which point there is an internal cycle (a three-sphere) that contracts smoothly, whereas the rest of the compact part of the geometry (a four-sphere) keeps finite. Thus, there is a mass gap in these theories, as we will argue when we discuss their spectrum in Chapter \ref{Chapter3_observables}. However, they are non-confining theories in general. As we illustrate in Figure \ref{fig:membreanesqq}, the ultimate reason stems from the computation of the quark potential in eleven dimensions. In this case, it is necessary to compute the area of a membrane that winds the M-Theory circle. As we will see, this circle also collapses smoothly at the IR, loosing its boundary. Because of that, there will be no issues related to charge conservation. As a result, a configuration consisting of two isolated membranes attached to the quarks at the boundary and hanging all the way down to the IR is now an allowed and competing configuration. Put differently, isolated quarks are allowed configurations. Moreover, the latter will be the energetically preferred configuration at large separations, cutting off the linear growth of the potential. 

This result makes sense, since having a mass gap has nothing to do, in principle, with the potential between quarks. There are theories with a mass gap and no area law \cite{Witten:1999ds} as well as the opposite \cite{Gubser:2004qj}. The latter is expected, for example, if a spontaneously broken symmetry leads to a massless Goldstone mode. There is no reason to expect the splitting of the flux tube between quarks in such scenario.

This is precisely the physics we want to investigate in this thesis. In the next Chapter, we will give all the details of the type IIA solutions that realise this mechanism. In Chapter \ref{Chapter3_observables}, we will show that generically they are non-confining, in the sense of a linear quark-antiquak potential. We will also show there their spectrum, and realise that there is indeed a mass gap. Moreover, we will examine whether entanglement entropy measures are sensitive to confining dynamics.

In Chapter \ref{Chapter4_thermo} we will discuss the finite temperature phase diagram of these theories. After that, in Chapter \ref{Chapter5_HoloCCFTs} we discuss the effect that flowing near a CFT can have, which is indeed realised in our family of theories. This discussion motivates the proposal of holographic complex conformal field theories, as we shall see.

Finally, in Chapter \ref{Chapter6_Transport} we will study transport in strongly coupled unpaired quark matter using the gauge/gravity correspondence.

\newpage
\thispagestyle{empty}

\chapter{Super Yang--Mills Chern--Simons--Matter theories}\label{Chapter2_B8family}

In the previous Chapter we have introduced holography and argued that it might turn useful in the study of non-perturbative phenomena of quantum field theories. In particular, we explained how confinement arises in holographic theories. We also elaborated on its relation to the shrinking of a cycle of the geometry.  

As we already advanced in the Introduction, in this thesis we will focus on a family of three-dimensional gauge theories, and we will study them by means of their gravitational duals. Let us emphasise that the supergravity solutions presented in this Chapter are not new \cite{Cvetic:2001ye}, but we gather them in what we hope is a user-friendly and comprehensive manner. All these solutions preserve two supercharges, corresponding to $\NN=1$ supersymmetry in three dimensions. Because the small amount of symmetry, it is difficult to determine the precise details of the dual gauge theories, although we can point out some properties they presumably have.
	
First, the interactions of the theories are Yang--Mills like, associated to a two-sites quiver of the form $\text{U}(N+M) \, \times \,\text{U}(N)$, similarly to the Klebanov--Witten theory \cite{Klebanov:1998hh}. The parameter distinguishing different members of the family will be denoted as $b_0\in[0,1]$, and is related to the difference between the inverse squared gauge couplings of each of the two gauge groups in the microscopic theory. The gravity duals preserve $\NN=1 $ supersymmetry and admit both a type IIA and an M-Theory description, though only the latter is regular in general. The theories are further supplemented with Chern--Simons--Matter terms (CSM) \cite{Loewy:2002hu,Hashimoto:2010bq}. We will refer to them as SYM--CSM theories for brevity.

The single SYM--CSM theory in the UV undergoes different RG flows depending on the value of $b_0$, leading to a variety of interesting IR phenomena. This is depicted in Figure \ref{fig:triangle}. For a generic value of $b_0\in(0,1)$, the supergravity solution caps of smoothly at the end-of-space. This is reminiscent of the Klebanov--Strassler solution \cite{Klebanov:2000hb} (see Section \ref{sec:KlebanovStrassler}), although the gauge theories that we will discuss are three-dimensional and with a better behaved UV. On the top of that, our SYM--CSM theories are gapped but non-confining for a generic value of the parameter.

In contrast, the two limiting values of the parameter are special. When $b_0=0$ the gap is lost and the theory flows to an infrared fixed point, dual to a conformal field theory (CFT). On the opposite $b_0=1$ limit, not only is the theory gapped, but also confining. On the gauge theory side, the existence of confinement seems to be a consequence of the absence of CSM interactions for this limiting case. As a result, we can think of these SYM--CSM theories as smoothly interpolating between conformal and confining regimes.

In order to understand the physics of these family of SYM--CSM theories, we review their supergravity dual solutions in this Chapter. Then, having the explicit solutions written down in this Chapter, we will discuss their physical properties afterwards.

\section{Supergravity solutions}

Let us review in this Section the main features of type IIA supergravity solutions dual to the family of gauge theories we consider in this thesis. Even though these solutions are regular only in eleven dimensions, we start their exposition from the ten-dimensional setup instead, since the geometric picture is cleaner. After that, we will explain how the uplift works.

We want to study SYM-like theories in three dimensions. Because of that, our starting point is the D2-brane ansatz \eqref{eq:D2brane}. Additionally, we want to get confining infrared physics with a similar mechanism to that of the KS model (see Section \ref{sec:KlebanovStrassler}). For that we need to replace the six-sphere by some compact manifold with non-trivial cycles. As the internal compact space we choose the squashed $\CP^3$, whose geometry we review in Appendix \ref{ap:CP3} and has the desired properties. The UV geometry will coincide with that induced by a stack of $N$ D2-branes sitting at the tip of a cone over $\CP^3$. This justifies the choice of the following ansatz for the metric (in string frame) and the dilaton field
\begin{equation}
\label{eq:10DansatzAP}\begin{aligned}
\dd s_{\rm st}^2 &=h^{-\frac12}\left[- \dd t^2 + \dd x_1^2 + \dd x_ 2^2\right]+h^{\frac12} \left[\dd r^2+e^{2f}\dd\Omega_4^2+e^{2g}\left[\left(E^1\right)^2+\left(E^2\right)^2\right] \right]\,, \\
e^\Phi&=h^{\frac14} \, e^\Lambda \,.
\end{aligned}
\end{equation}
The dilaton $\Phi$ and the functions $f$, $g$, $h$ and $\Lambda$ are functions solely of the radial coordinate $r$. Note also that $\CP^3$ is seen as the coset $\rm{Sp}(2)/\rm{U}(2)$, consisting of a two-sphere $\rm{S}^2$ (described by the vielbeins $E^1$ and $E^2$) fibred over a four-sphere $\rm{S}^4$, whose metric is given by $\dd\Omega^2_4$.

As explained in Appendix \ref{ap:CP3}, in this geometry there is a set of left-invariant forms, namely two 2-forms ($X_2$ and $J_2$), two 3-forms ($X_3$ and $J_3$), two 4-forms ($X_2\wedge J_2$ and $J_2\wedge J_2$) and the volume form of $\CP^3$ (which we denote by $\omega_6$). The fluxes of the geometry can be written in terms of these left-invariant forms in such a way that this symmetry ensures the consistency of the ansatz. By this we mean that all the internal angles will drop from the resulting equations, leading to dependence just on the radial coordinate. Therefore, we take the following forms 
\begin{equation}\label{eq:fluxesansatz}
\begin{array}{rlcrcl}
F_4 &=&  \dd^3x\wedge \dd(h^{-1}e^{-\Lambda})+ G_4 + B_2\wedge F_2\,, \qquad\qquad & F_2& =& Q_k (X_2 - J_2)\,, \\[2mm]
G_4 &=& \dd(a_J J_3) + q_c\left(J_2\wedge J_2 - X_2\wedge J_2\right)\,, \qquad\qquad & B_2& =& b_X X_2 + b_J J_2 \,,\\[1mm]
\end{array}
\end{equation}
where 
\begin{equation}\label{trunc}
a_J\,=\,\frac{e^{2g}\big( Q_kb_J-q_c\big)-2e^{2f+g-\Lambda}\big(b_J+b_X\big)+e^{2f}\left(q_c+Q_k\big(b_X-b_J\big)\right)}
{2\big(e^{2f}+e^{2g}\big)}\,.
\end{equation}
The parameters $Q_k$ and $q_c$ are constants which will be related to gauge theory parameters as we shall see later. Moreover, $b_J$ and $b_X$ are functions that depend on the radial coordinate which will satisfy first order equations as we argue next.

The system can preserve $\NN=1$ supersymmetry and the resulting BPS equations from the supergravity action can be solved in three consecutive steps. First, we can solve the equations for the metric functions, which follow directly from \cite{Conde:2011sw,Bea:2013jxa} and read
\begin{eqnarray}
\label{BPSsystem}
\Lambda'&=&2Q_k\,e^{\Lambda-2f}-Q_k\,e^{\Lambda-2g}\,, \nonumber\\[2mm]
f'&=&\frac{Q_k}{2}\,e^{\Lambda-2f}-\frac{Q_k}{2}\,e^{\Lambda-2g}+e^{-2f+g}\,, \nonumber\\[2mm]
g'&=&Q_k\,e^{\Lambda-2f}+e^{-g}-e^{-2f+g} \,.
\end{eqnarray}

Once the functions of the metric are known, one turns to solve the equations for the fluxes \cite{Faedo:2017fbv}
\begin{eqnarray} \label{BPSsystem_fluxes}
b_X' & = & 2\,e^{-4f+2g+\Lambda} \big( q_c+2\,a_J-Q_k\,b_J\big)\,, \nonumber \\[2mm]
b_J'&=&e^{-2g+\Lambda}\Big[ Q_k\left(b_J-b_X\right) +2\,a_J-q_c\Big]\,, 
\end{eqnarray}
where $a_J$ is given by \eqref{trunc}. Although there are many solutions in which these functions can be set to zero, we will generally need to keep them in order to get regular solutions. More precisely, they account for the presence of fractional branes in the system and prevent the eleven-dimensional warp factor from blowing up at the IR, which would result into a singularity, via the so called \textit{transgression mechanism}\cite{Cvetic:2000mh,Aharony:2008gk,Cvetic:2001ma}.

Finally, after solving the equations for the fluxes, the ten-dimensional warp factor $h$ is given by direct integration of
\begin{equation}\label{eq:10DwarpFactor_eq}
\left(e^\Lambda\,h\right)' = -e^{2\Lambda-4f-2g}\Big[Q_c+Q_k\,b_J \left(b_J- 2 b_X\right)+ 2 q_c \left(b_X- b_J\right) + 4 a_J\left(b_X+ b_J\right)\Big] \,. \\[1mm] 
\end{equation}
Here, $Q_c$ appears as an integration constant. Together with $Q_k$ and $q_c$, it is related to gauge theory parameters via the quantisation of the Page charges \cite{Marolf:2000cb}
\begin{eqnarray}\label{eq:charges_integral}
N&=& \frac{1}{2\kappa_{10}^2T_{D2}} \int_{\CP^3} \Big[\ -*F_4\ -B_2\wedge F_4\ +\frac{1}{2} B_2\wedge B_2\wedge F_2\ \Big] \,,\nonumber\\[2mm]
\bar M&=& \frac{1}{2\kappa_{10}^2T_{D4}} \int_{\CP^2} \Big[ \ F_4\ -B_2\wedge F_2\  \Big] \ = \ \frac{1}{2\kappa_{10}^2T_{D4}} \int_{\CP^2} G_4\,,\nonumber\\[2mm]
k&=& \frac{1}{2\kappa_{10}^2T_{D6}} \int_{\CP^1} F_2 \,,
\end{eqnarray}
which leads to \cite{Faedo:2017fbv,Hashimoto:2010bq}:
\begin{equation}\label{eq:gauge_parameters}
Q_c = 3\pi^2 \ls^5 g_s N,\qquad Q_k = \frac{\ls g_s}{2} k, \qquad q_c =\frac{3\pi \ls^3g_s}{4} \bar M =  \frac{3\pi \ls^3g_s}{4} \left(M-\frac{k}{2}\right)\ .
\end{equation}
In \eqref{eq:charges_integral} and \eqref{eq:gauge_parameters}, $N$ is the number of D2-branes and matches the rank of the gauge group\footnote{Following \cite{Aharony:2009fc}, the $k/2$ shift has been included in order to account for the Freed--Witten anomaly.} on the field theory. The parameter $Q_k$ gives rise to D4-brane charge, so $k$ is consequently expected to be the CS level of the dual gauge theory. Finally, $q_c$ is related to the number of fractional D2-branes that is needed to obtain a regular geometry \cite{Cvetic:2001ma}, where $M$ represents the shift in the gauge group due to fractional branes \cite{Aharony:2008gk}. In addition, recall that the string length $\ls$ and the string coupling constant $g_s$ are related to the \mbox{'t~Hooft coupling} via \eqref{eq:YMcoupling} and \eqref{eq:Hooft}, which lead to \begin{equation}
\label{eq:tHooft3D}
\lambda =\ls^{-1} g_s N\,,
\end{equation}
and has dimensions of $\sim$(length)$^{-1}$ in three dimensions.

Before getting into the details of the solutions we are interested in, let us show two particular simple solutions which are dual to superconformal field theories. Indeed, when $Q_k>0$ the BPS equations \eqref{BPSsystem} are solved by
\begin{equation}
\label{eq:ABJM}
e^{2f} = r^2\,,\qquad e^{2g} = r^2,\qquad e^{\Lambda} = \frac{r}{Q_k}\,,
\end{equation}
 which is the Aharony--Bergman--Jafferis--Maldacena (ABJM) fixed point \cite{Aharony:2008ug}. The supersymmetry of this solution is enhanced generically to $\NN=6$. Also, when $Q_k<0$ there is the Oogury--Park (OP) fixed point \cite{Ooguri:2008dk} at
 \begin{equation}\label{eq:OPsolution}
 e^{2f} = \frac{9}{5}r^2\,,\qquad  e^{2g} = \frac{9}{25}r^2\,,\qquad  e^{\Lambda} = \frac{3r}{5|Q_k|}\,. 
 \end{equation}
The gauge theory dual to the OP fixed point is an $\NN=1$ deformation of the ABJM model. It is easy to see that the uplifted eleven-dimensional geometry, which we discuss in the next Section, is AdS$_4$ times a round seven-sphere S$^7$ (orbifolded by $\Zinter_k$) in the ABJM case.  The unique difference in the OP case is that the sphere is not round but squashed.

\begin{figure}[t!]
	\begin{tikzpicture}[scale=3,very thick,decoration={markings,mark=at position .5 with {\arrow{stealth}}}]
	\node[above] at (0,0) {SYM--CSM $|$ D2};
	\node[below] at (0,-2.2) {Mass gap };
	\node[below] at (0,-2.35) {$\mathds{R}^7\times {\rm S}^4$};
	\node[left,red] at (-1,-.85) {OP CFT};
	\node[left,red] at (-1,-1) {AdS$_4\ \times \ $S$^7$};
	\node[right] at (1.6,-2.2) {Confinement};
	\node[right] at (1.6,-2.35) {$\mathds{R}^6\times{\rm S}^1\times {\rm S}^4$};
	\node at (-.6,-1) {$\Bp$};
	\node at (.63,-1) {$\Bm$};
	\draw[postaction={decorate},ultra thick,c4] (0,0) --  (0,-2) node[left,midway]{$\mathds{B}_8$};
	\draw[postaction={decorate},ultra thick, gray2] (-1,-1) -- (-1,-2) node[left,midway]{$\BOP$\,\,};
	\draw[postaction={decorate},ultra thick, c2] (0,0) .. controls (-.9,-.9) and (-1,-1) .. (-0.98,-2);
	\draw[postaction={decorate},ultra thick, c3] (0,0) .. controls (-.5,-.5) and (-.5,-1.5) .. (-.5,-2);
	\draw[postaction={decorate},ultra thick, c5] (0,0) .. controls (.5,-.5) and (.5,-1.5) .. (.5,-2);
	\draw[postaction={decorate},ultra thick, c6] (0,0) .. controls (.9,-.9) and (.95,-1.05) .. (1,-2);
	\draw[postaction={decorate},ultra thick, c7] (0,0) .. controls (.9,-.9) and (1.5,-1.6) .. (1.5,-2);
	\draw[postaction={decorate},ultra thick, c8] (0,0) .. controls (.95,-.95) and (1.8,-1.8) .. (1.9,-2);
	\draw[|-|] (-1,-2) -- (0,-2);
	\draw[-stealth] (0,-2) -- (2,-2);
	\node[left=5] at (-1,-2) {$b_0$};
	\node[below=5] at (-1,-2) {$0$};
	\node[below=5] at (0,-2) {$2/5$};
	\node[below=5] at (2,-1.95) {$1$};
	\draw[postaction={decorate},ultra thick,c1] (0,0) -- (-1,-1) node[left,midway]{$\Binf$\,};
	\draw[postaction={decorate},ultra thick,c9] (0,0) -- (2,-2) node[right,midway]{\, $\Bconf$};
	\draw [red, ultra thick,fill=red] (-1,-1) circle [radius=0.03]; 
	\draw [black, fill=black, ultra thick] (0,0) circle [radius=0.03];
	\end{tikzpicture}
	\caption{\small Pictorial representation of the $\B_8$ family of solutions. The asymptotic UV regime is given by the 3D super Yang--Mills theory with Chern--Simons interactions for the gauge fields (SYM--CSM). As we come down in energy (descending in the plot) the RG flow generically drives the theory to an IR regime with a mass gap. 
	For $b_0=0$ the IR is governed by the Ooguri--Park (OP) fixed point. The hue or the warmth of the curves will be roughly in one-to-one correspondence with the values of $b_0$ on the horizontal axis; in the Figures to follow we will try to maintain this mapping.
	}\label{fig:triangle}
\end{figure}

We are interested in a one-parameter family of solutions to \eqref{BPSsystem} and \eqref{BPSsystem_fluxes} which also have negative $Q_k$. In contrast to the previous examples, they are not superconformal in general. This family, as parametrised by $b_0\in[0,1]$, permits us to smoothly deform the IR properties of the dual theories. As a result, and depicted in Figure \ref{fig:triangle}, the single SYM-CSM theory at the UV dual to the D2-brane solution
\begin{equation}
	e^{2f}=2e^{2g} \sim r^2\,,\qquad e^{\Phi} \sim h^{\frac{1}{4}}\,,\qquad h\sim r^{-5}
\end{equation}
follows different trajectories along the RG flow down to a phase with a mass gap. The gap is lost when $b_0 = 0$, in which case the infrared of the theory is described by the OP fixed point \eqref{eq:OPsolution}. In the diagram we are also specifying the eleven-dimensional geometries that give raise to the different IR physics. As we will see in the next Chapter, the apparently innocent change of topology in the case $b_0=1$ will have dramatic consequences regarding the quark antiquark potential. In order to understand the difference between $b_0=1$ and $b_0\neq 1$, it is necessary to work out the uplift to eleven dimensions, where all these solutions enjoy a regular supergravity description.

\section{Uplift to 11 dimensions}

We have already pointed out that generically the metrics we are going to work with are not regular solutions of type IIA supergravity. The reason is that some curvature invariants blow up at the infrared. Mathematically, this is caused by a divergence of the warp factor $h$ at that point. Nevertheless, regular solutions are obtained when they are uplifted to eleven dimensions. This uplift can be performed via the usual ansatz 
\begin{equation}\label{eq:upliftansatz}
\dd s_{11}^2\,=\,e^{-\frac23\Phi}\,\dd s_{\rm s}^2+e^{\frac43\Phi}\,\ell_p^2 \,
\left(\dd\psi+C_1\right)^2\,,
\end{equation}
where $\psi$ parametrises the M-theory  circle, $\ell_p$ is the eleven-dimensional Planck length and $C_1$ is the RR one-form potential of type IIA. If we were considering the D2-brane solution, for which $C_1=0$, we would obtain an M-theory circle which would be trivially fibred over the rest of the directions. However, this is no longer going to be case for us, as $C_1$ is going to be given in terms of the angles and forms of $\CP^3$ (see Appendix \ref{ap:CP3}) through the following expression:
\begin{equation}
C_1 = -\parent{\cos\theta\ \dd \varphi\ - \ \xi \mathcal{S}^3}\,.
\end{equation}
This means that in terms of the vielbein
\begin{equation}
E^3\,=\,\dd\psi-\cos\theta\,\dd \varphi+\xi\,\mathcal{S}^3 \,,
\end{equation}
with $\psi\in\left[0,{4\pi}{|k|^{-1}}\right)$, the eleven-dimensional metric can be written in the M2-brane form 
\begin{equation}
\label{M2metric}
\dd s_{11}^2\,=\,H^{-2/3}\,\dd x_{1,2}^2+H^{1/3}\,\dd s_8^2\,,
\end{equation}
with\footnote{Note that the dependence on $k$ in the period of the angle $\psi$ will compensate with the charge $Q_k$.}
\begin{equation}\label{eq:8transverse}
\dd s_8^2\,=\,e^{-\Lambda}\,\Bigg[ \dd r^2+e^{2f}\,\dd\Omega_4+e^{2g}\,\left[\left(E^1\right)^2+\left(E^2\right)^2\right] \Bigg]+e^\Lambda\,Q_k^2\,\left(E^3\right)^2
\end{equation}
and
\begin{equation}
\label{HH}
H\,=\, h \, e^\Lambda \,.
\end{equation}
From \eqref{eq:10DwarpFactor_eq} it is straightforward to obtain the equation that $H$ has to fulfil, which is
\begin{equation}\label{eq:11DwarpFactor_eq}
H' = -e^{2\Lambda-4f-2g}\Big[Q_c+Q_k\,b_J \left(b_J- 2 b_X\right)+ 2 q_c \left(b_X- b_J\right) + 4 a_J\left(b_X+ b_J\right)\Big] \,. \\[1mm] 
\end{equation}

At this point, we can already understand how this uplift leads to a regular infrared: the divergence of $h$ will combine with the vanishing of $e^\Lambda$ in such a way that the eleven-dimensional warp factor $H$ acquires a finite value at the IR. Now that we know the form that our solutions take in eleven dimensions, it is easy to make contact with \cite{Cvetic:2001ye}, were the solutions we discuss were originally found. The functions $a, b$ and $c$ used there are related to ours through 
\begin{equation}
\label{abc}
a^2\,=\,e^{2g-\Lambda}\,,\qquad\qquad b^2\,=\,Q_k^2\,e^\Lambda\,,\qquad\qquad c^2\,=\,e^{2f-\Lambda}\,,
\end{equation}
and their equations for special holonomy are equivalent to our BPS system (\ref{BPSsystem}). To be precise, in order to recover the results of \cite{Cvetic:2001ye} we should set $Q_k=-1$. Presumably, the sign is  a choice of radial coordinate, and  absolute values different from one  correspond to orbifolds of the construction in \cite{Cvetic:2001ye}. We conclude that all the solutions of \cite{Cvetic:2001ye} are also solutions of our equations and that we have directly constructed their ten-dimensional description. In particular, each of the eleven-dimensional solutions (except for the one labelled $\Bconf$) is based on an eight-dimensional transverse geometry of Spin$(7)$ holonomy.

Now that the uplift is understood, let us list the actual solutions.

\section{Solution with vanishing CS level: $\Bconf$}

By first considering the rightmost solution represented in Figure~\ref{fig:triangle}, we can start realising how interesting these family of solutions are as far as our aim of understanding non-perturbative phenomena of quantum field theories is concerned. The reason is that, as we shall see in the next Chapter, this solution describes a confining field theory in three dimensions. For this particular case the value of the Chern--Simons (CS) level $k$ is set to zero. Changing coordinates in \eqref{10Dansatz} through 
\be
\dd r=\left(1-\frac{\rho_0^4}{\rho^4}\right)^{-1/2}\dd \rho 
\ee
with $\rho\in(\rho_0,\infty)$, we can find the following solution to the BPS system \eqref{BPSsystem}
\begin{equation}
e^{\Lambda}\,=\,1 \,, \qquad\qquad e^{2f}\,=\,\frac12\,\rho^2 \,, \qquad\qquad e^{2g}\,=\,\frac14\,\rho^2\,\left(1-\frac{\rho_0^4}{\rho^4}\right)\,.
\end{equation}

In the dimensionless coordinate
\begin{equation}
\zeta=\frac{\rho}{\rho_0}\in(1,\infty)\,,\qquad \zeta \in (0,\infty)
\end{equation}
the fluxes that solve \eqref{BPSsystem_fluxes} and regularise the solution are in this case
\begin{eqnarray}
b_J&=&\frac{Q_c}{4q_c}+\frac{2q_c}{3\rho_0}\left[\frac{\zeta\sqrt{\zeta^4-1}-\left(3\zeta^4-1\right)U(\zeta)}{\zeta^4-1}\right]\,,\nonumber\\[2mm]
b_X&=&-\frac{Q_c}{4q_c}-\frac{2q_c}{3\rho_0}\left[\frac{\zeta\sqrt{\zeta^4-1}-\left(3\zeta^4-1\right)U(\zeta)}{\zeta^4}\right]\,,
\end{eqnarray}
where the dimensionless function $U$ is defined as 
\begin{equation}
U(\zeta)\,=\,\int_1^\zeta\left(\sigma^4-1\right)^{-1/2}\dd \sigma\,.
\end{equation}
Since here $e^\Lambda=1$, the warp factors  in ten and eleven dimensions coincide, and are given by
\begin{equation}
\label{eq:B8confWarpFactor}
\begin{aligned}
h=H=\quad &\\[2mm]
\frac{128q_c^2}{9\rho_0^6}\int_\zeta^\infty&\left[\frac{2-3\sigma^4}{\sigma^3\left(\sigma^4-1\right)^2}+\frac{\left(4-9\sigma^4+9\sigma^8\right)U(\sigma)}{\sigma^4\left(\sigma^4-1\right)^{5/2}}+\frac{2\left(1-3\sigma^4\right)U(\sigma)^2}{\sigma^5\left(\sigma^4-1\right)^3}\right]\dd\sigma\,. \\[2mm]
\end{aligned}
\end{equation}

The vanishing of $k$ in this case causes the RR two-form $F_2$ from \eqref{eq:fluxesansatz} to vanish. This translates into the fact that the M-theory circle is trivially fibred and hence the eight-dimensional transverse metric in eleven dimensions is a direct product of the form $\mbox{M}_7 \times \mbox{S}^1$, where $\mbox{M}_7$ is the ${\rm G}_2$-manifold found in \cite{Bryand:1989mv,Gibbons:1989er}. In fact, this geometry is a regular solution of type IIA supergravity without the need of uplifting it to M-theory.

The uplift of this metric \eqref{eq:Full11Dmetric} is very simple and its transverse part reads 
\begin{equation}\label{G2holonomy}
\dd s_8^2\,=\,\frac{\dd \rho^2}{\left(1-\frac{\rho_0^4}{\rho^4}\right)}+\frac{1}{2}\,\rho^2\,\dd\Omega_4^2+\frac{1}{4}\,\rho^2\,\left(1-\frac{\rho_0^4}{\rho^4}\right)\left[\left(E^1\right)^2+\left(E^2\right)^2\right]+ \ell_p^2 \, \dd \psi^2\,.
\end{equation}
Consequently, the IR limit corresponds to $\mathds{R}^6\times{\rm S}^1\times{\rm S}^4$. The fact that the S$^1$ is trivially fibred will have noticeable consequences in the gauge theory side, as we will see in the next Chapter.

\section{Solutions with non-vanishing CS level}

Let us consider now non-vanishing CS level $|Q_k|\neq0$ (recall we stick to $Q_k<0$). This case gathers the rest of the solutions from Figure \ref{fig:triangle}. We will give all of them in this Section.

\subsection{The $\B_8^\pm$ solutions}
In order to solve the BPS equations \eqref{BPSsystem} and \eqref{BPSsystem_fluxes} and obtain supersymmetric backgrounds, we perform the change of radial coordinate 
\begin{equation}\label{eq:coordy}
\dd r = \frac{P}{|Q_k|\ v\ (1-y^2)} \dd y\,.
\end{equation}
Remarkably, the new $y$ coordinate allows us to find analytical expressions for all the functions of the metric but the warp factor. In the previous expression $v$ and $P$ are solutions of the following differential equations:
\begin{equation}\label{eq:eqvandP}
2\ (1-y^2)\ \frac{\dd v}{\dd y} = y\ v +2 \qquad, \qquad\frac{1}{P}\ \frac{\dd P}{\dd y} = \frac{v+1}{v\ (1-y^2)}\ .
\end{equation}

Moreover, the BPS system \eqref{BPSsystem} is solved by
\begin{equation}
e^{2f-\Lambda} = P \,, \qquad e^{2g-\Lambda} = \frac{P(v-2)}{v(1+y)}\,,\qquad e^{\Lambda} = \frac{4P(v-2)}{|Q_k|^2v^3(1+y)}\,,
\end{equation}
where we have chosen to show the combinations that appear in the eight-dimensional metric \eqref{eq:8transverse} transverse to the M2-branes in eleven dimensions. Its final form reads
\begin{equation}
\label{eq:eightB8}\begin{aligned}
\dd s_8^2\,=&\,\frac{v\,P\,\left(1+y\right)\dd y^2}{4\left(v-2\right)\left(1-y^2\right)^2}\ +\ P\,\dd \Omega_4^2\\[2mm]&+\ \frac{P\left(v-2\right)}{v\left(1+y\right)}\left[\left(E^1\right)^2+\left(E^2\right)^2\right]\ +\ \frac{4P\left(v-2\right)}{v^3\left(1+y\right)}\left(E^3\right)^2 \,.
\end{aligned}
\end{equation}

Despite permitting analytical expressions, choosing $y$ as the radial coordinate is not free from drawbacks. It forces us to split our family of solutions into two subfamilies $\Bp$ and $\Bm$, depending on the boundary condition imposed to the $v(y)$ function. More precisely, note that the three-sphere parametrised by $E^i$ will collapse for the particular value of the radial coordinate $y=y_0$ at which 
\begin{equation}
v(y_0)=2\,.
\end{equation}
Each choice of $y_0$ gives raise to a different theory. Put differently, each $y_0$ will be associated to a single value of the parameter $b_0$ in Figure \ref{fig:triangle}. Depending on the particular value of $y_0$, we need to distinguish between $v(y)=v^\pm(y)$ in the following two cases:

\begin{itemize}
	\item $\B_8^+$ when $-1<y_0<1$ (corresponding to $0<b_0<2/5$ in Figure \ref{fig:triangle} as we will see). In this case $v(y)$ reads	
	\begin{equation}
	\label{eq:solutionV1}
	v^+(y)\,=\,\frac{1}{\left(1-y^2\right)^{\frac{1}{4}}}\,\left(
	w_0^+ - \frac{\Gamma[\frac{1}{4}]^2}{2\sqrt{2\pi}}\, + \,_2F_1\left[\frac12,\frac34;\frac32;y^2\right] \times y \right)
	\end{equation}
	\item $\B_8^-$ when $1<y_0<\infty $ (corresponding to $2/5<b_0<1$ as we will see). In this case
	\begin{equation}
	\label{eq:solutionV2}
	v^-(y)\,=\,\frac{1}{\left(y^2-1\right)^{\frac{1}{4}}}\left(w_0^- - \frac{\Gamma[\frac{1}{4}]^2}{2\sqrt{\pi}}+\frac{2}{\sqrt{y}} \,\,_2F_1\left[\frac14,\frac34;\frac54; \frac{1}{y^2} \right]\right)
	\end{equation}
\end{itemize}
The cases $y_0=-1$ and $y_0=1$ are special and we will discuss them apart and the constant $w_0^\pm$ is fixed precisely such that $v(y_0)=2$ (see \eqref{eq:expressionsw_0} in Appendix~\ref{ap:expansions} for the explicit expressions). The collapse of the three-sphere happens smoothly at the IR, in such a way that the geometry will become $\R^7\times \text{S}^4$ and describes the IR of the theories. On the other hand, the UV is attained as $y\to 1$ in all cases. 

The two distinct choices of boundary condition in $v(y)$, lead to two different solutions for $P$
\begin{equation}\label{eq:solutionP}
P = P^{\pm} = P_0^{\pm} \frac{(1+y)^{\frac{3}{4}}}{(\pm 1 \mp y)^{\frac{3}{4}}}\ v^{\pm}(y)\,,
\end{equation}
where the upper (lower) sign choice corresponds to the $\Bp$ ($\Bm$) subfamily. We will keep this convention through the rest of this Section. Also, note $P_0^\pm$ appear like integration constants, which we can fix as follows. If we expand the eight-dimensional metric \eqref{eq:eightB8} about the UV region $y\approx1$, the change of coordinates at leading order gives
\begin{equation}
|Q_k|\, \big( \pm  1 \mp y\ \big)^{\frac{1}{4}} \, r \ =\  2^{\frac{7}{4}}\, P^{\pm}_0
\end{equation}
and we thus get
\begin{equation}
\dd s_8^2\,\propto\,\dd r^2+\frac12 \, r^2\Bigg( \dd\Omega_4^2+\frac12\left[\left(E^1\right)^2+\left(E^2\right)^2\right]\Bigg)+
\left(\frac{4P_0^\pm}{|Q_k| w_0^\pm}\right)^2\left(E^3\right)^2\,.
\end{equation}
Since the size of the $E^3$ circle becomes constant, we recognise this as the uplift of a D2-brane
metric. Given our parametrisation of the dilaton in \eqref{eq:10DansatzAP}, in order for the solution to asymptote to the D2-brane solution with the correct normalisation  of the gauge coupling we must impose the boundary condition $e^{\Lambda}\to1$.  This fixes the dimensionful constant $P_0^+$ to the value
\begin{equation}
\label{eq:P0choice}
P_0^\pm\,=\,\frac{|Q_k|^2\,w_0^\pm}{4}\,.
\end{equation}
Since $\psi$ has period $4\pi/|k|$, this is equivalent to normalising the asymptotic radius of the M-theory circle in the eight-dimensional transverse metric to the usual result
\be
R_{(11)} = \frac{2Q_k}{k} = g_s \ell_s \,.
\ee 
Note that $e^{\Lambda}\to1$ actually implies that we are setting $g_s=1$. Nevertheless, we will keep explicit factors of $g_s$ in our formulas in order to facilitate  comparison with the literature. 

To see that the collapse of the geometry is indeed smooth at the IR, we can examine the leading order of the eight-dimensional transverse part. If we perform a change of variables
\newcommand{\rIR}{{\bf r}}
\begin{equation}
4P_0^\pm (y-y_0) = (1-y_0)^\frac{5}{4} (1+y_0)^\frac{1}{4} \ {\rIR}^2\,
\end{equation}
the transverse metric at small $\rIR$ approaches 
\begin{eqnarray}\label{eq:8Dgapped}
\dd s_8^2\,=\,\dd\rIR^2+\frac14\rIR^2\left[\left(E^1\right)^2+\left(E^2\right)^2+\left(E^3\right)^2\right]+\frac{2P_0^\pm\left(1+y_0\right)}{\left(\pm  1\mp y_0^2\right)^{1/4}}\,\dd\Omega_4^2\,.
\end{eqnarray}
Since the $E^i$'s describe a three-sphere fibration over S$^4$, we find that in the IR the eight-dimensional transverse metric approaches locally $\R^4\times$S$^4$ at the IR. In order to get the full regular eleven-dimensional metric, it will be necessary to solve the BPS equations \eqref{BPSsystem}. We search for $b_J$ and $b_X$ with the appropriate boundary conditions that renders the warp factor regular. 

The solutions for the fluxes and the warp factor shall be computed numerically. For all the numerical we perform in this thesis, it turns useful to define dimensionless functions $\FF$, $\GG$, $\BB_X$, $\BB_J$, $\mathbf{h}$ and $\mathbf{\mathcal{H}}$ defined through
\begin{equation}\label{eq:dimlessfunctions}
\begin{aligned}
e^f &=  |Q_k|\ e^\FF\, , \qquad \qquad  e^g = |Q_k|\ e^\GG\,,  \qquad\qquad h= \frac{4q_c^2 +3 |Q_k| Q_c}{|Q_k|^6} \ \mathbf{h}\,, \\[2mm]
b_J &= - \frac{2q_c}{3 |Q_k|} - \frac{\sqrt{4 q_c^2 + 3 |Q_k| Q_c}}{3 |Q_k|}\ \BB_J\,, \\[2mm] 
b_X &=  \frac{2q_c}{3 |Q_k|} + \frac{\sqrt{4 q_c^2 + 3 |Q_k| Q_c}}{3 |Q_k|}\ \BB_X \ , \\[2mm]
H &=  \frac{4q_c^2 +3 |Q_k| Q_c}{\left(P_0^\pm\right)^3}\  \mathcal{H} \ ,
\end{aligned}
\end{equation} 

In terms of them, we will always be able to factor all charges out from the equations. In the present computation, we first write \eqref{BPSsystem_fluxes} in terms of these dimensionless functions and solve them perturbatively around the IR and around the UV. The requirement that the numerical integrations from the IR and the UV match at some intermediate value of the radial coordinate fixes the value of the undetermined parameters we will find in the expansions.

In the IR, defined by the condition $v(y_0)=2$, we find
\begin{eqnarray}\label{eq:fluxesExpansionsSUSYcoordyIR}
\mathcal{B}_J&=&1-\frac{1}{ 2\left(1-y_0^2\right)}\left(y-y_0\right)+\frac{2-3y_0}{8\left(1-y_0^2\right)^2}\left(y-y_0\right)^2+\mathcal{O}\left(y-y_0\right)^3\,, \nonumber\\[2mm]
\mathcal{B}_X&=&1-\frac{3}{4\left(1-y_0^2\right)^2}\left(y-y_0\right)^2+\mathcal{O}\left(y-y_0\right)^3 \,, \\[2mm]
\mathcal{H}&=& \mathcal{H}_{\text{\tiny{IR}}}\mp\frac{7}{48\left(1+y_0\right)^3\left(\pm 1\mp y_0^2\right)^{1/4}}\left(y-y_0\right)\\[2mm]&&-\frac{77(y_0-2)}{576\left(1+y_0\right)^3\left(\pm 1\mp y_0^2\right)^{5/4}}\left(y-y_0\right)^2+\mathcal{O}\left(y-y_0\right)^3 \,,
\nonumber
\end{eqnarray}
where we have already imposed regularity of the warp factor, which fixes the integration constants in $\mathcal{B}_J$ and $\mathcal{B}_X$. The only undetermined constant in the IR expansion is $\mathcal{H}_{\text{\tiny{IR}}}$, which will be fixed in the full numerical solution by requiring  D2-brane asymptotics in the UV with the correct normalisation. 

In the UV, located at $y\to1$, we find  the expansions
\begin{eqnarray}\label{eq:fluxesExpansionsSUSYcoordyUV}
\mathcal{B}_J&=&b_0\left[1+\frac{2^{9/4}}{w_0^\pm}\left|1-y\right|^{1/4}+\frac{2^{7/2}}{\left(w_0^\pm\right)^2}\left|1-y\right|^{1/2}-\frac{2^{19/4}}{\left(w_0^\pm\right)^3}\left|1-y\right|^{3/4}\right.\nonumber\\[2mm]
&&\left.\qquad+\frac{\beta_4}{\left(w_0^\pm\right)^4}\left|1-y\right|+\mathcal{O}\left|1-y\right|^{5/4}\right]\,,
\nonumber\\[2mm]
\mathcal{B}_X&=&b_0\left[1+\frac{2^{9/4}}{w_0^\pm}\left|1-y\right|^{1/4}+\frac{3\times2^{5/2}}{\left(w_0^\pm\right)^2}\left|1-y\right|^{1/2}+\frac{2^{23/4}}{\left(w_0^\pm\right)^3}\left|1-y\right|^{3/4}\right.\nonumber\\[2mm]
&&\left.\qquad-\frac{1}{\left(w_0^\pm\right)^4}\left(128+\frac{\beta_4}{2}\right)\left|1-y\right|+\mathcal{O}\left|1-y\right|^{5/4}\right] \,,\\[2mm]
\mathcal{H}&=&\mathcal{H}_{\text{\tiny{UV}}}+\frac{\left(1-b_0^2\right)}{15\times2^{3/4}\left(w_0^\pm\right)^2}\left|1-y\right|^{5/4}+\frac{2^{3/2}\left(1-2b_0^2\right)}{9\left(w_0^\pm\right)^3}\left|1-y\right|^{3/2}\\[2mm]
&&\qquad+\mathcal{O}\left|1-y\right|^{7/4}\,,\nonumber
\end{eqnarray}
The undetermined constants\footnote{Note that $\beta_4$ was denoted by $b_4$ in \cite{Faedo:2017fbv}.} in the UV are thus $b_0, \beta_4$ and 
$\mathcal{H}_{\text{\tiny{UV}}}$. The parameter $b_0$ is that of Figure \ref{fig:triangle}, which distinguishes between different theories in the family. It appears here as the leading coefficient in the expansions, and it is thus associated to a source in the field theory side. More precisely, varying $b_0$ corresponds to varying the UV asymptotic flux of the NSNS two-form through the $\CP^1$ cycle inside $\CP^3$. Since this asymptotic flux is expected to specify the difference between the gauge theory couplings \cite{Hashimoto:2010bq}, we can think of these family of theories as parametrised by this difference. Presumably, $\beta_4$ is related to the vacuum expectation value (VEV) of some operator in the gauge theory. Finally, $\mathcal{H}_{\text{\tiny{UV}}}$ must vanish in order to have the correct D2-brane asymptotics in the decoupling limit, \textit{i.e.}, in order for $H\to0$ in the UV.

\begin{figure}[t!]
	\begin{center}
		\begin{subfigure}{.45\textwidth}
			\includegraphics[width=\textwidth]{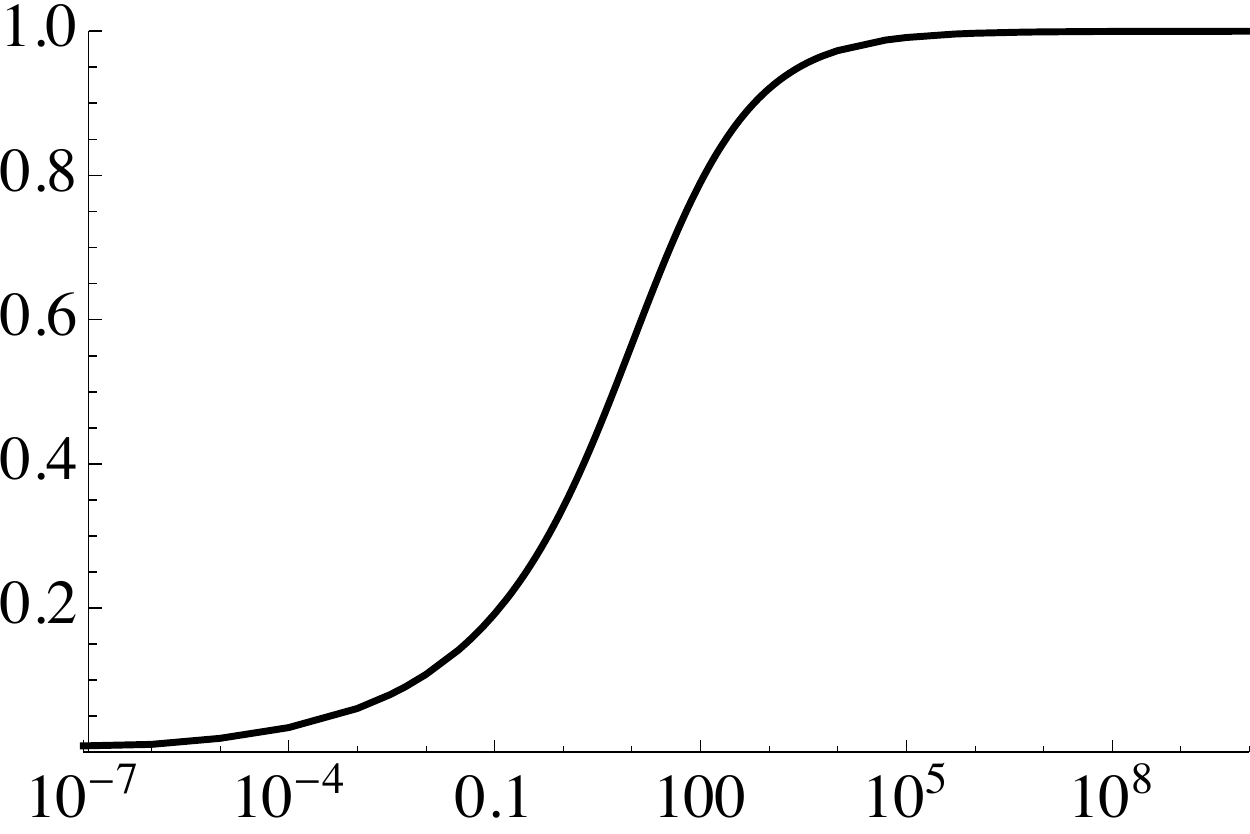} 
			\put(-140,110){$b_0$}
			\put(-80,-15){$y_0+1$}
		\end{subfigure}\hfill
		\begin{subfigure}{.45\textwidth}
			\includegraphics[width=\textwidth]{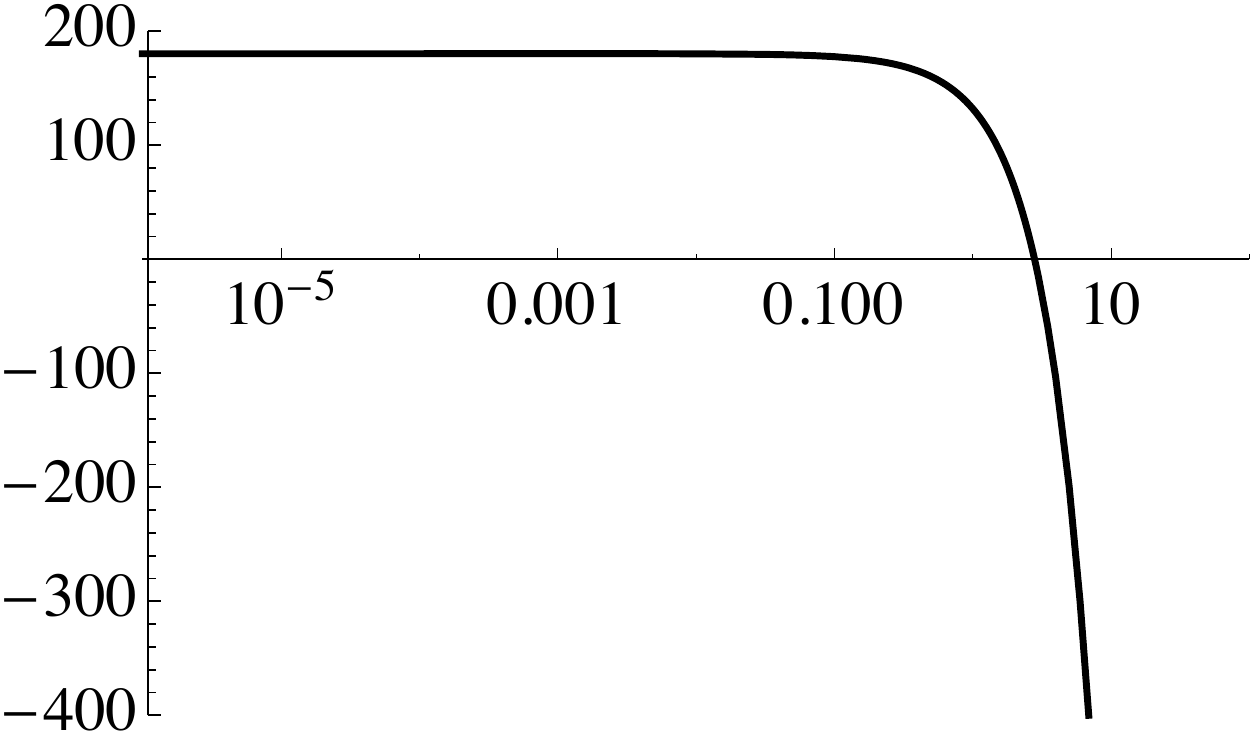} 
			\put(-140,110){$\beta_4$}
			\put(-80,-15){$y_0+1$}
		\end{subfigure}
		\caption{\small Values of the UV parameters $b_0$ (left) and  $\beta_4$ (right) from the numerical integration.}\label{fig.numparam}
	\end{center}
\end{figure}
\begin{figure}[t!]
	\begin{center}
		\includegraphics[width=.55\textwidth]{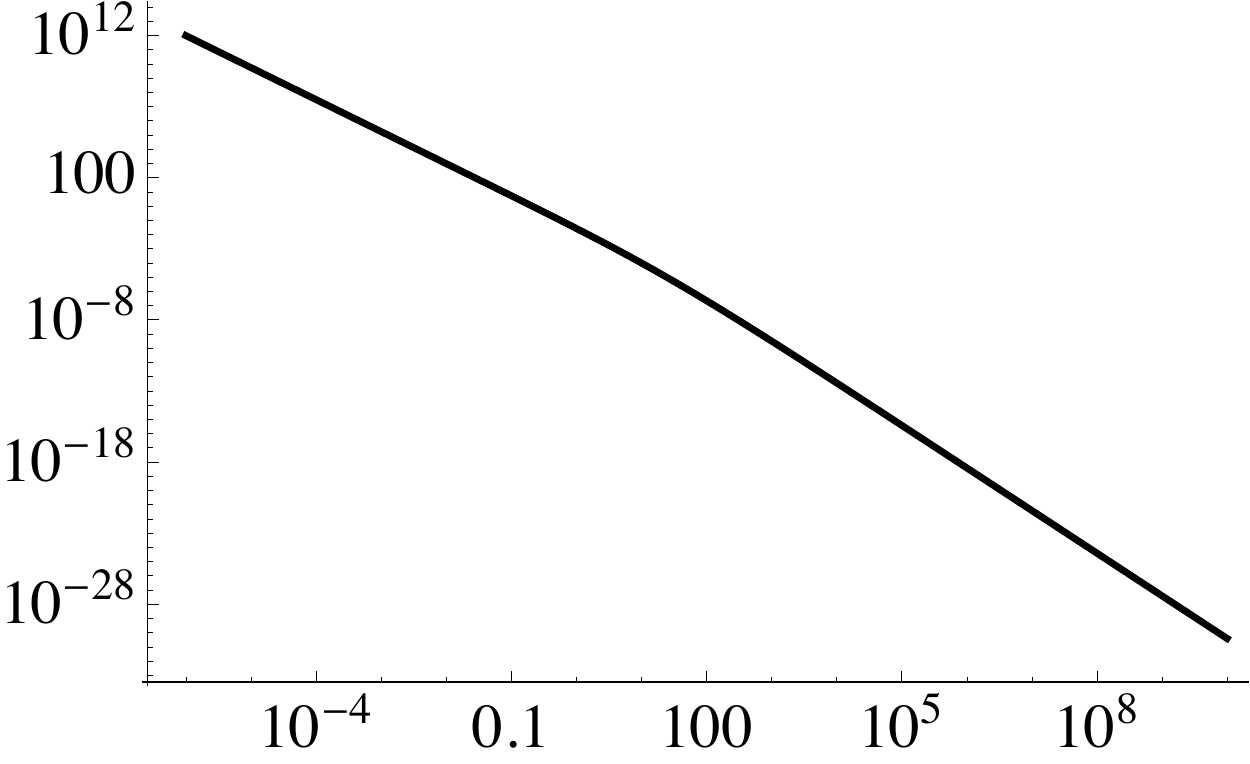} 
		\put(-210,35){\rotatebox{90}{$64 \, \mathcal{H}_{\text{\tiny{IR}}} / (w_0^{\pm})^3 $}}
		\put(-100,-15){$y_0+1$}
		\caption{\small Values of the parameter $\mathcal{H}_{\text{\tiny{IR}}}$ from the numerical integration.}\label{fig.numHIR}
	\end{center}
\end{figure}

Through numerical integration from both the IR and the UV, continuity in an intermediate point fixes the actual value of $\mathcal{H}_{\text{\tiny{IR}}}$, $b_0$ and $\beta_4$ for each choice of $y_0$ (i.e., for each theory). Once this is done there is a unique solution for each value of $y_0$ and the UV constants $b_0, \beta_4$ can be simply read off from the solution. The result is displayed in Figure~\ref{fig.numparam}, where we realise that $y_0$ is in one-to-one correspondence to the value of $b_0$, which takes values in the interval $(0,1)$ as we already mentioned. Also, Figure~\ref{fig.numHIR} shows the IR value of the warp factor.

\subsection{The $\mathds{B}_8$ solution}

Recall that the case $y_0=1$, which corresponds to $b_0=2/5$, was not covered by the analysis in the previous Section. There is nothing deep in this fact, but just that the coordinate $y$ given through \eqref{eq:coordy} is ill-defined in that case. Fortunately, this is not an issue, given that an analytical solution exists for this case in the original coordinate. It takes the simple form
\begin{equation}
e^{2f}\,=\,\frac12\,\frac{r^2\left(r-2r_0\right)}{\left(r-r_0\right)}\,,\quad e^{2g}\,=\,\frac14\,\frac{r^2\left(r-2r_0\right)^2}{\left(r-r_0\right)^2}\,,\quad e^{\Lambda}\,=\,\frac{r_0}{2|Q_k|}\,\frac{r\left(r-2r_0\right)}{\left(r-r_0\right)^2}\,,\\[1mm]
\end{equation}
with $r\in(2r_0,\infty)$ where
\be
\label{r0}
r_0 = 2|Q_k| \,,
\ee
so that $e^\Lambda \to 1$ asymptotically. Uplifting to eleven dimensions we get the transverse space
\begin{equation}\begin{aligned}
\dd s_8^2&=\frac{\left(r-r_0\right)^2\dd r^2}{r\left(r-2r_0\right)}+\frac{1}{2}r\left(r-r_0\right)\dd\Omega_4^2\\[2mm]
&+\frac{1}{4}r\left(r-2r_0\right)\left[\left(E^1\right)^2+\left(E^2\right)^2\right]+\frac{r_0^2}{4}\frac{r\left(r-2r_0\right)}{\left(r-r_0\right)^2}\left(E^3\right)^2\,.
\end{aligned}
\end{equation}
This metric was dubbed $\mathds{B}_8$ in \cite{Cvetic:2001ye, Cvetic:2001pga}. The geometry ends smoothly at $r=2r_0$ and has the same asymptotic behaviour as the  
$\Bpm$ solutions. In this case, completely analytic expressions can be found, even for the fluxes and the warp factor. In terms of a dimensionless coordinate 
\be
\rho=\frac{r}{r_0} \,,
\ee
the  fluxes take the form
\begin{equation}
\mathcal{B}_J\,=\,\frac{2\left(\rho^4+\rho^3-4\,\rho+4\right)}{5\rho^3\left(\rho-1\right)}\,,\qquad\qquad\qquad\mathcal{B}_X\,=\,\frac{2\left(\rho^5-10\,\rho+8\right)}{5\rho^3\left(\rho-1\right)^2}\,,
\end{equation}
where one integration constant was fixed to have D2-brane asymptotics in the UV while the other two were fixed by regularity.  The M2-brane warp factor can be found again in closed form and is simply
\begin{equation}\label{eq:Full11Dmetric}
\mathcal{H}\,=\,\frac{\left(1323\rho^6+924\rho^5+963\rho^4+510\rho^3-1340\rho^2-4340\rho+2800\right)}{47250\,\rho^9\left(\rho-1\right)^2}\,,
\end{equation}
which is perfectly regular at $\rho=2$. Notice that the boundary conditions have fixed all the integration constants, the only parameters being the quantised charges.

\subsection{$\Binf$ solution}
\label{binfsol}

In the particular case $b_0=0$, corresponding to $y_0=-1$, IR physics changes though. The eight-dimensional metric is still written as \eqref{eq:8transverse}, and the corresponding expressions for $v^+$ $w_0^+$, $P$ and $P_0^+$ can still be read of from \eqref{eq:solutionV1}, \eqref{eq:solutionP} and \eqref{eq:P0choice}. However, after performing the change of variables
\be \label{eq:IRofB8inf}
P_0^+\,\left(1+y\right)^{3/4}=\frac{9}{2^{11/4}\times5 }\,\varrho^{2} \,, 
\ee
one discovers that the  transverse space in the IR corresponds to the OP solution
\begin{equation}
\dd s_8^2\,=\,\dd \varrho^2+\frac{9}{20}\,\varrho^2\left[\dd\Omega_4^2+\frac{1}{5}\,\left[\left(E^1\right)^2+\left(E^2\right)^2+\left(E^3\right)^2\right]\right]\,,
\end{equation}
since one can recognise the metric inside the square brackets as the squashed seven-sphere. In fact, 
\begin{equation}\label{eq:rtovarrho}
20|Q_k|r=2\varrho^2
\end{equation}
relates \eqref{eq:IRofB8inf} to \eqref{eq:OPsolution}.
The full  geometry was denoted $\Binf$ in \cite{Hashimoto:2010bq} and its significance had been overlooked in studies prior to this reference. It interpolates between the theory on the D2-branes on the squashed $\CP^3$ and the OP fixed point, so it can be seen as an irrelevant deformation of the OP CFT whose UV completion is a SYM--CSM  theory.  

When $y_0\to-1$ the IR expansions \eqref{eq:fluxesExpansionsSUSYcoordyIR} are not well defined, reflecting the dramatic change in the IR, which in this case is a fixed point instead of a gapped phase. Indeed, we have that for this particular solution the fluxes are constant
\begin{equation}
b_J\,=\,-\frac{2q_c}{3|Q_k|}\,,\qquad\qquad\qquad b_X\,=\,\frac{2q_c}{3|Q_k|}\,,\qquad\qquad\qquad a_J\,=\,-\frac{q_c}{6}\,.
\end{equation}
Using this, it is easy to find the expansions for the warp factor. In the IR, around $y=-1$, we get
\begin{equation}
\mathcal{H}\,=\,\mathcal{H}_{\infty}+\left(y+1\right)^{-\frac{9}{4}}\frac{5}{9\times2^{\frac{5}{4}}}\left[\frac53-\frac{13}{8}\left(y+1\right)+\frac{815}{1664}\left(y+1\right)^2+\mathcal{O}\left(y+1\right)^3\right]\,.\\
\end{equation}
Notice that the constant term $\mathcal{H}_{\infty}$ is not the leading term in this case and this causes the metric to be AdS. On the other hand, the UV expansion gives again D2-brane asymptotics, as can be obtained from the general expansion of the $\Bpm$ family, setting $b_0=0$.

The only parameter to be found from the numerics is the $\mathcal{H}_{\infty}$ such that the warp factor has no constant piece in the UV. From our results we find $\mathcal{H}_{\infty}\simeq-0.0087$.

\subsection{$\BOP$ solution}\label{sec:BOP_solution}

Remarkably, the OP fixed point also admits a relevant deformation that can be solved for analytically. In our variables, the metric functions are
\begin{equation}\begin{aligned}
e^{2f}&=\frac95\,r^2\left[1-\left(\frac{r_0}{r}\right)^{\frac{5}{3}}\right],\qquad e^{2g}=\frac{9}{25}\,r^2\left[1-\left(\frac{r_0}{r}\right)^{\frac{5}{3}}\right]^2,\\[2mm] e^{\Lambda}&=\frac{3\,r}{5|Q_k|}\left[1-\left(\frac{r_0}{r}\right)^{\frac{5}{3}}\right]\,,
\end{aligned}
\end{equation}
with the radial direction ending at $r=r_0$.
Changing coordinates from $r$ to $\varrho$ through 
\eqref{eq:rtovarrho}
we see that this solution corresponds to the original Spin(7) manifold of \cite{Bryand:1989mv,Gibbons:1989er}, whose metric is
\begin{equation}\label{Spin7holonomy}
\begin{aligned}
\dd s_8^2&\,=\,\left[1-\left(\frac{\varrho_0}{\varrho}\right)^{\frac{10}{3}}\right]^{-1}{\dd \varrho^2}+\frac{9}{20}\,\varrho^2\,\dd\Omega_4^2\\[2mm]
&\quad +\frac{9}{100}\,\varrho^2\,\left[1-\left(\frac{\varrho_0}{\varrho}\right)^{\frac{10}{3}}\right]\left[\left(E^1\right)^2+\left(E^2\right)^2+\left(E^3\right)^2\right]\,.
\end{aligned}
\end{equation}
The UV of this flow is of course the OP fixed point while the IR, which lies at $\varrho(r_0)=\varrho_0$, is precisely of the form \eqref{eq:8Dgapped}, with the four-sphere radius proportional to $\varrho_0$.

The fluxes of this flow can also be solved for analytically. In terms of the dimensionless coordinate 
\be
\tilde r=\frac{r}{r_0} 
\ee
they are simply
\begin{equation}
\mathcal{B}_J\,=\,\frac{1}{\tilde r^{\frac{1}{3}}}\,,\qquad\qquad\qquad \mathcal{B}_X\,=\,\frac{6\tilde r^{\frac{5}{3}}-1}{5\tilde r^2}\,.
\end{equation}
Finally, the regular eleven-dimensional warp factor, for which there is also an analytical expression, reads
\begin{eqnarray}
\mathcal{H}&=&\frac{5}{243}\left[\frac{1}{\tilde r^2}-\frac{9}{\tilde r^{\frac{1}{3}}}-3\,\frac{\tilde r^{\frac{4}{3}}-\tilde r^{\frac{1}{3}}}{\tilde r^{\frac{5}{3}}-1}\right]\\[4mm]
&+&\frac{4\sqrt{2}}{81}\left[\sqrt{5+\sqrt{5}}\arctan\left(\mathbf{R}_-\right)+\sqrt{5-\sqrt{5}}\arctan\left(\mathbf{R}_+\right)\right] \,.\nonumber
\end{eqnarray}
where:
\begin{equation}
\mathbf{R}_{\pm} = \frac{\sqrt{10+2\sqrt{5}}}{4\tilde r^{\frac{1}{3}}+1 \pm \sqrt{5}}\,.
\end{equation}

Note that $\BB_J$ and $\BB_X$ contain the source of a dimension $8/3$ operator\footnote{The spectrum of scalar around the OP fixed point together with the dimension of the dual operators can be found in Table 1 of \cite{Faedo:2017fbv}.}, corresponding to the mode $\tilde r^{-1/3}$. Moreover, we have factored out the rest of scales in the solution, so all the VEVs and sources in the gauge theory are $\OO(1)$ in these units. This observation will play a crucial role in the next Chapter, when we discuss the spectrum of the theories.

Now we have given all the relevant expressions for the family of solutions whose RG flows where illustrated  in Figure~\ref{fig:triangle}. We would like to discuss how the two limiting solutions are reached, and address the question of what dynamics is imprinted on the flows close to them. We will come to that in Section~\ref{quasi}. Before, it is mandatory to study the energy range where our supergravity approach is reliable, which we discuss in the next Section.

\section{Range of validity}
\label{sec:validity_gs}

We now turn to the determination of the range of validity of the supergravity solutions above. Since in the UV the dilaton goes to zero, the correct description is the ten-dimensional one. This one extends up to the UV scale at which the curvature ceases to be small in string units. The Ricci scalar of the ten-dimensional solutions grows in the UV as 
\begin{equation}
\ell_s^2 R\sim \ell_s^2\left(\frac{|Q_k|}{4q_c^2+3Q_c|Q_k|}\frac{r}{\left(1-b_0^2\right)}\right)^{1/2}\,.
\end{equation}
Requiring this to be small and translating to a gauge theory energy scale $U$ via the usual relation $U=r/\ell_s^2$ \cite{Itzhaki:1998dd} we find the condition
\begin{equation}
U\ll\lambda\left(1+\frac{\bar{M}^2}{2N|k|}\right)\left(1-b_0^2\right)\,,
\end{equation}
where we recall that $\lambda$ is the 't~Hooft coupling with dimensions of energy, \eqref{eq:tHooft3D}. We observe that the usual result $U\ll\lambda$ for the D2-branes gets  modified due to the presence of the fractional branes. We have included the dependence on the choice of theory through the coefficient $1-b_0^2$,  which vanishes as $y_0^{-1/2}$ when $b_0\to1$. This is a manifestation of the fact that, in the limit $b_0\to 1$, $Q_k$ must scale as $|Q_k| \sim |k| \sim y_0^{-1/2}$ in order to obtain a valid supergravity description. The origin of this scaling together with more details will be given in the next Section.   

\begin{figure}[t]
	\begin{center}
		\includegraphics[scale=1]{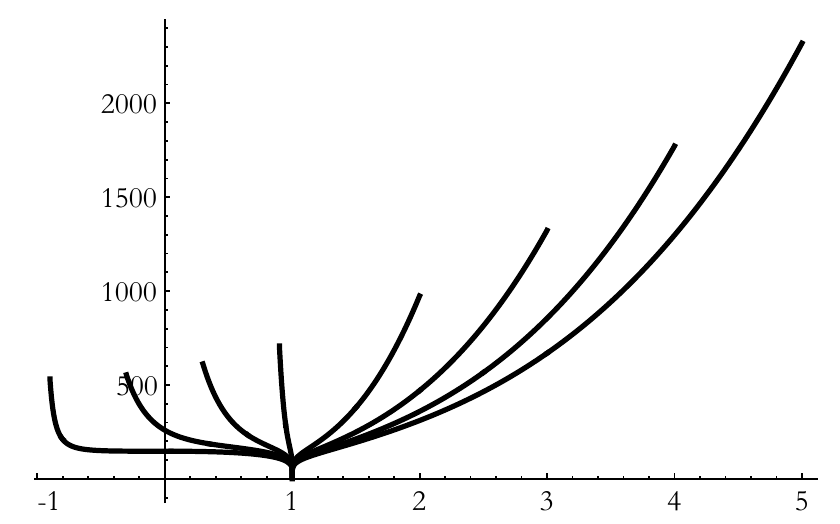} 
		\put(-250,155){\rotatebox{0}{$\left(4q_c^2+3Q_c|Q_k|\right)^{2/3} K$}}
		\put(-135,-10){$y$}
		\caption{\small Kretschmann scalar 
			$K=R_{\mu \nu \rho \sigma}  R^{\mu \nu \rho \sigma}$ as a function of $y$ for several $\B^{\pm}_8$ solutions with $y_0=-0.9,\, -0.3,\, 0.3,\, 0.9,\,2,\,3,\,4,\,5$ from left to right. We see that at $y=1$ all curves approach the same value since they all share the same UV asymptotics, whereas the curvature at the IR endpoint ($y=y_0$) diverges as $y_0^2$ as $y_0\to\infty$. }
		\label{Kret}
	\end{center}
\end{figure}

In the IR the ten-dimensional metrics are singular, so the correct description is given in terms of the eleven-dimensional solutions, in which the IR value of the Ricci scalar in units of the eleven-dimensional Planck length $\ell_p = g_s^{2/3} \ell_s$ is finite and scales  as
\begin{equation}
\ell_p^2\, R\sim\left(\frac{\bar{M}^2}{2}+N|k|\right)^{-1/3}\,.
\end{equation}
In order for this to be small we simply need to require that the combination 
\be
\frac{\bar{M}^2}{2}+N|k| \gg 1 \,.
\ee
For large $y_0$, however, the IR value of the Kretschmann scalar, $K=R_{\mu \nu \rho \sigma}  R^{\mu \nu \rho \sigma}$, shown in Figure \ref{Kret}, grows as 
\be
\ell_p^4 \, K \sim \left(\frac{\bar{M}^2}{2}+N|k|\right)^{-2/3} y_0^2  \,.
\ee
Thus in the limit $y_0 \to \infty$ we must impose the additional condition that 
\begin{equation}
|k|^6\left(\frac{\bar{M}^2}{2}+N|k|\right)\gg1\,,
\end{equation}
where again we have assumed that $k \sim y_0^{-1/2}$.

\section{Limiting Dynamics}
\label{quasi}
In this Section we will study the limits of the above metrics as $b_0\to 1$ and as $b_0 \to 0$.  In the first  case the solution approaches $\Bconf$ everywhere except in the deep IR. In the second case the solution approaches the combination of the $\Binf$ flow followed by the $\BOP$ flow. In this sense the solutions with generic $b_0$ continuously interpolate between quasi-confining and quasi-conformal  dynamics. We will verify this with explicit calculations of the physical quantities, such as the quark antiquark potential in Chapter \ref{Chapter3_observables} or some thermodynamic properties in Chapter \ref{Chapter4_thermo}.

\subsection{Quasi-confining dynamics}
\label{quasiconfining}

Consider the limit $b_0\to1$ (corresponding to $y_0\to \infty$) of the $\Bm$ solutions. Expanding the functions of the internal metric for large $y_0$ we find 
\begin{eqnarray}\label{largey0}
e^{2f}&=&\frac{4\left(P_0^-\right)^2}{|Q_k|^2}\left(\frac{y+1}{y-1}\right)^{1/2}\left[1-\frac{\left(y^2-1\right)^{1/4}}{\sqrt{y_0}}+\mathcal{O}\left(y_0^{-1}\right)\right]\,,\nonumber\\[2mm]
e^{2g}&=&\frac{4\left(P_0^-\right)^2}{|Q_k|^2}\frac{1}{\left(y^2-1\right)^{1/2}}\left[1-\frac{2\left(y^2-1\right)^{1/4}}{\sqrt{y_0}}+\mathcal{O}\left(y_0^{-1}\right)\right]\,.
\end{eqnarray}
Performing the change of variables
\begin{equation}
y\,=\,\frac{\rho^4+\rho_0^4}{\rho^4-\rho_0^4}
\end{equation}
we see  that,  to leading order, we recover the confining metric (\ref{G2holonomy}) with an internal scale given by
\begin{equation}
\rho_0^2 \,=\, \frac{8 \left(P_0^-\right)^2}{|Q_k|^2}  \,.
\end{equation}
Given that $P_0^-$ was fixed by the UV condition $e^\Lambda\to1$ as in (\ref{eq:P0choice}), to leading order in $y_0$ we have 
\begin{equation}
\label{rho0}
\rho_0^2\,=\,2\,|Q_k|^2\,y_0\,.
\end{equation}
Note that, since $y_0 \to \infty$, $\rho_0$ seems to grow without bound. 
One simple way in which we can think of this limit\footnote{See \cite{Faedo:2017fbv} for an alternative, equivalent way.} is that we scale the charge $|Q_k| \sim y_0^{-1/2}$ as we take $y_0 \to \infty$. This is intuitive since we know that the 
$\Bconf$ solution has $k=0$. By comparing with the analytic confining solution (\ref{G2holonomy}) it is possible to deduce how the parameters $b_0$, $\beta_4$ and $\mathcal{H}_{\text{\tiny{IR}}}$ must scale for large $y_0$, with the result
\begin{equation}
\left(1-b_0^2\right)\sim\frac{6K\left(-1\right)}{\sqrt{2}}\,y_0^{-\frac{1}{2}}\,,\qquad b_4\sim-2^{7/2}\,K\left(-1\right)\,y_0^{\frac{3}{2}}\,,\qquad \mathcal{H}_{\text{\tiny{IR}}}\sim\frac{h_{\rm conf}}{256}\,y_0^{-\frac{3}{2}}\,,\\[3mm]
\end{equation}
where  $K(m)$ is the complete elliptic integral of the first kind and $h_{\rm conf}$ is  the IR value of the warp factor for the confining solution given by (\ref{eq:B8confWarpFactor}) with $z=1$. We have verified these scalings with our numerical solutions. 

In terms of the $\rho$ coordinate, the first correction in (\ref{largey0}) (the second term inside the square brackets) takes the form
\begin{equation}
\frac{\rho_0}{\rho}\frac{1}{\sqrt{y_0}\left(1-\frac{\rho_0^4}{\rho^4}\right)^{1/2}}\,.
\end{equation}
We see that, no matter how large $y_0$ is, this first correction  competes with the  leading term (the 1 in (\ref{largey0})) sufficiently close to $\rho_0$. This was expected because we know that, sufficiently deep in the IR, the $\Bm$ and the $\Bconf$ metric differ dramatically from one another: in $\Bm$ the M-theory circle shrinks to zero size whereas in $\Bconf$ it does not. The intuitive picture is therefore that, by taking $y_0$ large enough, one can make the 
$\Bm$ and the $\Bconf$ metrics arbitrarily close to one another on an energy range that extends form the UV down to an IR scale arbitrarily close to the mass gap. Throughout this range the ${\rm S}^1$ of the internal metric has a constant and identical size in both metrics. Sufficiently close to the mass gap, however, the $\Bm$ metric abruptly deviates from the $\Bconf$ metric and the internal ${\rm S}^1$ closes off. Presumably  this fast change of the size of the circle is related  to the fact that the curvature in the deep IR diverges as $y_0 \to \infty$, as shown in Figure \ref{Kret}. 

\newpage

\subsection{Quasi-conformal dynamics}
\label{quasiconformal}

The $\Binf$ and the $\BOP$ solutions arise as two different limits of the $\Bp$ metrics. If the limit $y_0=-1$ of the $\Bp$ is taken at fixed $y$ then the result is the $\Binf$ solution, as we saw in \Sec{binfsol}. 

Instead, if we first focus on the IR of $\Bp$ by expanding around $y-y_0$, so that we see the $\mathds{R}^4\times{\rm S}^4$ region, and afterwards take the $y_0\to-1$ limit, then the $\BOP$ metric is reproduced. 
Indeed, for the size of the four-sphere in the eight-dimensional transverse space we have in the strict IR 
\begin{equation}
e^{2f-\Lambda}\,=\,2^{3/4}\,P_0\,\left(y_0+1\right)^{3/4}+\mathcal{O}\left(y_0+1\right)^{7/4}\,.
\end{equation}
Comparing with the IR expansion for $\BOP$ suggests the relation 
\begin{equation}\label{r0P0}
r_0\,=\,\frac{2^{3/4}P_0}{3|Q_k|}\left(y_0+1\right)^{3/4}\,.
\end{equation}
Note that, since the $\BOP$ metric does not share the same UV with the rest of the family, we do not expect \eqref{eq:P0choice} to be the right choice of $P_0$ so as to recover $\BOP$ when $y_0\to -1$. Rather, we may want to use \eqref{r0P0} to ensure that in this limit we approach $\BOP$, so
\begin{equation}
\label{eq:P0choiceIR}
P_0=\frac{3 |Q_k| }{2^{3/4} (y_0+1)^{3/4}} \, r_0
\end{equation}
Moreover, using this identification of parameters and integrating the change of coordinates \eqref{eq:coordy} in the IR and around $y_0+1$ we get
\begin{eqnarray}
y-y_0&=&\left[\frac{4\left(r-r_0\right)}{3r_0}+\frac{2\left(r-r_0\right)^2}{9r_0^2}+\mathcal{O}\left(r-r_0\right)^3\right]\left(y_0+1\right)\nonumber\\[2mm]
&-& \left[\frac{5\left(r-r_0\right)}{6r_0}+\frac{5\left(r-r_0\right)^2}{6r_0^2}+\mathcal{O}\left(r-r_0\right)^3\right]\left(y_0+1\right)^2\nonumber\\[2mm]
&+&\mathcal{O}\left(y_0+1\right)^3\,.
\end{eqnarray}
Finally, substituting this expansion together with (\ref{r0P0}) for the metric functions in the IR of the $\Bp$ family and taking the $y_0\to-1$ limit we arrive at
\begin{eqnarray}
e^{2f}&=&3r_0\left(r-r_0\right)+2\left(r-r_0\right)^2-\frac{\left(r-r_0\right)^3}{9r_0}+\mathcal{O}\left(r-r_0\right)^4\,,\nonumber\\[2mm]
e^{2g}&=&\left(r-r_0\right)^2-\frac{2\left(r-r_0\right)^3}{3r_0}+\mathcal{O}\left(r-r_0\right)^4\,,
\end{eqnarray}
which coincides, to this order, with the corresponding expansions for $\BOP$.

The intuitive picture is therefore that $\Bp$ solutions with $b_0 \gtrsim 0$  flow very close to the OP fixed point but eventually deviate from it. Put differently, the relevant deformation of the OP CFT makes these solutions exit the AdS region and develop a mass gap in the deep IR. After exiting the conformal region, their RG flow is well approximated by the metric $\BOP$. This quasi-conformal behaviour will manifest in many physical quantities we will compute in the coming Chapters.

Moreover, the existence of an exact moduli space raises the possibility of the spontaneous breaking of scale invariance, suggesting the possible presence of a light dilaton in the spectrum, as emphasised in e.g.~\cite{Gorbenko:2018ncu}. 
We will address whether a light dilaton is realised when we compute the spectrum of the theories and find the mass gap (see \cite{Elander:2021wkc} for a recent approach).

\newpage
\thispagestyle{empty}


\newpage
\chapter{Probing quasi-conformal and quasi-confining dynamics}
\label{Chapter3_observables}

In this Chapter we are going to compute some observables in the SYM--CSM theories we discussed in the previous Chapter. First of all, we will compute in Section \ref{sec:potential} the potential between a quark and an antiquark. This quantity will tell us that our theories are not going to be confining in general. However, computing their spectrum in Section~\ref{sec:spectrum} we argue that they are gapped. Hence, the $\B_8$ family provides a counterexample to the expectation that gapped gauge theories with a string dual are confining (see Section~\ref{sec:AdSsoliton} in the Introduction). Finally, in Section~\ref{sec:EE}, we examine whether entanglement entropy measures are sensitive to the confining nature of a gauge theory.

\section{Quark-antiquark potential}
\label{sec:potential}

We will now present a computation of the potential between an external quark and an external antiquark separated by a distance $L$ in the gauge theory directions. In the string description this would be extracted from the action of a string hanging from the quark and the antiquark. Instead, in M-theory we must consider a hanging membrane. An important point in the calculation is that, generically, the membrane action is UV divergent. We will renormalise away this divergence by subtracting the action of two disconnected membranes extending from the UV all the way to the IR end of the geometry. For all metrics except for $\Bconf$ this is in itself a physically acceptable configuration that competes with the connected configuration. In the case of $\Bconf$ the disconnected configuration is not a physically acceptable configuration to which the connected one can transition, but it can still be used as a mathematical well defined quantity that can be used  to regularise the membrane action.

Such computation is quite standard (see, for instance \cite{Maldacena:1998im,Rey:1998ik,Casalderrey-Solana:2019vnc}), and in our case reduces to performing the integrals
\begin{equation}
\label{eq:integrals_WilsonLoop}
\begin{aligned}
&\begin{aligned}
V= \frac{|Q_k|}{(2\pi)^2 \ell_p^3}\int_0^{\frac{4\pi}{|k|}}\dd\psi \ \times \ \Bigg[&2 \int_{r_*}^\infty\dd r\left[\left(\frac{H_*e^{-\Lambda_*}}{H_*e^{-\Lambda_*} - He^{-\Lambda}}\right)^{\frac{1}{2}}-1\right] - 2\int_{r_0}^{r_*}\dd r \, \Bigg]\\[2mm]
=\quad\frac{1}{2\pi \ell_ s^2}\quad\Bigg[&2 \int_{r_*}^\infty\dd r\left[\left(\frac{h_*}{h_* - h}\right)^{\frac{1}{2}}-1\right]- 2\int_{r_0}^{r_*}\dd r\ \Bigg]\,,\\[2mm]
\end{aligned}
\\[2mm]
&L= 2\int_{r_*}^{\infty} \dd r\  \frac{h\ }{\left(h_* - h\right)^{\frac{1}{2}}}\,,
\end{aligned}
\end{equation}
for the potential and the separation between quarks respectively. Note that, at the computational level, the integrals coming from the eleven-dimensional description coincide with those we would find in the type IIA picture.

\begin{figure}[t]
	\begin{center}
			\begin{subfigure}{.65\textwidth}
			\includegraphics[width=\textwidth]{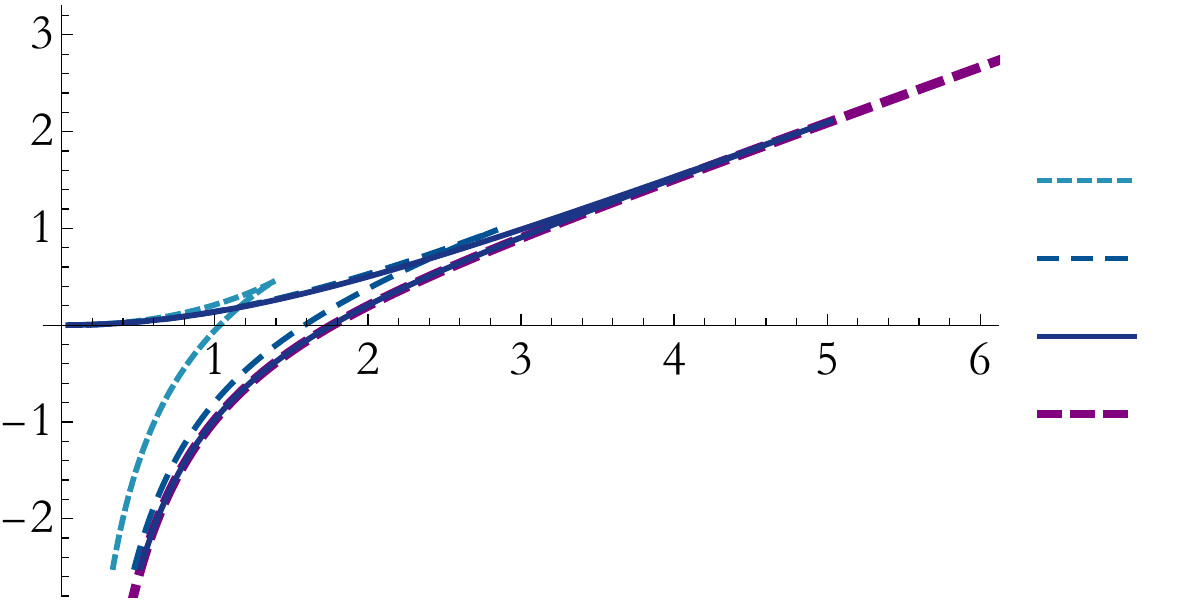} 
			\put(-220,120){$\overline{V}$}
			\put(-65,30){$\overline{L} $}
			\put(-7,75.5){$b_0 = 0.6835$}
			\put(-7,61){$b_0 = 0.9201$}
			\put(-7,46.5){$b_0 = 0.9972$}
			\put(-7,32){$b_0 = 1 $ $(\Bconf)$}
		\end{subfigure}\vspace{7mm}
		\begin{subfigure}{.65\textwidth}
			\includegraphics[width=\textwidth]{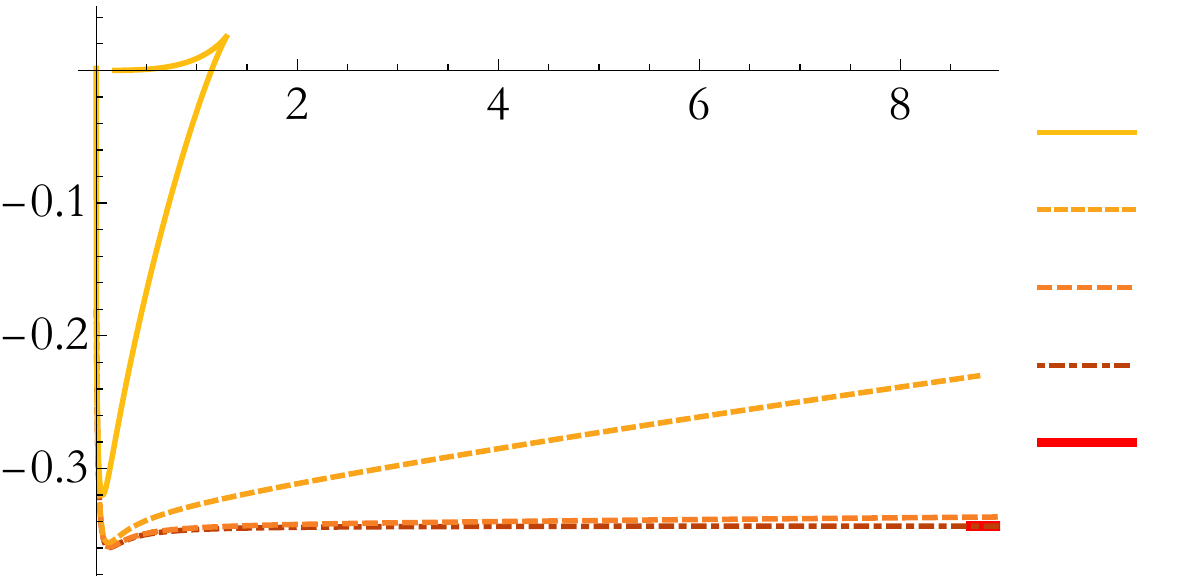} 
			\put(-220,110){$\overline{V}\cdot  \overline{L} $}
			\put(-65,70){$\overline{L} $}
			\put(-7,81){$b_0 = 0.1914$}
			\put(-7,66.5){$b_0 = 0.0604$}
			\put(-7,51){$b_0 = 0.0191$}
			\put(-7,36.5){$b_0 = 0 $ $(\B_8^\infty)$}
			\put(-7,21.5){OP | CFT}
		\end{subfigure}\hfill
		\caption{\small (Top) Quark-antiquark potential for several $\Bm$ solutions with $|Q_k|$ scaled as $|Q_k|^2=\rho_0^2/2y_0$, as dictated by \eqq{rho0}, where $\rho_0$ is the scale of the $\Bconf$ solution, whose quark-antiquark potential is shown as a dashed purple curve. (Bottom) Quark-antiquark potential for several $\Bp$ solutions and for the $\Binf$ solution. In these plots we use the dimensionless quantities defined in \eqref{norm}. The red, tiny segment at the right of the plot corresponds to the value of $VL$ for the OP fixed point. }\label{WLfig}
	\end{center}
\end{figure}

The results for the quark-antiquark potential $V$ as a function of the separation $L$ for the $\Bm$ and $\Bp$ solutions are shown in Figs.~\ref{WLfig}(top) and \ref{WLfig}(bottom), respectively, where
\begin{equation}
\label{norm}
\begin{aligned}
\overline{V} &\ =\  \frac{2\pi \ell_s^2}{|Q_k|}\, V\ =\ \frac{4\pi \, N}{\lambda \, |k|} \, V\,,\\[2mm]
\overline{L} &\ =\  \frac{|Q_k|^2}{\left(4q_c^2+3Q_c|Q_k|\right)^{1/2}}\, L \ =\  
\frac{\lambda \, |k|^2}{6\pi N \sqrt{\bar M^2+2|k|N}}\, L  \,.
\end{aligned}
\end{equation}\

The behaviour of these curves can be understood as follows. In the UV, \textit{i.e.}~in the limit $L\to 0$, the behaviour is the same for all curves, since it is dictated by their common D2-brane asymptotics, which implies $V L\sim - L^{1/3}$. Thus, as $L$ begins to increase from zero, the curves corresponding to the product $VL$ first go down ($VL$ becomes more negative) until they reach a turning point and start going up. 

This happens at the energy scale at which the Yang--Mills interaction ceases to dominate the dynamics and the CS interactions take over. This scale can be estimated from the radial position at which the first correction to the D2-brane metric is of the same order as the leading term, which yields  
\be
\lambda \frac{|k|}{N} \left( \frac{2b_0^2-1}{1-b_0^2} \right) \,.
\ee 
This is the usual result dressed with a function of the parameter $b_0$. 
After this point all curves except for the one corresponding to $\Binf$ reach $V=0$ and cross the horizontal axis at a separation that we call $l_{\text{qq,c}}$. When this happens  the disconnected configuration becomes energetically preferred. In the case of $\Binf$ the product $VL$ asymptotically approaches a negative constant  corresponding to the OP fixed point, as expected form the fact that this is the endpoint of the  $\Binf$ flow. In this case the preferred configuration is always the connected one. We see from Figure~\ref{WLfig}(bottom) that curves with $b_0$ closer and closer to $0$ become flatter and flatter and cross the horizontal axis at a larger and larger $l_{\text{qq,c}}$, manifesting quasi-conformal behaviour (see \cite{Faedo:2017fbv} for a quantitative argument of this fact). 

In the opposite limit, as $b_0\to 1$, we see from Figure~\ref{WLfig}(top) that the curves approach that of the $\Bconf$ solution, as expected from the discussion in Section~\ref{quasiconfining}. Note that the quark-antiquark potential corresponding to the $\Bconf$ theory grows linearly at large separations, and hence the theory is confining (see also the arguments in \cite{Herzog:2002ss}). This suggests that theories with $b_0\simeq 1$ exhibit quasi-confining dynamics. 

At distances $L$ slightly larger than $l_{\text{qq,c}}$ all curves but $\Bconf$ reach a cusp and ``turn back'', thus making the plots multivalued. We denote this maximum length at which we find the turning point by $l_{\text{qq,t}}$. The reason for the cusp is that, as  the penetration depth of the hanging membrane  inside the bulk increases beyond the point corresponding to the cusp,  $L$ begins to decrease.  This kind of behaviour also appears in e.g.~calculations of the quark-antiquark potential in solutions with horizons, \textit{i.e.}~in gauge theories at non-zero temperature. As in those cases, the part of the curve beyond the cusp is always energetically disfavoured 

We can conclude that the presence of confinement in the $\Bconf$ solution seems to be associated with the absence of CSM terms in the dual gauge theory. In fact, when CS are turned on it is a natural expectation \cite{Dunne:1998qy} that they give a mass to the gauge bosons, in which case colour charge is consequently screened \cite{Karabali:1999ef}.
	
Let us finish this Section mentioning that confinement in a theory with CS terms has  been previously considered in \cite{Witten:1999ds,Maldacena:2001pb} in the case of a single gauge group, as opposed to the product gauge group of our model. Despite this difference, our results are compatible with those of \cite{Witten:1999ds,Maldacena:2001pb}. Both of these references distinguish between the bare CS level occurring in the microscopic Lagrangian and the effective IR CS level, and both of them claim that confinement appears only when the latter is zero, \textit{i.e.}~when no CS terms are present in the IR theory. This suggests that, in our model, we should think of $k$ as the effective IR CS level, which is consistent with our identification of this parameter based on the analogy with the  ABJM IR fixed point. 

\section{Mass spectrum of the theories}\label{sec:spectrum}

Knowing that only one of our theories, $\Bconf$, is indeed confining, we undertake now the computation of their spectrum. We compute the masses of the spin-0 and spin-2 particles of the theory, wishing to convince ourselves that all the theories, but $\Binf$, are gapped. At the very end of this Section, we address the issue of whether quasi-conformal dynamics and the fact that the OP theory possesses an exact moduli space realise a light dilaton in theories flowing near the fixed point.

\subsection{Four-dimensional description}

Holographic spectra are conveniently computed using the gauge-invariant formalism discussed in \cite{Berg:2005pd,Elander:2010wd} (see also \cite{Bianchi:2003ug,Berg:2006xy,Elander:2009bm}). This has been used to compute spectra for a number of backgrounds with quasi-conformal dynamics obtained as deformations of the Maldacena--Nunez \cite{Maldacena:2000yy,Chamseddine:1997nm} and Klebanov--Strassler \cite{Klebanov:2000hb} backgrounds, leading to the identification of a light state when there is an operator that acquires a VEV that is parametrically larger than the scale of confinement \cite{Elander:2009pk,Elander:2012yh,Elander:2014ola,Elander:2017cle,Elander:2017hyr}. Quasi-conformal dynamics has also been explored holographically in a top-down context in  \cite{PremKumar:2010as,Anguelova:2010qh,Anguelova:2011bc}. 

This requires the existence of a lower-dimensional sigma model containing at least all the modes that are excited in the background solutions. Moreover, it must be a consistent truncation of ten or eleven-dimensional supergravity in order to ensure that the set of fields that one retains is closed, that is, they do not source additional modes outside the truncation.

Fortunately, all the backgrounds of interest can be obtained as a solution to a four-dimensional supergravity. The details of the full reduction from ten dimensions can be found in \cite{Cassani:2009ck}, from which we will keep only the scalars and follow the notation of \cite{Faedo:2017fbv}. The resulting sigma model contains gravity plus six scalars, three of which,  $\{ \Phi, U, V\}$, come from the metric together with the dilaton, while the rest, $\{a_J, b_J, b_X \}$, descend from the forms. The four-dimensional action is given in Appendix \ref{app:4Deffectivetheory}, and we write it here as\footnote{Note that  \eqref{eq:4daction} is related to \eqref{reduced_Action} using the following identifications $\tilde{\mathcal S}_4 ={\kappa_4^2}\cdot \mathcal S_4/2$, $G=\mathbf{G}/2$ and $\tilde{\mathcal V} = \mathcal V/4$.}
\begin{equation}
\label{eq:4daction}
\tilde {\mathcal S_4} = \int \dd\rho \, \dd^3 x \, \sqrt{-g}\left(\frac{R}{4}-\frac{1}{2}G_{ab} (\Phi^a) g^{MN} \partial_M\Phi^a\partial_N\Phi^b-\tilde {\mathcal V}(\Phi^a) \right) ,
\end{equation}
where $g_{MN}$ is the four-dimensional metric ($M,N = 0,1,2,3$), $G_{ab}$ is the sigma model metric ($a,b = 1, \cdots, 6$), whose explicit form is given by
\begin{equation}
\label{eq:sigmametric}\begin{aligned}
G_{ab}\partial_M\Phi^a\partial_N\Phi^b \, =\, & \frac{1}{4} \partial_M \Phi \partial_N \Phi +2 \partial_M U \partial_N U + 6 \partial_M V \partial_N V + 4  \partial_M U \partial_N V  \nonumber \\[2mm] 
&+ 16e^{-2U-4V+\frac{\Phi}{2}} \partial_M a_J \partial_N a_J \nonumber\\[2mm]
&+ 2e^{-4V-\Phi} \left( \partial_M b_J \partial_N b_J  + 2 \partial_M b_X \partial_N b_X \right) ,\nonumber
\end{aligned}
\end{equation}
and the potential $\tilde {\mathcal V}$ can be written in terms of a superpotential $\mathcal W$ as it is shown in the Appendix.

All the solutions preserve Poincar\'e invariance in three dimensions, so we restrict ourselves to backgrounds that only depend on a radial coordinate $\rho$, for which the metric has the form of a domain wall
\begin{equation}
\dd s_4^2 = \dd\rho^2 + e^{2A(\rho)} \dd x_{1,2}^2 .
\end{equation}
Moreover, they are $\mathcal{N}=1$ supersymmetric, so as usual they can be obtained from a set of BPS equations that read
\begin{equation}\label{eq:firstordereqs}
\Phi'^a =G^{ab} \frac{\partial \mathcal W}{\partial \Phi^b}, \hspace{1.5cm} A' = -\mathcal W ,
\end{equation}
where prime denotes derivatives with respect to $\rho$. 
The scalars $U$, $V$, $\Phi$ and $A$ are nothing but the convenient rewriting of the functions $e^{f}$, $e^{g}$, $e^{\Lambda}$ and $h$ from Chapter \ref{Chapter2_B8family} given in \eqref{eq:relation10Dto4D}, which we wish to reproduce here
\begin{eqnarray}
e^{\Phi}&=&h^{1/4}\,e^{\Lambda}\,,\nonumber\\[2mm]
e^{2U}&=&4\,h^{3/8}\, e^{2g-\Lambda/2}\,,\nonumber\\[2mm]
e^{2V}&=&2\,h^{3/8}\, e^{2f-\Lambda/2}\,,\nonumber\\[2mm]
e^{2A}&=&16\,h^{1/2}\, e^{4f+2g-2\Lambda}\,.
\end{eqnarray}
Also, the radial coordinate change from $\rho$ to $y$ is defined in terms of the functions we know via
\begin{equation}
\dd\rho = \dd y \, \frac{2 \sqrt{2} H^{3/4} P^{9/4} (v-2)^{1/4}}{|Q_k|^{1/2} v^{3/4} (1-y) (y+1)^{5/4}}\,,
\end{equation}
where $v(y)$ and $P(y)$ are given by (\ref{eq:solutionV1}$\, - \,$\ref{eq:solutionV2})  and \eqref{eq:solutionP} respectively, and $H(y)$ is the eleven-dimensional warp factor \eqref{HH}. As a consequence, the BPS equations following from equation~\eqref{eq:firstordereqs} coincide with \eqref{BPSsystem} and \eqref{BPSsystem_fluxes}.

\subsection{Fluctuations}
\label{sec:formalism}

In this Section, we summarise the gauge-invariant formalism that we used to compute the spin-0 and spin-2 glueball spectra of the dual theory. We follow closely \cite{Berg:2005pd,Elander:2010wd} to which the reader is referred for further details. 

Spectra are obtained by studying small fluctuations of the scalar fields and the metric around a given background. The fluctuations are taken to depend on both the radial coordinate $\rho$ as well the boundary coordinates $x^\mu$. After going to momentum-space (defining $m^2 = -q^\mu q^\nu \eta_{\mu\nu}$ where $q^\mu$ is the three-momentum and $\eta_{\mu\nu} = {\rm diag} (-1,1,1)$ is the boundary metric), one expands the equations of motion to linear order in the fluctuations and imposes the appropriate boundary conditions in the IR and UV. The spectrum is then given by those values of $m^2$ for which solutions exist.

More precisely, we expand in fluctuations $\{ \varphi^a, \nu, \nu^\mu, \mathfrak e^\mu{}_\nu, h, H, \epsilon^\mu \}$ as
\beqs
\Phi^a &=& \bar \Phi^a + \varphi^a, \nonumber\\
\dd s_4^2 &=& (1 + 2\nu + \nu_\sigma \nu^\sigma) \dd \rho^2 + 2 \nu_\mu \dd x^\mu \dd \rho + e^{2A} ( \eta_{\mu\nu} + h_{\mu\nu} ) \dd x^\mu \dd x^\nu, \\
h^\mu_{\ \nu} &=& \mathfrak e^\mu_{\ \nu} + i q^\mu \epsilon_\nu + i q_\nu \epsilon^\mu + \frac{q^\mu q_\nu}{q^2} H + \frac{1}{2} \delta^\mu_{\ \nu} h,\nonumber
\eeqs
where $\mathfrak e^\mu_{\ \nu}$ is transverse and traceless, $\epsilon^\mu$ is transverse, and the three-dimensional indices $\mu$, $\nu$ are raised and lowered by the boundary metric $\eta$.

The spin-2 fluctuation $\mathfrak e^\mu_{\ \nu}$ satisfies the linearised equation of motion
\beq
\label{eq:fluceoms2}
\left[\partial_\rho^2 + 3A' \partial_\rho + e^{-2A} m^2 \right] \mathfrak{e}^{\mu}_{\ \nu} = 0 .
\eeq
After forming the gauge-invariant combination \cite{Berg:2005pd,Elander:2009bm}
\beq
\mathfrak a^a = \varphi^a - \frac{\bar\Phi'^a}{4A'} h ,
\eeq
the linearised equation of motion for the spin-0 fluctuations can be written as

\begin{equation}
\label{eq:fluceoms}\begin{aligned}
&\left[ \mathcal D_\rho^2 + 3A' \mathcal D_\rho + e^{-2A} m^2\right] \mathfrak{a}^a \\
&\qquad - \Big[ \tilde{\mathcal V}^a_{\ |c} - \mathcal{R}^a_{\ bcd} \bar \Phi'^b \bar \Phi'^d + \frac{2 (\bar \Phi'^a \tilde{\mathcal V}_c + \tilde{\mathcal V}^a \bar \Phi'_c )}{A'} + \frac{4 \tilde{\mathcal V} \bar \Phi'^a \bar \Phi'_c}{A'^2} \Big] \mathfrak{a}^c = 0\,.
\end{aligned}
\end{equation}
The different quantities involved in this expression are
\beqs
\begin{array}{lcl}
	\mathcal G_{abc} = \frac{1}{2} \left( \partial_b G_{ca} +\partial_c G_{ab} - \partial_a G_{bc} \right),&\quad&\bar\Phi'_a = G_{ab} \partial_\rho\bar \Phi^b,\\[2mm]
	\mathcal R^a_{\ bcd}= \partial_c \mathcal G^a_{\ bd} - \partial_d \mathcal G^a_{\ bc} + \mathcal G^a_{\ ce} \mathcal G^e_{\ bd} - \mathcal G^a_{\ de} \mathcal G^e_{\ bc},&\quad&\tilde{\mathcal V}^a_{\ |b} = \frac{\partial \tilde{\mathcal V}^a}{\partial \Phi^b} + \mathcal G^a_{\ bc} \tilde{\mathcal V}^c,
\end{array}
\eeqs
that is, $\mathcal R^a_{\ bcd}$ is the Riemann tensor corresponding to the sigma model metric. On the other hand, the background covariant derivative is defined as $\mathcal D_\rho \mathfrak a^a = \partial_\rho \mathfrak a^a + \mathcal G^a_{\ bc} \bar \Phi'^b \mathfrak a^c$.

In order to obtain the spectrum, we first introduce IR (UV) cutoffs at $\rho_I$ ($\rho_U$). For backgrounds with an end-of-space in the IR located at $\rho = \rho_o$, the physical spectrum is obtained in the limit of $\rho_I \rightarrow \rho_o$, while similarly taking the limit of $\rho_U$ towards the location of the UV boundary. The boundary conditions at $\rho_I$ and $\rho_U$ are obtained by requiring that the variational problem be well-defined. This necessitates adding localised boundary actions in the IR and UV, which up to quadratic order in the fluctuations are determined by symmetry and consistency, except for a term quadratic in the scalar fluctuations. Taking the limit of this term corresponding to adding infinite boundary-localised mass terms for the scalar fluctuations, we obtain the boundary condition
\beq
\varphi^a |_{\rho_{I,U}} = 0,
\eeq
which in terms of the gauge-invariant variable $\mathfrak a^a$ becomes
\beqs
\label{eq:BCs1}
-\frac{e^{2A}}{m^2} \frac{\bar \Phi'^a}{A'} \left[ \bar \Phi'_b \mathcal D_\rho  - \frac{2\tilde{\mathcal V} \bar \Phi'_b}{A'} - \tilde{\mathcal V}_b \right] \mathfrak a^b \Big|_{\rho_i} = \mathfrak a^a \Big|_{\rho_{I,U}}.
\eeqs
Similarly, the boundary condition for the tensor fluctuations becomes
\beq
\label{eq:BCs2}
\partial_\rho \mathfrak{e}^{\mu}_{\ \nu} |_{\rho_{I,U}} = 0 .
\eeq
These boundary conditions assure that subleading modes are selected in the limit of taking $\rho_I$ towards the end-of-space and $\rho_U$ towards the UV boundary, in accordance with standard gauge-gravity duality prescriptions.

\subsection{Mass spectrum}

\begin{figure}[t!]
	\begin{center}
		\includegraphics[width=10cm]{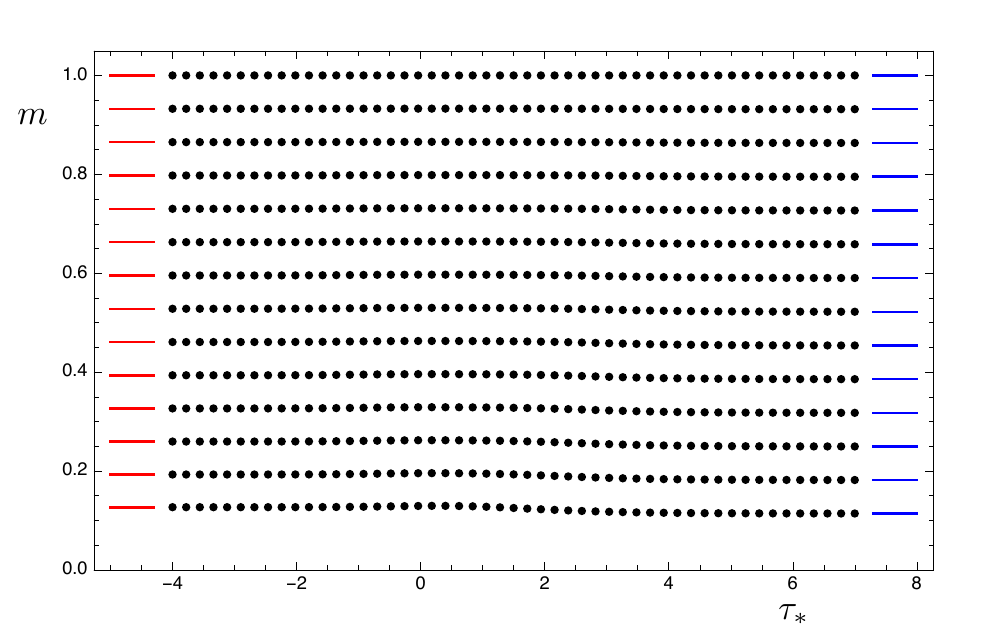}
		\caption{Mass spectrum $m$ of spin-2 states as a function of $\tau_*$ normalised to the heaviest state included in the plot, compared to the spin-2 spectrum of $\BOP$ (red, left) and $\Bconf$ (blue, right).}
		\label{Fig:SpectrumSpin2}
	\end{center}
\end{figure}

\begin{figure}[t!]
	\begin{center}
		\includegraphics[height=.85\textheight]{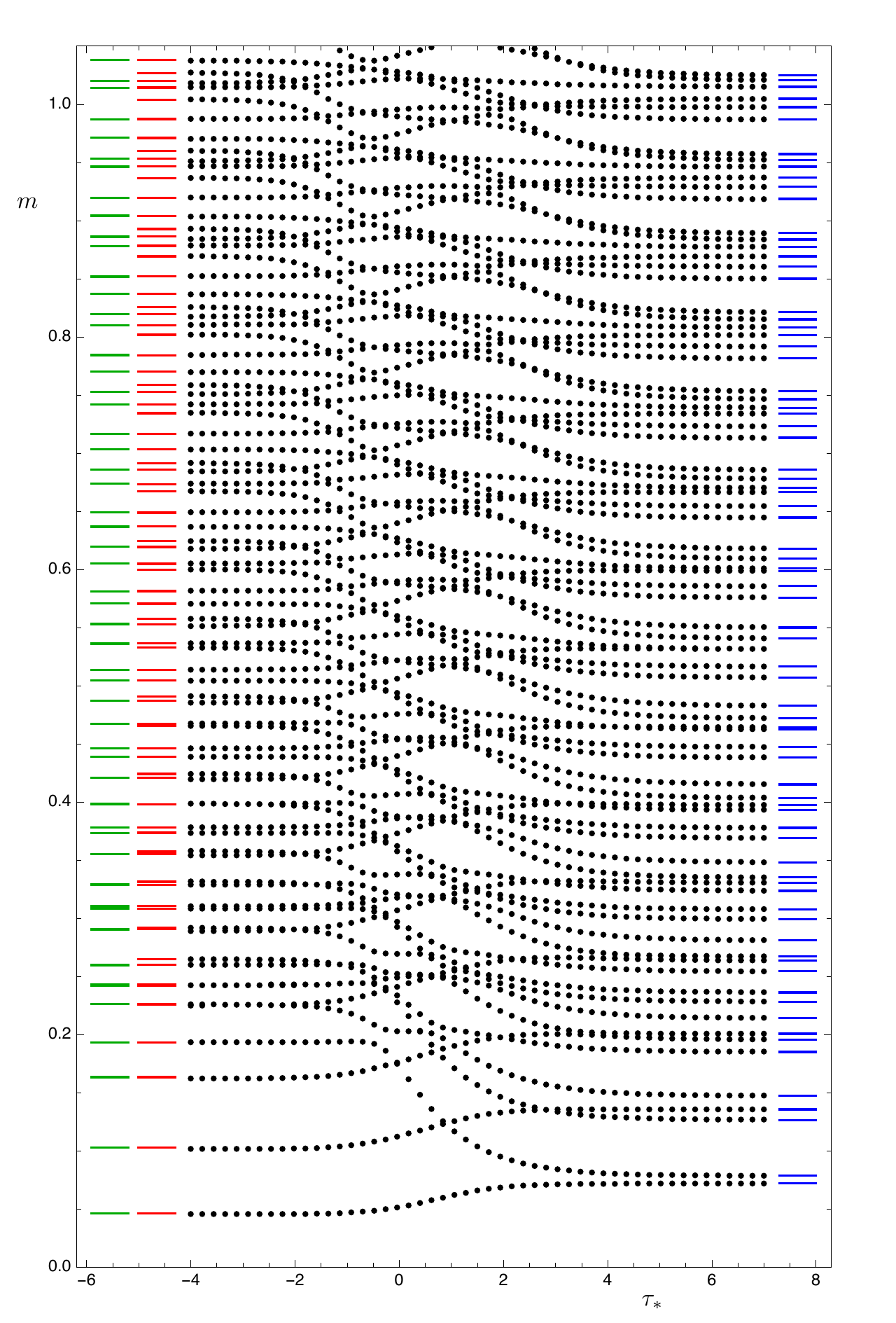}
		\caption{Mass spectrum $m$ of spin-0 states as a function of $\tau_*$ with the same normalisation as for Figure~\ref{Fig:SpectrumSpin2}. Also shown is the spin-0 spectrum of $\BOP$ (from left: in green for four scalars, in red for six scalars) and $\Bconf$ (blue, right).}
		\label{Fig:SpectrumSpin0}
	\end{center}
\end{figure}

Now that the procedure that we followed is clear, we will directly show the result for the spectrum of spin-0 and spin-2 particles. We refer to \cite{Elander:2018gte} for the details. Let us just stress a technical point concerning the boundary conditions used for the fluctuations. Following Section~\ref{sec:formalism}, we imposed equations~\eqref{eq:BCs1} and \eqref{eq:BCs2} for the spin-0 and spin-2 modes, respectively, at an IR cutoff, and checked that the calculation of the spectrum converges to the physical result when the cutoff is removed. In principle, we should have followed the analogous procedure in the UV, introducing a UV cutoff, and studied what happens when it is taken to the boundary of the spacetime. However, we found that reaching high enough values of the UV cutoff is numerically challenging, and because of this we took a different approach, namely to make use of the explicit UV expansions of the fluctuations. As noted in Section~\ref{sec:formalism}, the two approaches agree in the limit $\rho_U \rightarrow $ boundary.

Following \cite{Elander:2018gte}, instead of using $y_0$ or $b_0$, the different theories will be labelled in a new parameter $\tau_*$, given in terms of $y_0$ by 
\begin{equation}
\tau_*\,=\, \frac{1}{2}\log \left(1+y_0\right)\,.
\end{equation}
Note that the limit $\tau_*\to-\infty$ ($\tau_*\to+\infty$) corresponds to approaching the conformal (confining) region.

\begin{itemize}
	\item \textbf{Tensor modes}. The resulting spectrum for the different theories is shown in Figure~\ref{Fig:SpectrumSpin2}. As can be seen, there is a mass gap above which a tower of states appears. We have chosen to normalise the plot in such a way that the spacing of the heaviest modes remains the same for the different theories, in order to reflect the fact that the UV physics is the same for all the solutions. Conversely, the lowest lying states, which are sensitive to IR physics, show a slight dependence on the choice of theory, becoming lighter as we approach $\Bconf$. In the limits of $b_0 \to 0\,,\ 1$, the spectra corresponding to the $\BOP$ and $\Bconf$ backgrounds are recovered.
	
	\item \textbf{Scalar modes}. Figure~\ref{Fig:SpectrumSpin0} shows the scalar spectrum of different theories. The normalisation is the same as that for the spin-2 spectrum in Figure~\ref{Fig:SpectrumSpin2}. When $b_0\to 1$, the spectrum approaches that of the $\Bconf$ background. It is interesting that the heavier states in this limit seem to come in groups of six. This may be due to the fact that these states are mostly sensitive to the UV D2-brane geometry, which is simpler than the full solution. In the opposite limit, $b_0\to 0$, the spectrum approaches the one corresponding to the $\BOP$ solution. In this case, the theory flows close to the OP fixed point, and hence the $\BOP$ background is valid as an effective theory up to high energy scales. In this limit we observe that certain low-lying states become approximately degenerate. Presumably, this is due to an enhancement of symmetry that takes place in the IR of these flows. Indeed, the  S$^3$ of the internal geometry is not squashed in the $\BOP$ solution, which leads to an approximate SO$(4)$ symmetry enhancement in the IR of flows that pass very close to the OP fixed point. This same symmetry is the one that allows for a consistent truncation of the sigma model from six to four scalars that admits the $\BOP$ background as a solution. This truncation is described in Section 7.2 of \cite{Cassani:2011fu}, when the internal manifold is taken to be the seven-sphere.  Finally, we note that despite the fact that the theory can be made to flow arbitrarily close to the IR fixed point of OP, there is never a parametrically light state in the spectrum.
\end{itemize}

So, in spite of the OP fixed point possessing an exact moduli space and small $b_0$ flows exhibiting quasi-conformal behaviour, we did not find a light pseudo-dilaton. This can be understood as follows. The geometry of small $b_0$  RG flows consists of three parts. The first part takes the theory from the UV corresponding to the D2-branes down to the vicinity of the OP fixed point, following closely the solution $\Binf$. The second part consists of a region where the geometry is approximately AdS, and the dual theory remains close to the OP fixed point. From the point of view of this fixed point, how long it stays close is determined by the size of an irrelevant operator. The smaller $b_0$ is, the smaller this irrelevant operator is. The third and final part describes the flow from the vicinity of the OP fixed point to the deep IR where the theory develops a mass gap. For small $b_0$, this final part of the RG flow is very well captured by the $\BOP$ solution. The key point is that, as mentioned in Section~\ref{sec:BOP_solution}, the RG flow described by the $\BOP$ solution is triggered by both a source and a VEV of comparable sizes. In other words, the magnitudes of the explicit and the spontaneous breaking of scale invariance are similar. It would be interesting to investigate whether more general flows exist that start at the OP fixed point and for which the ratio between the spontaneous and the explicit breaking of scale invariance can be made arbitrarily large. 

\section{Entanglement entropy}\label{sec:EE}

Entanglement entropy is a non-local quantity that permits us to study non-perturbative phenomena of quantum field theories. Despite of its simple definition, it turns out to be very difficult to compute in interacting field theories. The situation is also conceptually challenging in gauge field theories due to the lack of local tensor product decomposition of the physical Hilbert space of gauge invariant states ${\mathcal{H}}_\text{phys}\bcancel{\rightarrow} {\mathcal{H}}_A\otimes {\mathcal{H}}_B$ \cite{Ghosh:2015iwa}. In holography, the Ryu--Takayanagi (RT) formula \cite{Ryu:2006bv} is conjectured to provide us with the entanglement entropy for a given field theory at strong coupling and in the limit of large-$N$. The RT formula states that the entanglement entropy associated with a region $A$ is given by the minimal area of a co-dimension two bulk surface exploring the dual 10-dimensional classical background geometry, anchored onto $\partial A$ at the boundary of the bulk spacetime. 

The aim of this Section is to examine whether the holographic entanglement entropy as given by the RT prescription can reveal if a gauge field theory is confining. This question was raised by several works which found evidence for phase transitions in entanglement entropies as functions of relevant length scales in different confining backgrounds at large-$N$ \cite{Klebanov:2007ws,Nishioka:2006gr,Nishioka:2009un,Kol:2014nqa}. The phase transitions additionally share some resemblance with the deconfinement phase transitions happening in the same models at finite temperature. This pronounces the expectation that entanglement entropy probes confinement. Interestingly, a number of studies have scrutinised the AdS/CFT results by a comparison to entanglement entropies calculated from 4D pure glue non-Abelian Yang--Mills theories on a lattice with a small number of colours \cite{Buividovich:2008gq,Buividovich:2008kq,Itou:2015cyu,Rabenstein:2018bri}.

Additionally, some works have emphasised that the phase transition of the entanglement entropy at large-$N$ signals the deconfinement transition of the underlying gauge theory and
can be used as a diagnostic for gauge theories with varying degree of resemblance to QCD, see, \emph{e.g.}, \cite{Lewkowycz:2012mw,Kol:2014nqa,Mahapatra:2019uql,Knaute:2017lll,Arefeva:2020uec}. However, 
sometimes the mere presence of an energy scale leads to a phase transition of the entanglement entropy \cite{Klebanov:2012yf,Jokela:2019tsb} in phases with no link to confinement. 

\begin{figure}[t]
	\begin{center}
		\begin{subfigure}{0.49\textwidth}
			\includegraphics[width=1.\textwidth]{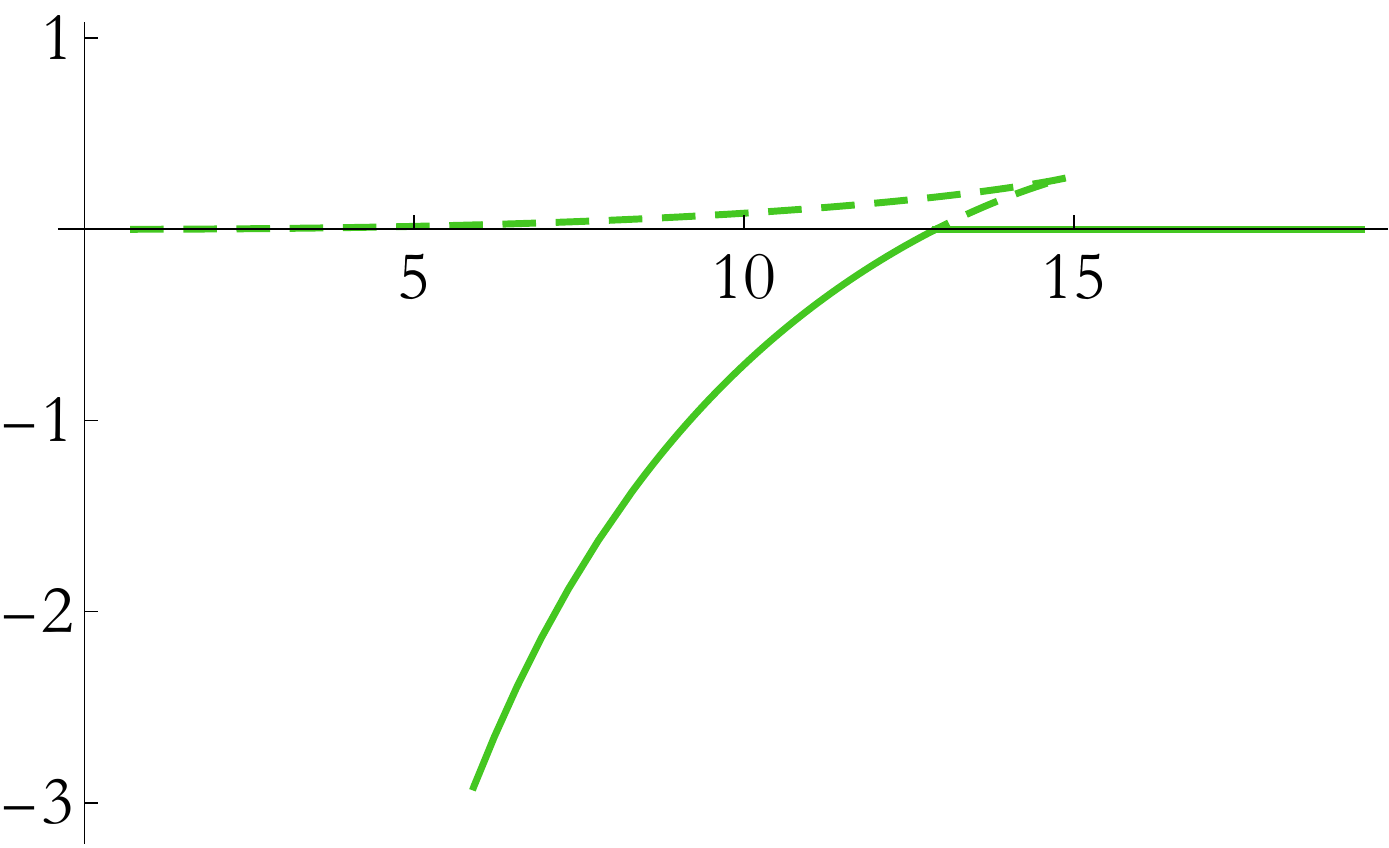} 
			\put(-150,90){$\overline{V}$}
			\put(-22,60){$10^2 \overline{L}$}
		\end{subfigure}\hfill
		\begin{subfigure}{.48\textwidth}
			\includegraphics[width=1.\textwidth]{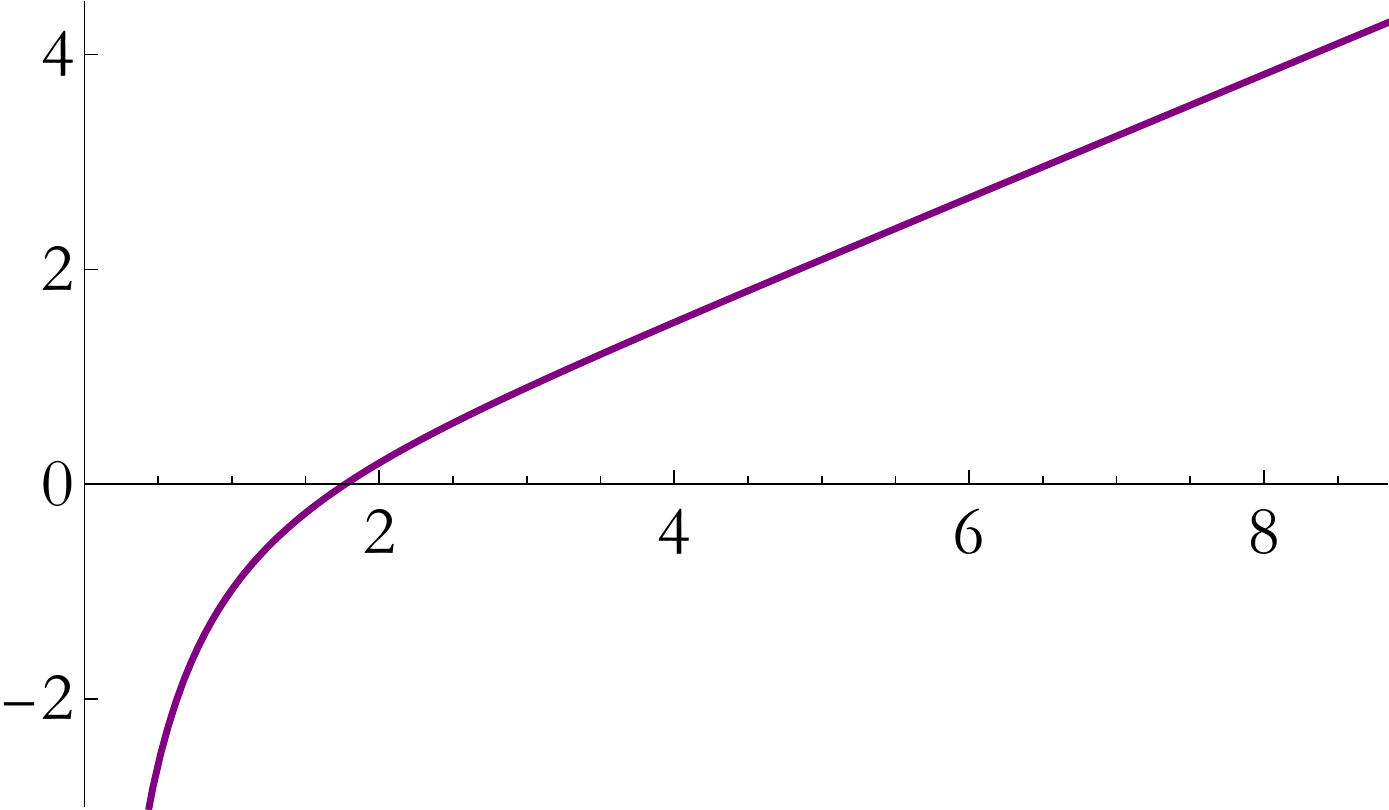} 
			\put(-150,90){$\overline{V}$}
			\put(-10,15){$\overline{L}$}
		\end{subfigure}
		\caption{\small Quark-antiquark potential for a non-confining theory with a mass gap $\B_8$ (Left) and for the confining theory $\Bconf$ (Right). Solid curves stand for the value of the potential for the dominant configuration whereas dashed curves depict those of unstable configurations. The string breaks in the $\B_8^0$ theory when the curve crosses zero, signalling the splitting of a meson into two quarks in the gauge theory. As explained in Section~\ref{sec:potential}, splitting cannot happen in the confining case $\Bconf$, where consequently the connected configuration is always the dominant one, leading to the linear growth of the potential for large values of the separation between the quark and the antiquark.}\label{fig.WilsonLoop}
	\end{center}
\end{figure}

In order to address the question of whether the entanglement entropy is sensitive to the fact that a theory is confining, we are going to consider different entangling surfaces in the family of solutions we constructed in Chapter~\ref{Chapter2_B8family}. We will compare them to the results we obtained in Section~\ref{sec:potential} for the quark antiquark potential. Specifically, we will focus in the non-confining gapped theory with $b_0=2/5$ and the confining $\Bconf$, whose plots for the potential we reproduce in Figure~\ref{fig.WilsonLoop}.

One of the main results of this Section will be that the entanglement entropy measures we consider, while being sensitive to the presence of the mass gap, do not distinguish confining from non-confining theories contrary to the quark-antiquark potential. This means that the mass gap fixes the maximum typical distance between entangled states, which \textit{a priori} is independent from the interactions between infinitely massive and very distant quarks. None of the entanglement measures we investigate is able to distinguish between the confining theory from those that are only gapped. Similarly, we can assert that entanglement measures are not sensitive to the presence of CS interactions. On the contrary, we will show that the near-proximity of the OP CFT at some intermediate energy scale is clearly captured, in addition to by the entanglement entropy itself, by the mutual information and the $F$-functions counting the numbers of degrees of freedom in $(2+1)$-dimensional field theories.

Although in the end we want to consider entanglement entropy of strips and disks, we plan to explain first the general setup and specialise to those cases afterwards. Following \cite{Ryu:2006bv}, the entanglement entropy of a QFT region $A$ bounded by $\partial A$ is given by the area of a minimal surface $\Sigma_A$ anchored on $\partial A$ at the spacetime boundary. This minimal surface $\Sigma_A$ is co-dimension two, {\emph{i.e.}}, an eight-dimensional submanifold embedded in our ten-dimensional type IIA geometry \eqref{10Dansatz}, wrapping completely the internal part of the geometry. For simplicity, we will refer to this embedding as the Ryu--Takayanagi (RT) surface associated to $A$. The holographic entanglement entropy in string frame reads
\begin{equation}\label{eq:EE_formula}
S_A = \frac{1}{4 G_{10}}\ \int_{\Sigma_A} \dd^8\sigma \ e^{-2\Phi}\sqrt{\det \mathsf{g}} \ ,
\end{equation}
where the $\sigma$'s are coordinates on $\Sigma_A$, the constant $G_{10} = (16\pi)^{-1} (2\pi)^7 g_s^2 \ls^8$ is the ten-dimensional Newton's constant\footnote{In particular, $16\pi G_{10} = 2\kappa_{10}^2$, see \eqref{eq:newtons_tension}.}, the function $\Phi$ is the dilaton, and $\mathsf{g}$ is the induced metric on $\Sigma_A$ in string frame:
\begin{equation}\label{eq:embedding}
\mathsf{g}_{\alpha\beta} = \frac{\partial x^{\mu}}{\partial \sigma^\alpha}\frac{\partial x^{\nu}}{\partial \sigma^\beta} \ g_{\mu\nu} \ .
\end{equation}
Since we require that $\Sigma_A$ wraps the whole internal space, integration over six of the eight $\sigma$'s coordinates gives a factor of $V_6 = 32\pi^3 /3$ in \eqref{eq:EE_formula}, which is the volume of $\CP^3$. The embedding of the surface is determined by the equations of motion for the scalar fields that we will choose to present as follows
\begin{equation}
\label{eq:general_embedding}
t= \text{constant} \quad , \quad x_1 = x_1(\sigma_1,\sigma_2) \quad , \quad x_2 = x_2(\sigma_1,\sigma_2) \quad , \quad r = r(\sigma_1,\sigma_2) \ .
\end{equation}
We then vary \eqref{eq:EE_formula} with respect to the fields and obtain the Euler--Lagrange equations
\begin{equation}
\label{eq:EulerLagrange}
\frac{\partial\LL}{\partial\phi^i} - \partial_\mu\left(\frac{\partial \LL}{\partial(\partial_\mu\phi^i)}\right) = 0 \, , \quad \mbox{for }\phi^i\in\{x^1,\ x^2,\ r\}\mbox{ and }\mu=\sigma^1, \ \sigma^2 \ ,
\end{equation}
which comprise three second order partial differential equations. These equations are the equations of motions that a surface has to fulfil in order to be extremal. We will stick to the entanglement entropy between strips and disks because in that case the problem simplifies and we just have to solve second order ordinary differential equations. Our expectation is that these two cases capture the main features one might encounter considering other shapes for $A$.

This system is quite similar to that of \cite{Bea:2013jxa}, which we followed to perform our computations. We relegate the most technical details to the Appendixes and refer there for further details. 

\subsection{Entanglement entropy of the strip}\label{sec:strip}

\begin{figure}[t]
	\begin{center}
		\includegraphics[width=\textwidth]{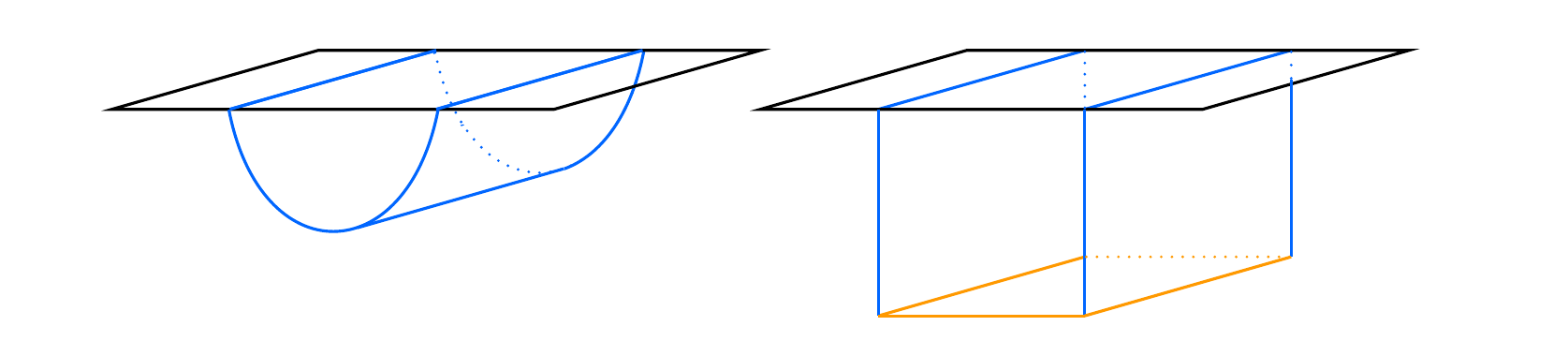} 
		\put(-140,-10){$\sqcup$ configuration}
		\put(-285,-10){$\cup$ configuration}
		\caption{\small  The two possible and competing configurations of the RT surface we have to consider when computing entanglement entropies of strips: for small widths of the strip, a ``connected'' configuration $\cup$, which does not reach the bottom of the geometry (left), competes with the ``disconnected'' configuration $\sqcup$, which reaches the end-of-space (right).
		}\label{fig:strip_conf}
	\end{center}
\end{figure}

Let us now specialise our setup and pick a particular boundary region $A$. Let $A$ be a strip of width $l$. For this choice, we will find two possible embeddings that extremise the area of the RT surface, depicted in Figure~\ref{fig:strip_conf}. First, there is the configuration denoted by $\cup$, which is specified by the choice
\begin{equation}
\label{eq:embedding_strip1}
t= \text{constant}, \quad x^1 = \sigma^1\in[-l/2,l/2],\quad x^2 = \sigma^2\in \R,\quad r = r(\sigma^1)\in [r_*,\infty)\,,
\end{equation}
where $r_* \ (> r_s)$ is the value of the radial coordinate at which the RT surface has a turning point. By $r_s$ we denote the position of the end-of-space in the $r$ coordinate, $r_s=r(y_0)$ with the change of coordinates \eqref{eq:coordy}. 

When we substitute the ansatz \eqref{eq:embedding_strip1} in the equations of motion \eqref{eq:EulerLagrange}, we are left with a unique second order differential equation. Focusing on the induced metric obtained by substituting \eqref{eq:embedding_strip1} into the background metric \eqref{eq:10DansatzAP}, and taking the squared root of its determinant as specified by \eqref{eq:EE_formula}, we obtain an expression for the entanglement entropy in this case
\begin{equation}
\label{eq:EE_strip}
S_\cup(l) = \frac{V_6 L_y}{4 G_{10}} \int_{-\frac{l}{2}}^{\frac{l}{2}}\dd \sigma^1\  \Xi^{\frac{1}{2}} \left(1+h\ \dot r^2\right)^{\frac{1}{2}}\ ,
\end{equation}
where the dot stands for differentiation with respect to $\sigma^1$, $L_y = \int_{\R}\dd\sigma^2$, and 
\begin{equation}\label{eq:def_XI}
\Xi = h^2 e^{8f+4g-4\Phi}\ .
\end{equation} 
In \eqref{eq:EE_strip} there is a conserved quantity that allows us to solve the remaining second order equation and will help us to perform the integrals involved in the computation of the entanglement entropy. We relegate these details to Appendix~\ref{ap:strip}.

There is also a second possible configuration we denoted by $\sqcup$ in Figure~\ref{fig:strip_conf}. This extremal surface consists of three pieces: two of them extend from the UV all the way down to the IR, whereas the third one lays at the bottom of the geometry connecting the two. The embeddings for the two former ones are specified by demanding
\begin{equation}
\label{eq:embedding_strip2}
t= \text{constant}\, , \quad x^1 = \pm \frac{l}{2} = \text{constant}\, , \quad x^2 = \sigma^2\in \R\,  , \quad r =\sigma^1 \in [r_s,\infty) \ ,
\end{equation}
which automatically fulfil the equations of motion \eqref{eq:EulerLagrange}. As we pointed out, these embeddings are describing two submanifolds that hang from the UV, where they are attached to one of the edges of the strip; towards the IR, where they end at the regular bottom of the geometry at $r=r_s$. Note that on their own they do not yet constitute a valid RT surface for the strip, since 
they are not homologous to the boundary region $A$ unless they are connected at the bottom with the third piece  mentioned above. This last piece is located at the bottom of the geometry and specified in the following way:
\begin{equation}
\label{eq:bottom}
t= \text{constant} , \quad x^1 = \sigma^1 \in[-l/2,l/2] \, ,\quad x^2 = \sigma^2 \in \R \, ,\quad r = r_s = \text{constant}\ .
\end{equation}
It also fulfils equations \eqref{eq:EulerLagrange}. Additionally, it has zero area due to the fact that it is wrapping the three-cycle which  contracts smoothly at the end-of-space. Thus, the action \eqref{eq:EE_formula} in this second RT surface $\sqcup$ consequently leads to the expression 
\begin{equation}
\label{eq:EE_disconnected}
S_\sqcup =  2 \ \frac{V_6 L_y}{4 G_{10}} \int_{r_s}^{\infty}\dd  r\  \Xi^{\frac{1}{2}} h^{\frac{1}{2}}\ 
\end{equation}
for the entanglement entropy. Notice that (\ref{eq:EE_disconnected}) does not {\emph{explicitly}} depend on $l$. The dependence on $l$ is only through the boundary conditions (\ref{eq:embedding_strip2}) by anchoring the surface to the boundary region $A$.  
The configuration of a single dipping surface of the form (\ref{eq:embedding_strip2}), connected to a semi-infinite bottom embedding, would lead to the entanglement entropy of spacetime divided in half. It is immediate that the corresponding entanglement entropy cannot depend on the position of the domain wall; similar argument applies to (\ref{eq:EE_disconnected}).

\begin{figure}[t]
	\begin{center}
		\begin{subfigure}{0.47\textwidth}
			\includegraphics[width=\textwidth]{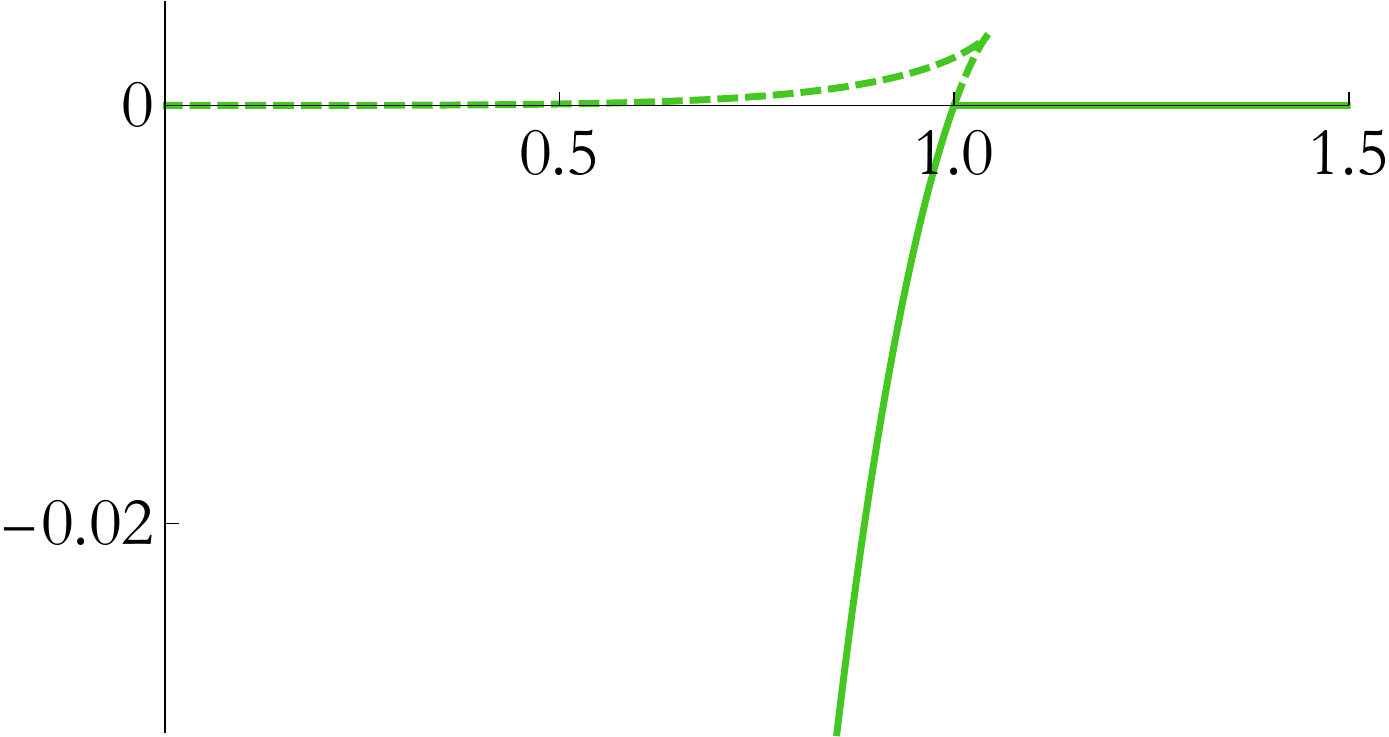} 
			\put(-160,90){$\Delta \overline S$}
			\put(-30,-10){$l/l_c$}
		\end{subfigure}\hfill
		\begin{subfigure}{.47\textwidth}
			\includegraphics[width=\textwidth]{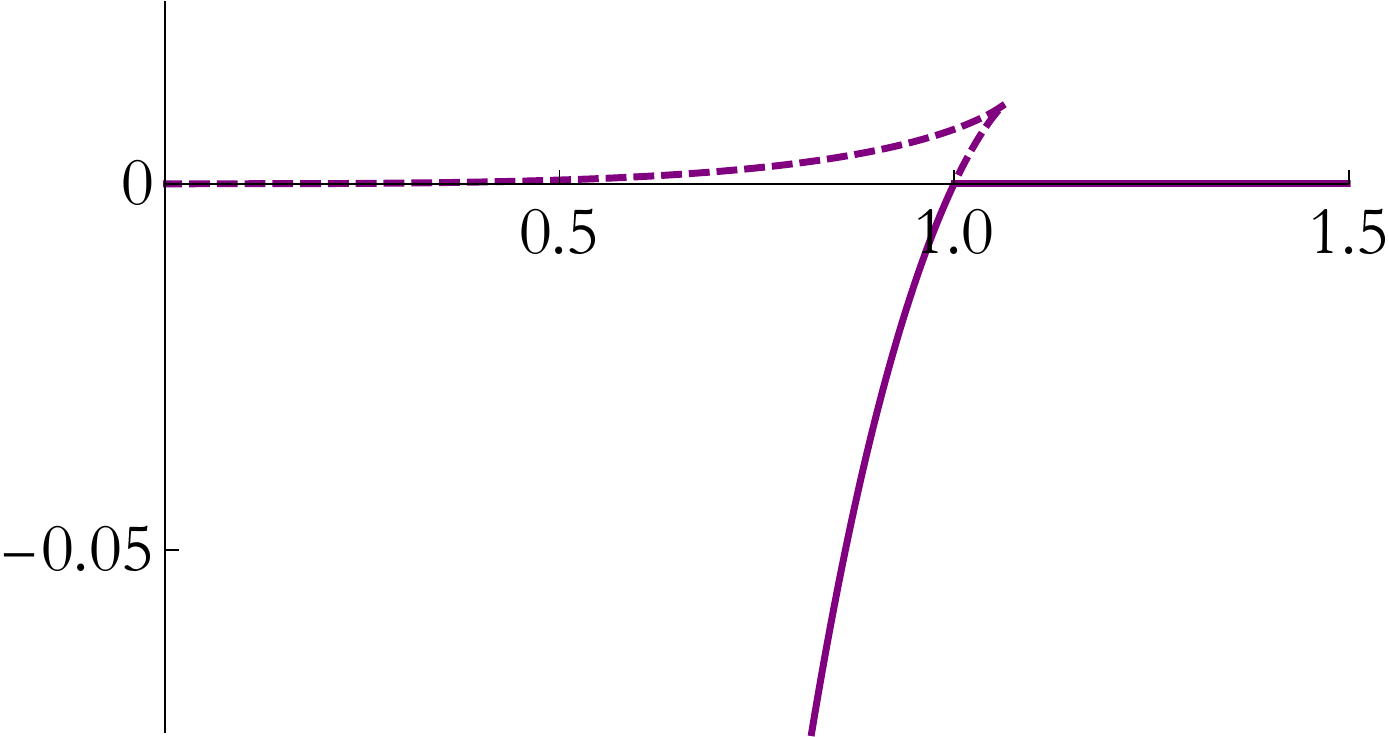} 
			\put(-160,90){$\Delta \overline S$}
			\put(-30,-10){$l/l_c$}
		\end{subfigure}
		\caption{\small Entanglement entropy of a single strip as a function of the width of the strip in the gapped non-confining theory $\B_8$ (Left) and in the confining one $\Bconf$ (Right). We plot the rescaled quantity \eqref{eq:dimensionlessEEstrip} defined in Appendix~\ref{ap:strip} as a function of the strip width normalised to the value above which the ``disconnected'' configuration $\sqcup$ becomes the dominant one.
		}\label{fig.EEplotsStrip}
	\end{center}
\end{figure}

Notice that for a given strip of width $l$ we have two candidate RT surfaces (although we will see that the configuration $\cup$ is only present below some critical value of $l$), so we need to infer which one is the minimal one. The correct choice can be expressed as:
\begin{equation}
S_{\text{strip}}(l) = \min\{S_\cup(l),\ S_\sqcup\} \ .
\end{equation}
Notice that both \eqref{eq:EE_strip} and \eqref{eq:EE_disconnected} are UV divergent. However, instead of regulating the entropy functionals for each case, which comes with its own subtleties, we are content with comparing the on-shell actions. To be more precise, we will consider the following object
\begin{equation}
\label{eq:EE_strip_reg}
\Delta S (l) = S_\cup (l) - S_{\sqcup}\ .
\end{equation}
Then, while \eqref{eq:EE_strip} and \eqref{eq:EE_disconnected} are UV divergent quantities, since their divergence structure is the same due to homogeneity, \eqref{eq:EE_strip_reg} is a finite quantity. 
Moreover, there will be values for the width of the strip at which $\Delta S (l)$ will flip sign, signalling a critical value above which the ``disconnected'' configuration $\sqcup$ becomes preferred. In fact, we can see that happening in Figure~\ref{fig.EEplotsStrip}, where we show the value of a dimensionless version of \eqref{eq:EE_strip_reg} defined in Appendix~\ref{ap:strip} as a function of the width of the strip. Above a certain critical value of the width of the strip $l=l_c$ the disconnected configuration becomes the dominant one. On top of that, $\Delta S(l)$ has a turning point causing the disconnected configuration to be the only existing one for large enough widths $l$.

\begin{figure}[t]
	\begin{center}
		\begin{subfigure}{0.45\textwidth}
			\includegraphics[width=\textwidth]{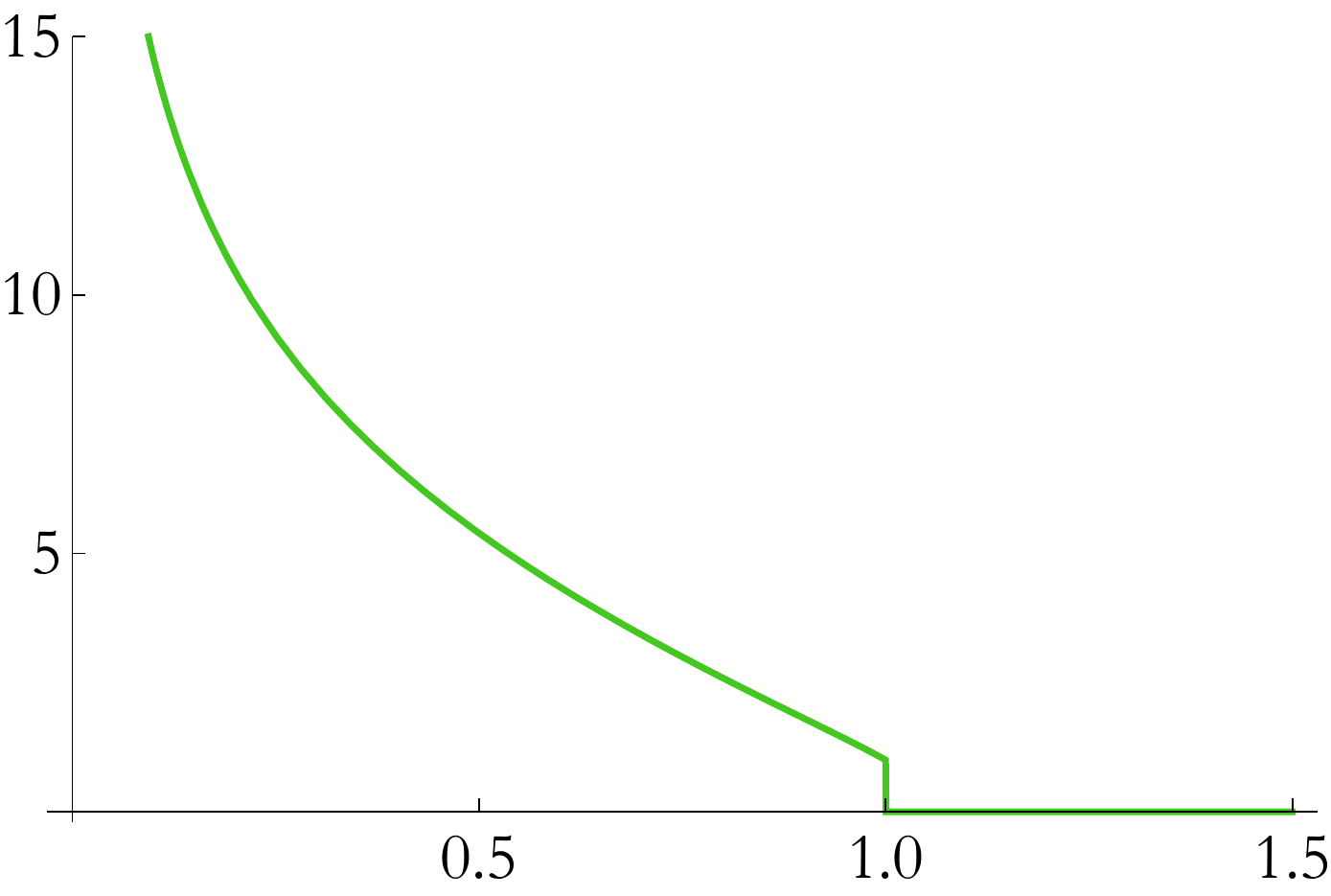} 
			\put(-160,110){$\ffunc(l)/\ffunc(l_c)$}
			\put(-30,-10){$l/l_c$}
		\end{subfigure}\hfill
		\begin{subfigure}{.45\textwidth}
			\includegraphics[width=\textwidth]{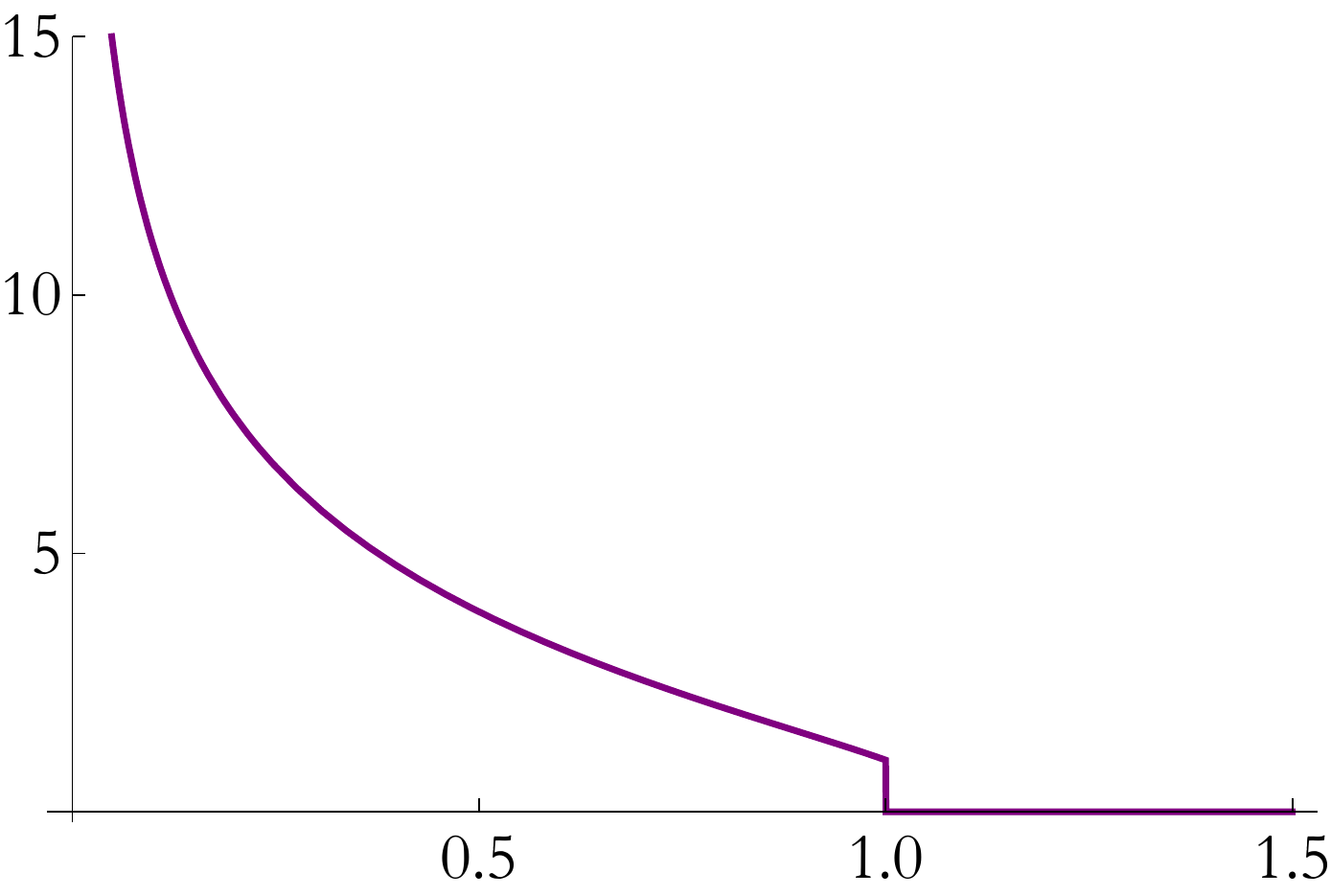} 
			\put(-160,110){$\ffunc(l)/\ffunc(l_c)$}
			\put(-30,-10){$l/l_c$}
		\end{subfigure}
		\caption{\small Function $\ffunc(l)$ for a single strip as a function of strip width in the gapped non-confining theory $\B_8$ (Left) and the confining one $\Bconf$ (Right). Both quantities are normalised to their value at the point where the ``disconnected'' configuration $\sqcup$ becomes dominant, above which it is strictly zero.
		}\label{fig.FRplotsStrip}
	\end{center}
\end{figure}

\begin{figure}[t]
	\begin{center}
		\begin{subfigure}{0.45\textwidth}
			\includegraphics[width=\textwidth]{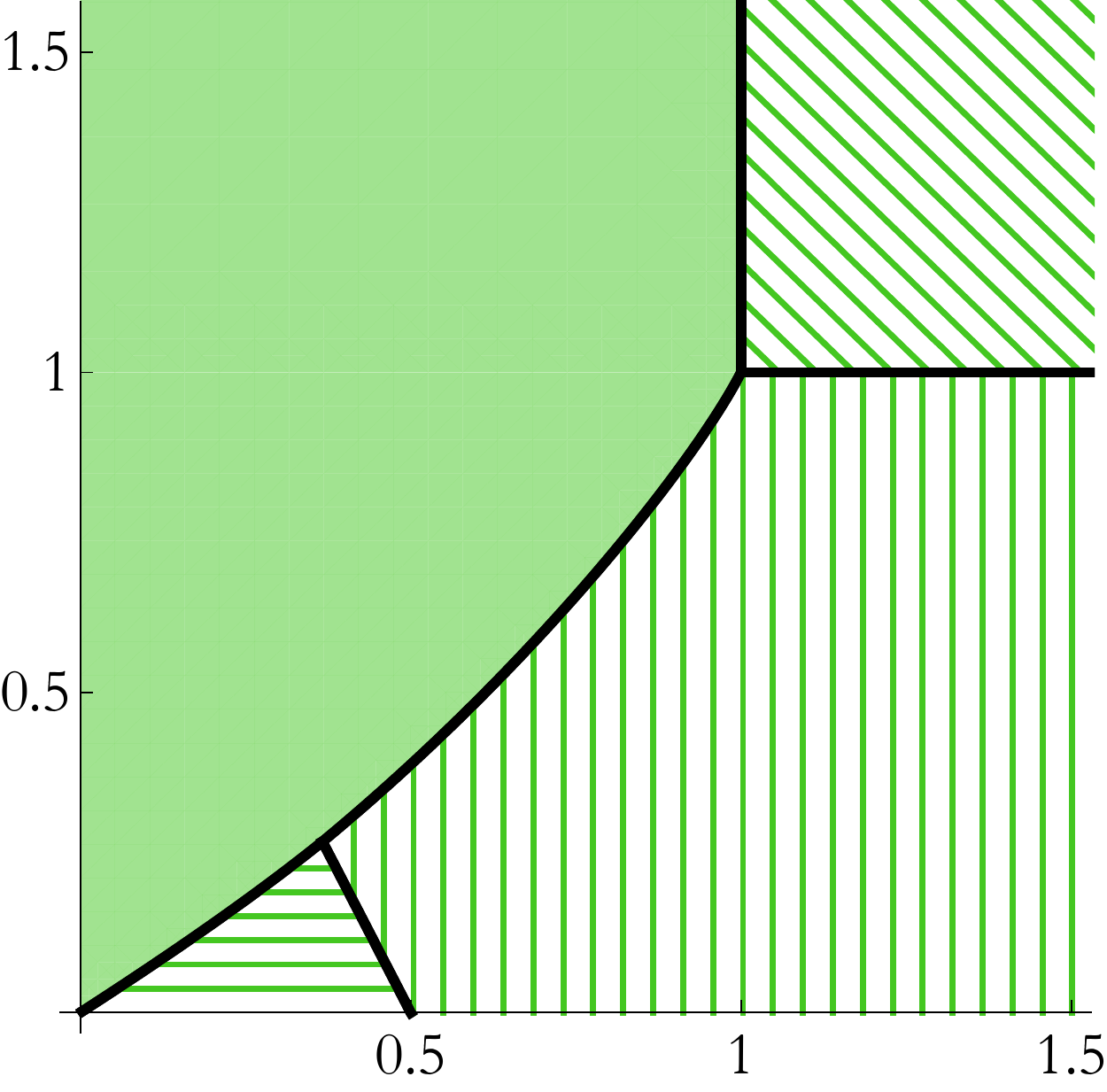} 
			\put(-150,160){$s/l_c$}
			\put(-30,-10){$l/l_c$}
			\put(-30,-35){$ $}
		\end{subfigure}\hfill
		\begin{subfigure}{.45\textwidth}
			\includegraphics[width=\textwidth]{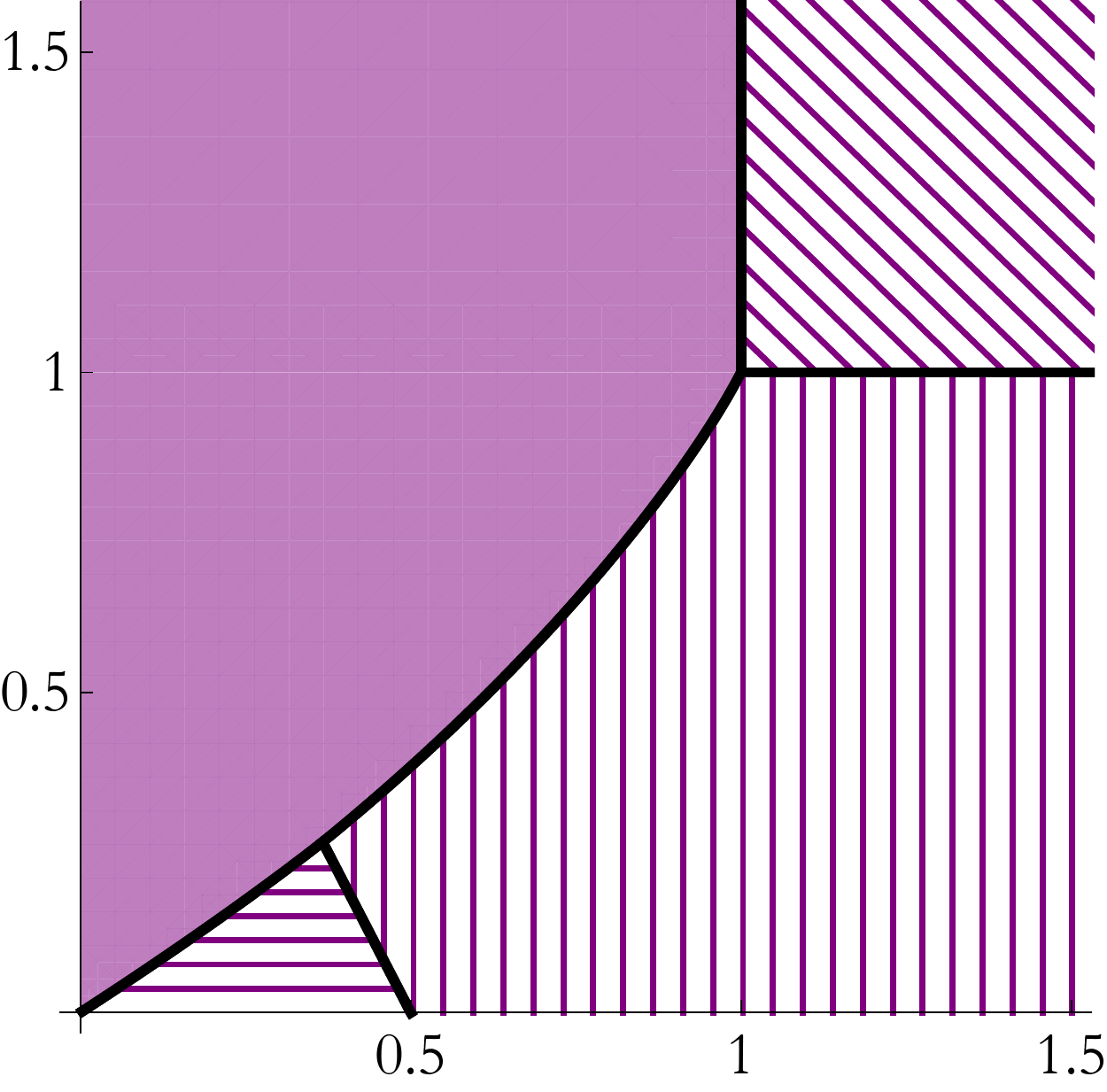} 
			\put(-150,160){$s/l_c$}
			\put(-30,-10){$l/l_c$}
			\put(-30,-35){$ $}
		\end{subfigure}
		\includegraphics[width=\textwidth]{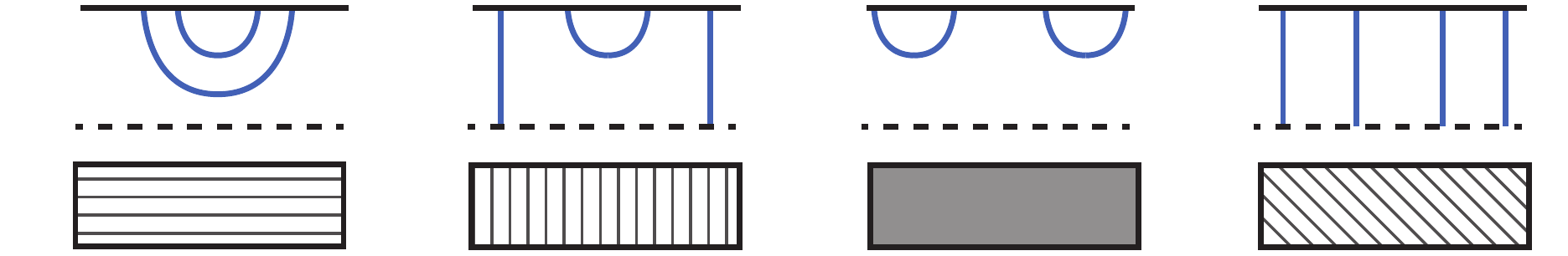} 
		\caption{\small Phase diagram of mutual information of entanglement entropy of strips in the gapped non-confining theory $\B_8$ (Left) and the confining one $\Bconf$ (Right). Solid black curves consist of points where mutual information \eqref{eq:mutual_info} vanish and a phase transition takes place. Separation between strips $s$ is shown in the vertical axes and their width $l$ is shown in the horizontal axes, both quantities normalised to the value of the width $l_c$ at which the ``disconnected'' configuration $\sqcup$ becomes dominant in each case.
		}\label{fig.mutualInfostrips}
	\end{center}
\end{figure}

Once the entanglement entropy of a single strip as a function of its width is known, it is straightforward to compute an entropic c-function that is conjectured to measure the number of degrees of freedom at scale  $l$ \cite{Casini:2004bw}. Generalisations to higher dimensions, within holography, include the original $(3+1)$-dimensional proposal of \cite{Nishioka:2006gr} as well as those conjectured in arbitrary dimensions \cite{Myers:2010tj,Liu:2012eea,Liu:2013una}.
Our field theory is $(2+1)$-dimensional, in which case our function is linked with the F-theorem and we compute an object that we denote by $\ffunc$:
\begin{equation}
\label{eq:Ffunction_strip}
\ffunc_{\text{strip}}(l) = \frac{l^2}{L_y}\frac{\partial S_{\text{strip}}}{\partial l} = \frac{l^2}{L_y}\frac{\partial \Delta S}{\partial l}\ ,
\end{equation}
where in the last equality we took into account \eqref{eq:EE_strip_reg} and the fact that $S_\sqcup$ does not explicitly depend on $l$. We plot this quantity in Figure~\ref{fig.FRplotsStrip}. The behaviour of this quantity is an immediate consequence of the behaviour of $\Delta S $: it decreases continuously until the point where the connected configuration is disfavoured, where it suddenly jumps to zero. For values of the widths bigger than $l_c$, $\ffunc_{\text{strip}}(l)$ is identically zero. 
In this case the RT surface is probing IR scales and is not sensitive to massive degrees of freedom which are gapped. 

Let us pause to make a technical comment. Taking numerical derivatives in (\ref{eq:Ffunction_strip}) can be demanding. One can completely circumvent this procedure by using the chain rule explained in \cite{Jokela:2020auu} to find that\footnote{Note the usage of the rescaled dimensionless quantities in favour of $\Delta S$, $l$ and $\Xi$ as defined in Appendix~\ref{ap:strip} and \ref{ap:disk}.}
\begin{equation}
\label{eq:derivative}
\frac{\partial \Delta \overline S}{\partial \overline l} = \overline \Xi_*^{\frac{1}{2}} \ .
\end{equation}
From this expression it is manifest that the computation of $\ffunc_{\text{strip}}$ only involves finite quantities and no UV regularisation is invoked. We have explicitly checked that results following from (\ref{eq:derivative}) and (\ref{eq:Ffunction_strip}) agree to great accuracy.

Another interesting quantity to compute is the mutual information between two entangling strips $A$ and $B$, given by 
\begin{equation}\label{eq:mutualpreinfo}
I(A,B) = S_A + S_B - S_{A\cup B} \ .
\end{equation}
The mutual information characterises the amount of information shared by the two domains \cite{Headrick:2010zt}. 
If we consider mutual information between two strips of the same width $l$ which are separated by a distance $s$, the expression (\ref{eq:mutualpreinfo}) can be written as
\begin{equation}
\label{eq:mutual_info}
I(A,\tilde A) = 2 \Delta S(l) - \Delta S (2l+s) - \Delta S (s) \ .
\end{equation}
It is possible then to draw a phase diagram of the dominant phases, the phase boundaries given by the loci where the mutual information vanishes. This is shown in Figure~\ref{fig.mutualInfostrips}, where it can be seen that four different regions arise, depending on which is the dominant configuration in each case. The result is analogous to those of \cite{Ben-Ami:2014gsa,Jokela:2019ebz} discussing cigar-like geometries. Interestingly, similar four-zone phase diagram is present in anisotropic non-confining geometries \cite{Jokela:2019tsb,Hoyos:2020zeg}, pronouncing the fact that it is the presence of an internal mass scale behind the entanglement entropy phase transitions and not necessarily the confinement. 

\subsection{Entanglement entropy of the disk}\label{sec:disk}

\begin{figure}[t]
	\begin{center}
		\includegraphics[width=\textwidth]{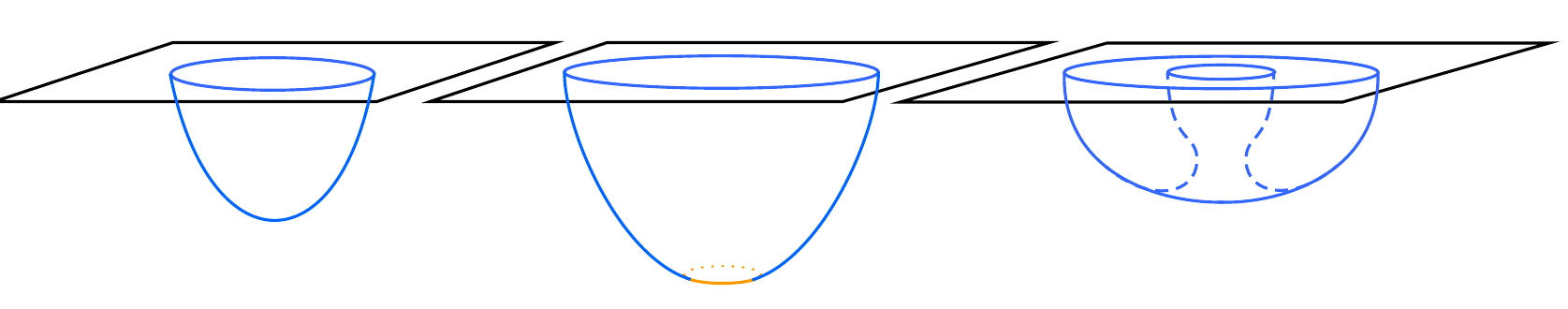} 
		\put(-84,-10){(c)}
		\put(-193,-10){(b)}
		\put(-292,-10){(a)}
		\caption{\small Three types of embeddings found when solving the equation of motion found when computing entanglement entropy on disks. When the radius of the disk is sufficiently small, the embedding does not reach the IR part of the geometry (a). For large values of the radius of the disk, the embedding reaches the end-of-space (b). Additionally, when considering mutual information between disks, embeddings like (c) should be taken into account.
		}\label{fig:disk_conf}
	\end{center}
\end{figure}

Similarly, we can study the case in which the entangling region is a disk of radius $R$ to inspect which lessons from the preceding Subsection are intact. In the case of a disk, we perform a change of coordinates to polar coordinates in the gauge theory directions, namely $x^1 = \rho \cos \alpha$, $x^2 = \rho \sin \alpha$. This choice will reduce our system of three second order differential equations obtained from \eqref{eq:EulerLagrange} to a single second order differential equation. Taking that into account, the embedding of the RT surface in our background solutions is going to be determined by the choice
\begin{equation}
\label{eq:embedding_disk}
t= \text{constant} \, , \quad \rho =\sigma^1 = [0,R] \, , \quad \alpha = \sigma^2 \in [0,2\pi] \, , \quad r = r(\sigma^1) \in [r_*,\infty)\ ,\\[2mm]
\end{equation}
and such that 
\begin{equation}\label{eq:radiusat infty}
\lim_{\rho\to R} r(\rho) = \infty \ ,
\end{equation} 
which essentially tells us that the embedding is attached to the circumference of the corresponding disk. Satisfying these conditions, there are two distinct types of embeddings, depicted in Figure~\ref{fig:disk_conf} (a) and (b). One possibility, happening when the radius of the disk is sufficiently small, is that the RT surface does not reach the bottom of the geometry, leading to the condition 
\begin{equation}
\label{eq:disk_BC1}
r(0) = r_* \qquad , \qquad  \dot r(0) = 0 \ .
\end{equation}
This is what happens in Figure~\ref{fig:disk_conf}(a). Another option is that the embedding does enter all the way down to the IR part of the geometry, in which case the condition is that
\begin{equation}
\label{eq:disk_BC2}
\lim_{\rho\to \rho_*} r(\rho) = r_s \ .
\end{equation}
Recall $r_s$ is the value of the radial coordinate $r$ at the end-of-space. In this case, similarly to the strip configuration, as represented in Figure~\ref{fig:disk_conf}(b), we need an extra piece of surface laying at the bottom of the geometry in order to complete the RT surface. We already argued above that this piece, specified by \eqref{eq:bottom}, fulfils the equations of motion.

\begin{figure}[t]
	\begin{center}
		\begin{subfigure}{0.45\textwidth}
			\includegraphics[width=\textwidth]{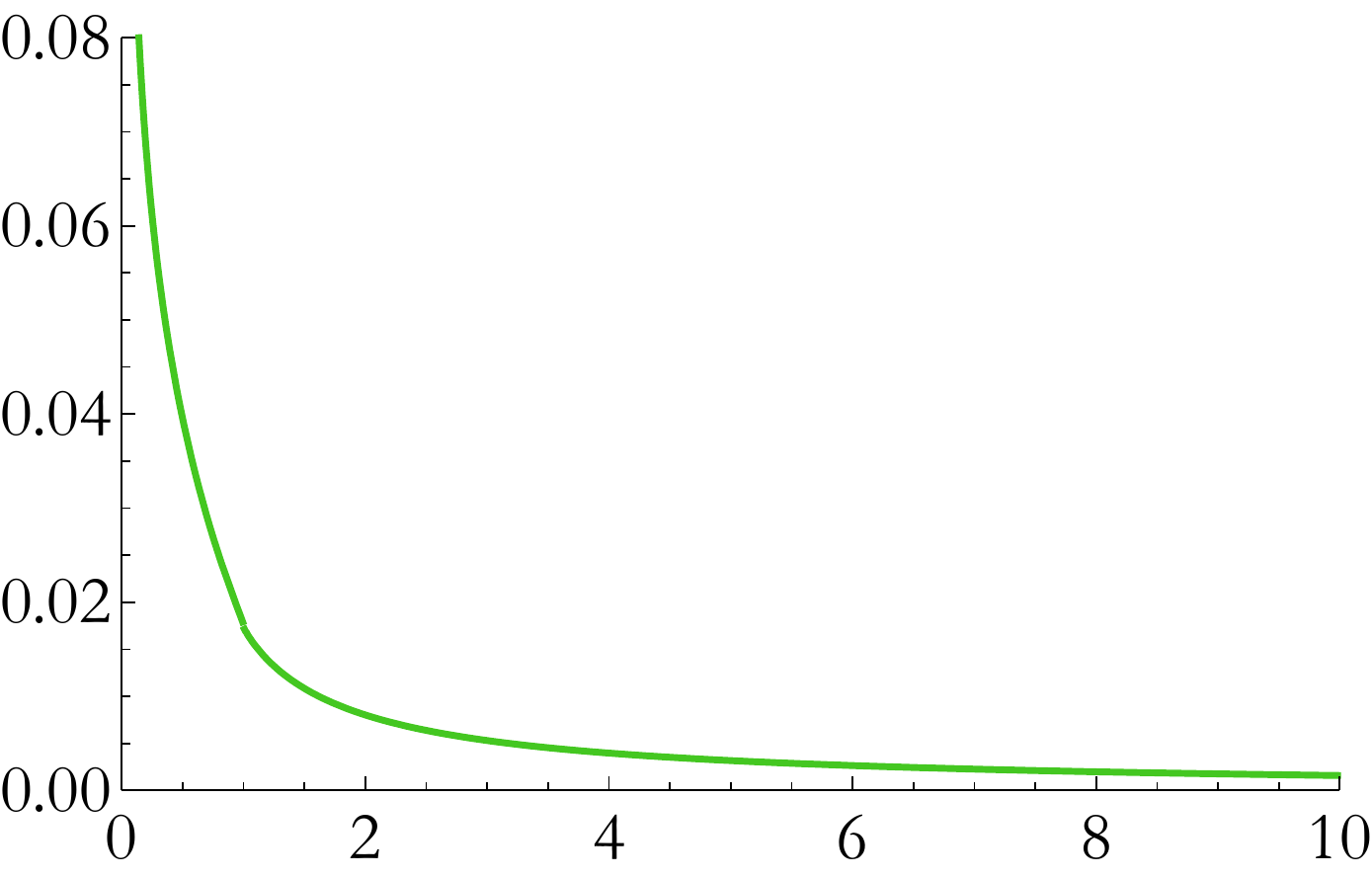} 
			\put(-150,110){$\overline\ffunc_D(R)$}
			\put(-30,-10){$R/R_c$}
		\end{subfigure}\hfill
		\begin{subfigure}{.45\textwidth}
			\includegraphics[width=\textwidth]{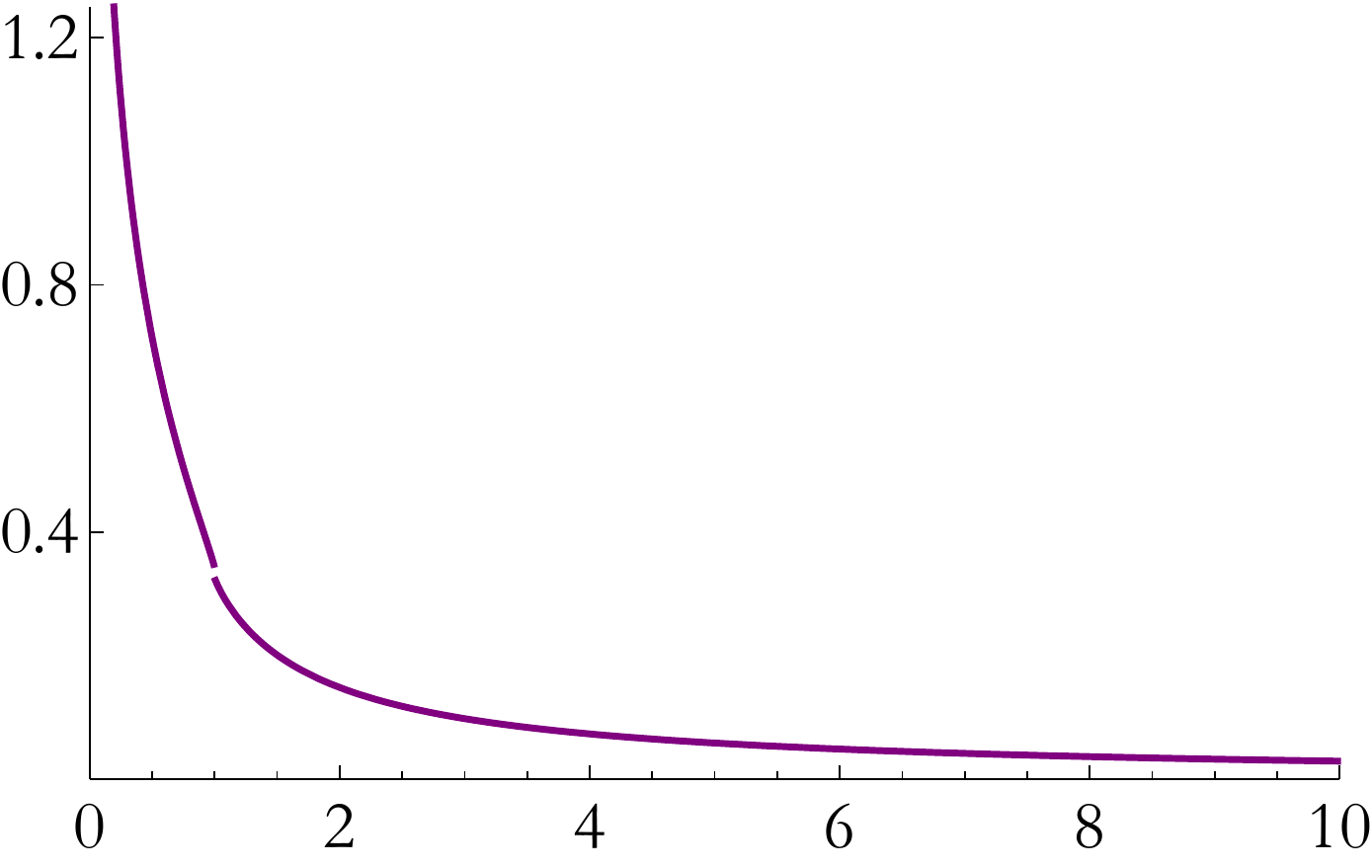} 
			\put(-150,110){$\overline\ffunc_D(R)$}
			\put(-30,-10){$R/R_c$}
		\end{subfigure}
		\caption{\small Measure of the flow of the degrees of freedom in the gapped non-confining theory $\B_8^0$ (Left) and in the confining one $\Bconf$ (Right). Notice that we plot rescaled quantities as explained in Appendix~\ref{ap:disk}, as a function of the radius of the disk normalised to the smallest radius for which the RT surface reaches the end-of-space.
		}\label{fig.FofR_Disk}
	\end{center}
\end{figure}

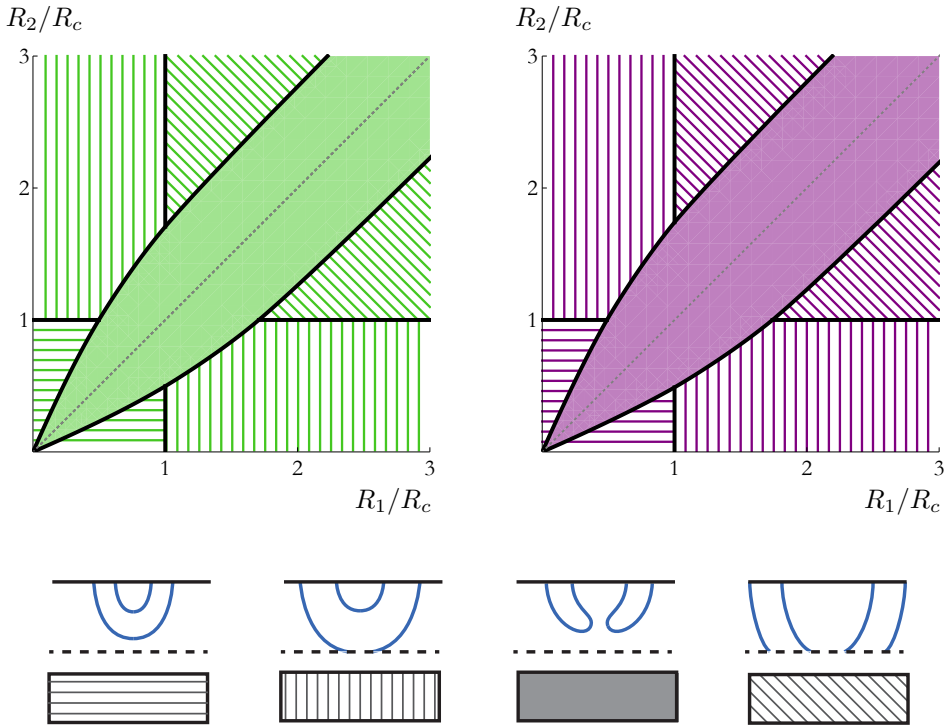
\begin{figure}[t]
	\begin{center}
		\begin{subfigure}{0.45\textwidth}
			\includegraphics[width=\textwidth]{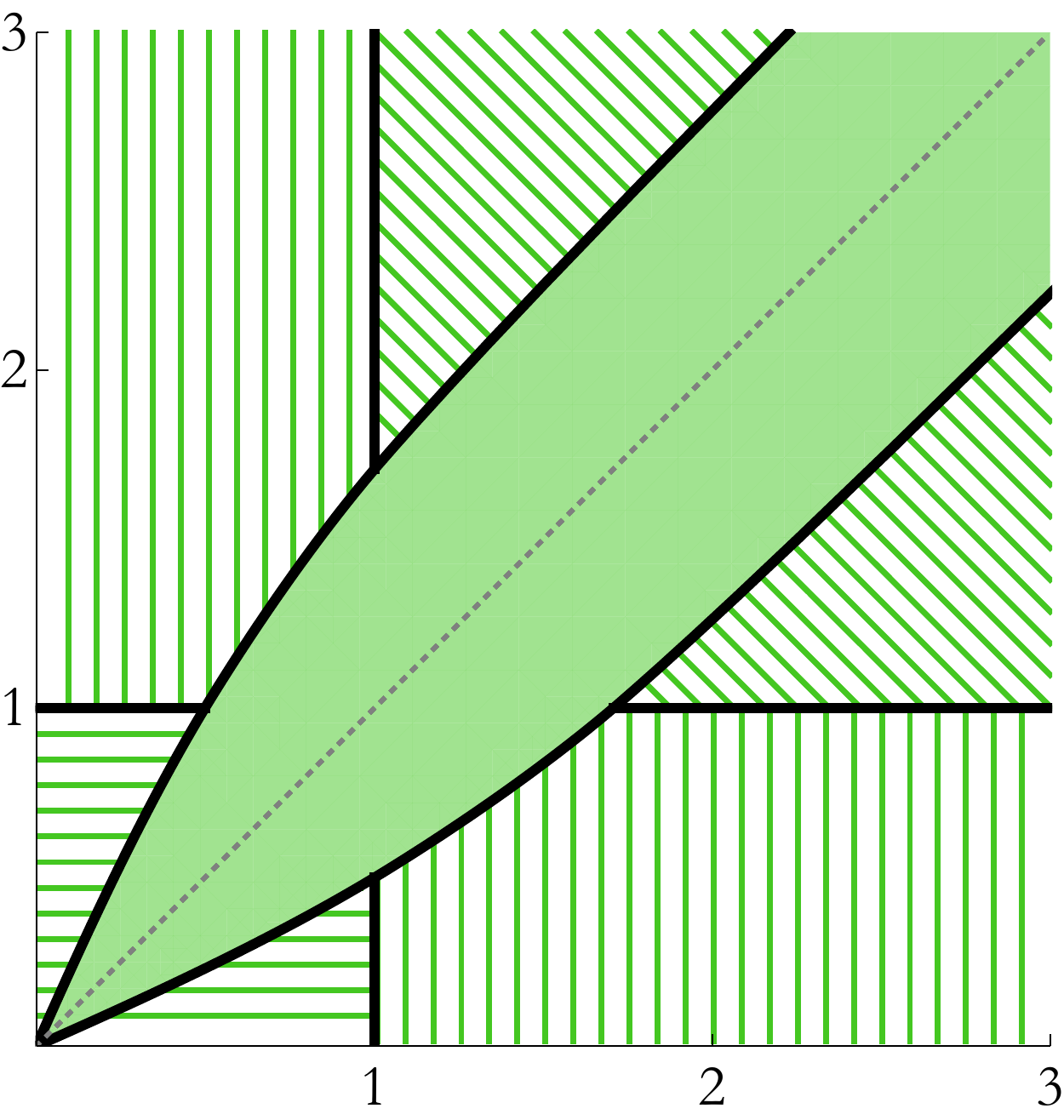} 
			\put(-160,170){$R_2/R_c$}
			\put(-30,-10){$R_1/R_c$}
			\put(-30,-35){$ $}
		\end{subfigure}\hfill
		\begin{subfigure}{.45\textwidth}
			\includegraphics[width=\textwidth]{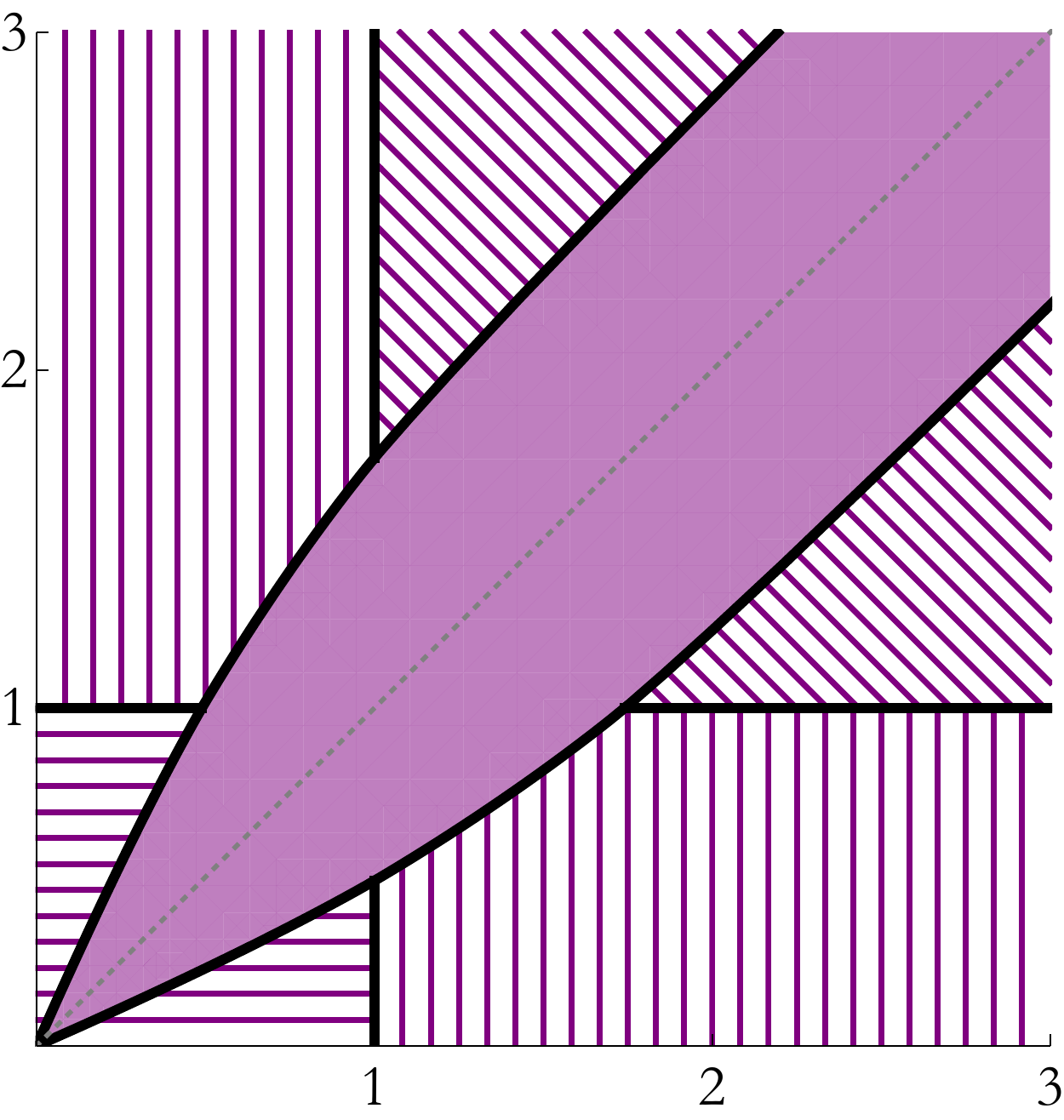} 
			\put(-160,170){$R_2/R_c$}
			\put(-30,-10){$R_1/R_c$}
			\put(-30,-35){$ $}
		\end{subfigure}
		\includegraphics[width=\textwidth]{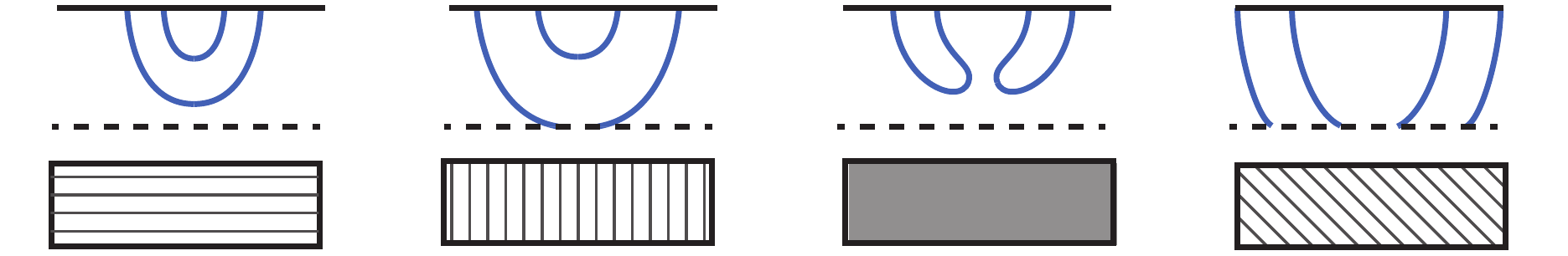} 
		\caption{\small Phase diagram stemming from the analysis of mutual information of entanglement entropy of disks in the gapped non-confining theory $\B_8^0$ (Left) and in the confining one $\Bconf$ (Right). On the axes we show the radius $(R_1,R_2)$ of the corresponding disks normalised to the value $R_c$ of the minimum radius for which the RT surface reaches the end-of-space.
		}\label{fig.mutualInfo}
	\end{center}
\end{figure}

When substituting \eqref{eq:embedding_disk} into \eqref{eq:EE_formula} we get
\begin{equation}\label{eq:EE_disk}
S_{\text{disk}}=\frac{V_6}{4G_{10}} \int_0^{2\pi}\dd\sigma^2  \int_0^R \dd \rho \ (1+h\ \dot r ^2)^{\frac{1}{2}}\ \rho\ \Xi^{\frac{1}{2}}
\end{equation}
and the first integral immediately gives us a factor of $2\pi$. The (\ref{eq:EE_disk}) is UV divergent and we will need to regularise it using counterterms, as shown in Appendix \ref{ap:disk}. Having done so, we will refer to the renormalised quantity as $S_{\text{disk}}^{\text{reg}}$. An analogous quantity to \eqref{eq:Ffunction_strip} can be defined for the disk to measure the change in the degrees of freedom \cite{Liu:2012eea}
\begin{equation} \label{eq:Ffunction_disk}
\ffunc_{\text{disk}}(R) = R\cdot \frac{\dd S^{\text{reg}}_{\text{disk}}}{\dd R} - S^{\text{reg}}_{\text{disk}} \ .
\end{equation}
In Figure~\ref{fig.FofR_Disk} we plot a rescaled version of this quantity, again for the two theories $\B_8^0$ and $\Bconf$. We cannot rule out the possibility that $\ffunc_{\text{disk}}$ is discontinuous: from our numerical computations it seems there is a tiny jump when the transition between the two configurations happens. This could in principle happen when the embedding starts reaching the IR bottom of the geometry, 
where the altered boundary conditions can lead to a discontinuity in the derivative of the entropy with respect to the radius. A discontinuity would, however, be in conflict with statements in \cite{Klebanov:2012yf}, where this analysis was done in a different albeit similar system to ours and it was claimed that this function has a second-order phase transition. Numerics are quite involved when the transition occurs, so it is difficult for us to distinguish small discontinuities from numerical errors. It would be interesting to generalise the chain rule argument of \cite{Jokela:2019tsb} that was used in the case of strip configurations to bypass numerical artefacts

Apart from the embeddings we just mentioned, there is another profile for the embedding that satisfies the equation of motion and is important to be taken into account. We depict this in Figure~\ref{fig:disk_conf}(c). It naturally emerges when one imposes the boundary conditions
\begin{equation}
\label{eq:embeddding_teo_disks}
r(\rho_*) =  r_*\qquad , \qquad  \dot r(\rho_*) = 0 \ ,
\end{equation}
with $r_*\neq r_s$ and $\rho_*\neq 0$. Although the picture is self-explanatory, the way in which the equation of motion with the boundary conditions (\ref{eq:embeddding_teo_disks}) is solved is slightly involved and hence explained in detail in Appendix~\ref{ap:disk}. The relevant outcome is that this embedding is attached to two disks at the boundary, thus allowing us to compute mutual information as we did in \eqref{eq:mutual_info}. We can therefore map out the phase diagram in terms of the two radii $(R_1,R_2)$ where the mutual information (\ref{eq:mutualpreinfo}) associated with disk configurations vanishes and showing the regions where each configuration is dominant. This phase diagram is plotted in Figure~\ref{fig.mutualInfo}. We note that the result is on par with the work in \cite{Nakaguchi:2014pha}, where a similar study for concentric circles was carried out in the gapped solution from \cite{Cvetic:2000db}.

\subsection{Limiting cases}\label{sec:limiting}

Having discussed the generic features and compared the confining and the non-confining theories, we now narrow down the scope and discuss specific cases close to the limiting values for the parameter $b_0$.
As described in Chapter \ref{Chapter2_B8family}, these limiting values of $b_0$ are interesting because they lead to radically different IR dynamics. In this Section we want to study these different limits from the point of view of information probes and illustrate their power in revealing interesting physics.

\vspace{.3cm}
\noindent$\blacksquare\ $ \textbf{Quasi-confining regime}
\vspace{.1cm}

\begin{figure}[t]
	\begin{center}
		\begin{subfigure}{.65\textwidth}
			\includegraphics[width=\textwidth]{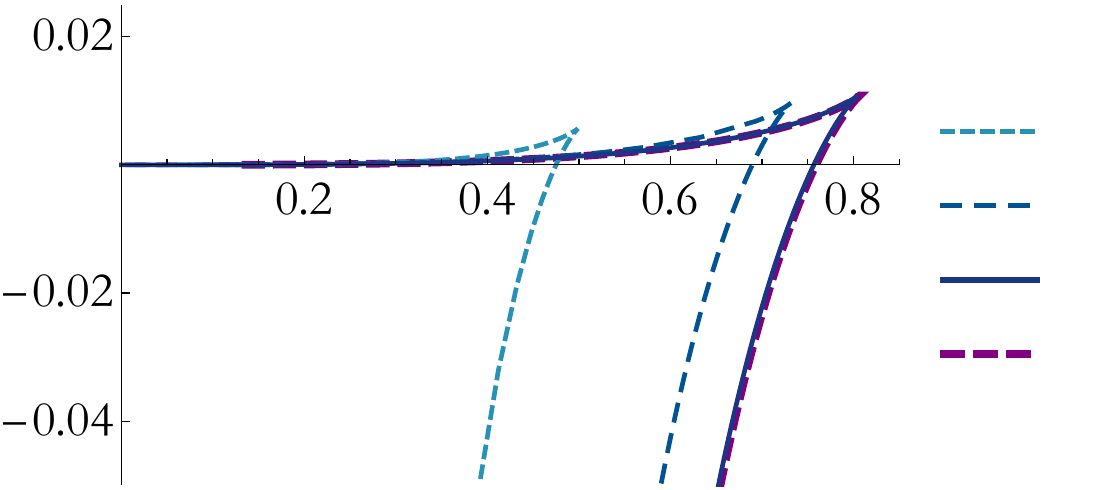} 
			\put(-240,90){$\overline S^{\text{reg}}$}
			\put(-45,0){$\overline l$}
			\put(-7,88){$b_0 = 0.6835$}
			\put(-7,69){$b_0 = 0.9201$}
			\put(-7,50){$b_0 = 0.9972$}
			\put(-7,31){$b_0 = 1 $ $(\Bconf)$}
		\end{subfigure}\vspace{7mm}
		\begin{subfigure}{0.65\textwidth}
			\includegraphics[width=\textwidth]{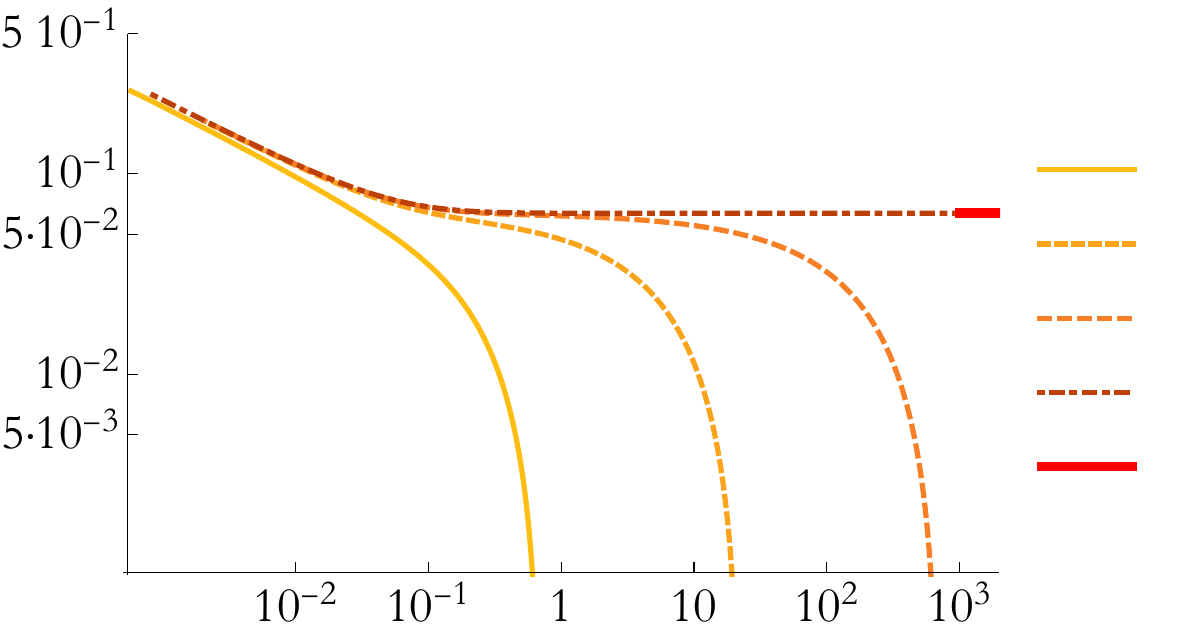} 
			\put(-260,110){$\overline S^{\text{reg}} \cdot \overline l$}
			\put(-30,5){$\overline l$}
			\put(-7,106){$b_0 = 0.1914$}
			\put(-7,88){$b_0 = 0.0604$}
			\put(-7,70){$b_0 = 0.0191$}
			\put(-7,52.5){$b_0 = 0 $ $(\B_8^\infty)$}
			\put(-7,36){OP | CFT}
		\end{subfigure}
		\caption{\small  (Bottom) Entanglement entropy of a single strip (multiplied by width $l$) as we dial the parameter $b_0$ to zero. In all cases for non-vanishing $b_0$ we find that this quantity vanishes, contrary to the case $b_0=0$ where we find precisely the CFT result. Notice, that closer the RG flow passes the OP fixed point, the wider the entanglement plateau, before the entanglement entropy plunges.  (Top) Entanglement entropy of a strip as we increase $b_0$ and hence creeping towards the confining theory. Note that for the penultimate value for $b_0$, the curve is overlapping with that of $\Bconf$.
		}\label{fig.limiting_EEstrip}
	\end{center}
\end{figure}

\begin{figure}[t]
	\begin{center}
		
		\begin{subfigure}{.45\textwidth}
			\includegraphics[width=\textwidth]{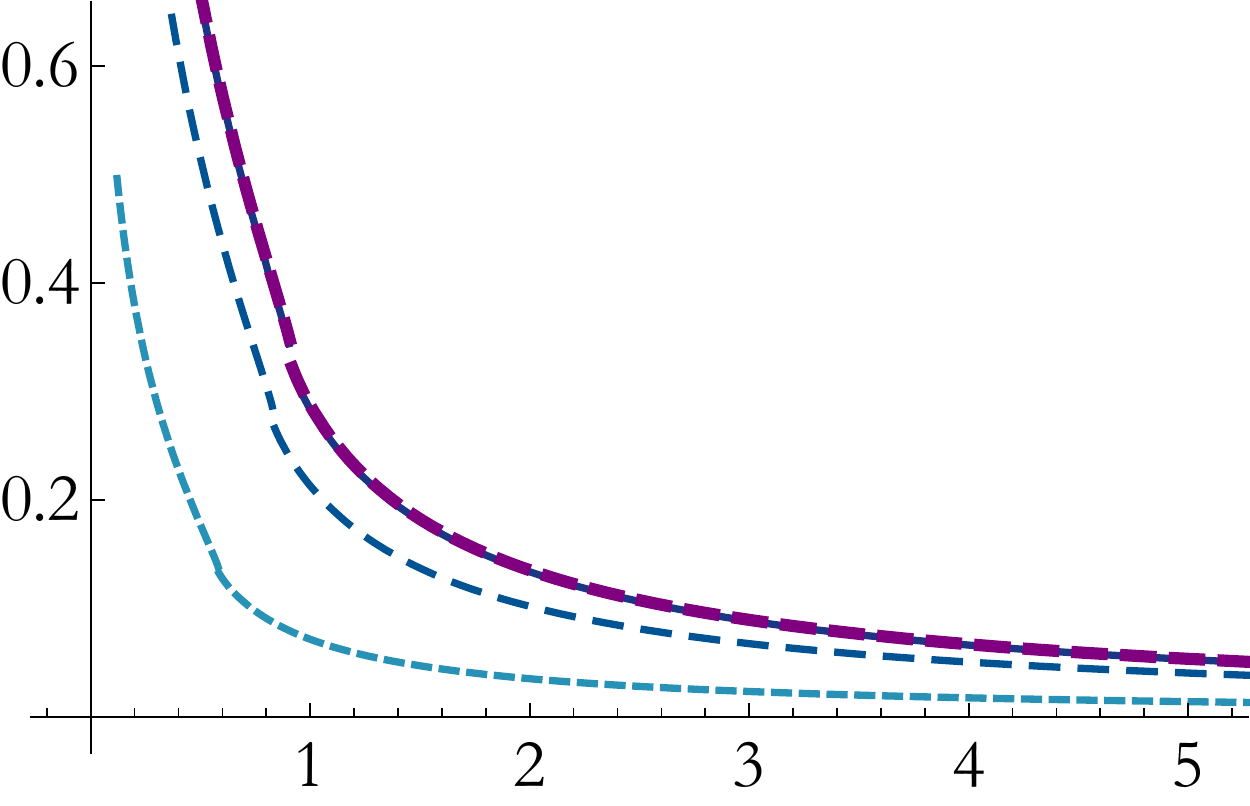} 
			\put(-150,110){$\overline  \ffunc_D(R)$}
			\put(-10,-10){$\overline R $}
		\end{subfigure}
		\hfill
		\begin{subfigure}{0.45\textwidth}
			\includegraphics[width=\textwidth]{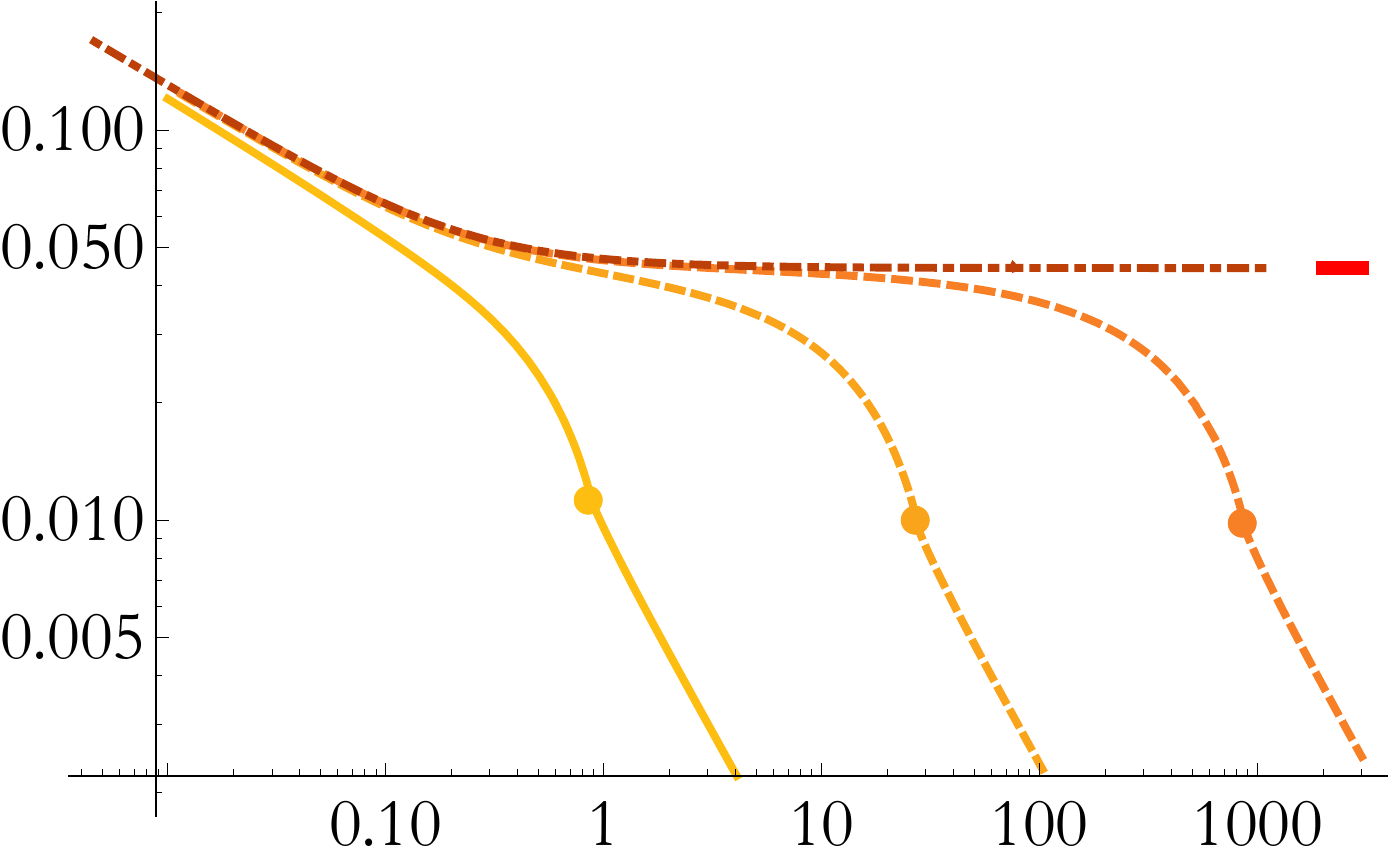} 
			\put(-150,110){$\overline  \ffunc_D(R)$}
			\put(-10,-10){$\overline  R$}
		\end{subfigure}
		\caption{\small (Right) Function $\overline \ffunc_D$ from the entanglement entropy of a disk as a function of its radius $R$. Dots stand for the transition of the embedding, namely those represented in Figure~\ref{fig:disk_conf}(a) to \ref{fig:disk_conf}(b). (Left) Same quantities in linear scale as we approach the confining theory. In both cases, when the transition happens, there is a visible change in the behaviour of the curves. This signals the proximity of the IR gapped phase. See legends in Figure~\ref{fig.limiting_EEstrip}.
		}\label{fig.limiting_FofRdisk}
	\end{center}
\end{figure}

\begin{figure}[t]
	\begin{center}
		\begin{subfigure}{.45\textwidth}
			\includegraphics[width=\textwidth]{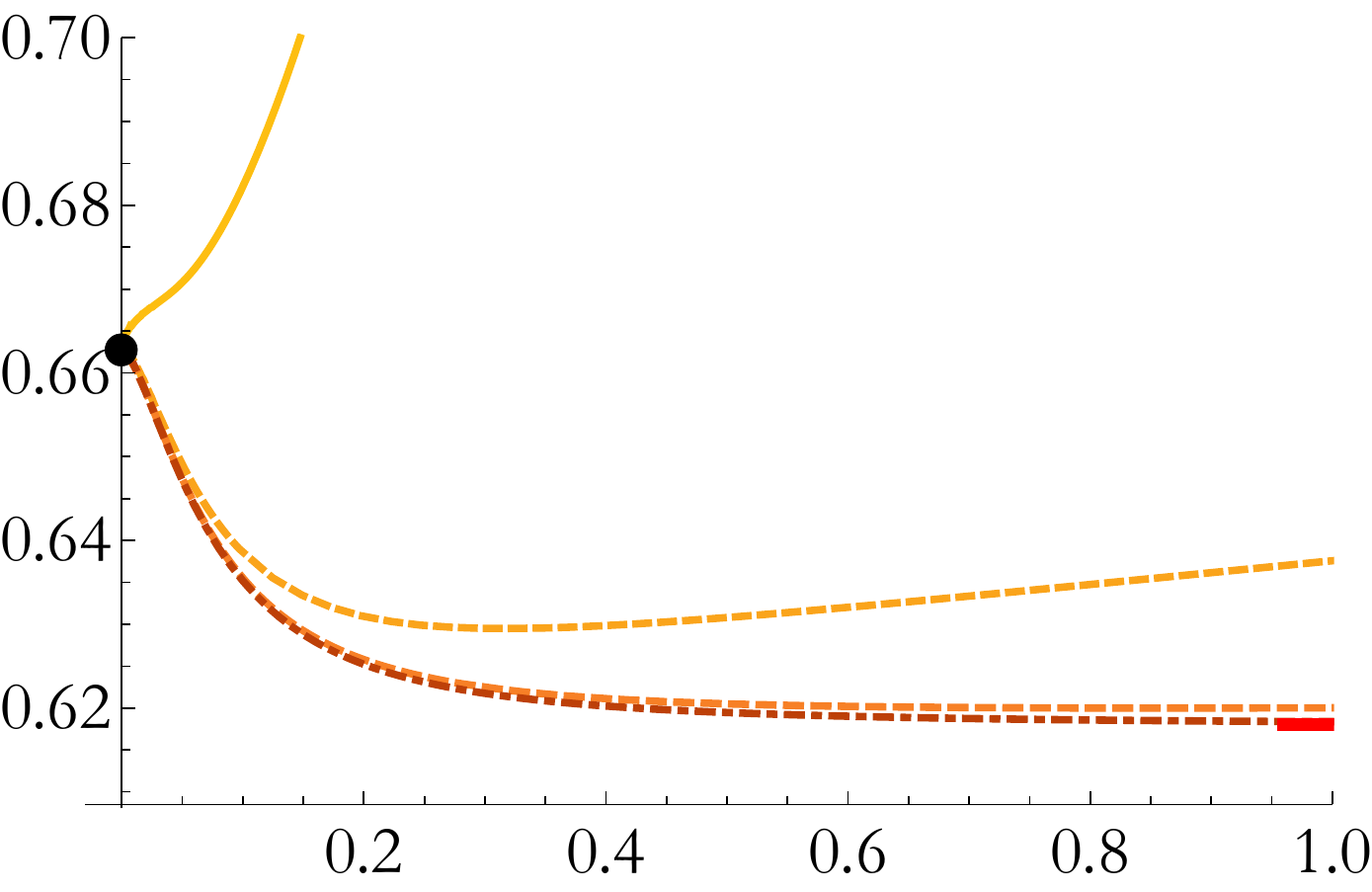} 
			\put(-150,110){$s/l$}
			\put(-30,-10){$\overline l$}
		\end{subfigure}\hfill
		\begin{subfigure}{0.45\textwidth}
			\includegraphics[width=\textwidth]{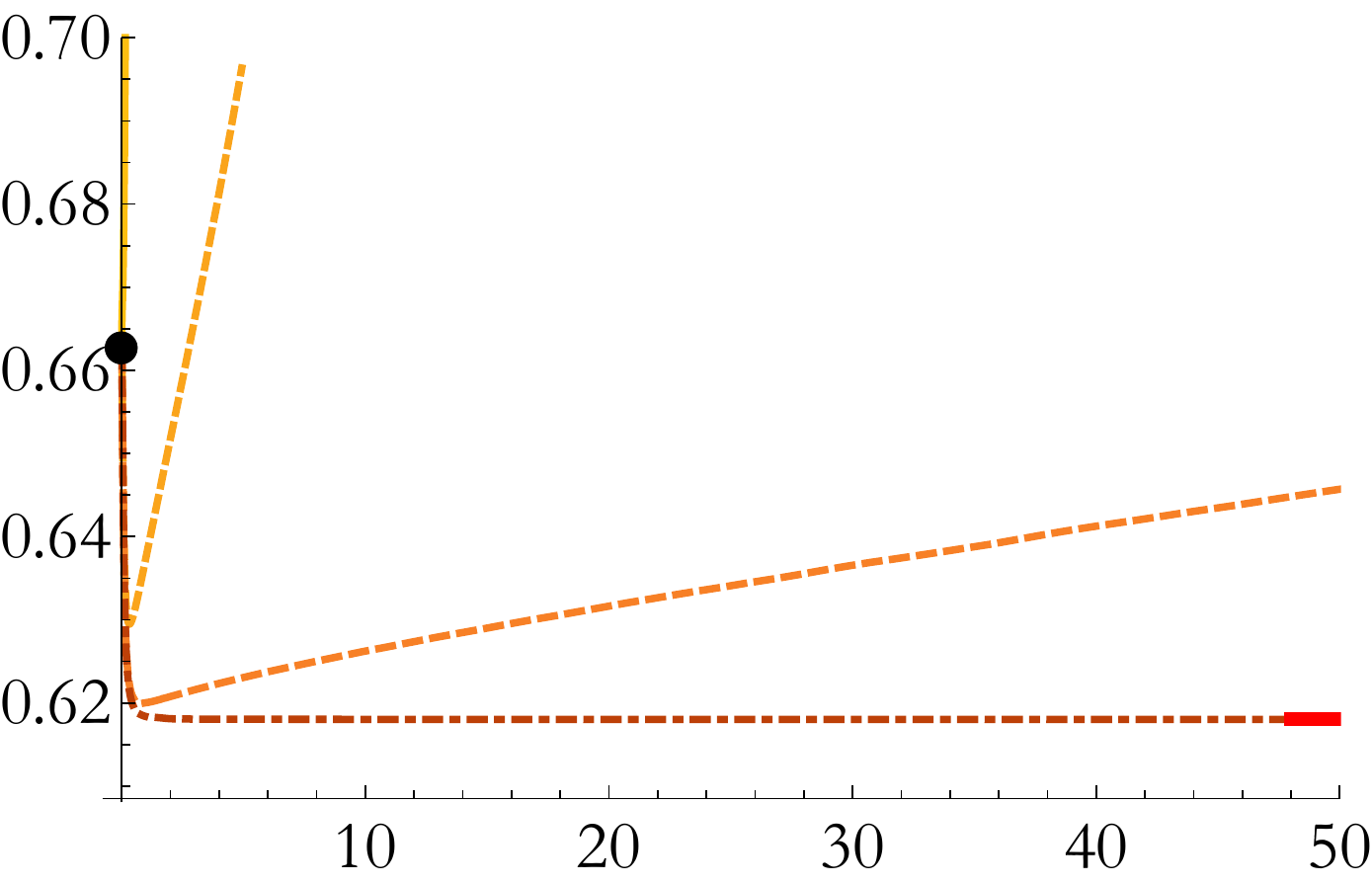} 
			\put(-150,110){$s/l$}
			\put(-30,-10){$\overline  l$}
		\end{subfigure}
		\caption{\small Ratio of the values of the width of strips $l$ which are separated by a distance $s$ at the point, where mutual information vanishes. Right plot extends the range of the left plot to larger strip widths. Notice that the numerics match analytic results both for small $l$ at the UV (\ref{eq:mutualD2}) and large $l$ at the IR (\ref{eq:mutualCFT}). See legends in Figure~\ref{fig.limiting_EEstrip}.
		}\label{fig.golden}
	\end{center}
\end{figure}

Let us start with the case $b_0=1$, which corresponds to the confining theory $\Bconf$. If we aim to understand this solution as a limit of the $\B_8$ family of theories, the charge $Q_k$ -- related to the CS level by \eqref{eq:gauge_parameters} -- must be rescaled. This is needed in order to obtain vanishing CS level (see Section~\ref{quasi}). After performing this rescaling, in the limit $b_0\to 1$, all quantities will approach the values that can be obtained directly in the case $\Bconf$. This fact is illustrated in Figure~\ref{fig.limiting_EEstrip} (Top), where we find that the entanglement entropy of the strip smoothly approaches the one for the confining case as $b_0$ is gradually raised towards unity.
We also show the same expected behaviour for the number of degrees of freedom, {\emph{i.e.}}, for $\ffunc_D$ function in Figure~\ref{fig.limiting_FofRdisk} (Left) for disk configurations.

\vspace{.3cm}
\noindent$\blacksquare\ $ \textbf{Quasi-conformal regime}
\vspace{.1cm}

On the opposite limit, for $b_0=0$ the ground state RG flow, denoted by $\B_8^\infty$, ends at a fixed point, as indicated by the leftmost arrow in Figure~\ref{fig:triangle}. Flows with $b_0 > 0$ approach the fixed point but never reach it. We will investigate the imprints that this passage close to the conformal fixed point leaves on information theoretic quantities. 

To start with, the OP conformal point has the standard expression for the finite part of the entanglement entropy, which is readily available by analytic methods. For strips, the full expression reads
\begin{equation}\label{eq:OPconformal}
\Delta S_{\text{OP}}(l) =  - \ \frac{9 \ \lambda\ L_y\ V_6}{2^8 \pi^4}\cdot \frac{|k|\left(\bar{M}^2+2|k|N\right)}{N} \ 
\times \
\frac{200\ \pi^3}{729\cdot \Gamma\left[\frac{1}{4}\right]^4}\sqrt{\frac{5}{3}} \cdot \frac{1}{l} \ .
\end{equation}
Note that $ \Delta S_{\text{OP}}(l) \cdot l $ is a constant depending on the number of degrees of freedom and we denote this as a plateau in the following.
As a consequence, flows with small $b_0$ pass close to the CFT also induce plateaux, which are wider the closer we are to the fixed point. Interestingly, it has been noted that 
the proximity to fixed points causes ``walking regime'', {\emph{i.e.}}, a regime of energy where some dimensionless quantities are approximately constant, even when such fixed points reside in the complex plane \cite{Gorbenko:2018ncu}. We will investigate this in Chapter \ref{Chapter5_HoloCCFTs}.

The entanglement entropy of a single strip times its width for different values of $b_0$ approaching zero is shown in Figure~\ref{fig.limiting_EEstrip} (Bottom). Note again that we plot the rescaled quantities as defined in Appendix~\ref{ap:strip}. It is easy to see that the closer we are to $b_0=0$, the wider the range where $ \Delta \overline S \cdot \overline l$ traces the conformal value given by (\ref{eq:OPconformal}). As soon as the quasi-conformal regime is departed, the curve abruptly decreases towards zero.

The near-proximity of the conformal fixed point will leak to the behaviour of many other quantities. Perhaps most prominently this fact is captured by the $c$-functions. Indeed, let us consider the function that measures the degrees of freedom for the disk configurations. Quasi-conformal regime is clearly visible in Figure~\ref{fig.limiting_FofRdisk} (Right). Note, however, that $\overline  \ffunc_D$ vanishes asymptotically: the way we see this in that plot is that, once the embedding reaches the IR of the geometry, the behaviour of $\overline  \ffunc_D$ rapidly renders into a straight line in the log-log plot. From its slope we can conclude that, for large values of $R$, the function $\overline \ffunc_D$ decreases  as $1/R$ for any $b_0$, which agrees with \cite{Klebanov:2012yf}.

Finally, we would like to show another quantity where quasi-conformal regime becomes manifest, arising from the computation of mutual information. Recall from above the computation of the mutual information between two parallel strips of the same width $l$ which are separated by a distance $s$ in \eqref{eq:mutual_info}. The values $(s,l)$ for which it vanishes are represented in plot Figure~\ref{fig.mutualInfostrips}. Interestingly, this critical ratio in a CFT is universal and given by the golden ratio \cite{Ben-Ami:2014gsa,Balasubramanian:2018qqx}
\begin{equation} \label{eq:mutualCFT}
\frac{s}{l}\Big |_{\text{CFT}}= \varphi^{-1} = \frac{-1+\sqrt{5}}{2} \approx 0.618 \ .
\end{equation} 
Moreover, in the UV this ratio is also fixed in all cases by D2-brane asymptotics \cite{vanNiekerk:2011yi}, leading to 
\begin{equation}\label{eq:mutualD2}
\frac{s}{l}\Big|_{\text{D2}} = -1+\sqrt{1+\beta}\approx 0.663 \ ,
\end{equation} 
where $\beta$ is the (unique) real root of the polynomial 
\begin{equation}\begin{aligned}
b(x)\, =\ &64 x^{11}-64 x^{10}+16 x^9-400 x^8+x^7+191 x^6+768 x^5+744 x^4\\&-192 x^3-704 x^2-1024 x-512 \ .
\end{aligned}
\end{equation}
As we analyse the mutual information $I(s,l)$, and in particular the boundary where it vanishes and defines the critical ratio $s/l$ for different theories, we find rich behaviour revealed in Figure~\ref{fig.golden} that can be interpreted as follows. Let us consider small $b_0$ values, so that the flows approach the OP fixed point. At the UV, {\emph{i.e.}} for small widths $l$, the critical ratio $s/l$ starts from the UV value (\ref{eq:mutualD2}), decreasing and developing a global minimum before increasing again towards IR (large $l$). For even smaller values of $b_0$ the critical $s/l$ curve can get arbitrarily close to the CFT value (\ref{eq:mutualCFT}), eventually diverging from it. However, only in the strict $b_0=0$ case, corresponding to $\B_8^\infty$ will we reach (\ref{eq:mutualCFT}) in the asymptotic IR regime. Our results are suggestive that the CFT value acts as a lower bound on critical $s/l$ and it would be interesting to understand the reason behind this. 

\section{Discussion}\label{sec:ChapterObservables_Discussions}

In this Chapter we have computed the quark antiquark potential in our family of theories. Moreover, we computed the mass of the first spin-0 and spin-2 particles and realised that they posses a mass gap. As a consequence, we can conclude that these theories are not confining in general, despite being gapped. The limiting cases are special though, since $\Bconf$ is both gapped and confining, whereas $\Binf$ has none of these properties.

One important further message of this Chapter is that holographic entanglement entropy and Wilson loops cannot be considered interchangeably as mediators of the fact whether the theory is confining. Whereas the quark-antiquark potential is sensitive to the fact that the flux tube between two infinitely massive quarks cannot break apart, entanglement entropy seems to be signalling the presence of the mass gap, capping off the flow of information. An observation worth highlighting, in the current context, is that the entanglement measures are insensitive to the CS interactions, which may be of relevance. This aspect deserves to be more properly understood. To this end, we hope to make closer contact with recent important studies of entanglement entropies in $(2+1)$-dimensional Chern-Simons field theories \cite{Agarwal:2016cir}. 

\begin{figure}[t]
	\begin{center}
		\includegraphics[width=.7\textwidth]{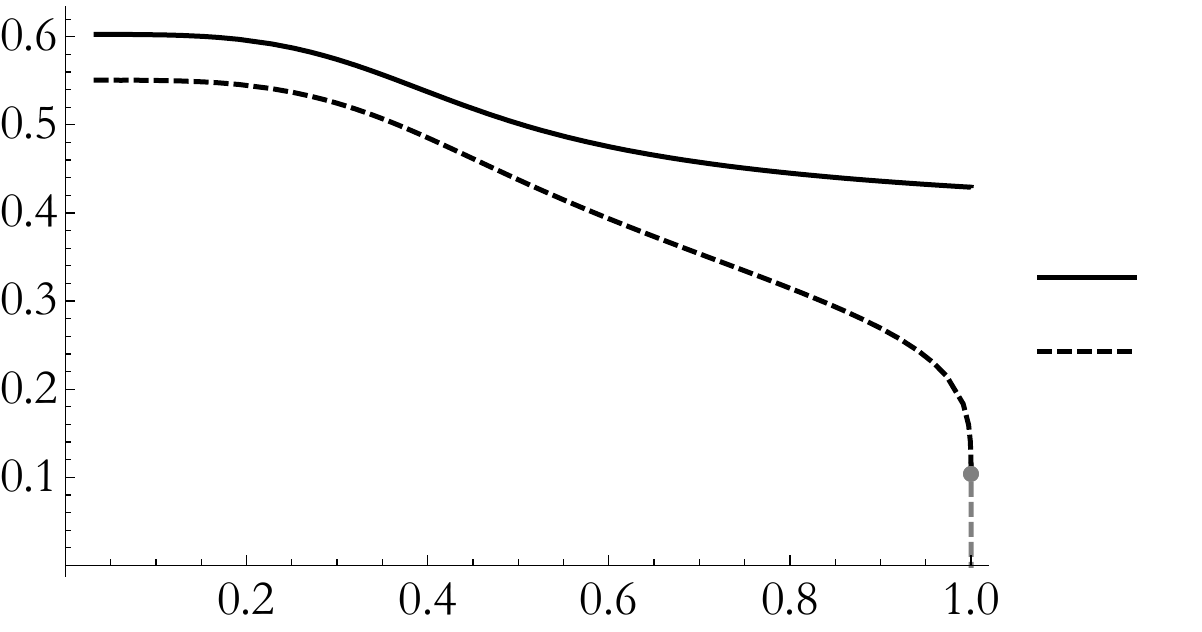} 
		\put(-7,84){$l_c/l_{\text{qq},c}$}
		\put(-7,65){$l_t/l_{\text{qq},t}$}
		\put(-30,10){$b_0$}
		\caption{\small  Ratios between scales stemming from quark-antiquark potentials and entanglement entropy of strips as a function of $b_0$. The solid curve is the ratio between the critical lengths where the disconnected strings and $\sqcup$-RT-surface become dominant for the Wilson loop and for entanglement entropy of a strip, respectively. The dashed curve stands for the ratio between the turning points of hanging string configurations and the RT embeddings. The last part of this curve, drawn fainter and below the grey dot, stands for extrapolation.}
		\label{fig:compare_scales}
	\end{center}
\end{figure}

A natural question to address is how the scales that we identify in entanglement entropy measures relate to the physical scales given by the mass gap. For that, one possibility would be to compare entanglement entropy scales to the masses of the lightest states in the spectrum of spin-2 and spin-0 particles obtained Section~\ref{sec:spectrum}. It is not completely clear, however, how to make this comparison and it is furthermore not evident if the states found there actually are the lightest ones. We are content with comparing to the scales that arise from quark-antiquark potentials. Taking Figures~\ref{fig.WilsonLoop} and \ref{fig.EEplotsStrip} into consideration, there are several scales that could be considered. The first one is the one we denoted by $l_c$, the width of the strip at which the $\sqcup$ configuration becomes preferred. This scale roughly indicates above which width there is some region (in the middle) of the strip which is not entangled with the complement. It is natural to compare this scale to the length at which the fluxtube between quarks breaks apart, $l_{\text{qq},c}$, which sets the length above which the quarks become screened by CS interactions. It is therefore expected to be related to some correlation length of the ambient field theory. 

In Figure~\ref{fig:compare_scales} we depict the ratio set by the entanglement measure to the one resulting from quark potential. We find that this ratio is around $l_c/l_{\text{qq},c}\approx 0.5$ for all theories. The fact that both scales are of the same order is a manifestation of the fact that the correlation length induced by the screening of the quark is linked with the separation between entangled states. An interesting question is in how general this statement holds. One can address this in other theories possessing phase transitions for entanglement measures by extracting the corresponding correlation lengths, and comparing them to the critical length resulting from string breaking.

There is another useful comparison to be discussed. The turning points of the RT surfaces and the string embeddings will give rise two additional numbers. We refer to them as $l_t$ and $l_{\text{qq},t}$, respectively. These are defined as the loci of the maxima as depicted in Figure~\ref{fig.WilsonLoop}~(left) and Figure~\ref{fig.EEplotsStrip}, respectively. As one dials $b_0\to 1$, {\emph{i.e.}}, we approach the confining $\Bconf$, the value of $l_{\text{qq},t}$ diverges; see Figure~\ref{WLfig}. As shown in Figure~\ref{fig.limiting_EEstrip}~(top), $l_{t}$ however saturates to the value given by the turning point of the confining theory. This causes the ratio $l_{t}/l_{\text{qq},t}$ to vanish in the limit $b_0\to 1$. This yields another  manifestation that entanglement entropy measures are not sensitive to confining dynamics.

A motivation for the claim that entanglement entropy probes confinement in \cite{Klebanov:2007ws} was the similarity between the phase transition from connected to disconnected configuration of strip entanglement entropy and the first-order finite temperature deconfinement transitions which is typically found in gravitational duals of confining gauge theories. We feel that this proposal should be viewed with caution. To be concrete, although all the SYM--CSM theories of the family we are studying have the same qualitative behaviour as far as entanglement entropy measures are considered (except for $\B_8^\infty$), their finite temperature physics will appear richer. In the next Chapter, the thermodynamic phase diagram will be mapped out with lavish transitions of first and second order, including a triple point and a critical point. In fact, it would be interesting to extend our analysis to finite temperature and similarly map out the phase diagram following from entanglement thermodynamics. It would be particularly compelling to make a thorough study of the entanglement phase diagram around the thermodynamic critical point of these theories. Such a top-down computation could further be compared to the analogous computation carried out in a holographic model of QCD \cite{Knaute:2017lll}.

Since the entanglement entropy does not seem to single out confining geometries from those that are not, one could raise to concern if some other set of boundary data would be better suited to bulk reconstruction. A particularly compelling option would be to lean on Wilson loops \cite{Jokela:2020auu,Hashimoto:2020mrx}, for example. In fact, in \cite{Hashimoto:2020mrx} (see also \cite{Sonnenschein:1999if}) it was argued that the linear quark-antiquark potential necessary leads to an ``IR bottom'' in the in deep interior of bulk spacetime. This, however, is not yet completely satisfactory due to a technical reason: having an regular ``IR bottom'' (typically realised by cigar-like geometries) is not in one-to-one correspondence with linear potential. This statement follows from the models we discussed. Then, the obstacle present at finite temperature is faced here as well, whence the hanging string breaks due to pair creation and therefore reconstruction of the IR part of the geometry is not possible. We feel that these issues can be resolved upon closer inspection and hope to address them in future works. 

A natural extension of our work would be to consider multiparty entanglement, between strips \cite{Ben-Ami:2014gsa}, or disks, for example, and analyse more refined probes of information flows. It would be particularly interesting to analyse how the extensivity of the mutual information \cite{Balasubramanian:2018qqx} behaves under the RG flow in our family of geometries and whether some intermediate length scales are more super-extensive than the CFT at the OP fixed point. A further investigation of entanglement measures for mixed states \cite{Takayanagi:2017knl,Kudler-Flam:2018qjo,Tamaoka:2018ned,Jokela:2019ebz,Lala:2020lcp,1822901}, contrary to pure states as studied in this Chapter, might shed more light on why entanglement wedge cross sections behave non-monotonically under RG flows \cite{Jokela:2019ebz}, specifically in confining geometries. Moreover, a particularly compelling scenario would be to add a magnetic (Kondo) impurity \cite{Benincasa:2012wu,Erdmenger:2013dpa,Erdmenger:2015spo} and study how the entanglement entropy behaves in our family of solutions, and in particular if some form of g-theorem holds. 
Finally, it would be interesting to make some contact with a phenomenon called partial deconfinement \cite{Watanabe:2020ufk,Hanada:2018zxn,Hanada:2019czd} and investigate if entanglement measures can lead to some new interesting insights.

\newpage
\thispagestyle{empty}


\newpage
\chapter{Thermodynamic phase diagram}
\label{Chapter4_thermo}

The aim of this Chapter is to study the thermodynamics of the SYM--CSM theories whose ground state we have been studying. Generically, as the temperature increases gradually from zero, the system undergoes a phase transition from the gapped state to an ungapped phase, as we will see. The ungapped nature of this phase follows  from the fact that, on the gravity side, it is  characterised by the presence of a black brane horizon, which supports excitations with arbitrarily low energy such as hydrodynamic modes. Therefore  we will refer to this phase transition as a \textit{degapping transition}. Its details and the behaviour of the system at even higher temperatures depend on the value of $b_0$ in a non trivial way, as we shall see.

The computation of the black brane solutions will be described in Section~\ref{HighT}. Before going into that, in the next Section we wish to discuss how the low-temperature phase of the system is straightforwardly obtained from the supersymmetric solutions constructed in Chapter~\ref{Chapter2_B8family}. 

\section{Low-temperature phases}
\label{LowT}
The family of solutions constructed in Chapter~\ref{Chapter2_B8family} can be straightforwardly heated up by going to Euclidean space and declaring the time direction to be compact. As usual, the period $\beta$ of the Euclidean time is related to the temperature $T$ in the dual gauge theory as
\begin{equation}\label{beta}
\beta\,=\,\frac{1}{T}\,.
\end{equation}

The thermodynamic properties are very simple. The ground states are supersymmetric, so in the appropriate renormalisation scheme (see Appendix~\ref{app:4Deffectivetheory}) the free energy $F$, computed as the bulk on-shell action,  vanishes. Since formally these 
finite-temperature solutions coincide with the supersymmetric ones (they only differ globally in the time direction), their free energy will also vanish in the same scheme. This happens independently of the temperature, and therefore the entropy $S$ is also zero. 

These thermal states are continuously connected to the ground state, so we expect them to correspond to a phase still exhibiting a mass gap. On the other hand, we know that replacing the regular IR with a horizon also introduces temperature into the system while removing the gap. These solutions, describing black branes, can have the same UV asymptotics and therefore correspond to other finite-temperature states of the same gauge theories. In the following we construct the black brane solutions and show the existence of several phase transitions.

\section{High-temperature phases}
\label{HighT}

In this Section we construct high-temperature phases of our SYM--CSM gauge theories. On the gravity side of the duality this corresponds to the replacement of the regular IR by a horizon, giving rise to a black brane. These black branes need to satisfy, at leading order, identical D2-brane UV boundary conditions as the zero-temperature solutions in order to correspond to states in the same gauge theory duals. Therefore we consider the 10D ansatz
\begin{equation}
\label{10Dansatz}\begin{aligned}
\dd s_{\rm st}^2 &=h^{-\frac12}\left(-\mathsf{b}\dd t^2 + \dd x_1^2 + \dd x_ 2^2\right)+h^{\frac12} \left(\frac{\dd r^2}{\mathsf{b}}+e^{2f}\dd\Omega_4^2+e^{2g}\left[\left(E^1\right)^2+\left(E^2\right)^2\right] \right)\,\\[2mm]
e^\Phi&=h^{\frac14} \, e^\Lambda \,,
\end{aligned}
\end{equation}
which differs from that in \eqref{eq:10DansatzAP} only in the presence of a blackening factor $\mathsf{b}$. This function must have a simple zero at the position of the horizon for the solution to describe a black brane. 

In this case, the ansatz for the different fluxes takes a very similar form to the one for supersymmetric solutions \eqref{eq:fluxesansatz}, namely
\begin{equation}\label{eqfluxesansatzFT}
\begin{array}{rclcrcl}
F_4 &=&  \mathbf{f}_4 *\omega_6 + G_4 + B_2\wedge F_2 \,,&\qquad& F_2& =& Q_k (X_2 - J_2)\,,\\[2mm]
G_4 &=& \dd(a_J J_3) + q_c\left(J_2\wedge J_2 - X_2\wedge J_2\right)\,,&\qquad& B_2& =& b_X X_2 + b_J J_2\,,
\end{array}
\end{equation}
where we have defined the quantity 
\begin{equation}
\mathbf{f}_4= Q_k b_J^2 +2(q_c+2a_J)b_X - 2 b_J(q_c-2a_J+Q_kb_X) +Q_c\,.
\end{equation}
This ansatz reduces to that of Chapter \ref{Chapter2_B8family} when the blackening factor $\mathsf{b}$ is set to one. An important point we should mention is that black brane solutions are not expected to solve the BPS equations \eqref{BPSsystem} and \eqref{BPSsystem_fluxes} in general, since finite temperature breaks supersymmetry. Rather, we will seek for solutions of the equations of motion of type IIA supergravity. In string frame, the equations of motion for the forms read
\begin{eqnarray}\label{formeqs}
\dd *F_4 + \mathsf{H}\wedge F_4 &=&0\,,	\nonumber\\[2mm]
\dd  *F_2 + \mathsf{H}\wedge* F_4 &=&0\,,\\[2mm]
\dd\left(e^{-2\Phi}  * \mathsf{H}\right)-F_2\wedge *F_4 -\frac{1}{2} F_4\wedge F_4 &=&0\,,\nonumber
\end{eqnarray}
where $\textsf{H} = \dd B_2$. The field strengths are such that the following Bianchi identities must be satisfied
\begin{equation}
\dd \mathsf{H}=0\,, \qquad\qquad \dd F_2 =0\,,  \qquad\qquad \dd F_4 = \mathsf{H}\wedge F_2\,.
\end{equation}
On top of that, we have the equations governing  the dilaton
\begin{eqnarray}
R+4\nabla_M\nabla^M \Phi - 4 \nabla^M\Phi \nabla_M \Phi-\frac{1}{12} \mathsf{H}^2 = 0\,,
\end{eqnarray}
and the metric
\begin{equation}\begin{aligned}
&R_{MN} + 2 \nabla_M\nabla_N \Phi -\frac{1}{4} \mathsf{H}_{MN}^2 \\[2mm]&\qquad= e^{2\Phi}\left[
\frac{1}{2} (F_2^2)_{MN} + \frac{1}{12}(F_4^2)_{MN} - \frac{1}{4}g_{MN}\left(
\frac{1}{2}F_2^2 +\frac{1}{24}F_4^2\right)\right]\,,
\end{aligned}
\end{equation}
in a self-explanatory notation, where $R$ and $R_{MN}$ are the Ricci scalar and the Ricci tensor respectively.

For the sake of simplification, we will again consider the dimensionless rescaled functions defined in \eqref{eq:dimlessfunctions}, together with 
\begin{equation}\label{eq:dimlessAJ}
a_J = -\frac{q_c}{6} - \sqrt{4q_c^2 + 3|Q_k| Q_c} \ \AAA_J\,,
\end{equation}
which cannot be given in terms of $b_J$ and $b_X$ when the system is considered at a finite temperature. Furthermore, working with the dimensionless radial coordinate 
\begin{equation}\label{ucoord}
u\,=\,\frac{|Q_k|}{r}
\end{equation}
all the charges drop from the equations. This means that, up to simple rescalings, the only parameter distinguishing one theory from another is again $b_0$. In particular, the mass gap at zero temperature (in units of the 't~Hooft couplings) is fixed by this parameter. The thermal phase transitions that we will exhibit will  take place at specific values of the ratio $T/M_\textrm{gap}$ that are determined by $b_0$. 

\subsection{Boundary conditions} 

In the radial coordinate defined in \eqref{ucoord} the UV is located at $u=0$. The leading behaviour of the functions must coincide with that of the ground state. It is then possible to solve the equations order by order in the radial coordinate around this point, obtaining an expansion of the form 
\begin{eqnarray}\label{UVexpansions}
e^\FF&=& \frac{1}{u\sqrt{2}}  \Big[1+{ f_1}{u}+\cdots+{ f_4}{u^4}+ { f_5}{u^5}+\cdots+{ f_{10}}{u^{10}}+\OO(u^{11})\Big]\,,\nonumber\\[2.2mm]
e^\GG&=& \frac{1}{2u}  \Big[1+\OO(u)\Big]\,, \qquad \qquad \qquad\mathbf{h} \,=\,\frac{16}{15}(1-b_0^2)\ u^5\   \Big[1+\OO(u) \Big]\,,\nonumber\\[2.2mm]
\mathsf{b}&=& 1+{\mathsf{b}_{5}}{u^{5}}+\OO(u^6)\,,\qquad \qquad \,e^\Lambda\,=\, 1+\OO(u)\,,\\[2.2mm]
\BB_J&=& b_0 +\cdots + { b_4} {u^4} +\cdots + { b_6} {u^6} + \cdots + { b_9} {u^9} +\OO(u^{10})\,, \nonumber\\[2.2mm]
\quad \BB_X&=&b_0+ \OO(u)\,,\qquad\qquad\qquad\quad\,\AAA_J\,=\,\frac{b_0}{6}+ \OO(u)\,, \nonumber
\end{eqnarray}
where we have made explicit the parameters that remain undetermined by the equations of motion. The first coefficients not explicitly shown in this expansion, written in terms of the ones that are shown, can be found in Appendix \ref{ap:UVexp}. As we already argued, $b_0$ selects a particular member of the family of gauge theories. The radial gauge chosen in equation~\eqref{10Dansatz}, $g_{tt} g_{rr}=-1$, still leaves a residual freedom to shift the radial coordinate, $r\to r + \mbox{constant}$. This allows us to fix the value of $f_1$ without loss of generality. We choose $f_1=-1$ in order to facilitate the comparison with the supersymmetric ground state.  We are thus left with seven subleading, undetermined parameters in the UV. They correspond to normalisable modes and are hence related to the temperature (see equation~\eqref{b5_conserved} below) and to VEVs for different operators in the dual gauge theory, including the energy-momentum tensor. These parameters are fixed dynamically in the complete solution once we impose the presence of a regular horizon.  

The existence of a horizon  is encoded in a simple zero of the blackening factor $\mathsf{b}$. At the horizon we demand regularity for the rest of the functions, so that they reach a finite value. We denote the position of the horizon as $u=u_h$ in the dimensionless radial coordinate introduced in equation~\eqref{ucoord}. In this way the functions enjoy an expansion of the form
\begin{equation}\label{Horizon_expansions}
\begin{array}{rclrcl}
e^\FF &=&  f_h  +\OO(u-u_h)  \ ,&
\quad e^\GG&=&  g_h + \OO(u-u_h)  , \\[2mm]
e^\Lambda&=& \lambda_h + \OO(u-u_h)\,, \ &
\quad\mathbf{h}& =& h_h +\OO(u-u_h)\,,\\[2mm]
\mathsf{b}&=&  \mathsf{b}_h (u-u_h)+\OO(u-u_h)^2\,,&
\quad\BB_J&=& \xi_h +\OO(u-u_h)\,,  \\[2mm]
\BB_X&=& \chi_h +\OO(u-u_h)\,,  &
\AAA_J&=& \alpha_h+\OO(u-u_h)   \ ,\\[2.5mm]
\end{array}
\end{equation}
All the subleading coefficients are determined in terms of these eight leading-order ones (see Appendix \ref{ap:horizonexp} for the first coefficients). The position of the horizon $u_h$ controls the temperature of the black brane, so the phase diagram can be explored by changing its value. 

\subsection{Numerical integration}

The perturbative analysis of the equations of motion with the desired boundary conditions in the UV and IR leaves us with the following undetermined parameters 
\begin{eqnarray}
\text{in the UV:}&\qquad&  f_4,\ f_5,\ f_{10},\ b_4,\ b_6,\ b_9,\ \mathsf{b}_5\nonumber\\
\text{at the horizon:}&\qquad& g_h,\ f_h,\ \lambda_h,\ h_h,\ \mathsf{b}_h,\ \alpha_h,\ \xi_h,\ \chi_h \,.
\end{eqnarray}
For each value of $b_0$ and $u_h$ we must find these fifteen parameters dynamically in the complete solution. Since we are solving eight second-order equations subject to one first-order (Hamiltonian) constraint, the problem is well posed. It should be mentioned that, as we note in equation~\eqref{conserved_current}, there is a conserved quantity along the radial coordinate. When substituting the expansions, this conserved quantity relates one of the UV parameters with horizon data as
\begin{equation}\label{b5_conserved}
\mathsf{b}_5 = \frac{16}{5}\, \frac{ \mathsf{b}_h\ f_h^4\ g_h^2}{\lambda_h^2}\ u_h^2\,.
\end{equation}
This equation can be used as a check for the numerical solutions. Alternatively, it can be imposed {\it ab initio}, reducing the number of parameters to be found numerically to fourteen.  

The solutions were constructed using a shooting method. The boundary conditions \eqref{UVexpansions} and \eqref{Horizon_expansions} were imposed close to the UV and the horizon, respectively, with an initial guess for the unfixed parameters, giving us a seed solution. The full equations were then integrated numerically both from the UV and from IR up to a designated intermediate matching point. For arbitrary values of the parameters the functions would be discontinuous at that point. Starting from this seed, we then used the Newton--Raphson algorithm to find the values of the parameters such that the functions and their derivatives are continuous at the matching point with the desired precision.    

The main challenge when solving a system of equations via the shooting method is to find a good initial seed. Ultimately, we are using a Newton--Raphson algorithm on a fourteen-dimensional parameter space (after making use of equation~\eqref{b5_conserved}), so we need to start the search close enough to the correct values of the unknown constants for it to converge. A good seed for a solution can be provided by another solution close enough in parameter space since, by continuity, the values of the unknown constants will be similar. In other words, once we have a black brane with given control parameters $b_0$ and $u_h$, we can use the values of the rest of the parameters as the initial guess for another solution with slightly corrected control parameters. The problem thus reduces to finding the first black brane.  

Fortunately the system admits one analytic solution describing a black brane. As we already mentioned, the RG flow corresponding to $b_0=0$, denoted by $\B_8^\infty$ at the left of Figure~\ref{fig:triangle}, ends at an IR fixed point. This is located at $r=\pi\, |Q_k|$ (equivalently, $u= \pi^{-1}$) so that near this point the functions in the metric read 
\begin{equation}\label{AdS-OP}
\begin{array}{rclcrcl}
 e^{2f}&=& 
 \displaystyle\frac{9}{5} \left(r+\pi Q_k\right){}^2\,, &\qquad\qquad& e^{2g}& =&\displaystyle\frac{9}{25} \left(r+\pi Q_k\right){}^2\,,\\[3mm]
e^\Lambda& =& \displaystyle\frac{3}{5 |Q_k|} \left(r+\pi Q_k\right)\,,&\qquad\qquad&h&=&\displaystyle\frac{125 \,Q_k^4}{2187\left(r+\pi Q_k\right)^4}\,,
\end{array}
\end{equation}
giving rise to an ${\rm AdS}_4$ geometry, as expected. It is then straightforward to include a horizon by considering the blackening factor
\begin{equation}\label{AdSblackening}
\mathsf{b}(r)=1-\frac{(r_h+\pi Q_k)^3}{(r+\pi Q_k)^3}\,,
\end{equation}
which describes an AdS-Schwarzschild black brane. This is an exact solution to the equations of motion when $b_0=0$. 

Based on this observation, the strategy to find the first black brane with the desired asymptotics is the following. We put a horizon in the region where the geometry is close to AdS-Schwarzschild and we choose as horizon data the values of the functions for $\B_8^\infty$ at that point. Since this should correspond to a solution at low temperature, the UV parameters are then taken as the ones for the metric without a horizon. Also, we use equation~\eqref{AdSblackening} as the seed for the blackening factor both at the UV and at the horizon. As a final simplification, we note that for $b_0=0$ the dimensionless functions $\BB_J$, $\BB_X$ and $\AAA_J$ from equations \eqref{eq:dimlessfunctions} and \eqref{eq:dimlessAJ} vanish exactly and, consequently, there are three equations which are trivially satisfied and six fewer parameters  to be found. In this way we were able to find a black brane solution in the $\B_8^\infty$ RG flow, with the horizon located at $u_h\approx \pi^{-1} -4\times 10^{-4}$. 

Once we have the first black brane we slightly decrease $u_h$ and use as seed the values of the parameters we have just obtained, finding in this way another solution. Thus, we can iteratively find hotter black branes by interpolating the values of the parameters of the previous solutions to obtain a good seed for the next one. The resulting family of black branes has the expected properties. At low temperatures the dependence of the free energy on the temperature is dictated by conformal invariance,
\begin{equation}\label{FreeECFT}
F\,\sim\, \frac{1}{|k|}\left(\frac{\overline{M}^2}{2} +N|k|\right)^{3/2}T^3\,.
\end{equation}
In the opposite, UV-limit, for temperatures above 
\begin{equation}\label{Tconformal}
\lambda\,\frac{k^2}{N}\left(\frac{\overline{M}^2}{2} +N|k|\right)^{-1/2}
\end{equation}
(recall $\lambda=g_s\ell_s^{-1}N$ is the 't~Hooft's coupling \eqref{eq:tHooft3D}), we  recover the D2-brane solution and the  free energy behaves as in a three-dimensional Yang--Mills theory at strong coupling:
\begin{equation}
F\,\sim\, \lambda^{-1/3}\left(\frac{N}{|k|^5}\right)^{1/3}\left(\frac{\overline{M}^2}{2} +N|k|\right)^{5/3}T^{10/3}\,.
\end{equation}
This UV behaviour will be shared by the entire family of solutions. Note that in the 
$N\to \infty$ limit the free energy scales with the number of colours as $N^2$, as expected on a deconfined phase. 

Taking into account that all the flows share the same UV asymptotics except for the value of $b_0$, we can use the solutions we just constructed as a seed to find black branes with non-vanishing $b_0\gtrsim0$. The seed will be better the higher the temperature is, since the flows differ mostly in the IR. Now $\BB_J$, $\BB_X$ and $\AAA_J$ are also non-vanishing, hence we have to solve the full system of equations and shoot for the complete set of parameters. In this way we obtained the first high-temperature black brane for a small non-vanishing value of $b_0$, whose ground state is gapped. Again, using this first solution as a seed it is possible to gradually increase the value of $u_h$ and find colder black branes. The process ends when some of the functions, namely $e^\FF$, $e^\GG$ and $e^\Lambda$, vanish at the horizon within numerical precision.  We will refer to the value of $u_h$ where this happens as $u_N$,  which is a meaningful quantity because we have fixed the radial gauge completely. In the case of the zero-temperature, regular solutions the vanishing of these metric functions is precisely what produces the end-of-space and the mass gap. Below we will investigate whether this happens in the current context in a smooth or in a singular manner. 

Once we have the entire family of black branes for this particular $b_0$, we slightly increase its value. The hottest black branes would again be similar to the ones we had constructed, that may then be used as a seed. With this strategy it is possible to explore the entire space of parameters in the $\left(b_0,u_h\right)$-plane. Plots for some of the solutions can be found in Appendix~\ref{ap:solutions}. 

\subsection{Zero-entropy limit} \label{sec:zeroEntropyLimit}

Before moving into the  study of the phase diagram, let us discuss in more detail the solutions that we obtain in the limit in which we remove the horizon for the black brane solutions with $0 < b_0 < 1$. This is achieved when the values of the functions $e^\FF$, $e^\GG$ and $e^\Lambda$ become zero at $u_h=u_N$, since the vanishing of these functions implies that the area of the black branes, and therefore their entropy, goes to zero. As we will see, in this limit the temperature is still non-zero. 

\begin{figure}[t]
	\begin{center}
		\includegraphics[width=.58\textwidth]{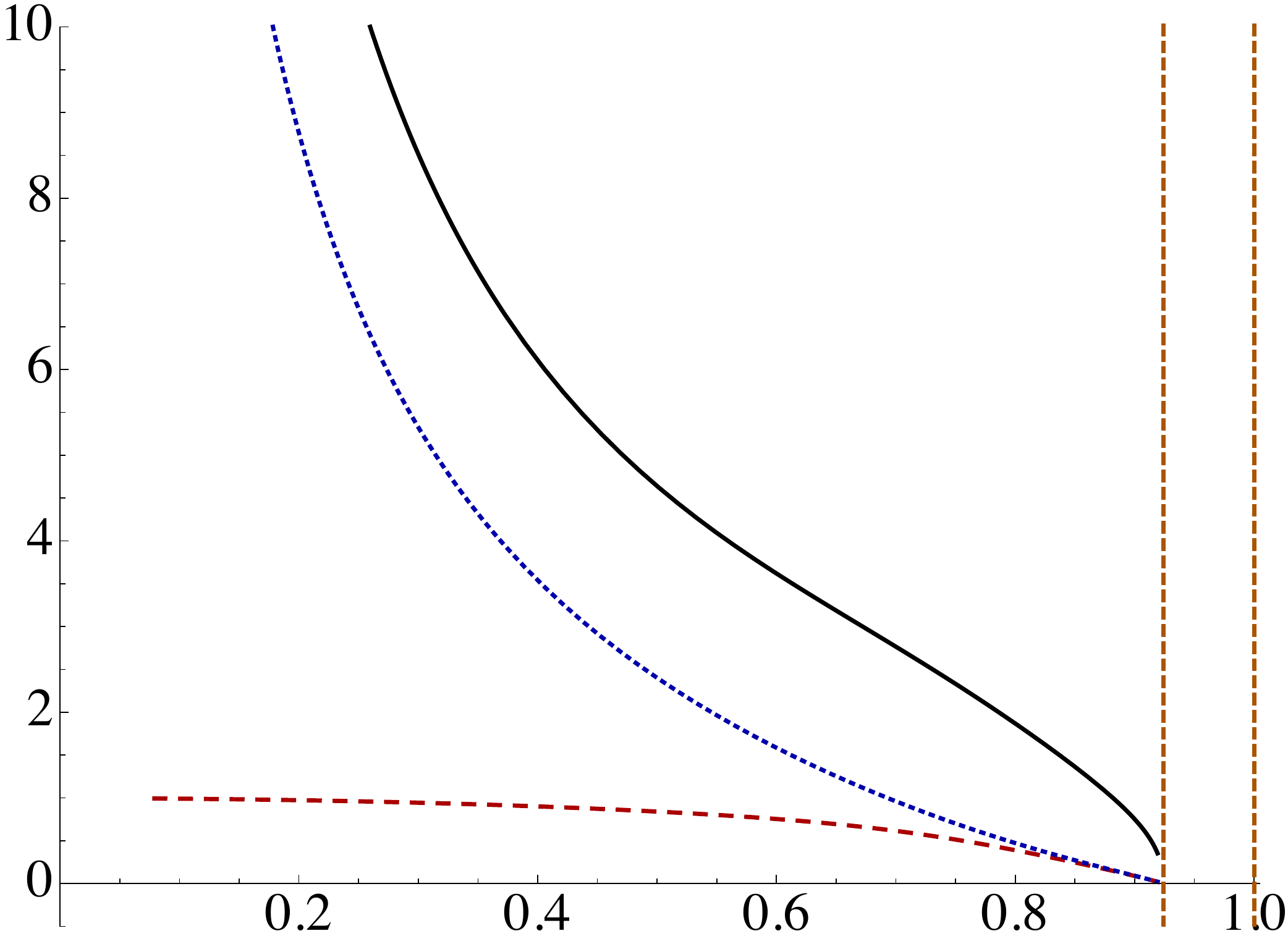} 
		\put(-220,160){\rotatebox{0}{$f_h,\ g_h,\ \lambda_h$}}
		\put(9,9){$u_h/u_s$}
		\caption{\small Horizon values of $e^\FF$, $e^\GG$ and $e^\Lambda$ in solid black, dotted blue and dashed red, respectively, as a function of the position of the horizon normalised to $u_s$, which is the where the supersymmetric regular ground-state solution ends. The dashed orange vertical line to the left corresponds to $u_N$, while the one to the right represents $u_s$. Between those two lines we have not found any black brane solutions. In this plot, we have fixed $b_0=2/5$.}
		\label{fig:HorizonValues}
	\end{center}
\end{figure}

As we have defined it, $u_N$ is the maximum value of $u_h$ for a given $b_0$. The first observation is that this does not coincide with $u_s(b_0)$, defined\footnote{The parameter $u_s$ gives the location of the end-of-space in the $u$ coordinate defined in \eqref{ucoord}. Hence, it plays the same role as that of $y_0$ in the $y$ coordinate \eqref{eq:coordy} used in Chapter~\ref{Chapter2_B8family}, which we know is in a one-to-one relation with $b_0$, as shown in Figure~\ref{fig.numparam} and discussed there. The precise relation between $y_0$ and $u_s$ is given in \eqref{eq:expression_us}.} as the position of the end-of-space for the supersymmetric, regular, zero-temperature solution with the same $b_0$. Note that it is meaningful to compare these two coordinate values because we have fixed the radial gauge completely. The discrepancy can be seen for the particular example of $b_0=2/5$ in Figure~\ref{fig:HorizonValues}, where we show the value of $e^\FF$, $e^\GG$ and $e^\Lambda$ at the horizon as a function of the position of the horizon normalised to $u_s$. All three values vanish at the same point $u_N<u_s$. The same phenomenon can be seen for different values of $b_0$ in Appendix~\ref{ap:solutions}.

As a consequence, when we take the area of the horizon to zero, which we refer to as the ``zero-entropy limit,'' we do not recover the  ground state solutions discussed in Chapter~\ref{Chapter2_B8family}. This is in sharp contrast with well known examples such as AdS-Schwarzschild or black Dp-branes, where the usual AdS and Dp-brane metrics are recovered as one removes the horizon. It is therefore interesting to understand the nature of the solution obtained in this regime.

\begin{figure}[t]
	\begin{flushleft}
		\hspace{1cm}\includegraphics[width=.72\textwidth]{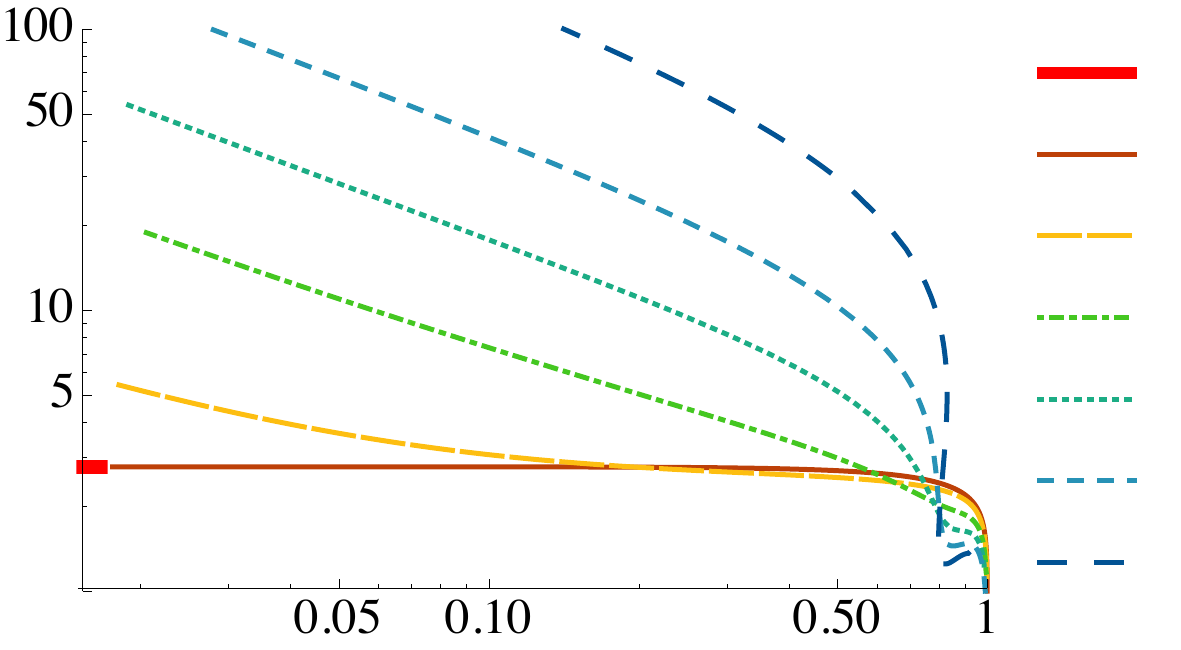} 
		\put(-270,150){\rotatebox{0}{$-\left(4q_c^2+3Q_c|Q_k|\right)^{1/3} R_h$}}
		\put(-45,-13){$\lambda_h$}
		\put(-2, 119){OP | CFT}
		\put(-2, 102.5){$b_0 = 0 $ $(\Binf)\qquad$}
		\put(-2,85){$b_0 = 0.1914$}
		\put(-2,67.5){$b_0 = 0.4000$}
		\put(-2,50){$b_0 = 0.5750$}
		\put(-2,32.5){$b_0 = 0.7109$}
		\put(-2,15){$b_0 = 0.8400$}
		\caption{\small Eleven-dimensional Ricci scalar as a function of $e^\Lambda$, both evaluated at the horizon, for different black brane solutions. Notice that as $\lambda_h$ goes to zero (\textit{i.e.}~in the zero-entropy limit), the Ricci scalar diverges for a generic value of $b_0>0$, whereas for $\B_8^\infty$ it approaches the constant value corresponding to the IR AdS. The curvature invariants obtained by squaring the Ricci and Riemann tensors have an analogous behaviour.}
		\label{fig:Ricci}
	\end{flushleft}
\end{figure}
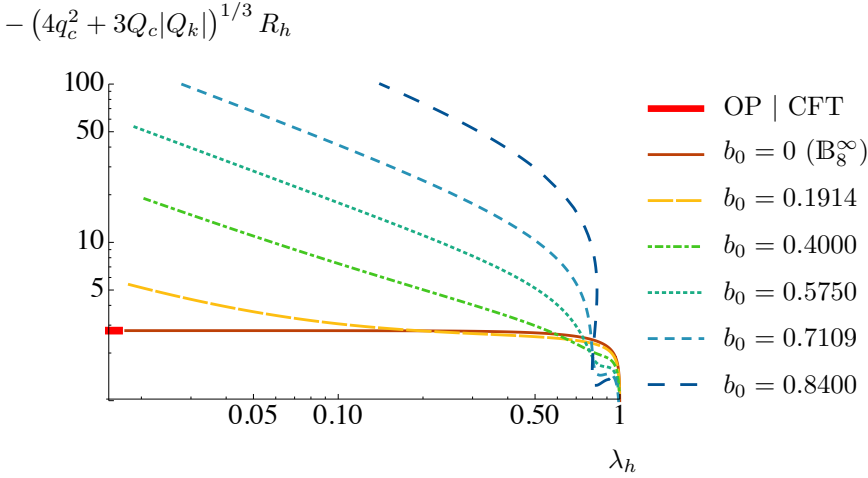

\begin{figure}[t!]
	\begin{center}
		\begin{subfigure}{0.45\textwidth}
			\includegraphics[width=\textwidth]{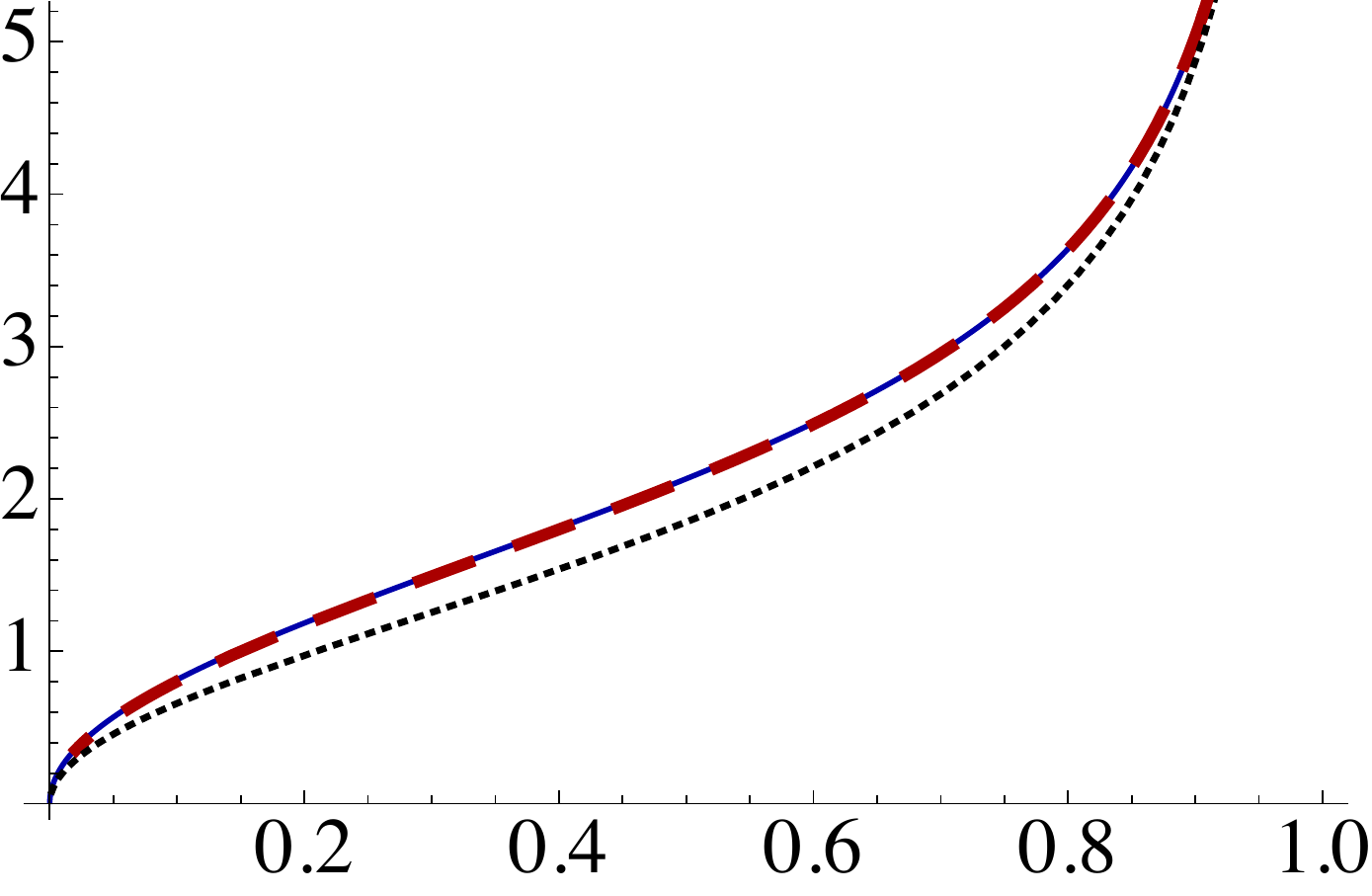} 
			\put(-150,105){$e^{\FF}$}
			\put(-5,15){$e^{\Lambda}$}
		\end{subfigure}\hfill
		\begin{subfigure}{.45\textwidth}
			\includegraphics[width=\textwidth]{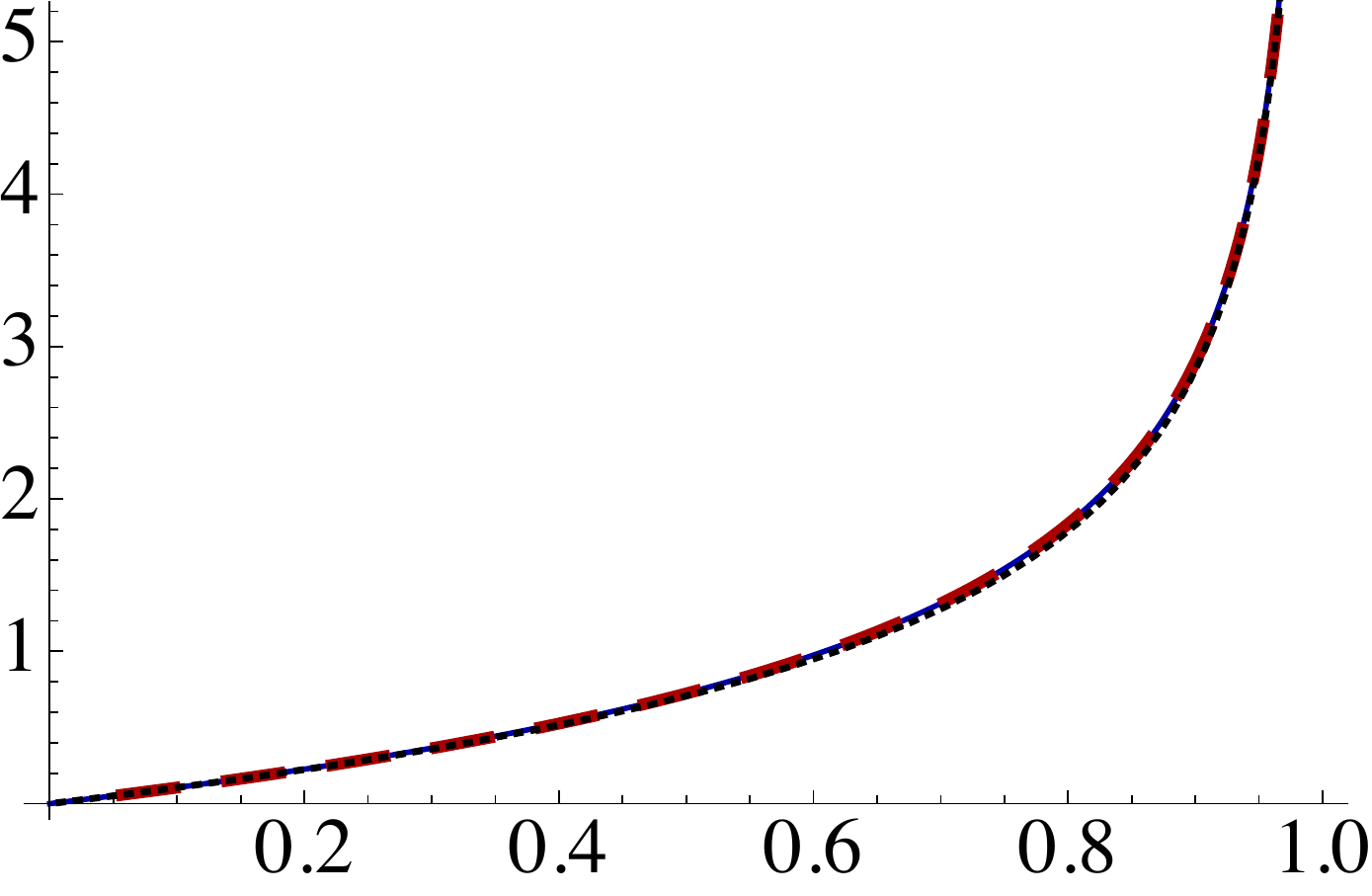} 
			\put(-150,105){$e^{\GG}$}
			\put(-5,15){$e^{\Lambda}$}
		\end{subfigure}\vspace{7mm}
		\begin{subfigure}{0.45\textwidth}
			\includegraphics[width=\textwidth]{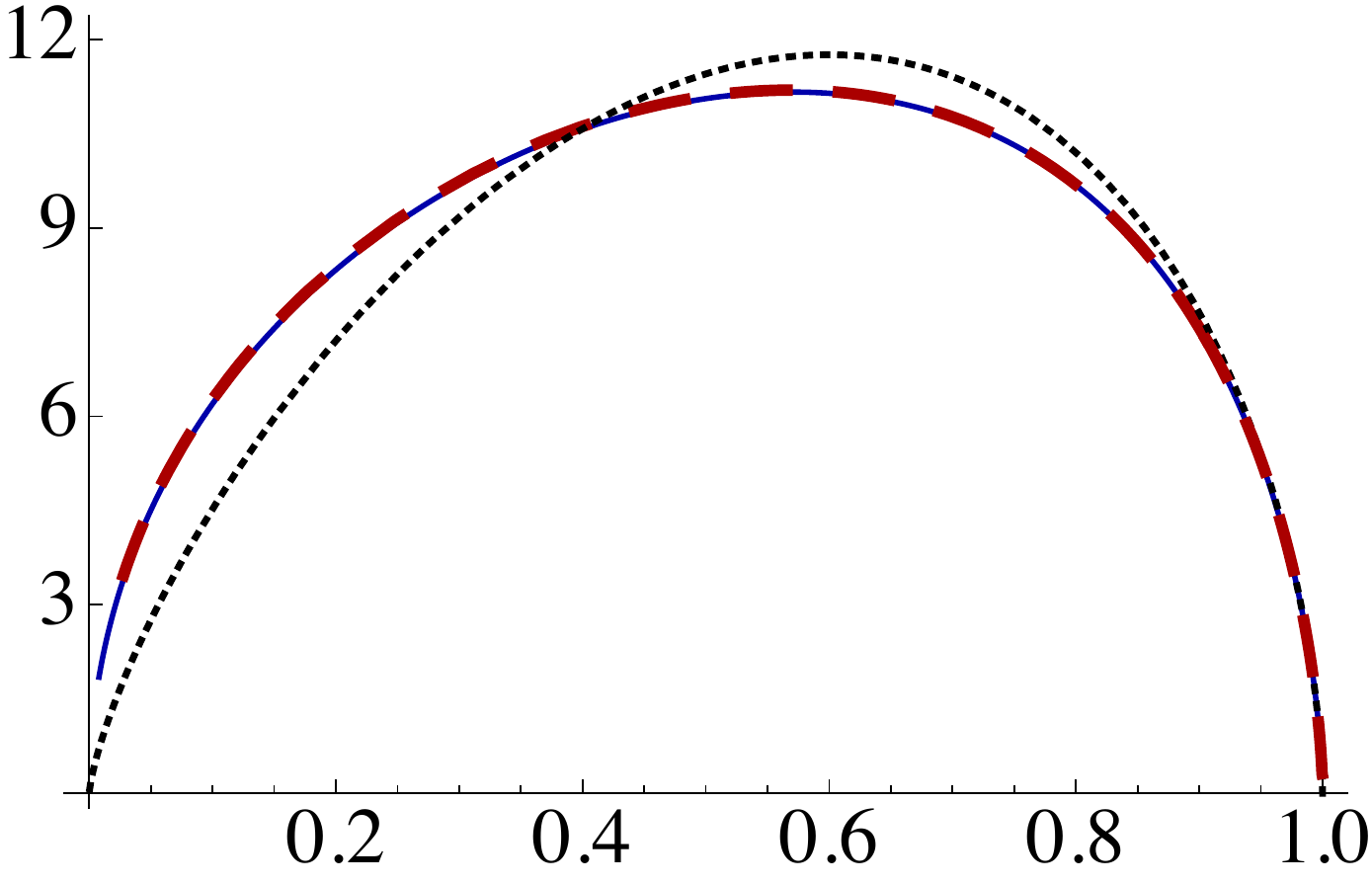} 
			\put(-150,115){$\frac{|Q_k|^{\frac{3}{2}}}{(4 q_c^2 + 3Q_c |Q_k|)^{\frac{1}{4}}}10^2e^\Phi$}
			\put(-5,15){$e^{\Lambda}$}
		\end{subfigure}\hfill
		\begin{subfigure}{.45\textwidth}
			\includegraphics[width=\textwidth]{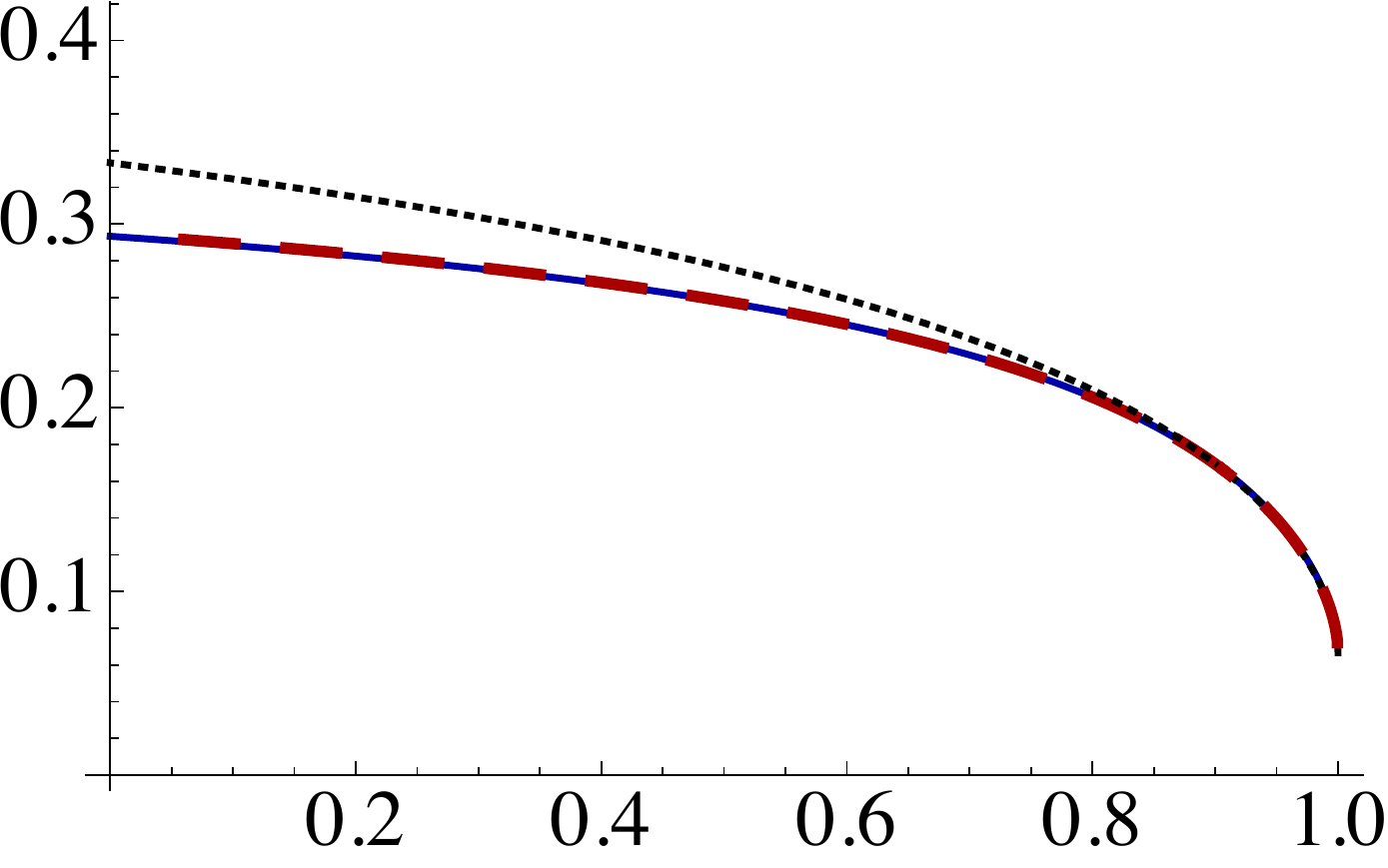} 
			\put(-150,105){$\AAA_J$}
			\put(-5,15){$e^{\Lambda}$}
		\end{subfigure}\vspace{10mm}
		\begin{subfigure}{0.45\textwidth}
			\includegraphics[width=\textwidth]{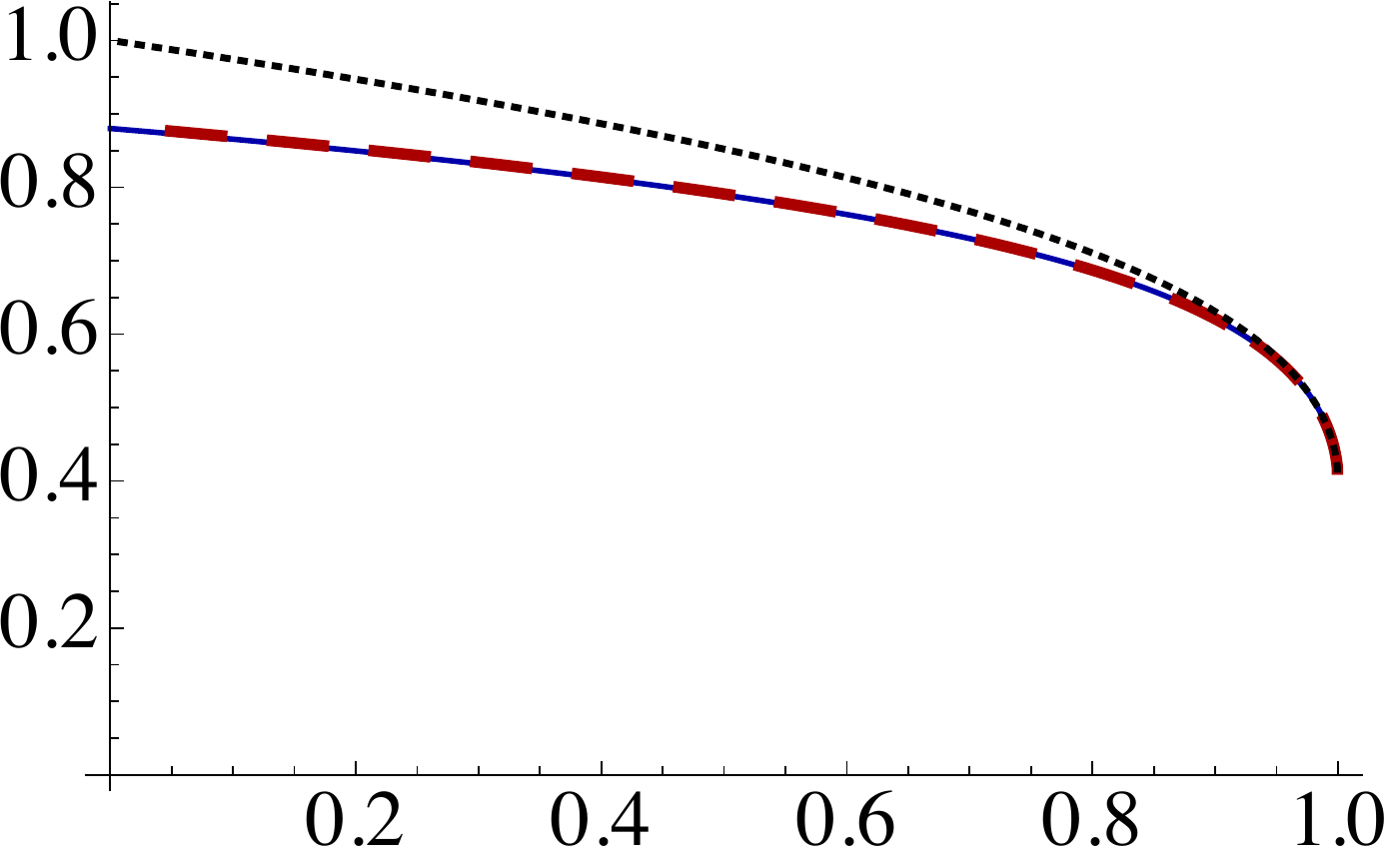} 
			\put(-150,105){$\BB_J$}
			\put(-5,15){$e^{\Lambda}$}
		\end{subfigure}\hfill
		\begin{subfigure}{.45\textwidth}
			\includegraphics[width=\textwidth]{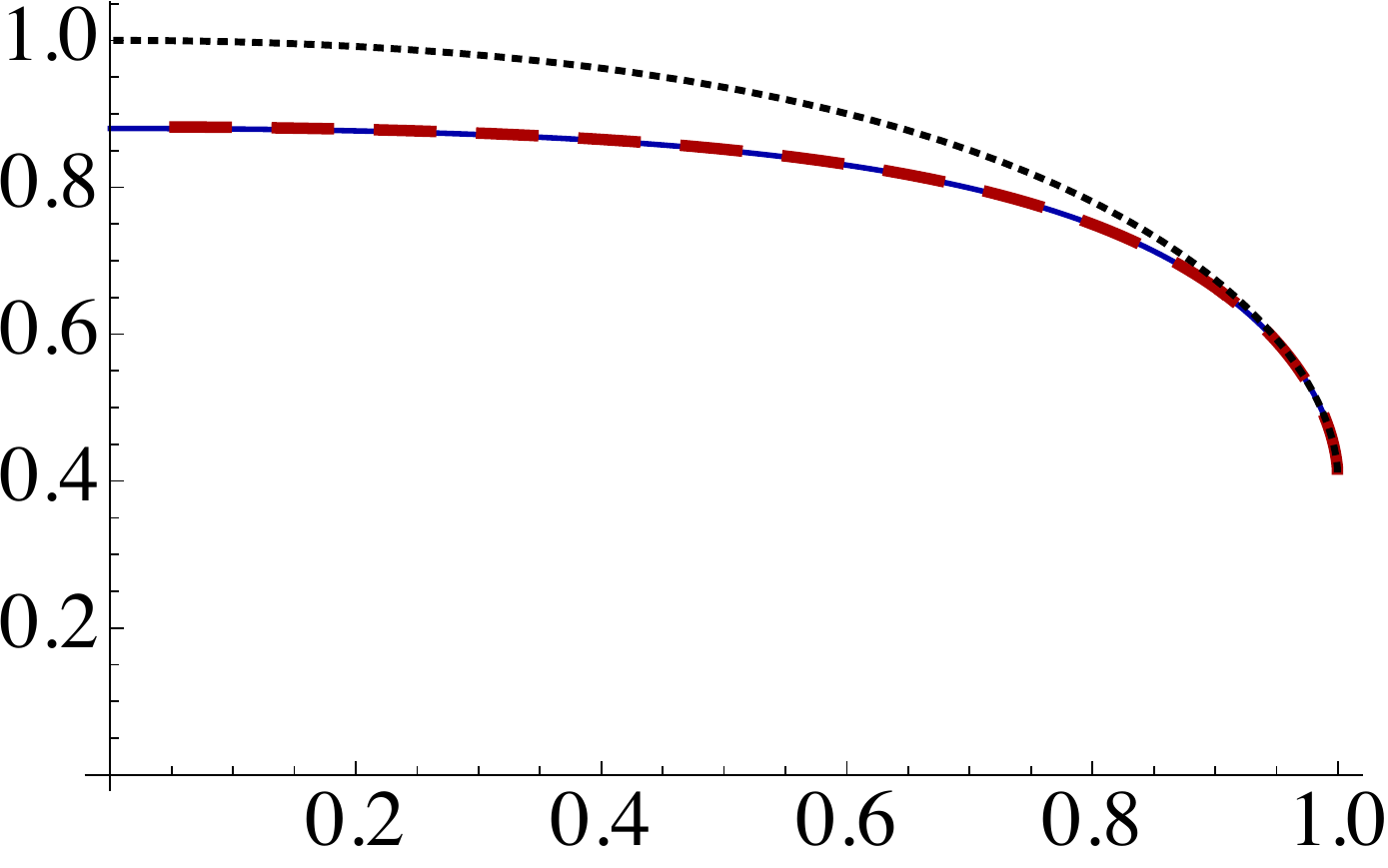} 
			\put(-150,105){$\BB_X$}
			\put(-5,15){$e^{\Lambda}$}
		\end{subfigure}
		\caption{\small Comparison of three solutions with $b_0=2/5$. The red, dashed curves correspond to the lowest-entropy solution that we were able to construct. The solid, blue curves show the horizonless, supersymmetric singular solution. These two solutions only differ in a tiny IR region. The dotted, black curves correspond to the horizonless, regular solution.}
		\label{fig.ZeroEntropyLimit}
	\end{center}
\end{figure}
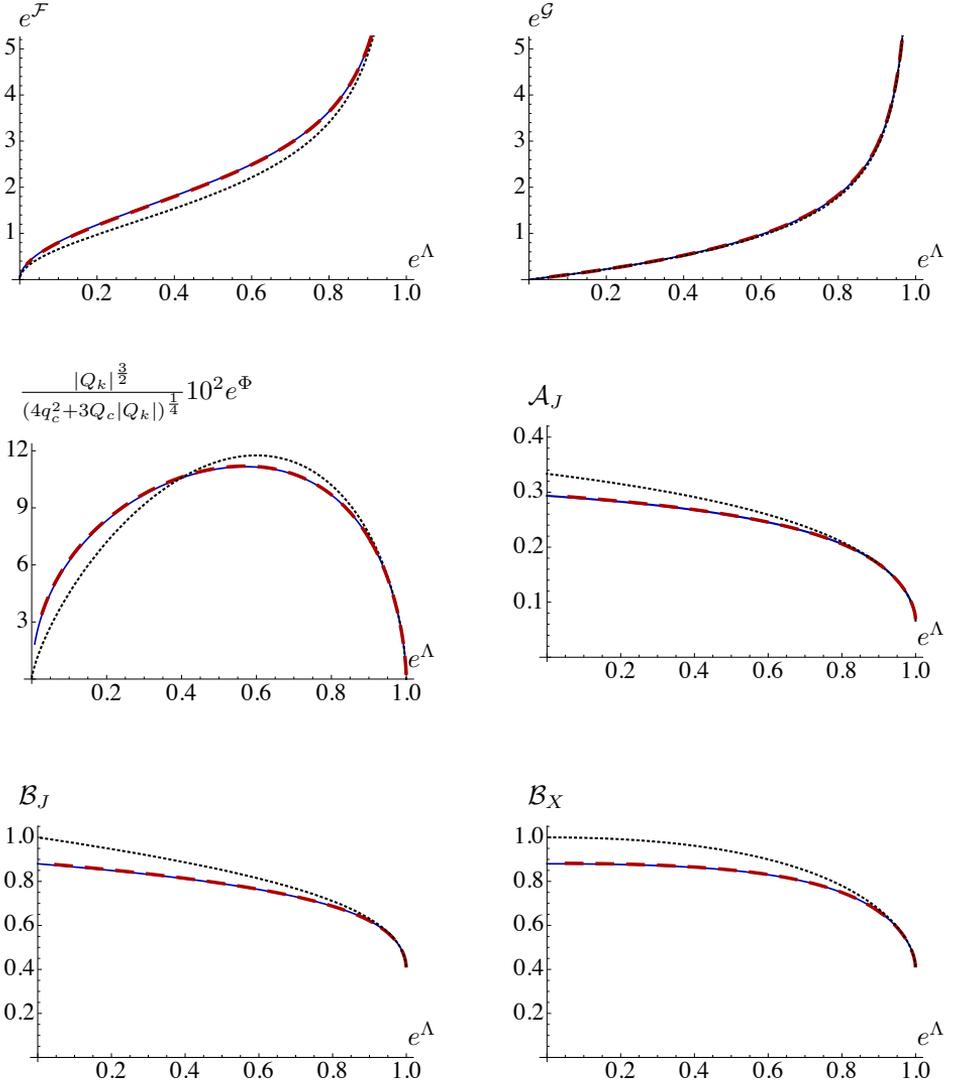

One indication comes from the fact that the eleven-dimensional curvature at the horizon diverges as we remove it. This suggests that a naked singularity is uncovered in this limit. This is illustrated  in Figure~\ref{fig:Ricci}, where we show the Ricci scalar in terms of the horizon value of $e^\Lambda$,  $\lambda_h$, in order to obtain a manifestly gauge-invariant plot.  It can be seen that the scalar curvature at the horizon diverges as $R\sim \lambda_h^{-2/3}$ in the limit $\lambda_h\to0$, that is, as $u_h\to u_N$.
This behaviour is of course inherited from that of the metric functions. Indeed, it is possible to infer their dependence on $\lambda_h$ near zero from the numerics. For the internal components we obtain
\begin{equation}
e^\FF \sim \lambda_h^{1/2}\,,\qquad \qquad\qquad e^\GG \sim \lambda_h\,. 
\end{equation}
It turns out that this is precisely their behaviour in the IR of the $\mathds{B}_8$ family, which is generically $\mathds{R}^4\times{\rm S}^4$ (see equation~\eqref{eq:8Dgapped}). The transverse geometries in eleven dimensions\footnote{As given by \eqref{eq:8transverse}, with the metric functions found numerically.} obtained by removing the horizon are thus perfectly regular, capping off smoothly at a certain $u_N$. Furthermore, the coincidence with a $\mathds{B}_8$ metric goes beyond the IR and extends to the entire solution. In Figure~\ref{fig.ZeroEntropyLimit} (top) we compare the metric functions that describe the transverse metric for the solution with lowest entropy that we constructed for $b_0=2/5$ with those of the $\mathds{B}_8$ metric that ends at the same value of $u_N$. They coincide within numerical precision except in a tiny IR region due to the small horizon that still exists in the thermal solution. We conclude that, as far as the eight-dimensional transverse geometry is concerned, the zero-entropy limit $u_h\to u_N$ describes a regular $\mathds{B}_8$ metric with the end-of-space at $u_N$ instead of the $u_s$ of the corresponding ground state.

The singularity in the full eleven-dimensional metric must then come from the warp factor. As we approach the zero-entropy limit it goes as
\begin{equation}\label{divh}
\mathbf{h}\sim \lambda_h^{-2}\,.
\end{equation}
Again, this behaviour can be identified with that of a known solution. As explained in Chapter~\ref{Chapter2_B8family}, for the horizonless solutions to be regular it is necessary to include internal fluxes (describing fractional branes) with fine-tuned values. If these values are not the correct ones the warp factor is IR-singular, with a divergence given exactly by Eq.~\eqref{divh}. We thus conclude that the zero-entropy limit of our black-brane solutions is a supersymmetric solution with fluxes, but the limiting values of these fluxes are not the correct ones to render the solution regular in the IR.  This can be seen in Figure~\ref{fig.ZeroEntropyLimit}, where we compare, for $b_0=2/5$, the lowest-entropy black brane solution that we constructed  (i) with the supersymmetric solution to the BPS equations in \eqref{BPSsystem_fluxes} with the end-of-space at $u_N(2/5)$, and  (ii) with the regular supersymmetric ground state. The first two backgrounds agree within numerical precision except in a tiny IR region due to the impossibility of reaching $u_h\to u_N$ numerically, while the regular one is clearly distinct.  

In summary, in the limit in which we remove the horizon and the entropy vanishes we find a supersymmetric solution which is nevertheless singular, since the subleading coefficients that we recover are not the ones that ensure regularity of the ground state. The singularity is good in the sense of \cite{Gubser:2000nd} because it can be hidden behind a horizon. It is interesting that this limit of zero entropy is reached at a finite temperature, as we will see  in detail in the next Section. 

\vfill

\section{Thermodynamics and the phase diagram}
\label{sec:thermodynamics}

In this Section we will discuss the main thermodynamic properties of the black branes we have constructed and confirm the presence of phase transitions between themselves and between them and the low-temperature phases described in Section~\ref{LowT}. In this way we will construct   the phase diagram in the $\left(b_0,T\right)$-plane for the entire family of theories. 

\subsection{Thermodynamic quantities}

The renormalised four-dimensional bulk action $S_{\mbox{\footnotesize ren}}$ describing the system is obtained in Appendix~\ref{app:4Deffectivetheory}. From this it is possible to extract the different thermodynamic quantities. The free energy density is given by
\begin{equation}
F= -\frac{S_{\mbox{\footnotesize ren}}}{\beta V_2}\,,
\end{equation}
with $V_2$ the (infinite) volume in the spatial directions and $\beta$ the period of the compact Euclidean time. This period, which is related to the temperature through equation~\eqref{beta}, is fixed in the black-brane  geometries by requiring the absence of conical singularities. As usual the Bekenstein--Hawking entropy is given by the area of the horizon. Using the known UV and horizon expansions \eqref{UVexpansions} and \eqref{Horizon_expansions} to evaluate these expressions, we obtain the following dimensionless values for the free energy, entropy and temperature 

\vfill\newpage

\begin{eqnarray}\label{themo_quantities}
\overline{F}&=&\frac{2\kappa_4^2 }{|Q_k|^5}\ F =-\frac{411}{2}-6 f_4-2 f_5+\frac{3}{2}  \mathsf{b}_5\,,\nonumber\\ [2mm]
\overline{S}&=&\frac{2\kappa_4^2}{|Q_k|^3(4q_c ^2+3Q_c|Q_k|)^{1/2}} \ S =\frac{64 \pi  f_h^4 g_h^2 \sqrt{h_h}}{\lambda _h^2}\,,\\ [2mm]
\overline{T}&=&\frac{(4q_c^2 + 3Q_c|Q_k|)^{1/2}}{|Q_k|^2}\ T =- \frac{1}{4\pi} \frac{\mathsf{b}_hu_h^2}{\sqrt{h_h}}\,,\nonumber
\end{eqnarray}
where $\mathsf{b}_5$ is understood to be given by \eqref{b5_conserved} in terms of horizon data. Similarly, the internal energy density $E$  and pressure $P$, computed from the energy-momentum tensor in \eqref{EMtensor}, are given by
\begin{equation}
\frac{2\kappa_4^2 P}{|Q_k|^5} = \frac{411}{2} +6 f_4+2 f_5-\frac{3}{2}  \mathsf{b}_5\,, \qquad
\overline{E}=\frac{2\kappa_4^2 E}{|Q_k|^5} =-\frac{411}{2}-6 f_4-2f_5-\frac{7}{2}  \mathsf{b}_5\,,
\end{equation}
which fulfil the expected thermodynamic relations $F=-P$ and $F=E-TS$. Finally, it must be verified that 
\begin{equation}
S= - \frac{\dd F}{\dd T}\,,
\end{equation}
which can be used as a check of the numerical results. Other quantities like the specific heat and the speed of sound can be straightforwardly computed from these thermodynamic potentials as:
\begin{equation}
\label{cs}
c_v = \frac{\dd E}{\dd T} \sac c_s^2 = \frac{\dd P}{\dd E} =\frac{S}{c_v} = \frac{S}{T}\frac{\dd T}{\dd S} \,.
\end{equation}

\subsection{Phase diagram}
\label{sec:phase_diagrams}

For each value of  $b_0$, namely for each gauge theory in the family, we must determine the preferred solution at each temperature. The first type of solutions that compete are those of Chapter~\ref{Chapter2_B8family}, which are obtained from the horizonless, supersymmetric, regular solutions simply by compactifying the Euclidean time. The period of this thermal circle is arbitrary, so these geometries exist for any  temperature and, in our renormalisation scheme, they have vanishing  free energy and entropy. The second type of solutions are the black branes found in Section~\ref{HighT}. Their free energy and entropy are given in terms of UV and horizon data by Eq.~\eqref{themo_quantities} and have a non-trivial dependence on the temperature. If several solutions exist at a given temperature, the one with lower free energy will be thermodynamically preferred. Since the free energy of the first type of solutions vanishes, the black branes will dominate if their free energy is negative. At points where the free energy of black branes crosses zero there is a phase transition to the gapped state.   The nature of the phase transition depends on the value of $b_0$. We identified two special values of $b_0$ at which the qualitative features change: 
\begin{equation}
\label{below}
b_0^\textrm{critical}\approx 0.6815 \, ,\qquad b_0^\textrm{triple}\approx 0.6847 \,.
\end{equation}
Note that these are very close to one another. We will come back to this in Section~\ref{sec:conclusions}. These two values give rise to the following three regions:  
\begin{figure}[t!]
	\begin{center}
		\begin{subfigure}{.45\textwidth}
			\includegraphics[width=\textwidth]{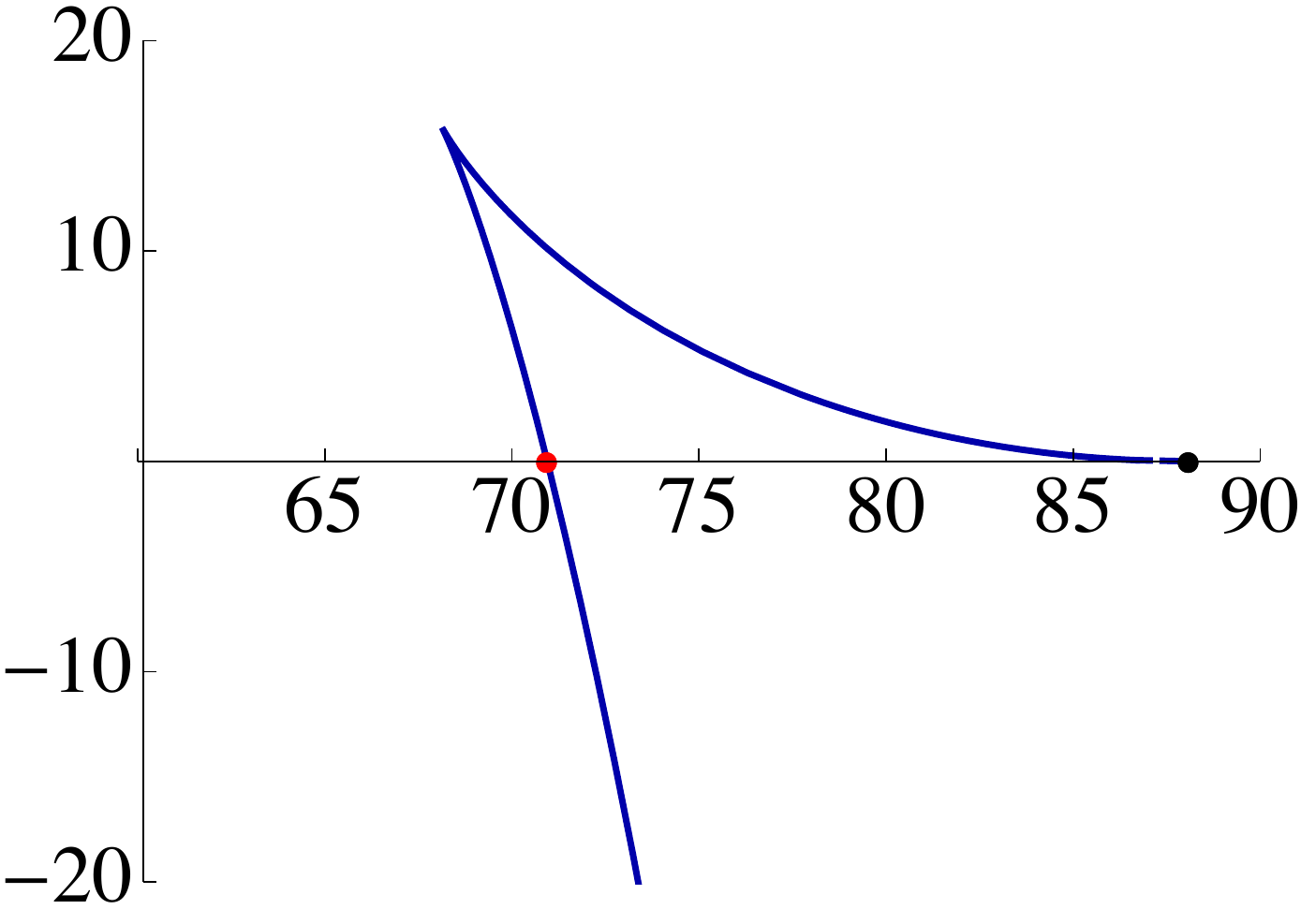} 
			\put(-160,120){$10^{-4}\ \overline F$}
			\put(-10,65){$\overline T$}
		\end{subfigure}\hfill
		\begin{subfigure}{.45\textwidth}
			\includegraphics[width=\textwidth]{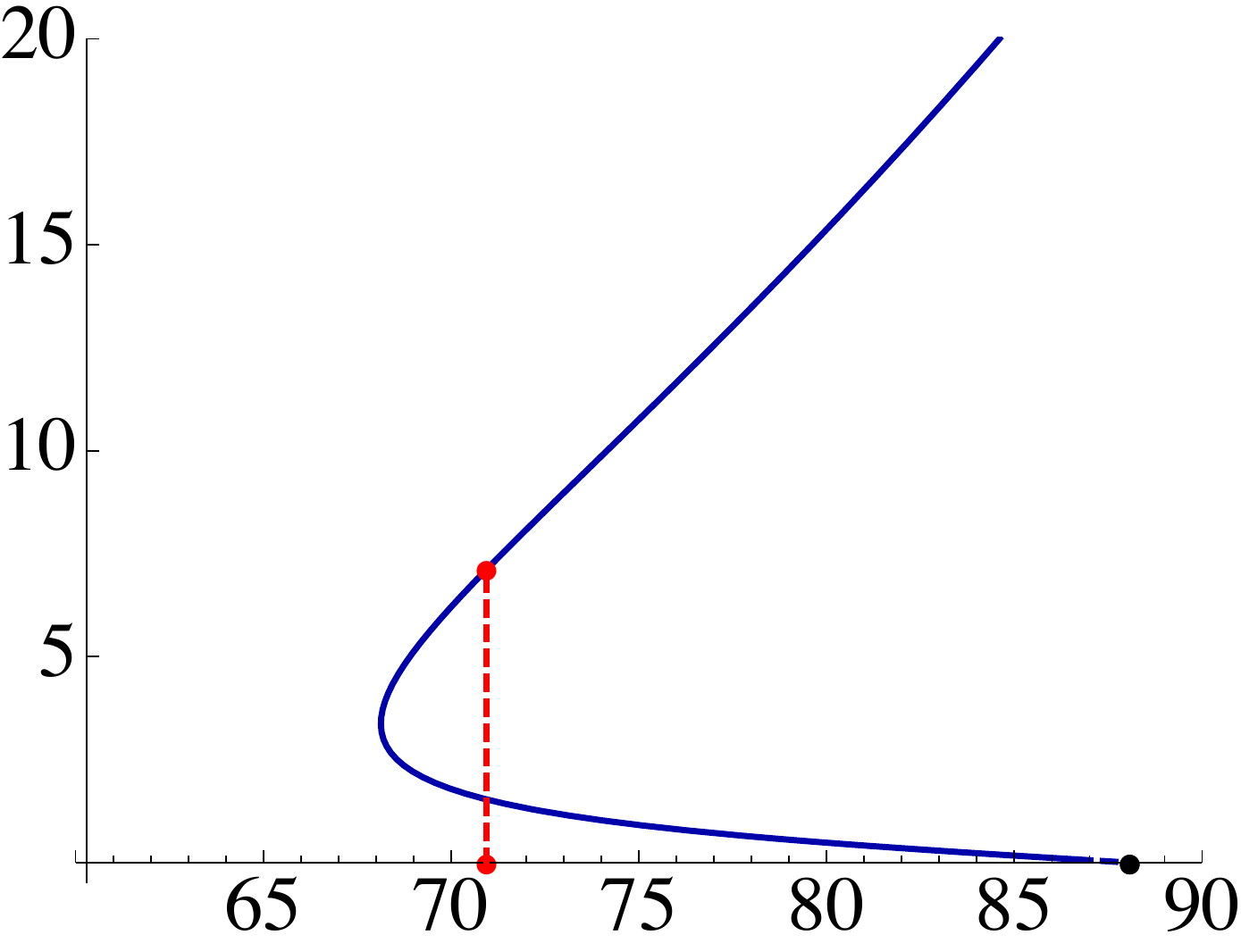} 
			\put(-160,125){$10^{-4}\  \overline S$}
			\put(-10,20){$\overline T$}
		\end{subfigure}
		\caption{\small Free energy density (left) and entropy density (right) as a function of the temperature for black branes in the theory with $b_0=0.7902$ (Case A). When the free energy crosses the axis, there is a first-order phase transition at which the entropy changes discontinuously. The dashed, red line indicates  the critical temperature. The region close to the horizontal axis of the blue curves, shown with dashes, is the result of an extrapolation to zero entropy.}\label{fig.TypeI_PT}
	\end{center}
\end{figure} 

\begin{figure}[t]
	\begin{center}
		\begin{subfigure}{.45\textwidth}
			\includegraphics[width=\textwidth]{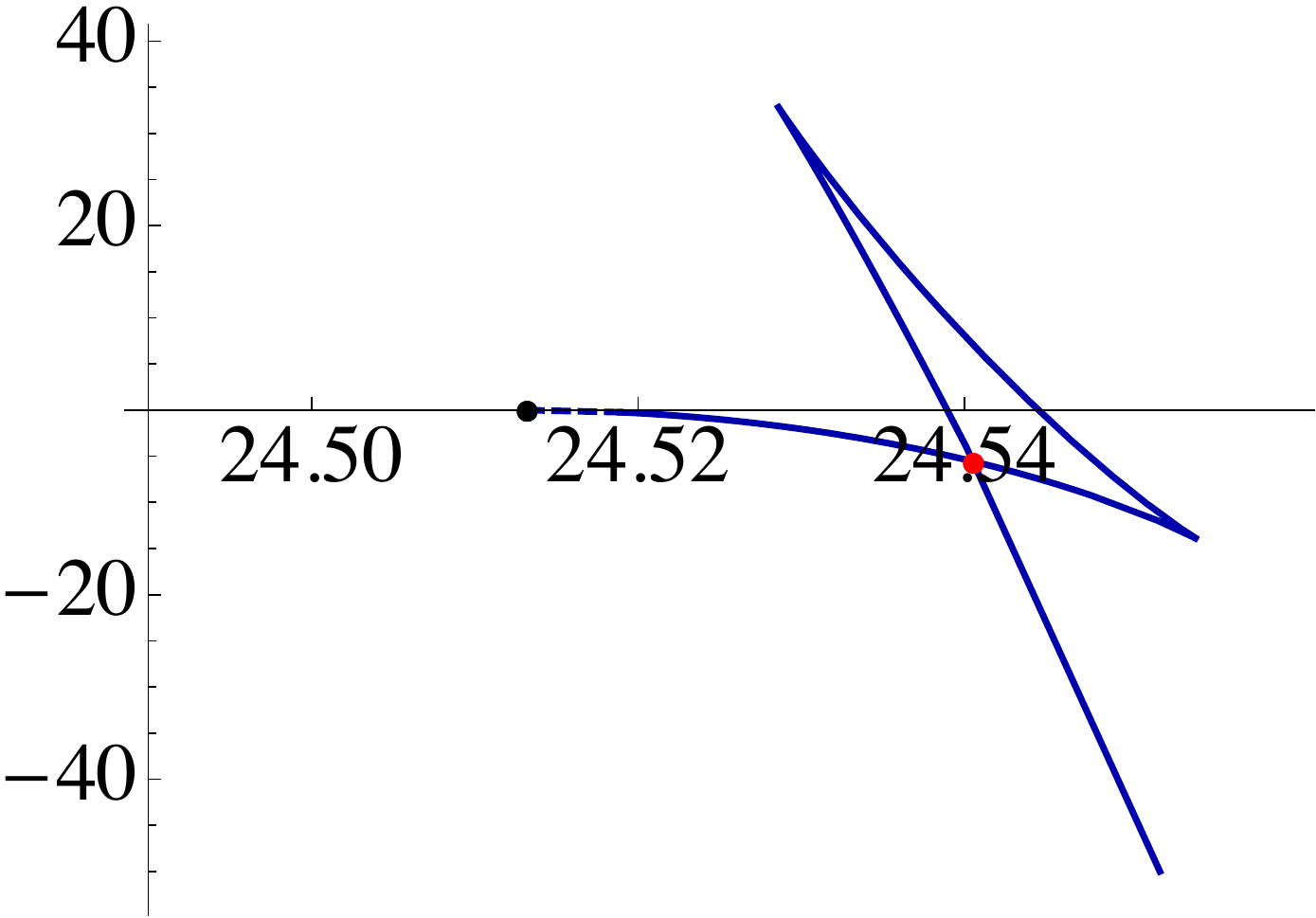} 
			\put(-145,115){$\overline F$}
			\put(-5,65){$\overline T$}
		\end{subfigure}\hfill
		\begin{subfigure}{.45\textwidth}
			\includegraphics[width=\textwidth]{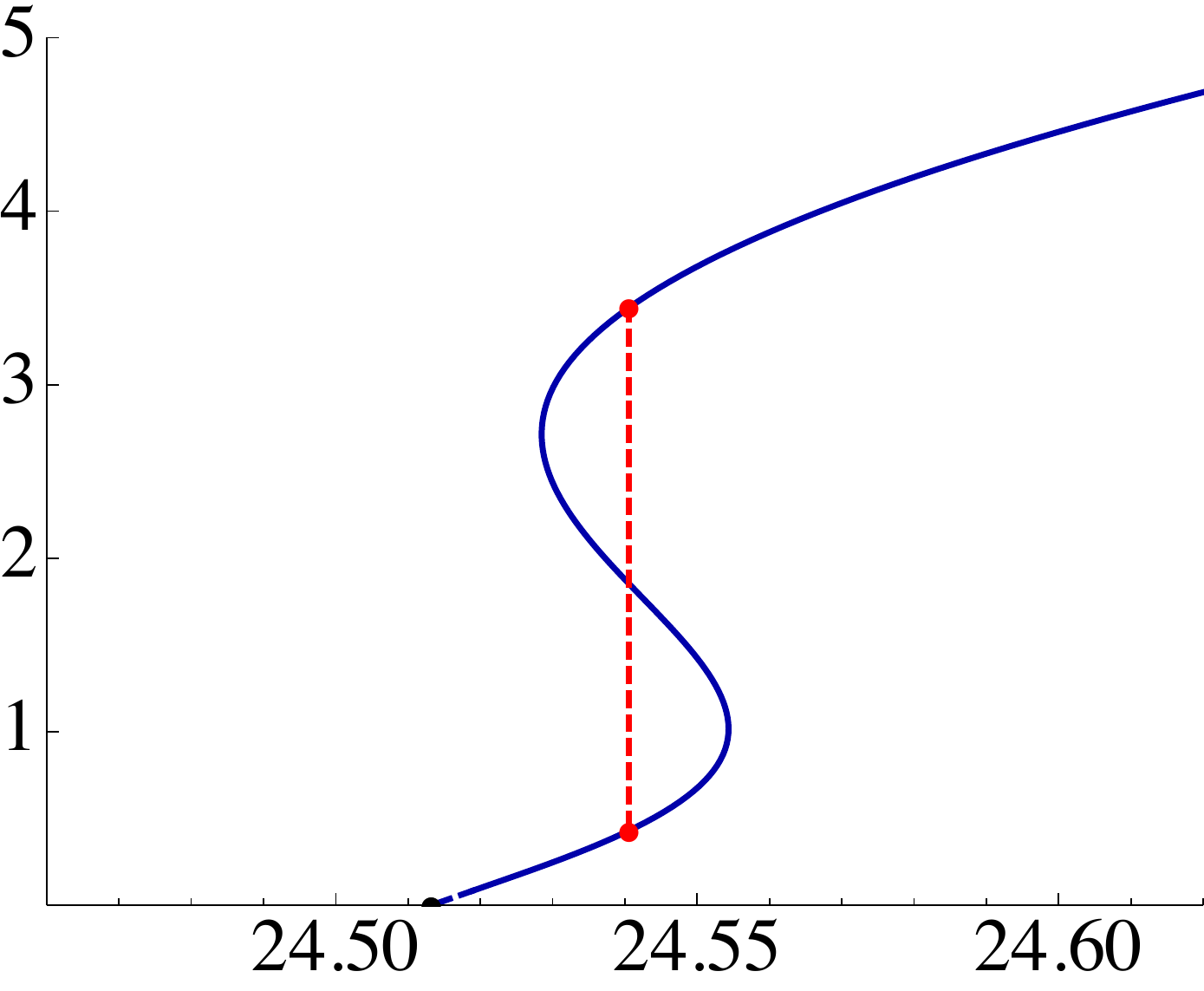} 
			\put(-160,135){$ 10^{-3}\ \overline S$}
			\put(-10,15){$\overline T$}
		\end{subfigure}
		\caption{\small Free energy (left) and entropy (right) as a function of the temperature for black branes in the theory with $b_0=0.6835$ (Case B). As the temperature decreases, there is a first order phase transition between two branches of black branes.
			At the transition the entropy changes from some finite value to another finite value, represented by the dashed red line. Decreasing further the temperature, one would find another phase transition from the second black brane phase to the regular phase without a horizon. This happens at some finite value of the temperature and vanishing entropy. The region close to the horizontal axis of the blue curve, shown with dashes, is the result of an extrapolation to zero entropy.}\label{fig.TypeII_PT}
	\end{center}
\end{figure}

\begin{figure}[t]
	\begin{center}
		\begin{subfigure}{.45\textwidth}
			\includegraphics[width=\textwidth]{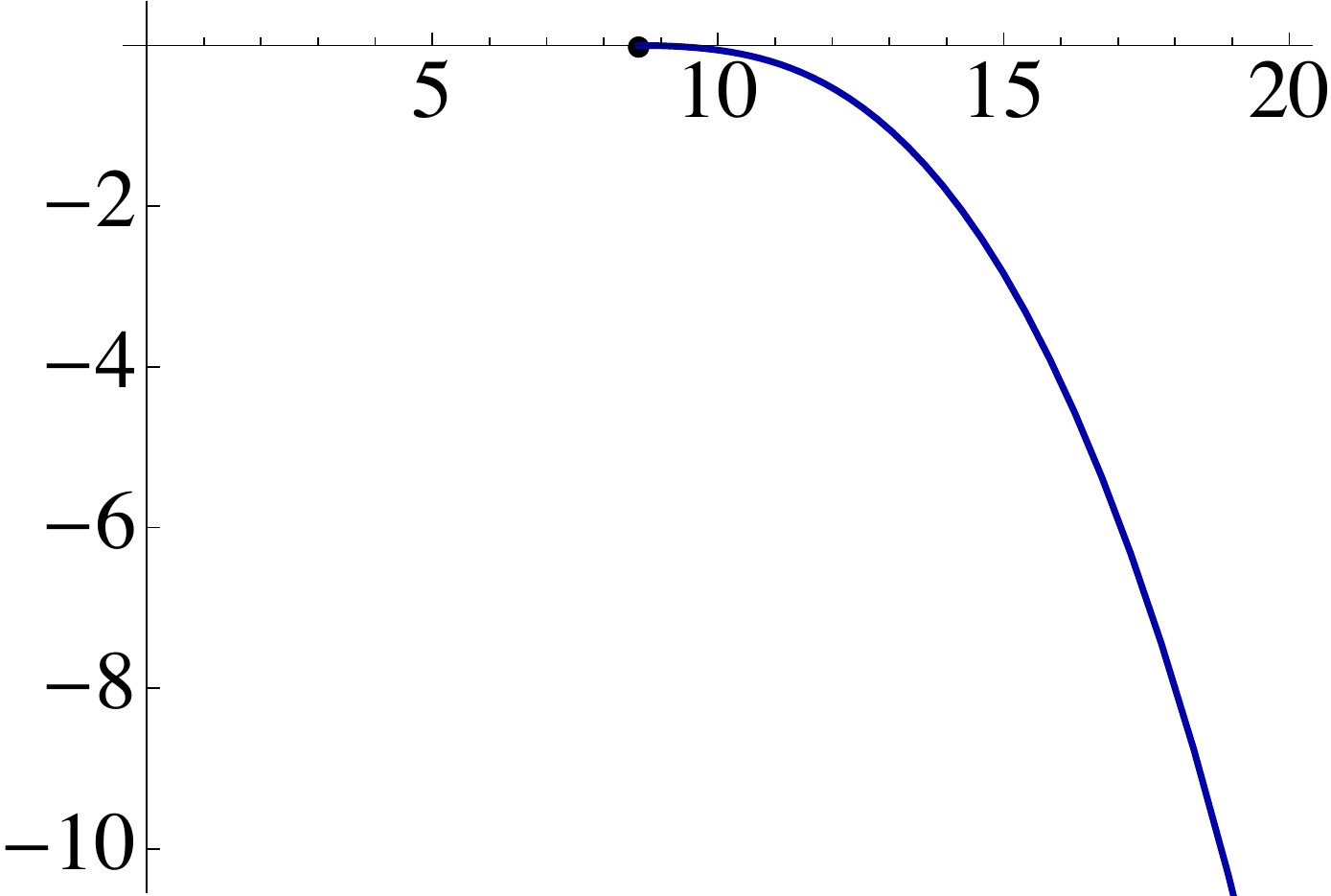} 
			\put(-150,120){$10^{-4}\ \overline F$}
			\put(-10,80){$\overline T$}
		\end{subfigure}\hfill
		\begin{subfigure}{.45\textwidth}
			\includegraphics[width=\textwidth]{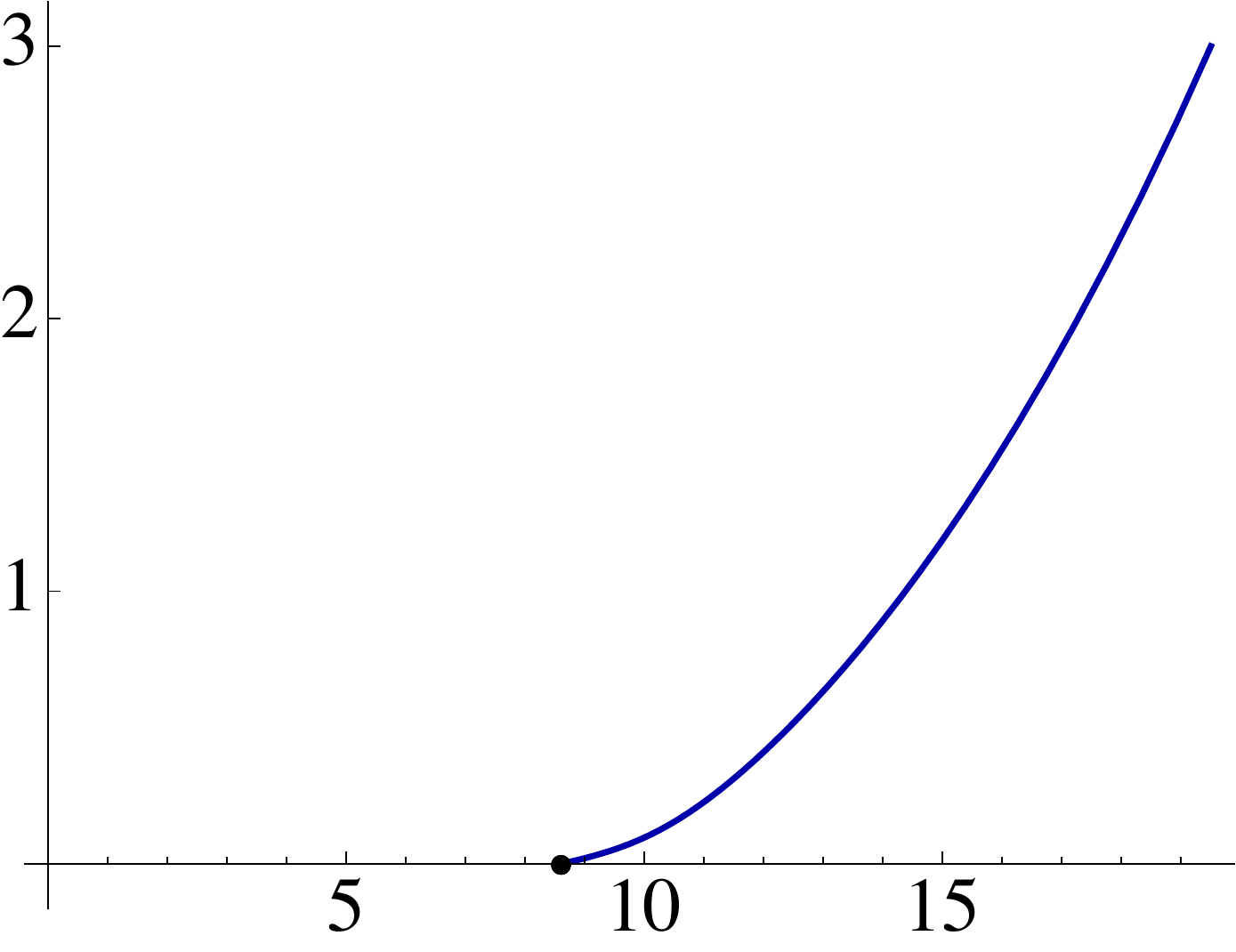} 
			\put(-160,130){$10^{-4} \ \overline S$}
			\put(-10,15){$\overline T$}
		\end{subfigure}
		\caption{\small Free energy density (left) and entropy density (right) as a function of the temperature for black branes in the theory with $b_0 = 0.5750$ (Case C). In this example, there is a single phase transition that takes place at finite temperature and zero entropy. The region close to the horizontal axis of the blue curve, shown with dashes, is the result of an extrapolation to zero entropy.}\label{fig.TypeIII_PT}
	\end{center}
\end{figure}
\begin{figure}[t!]
	\begin{center}
		\begin{subfigure}{.92\textwidth}
			\includegraphics[width=\textwidth]{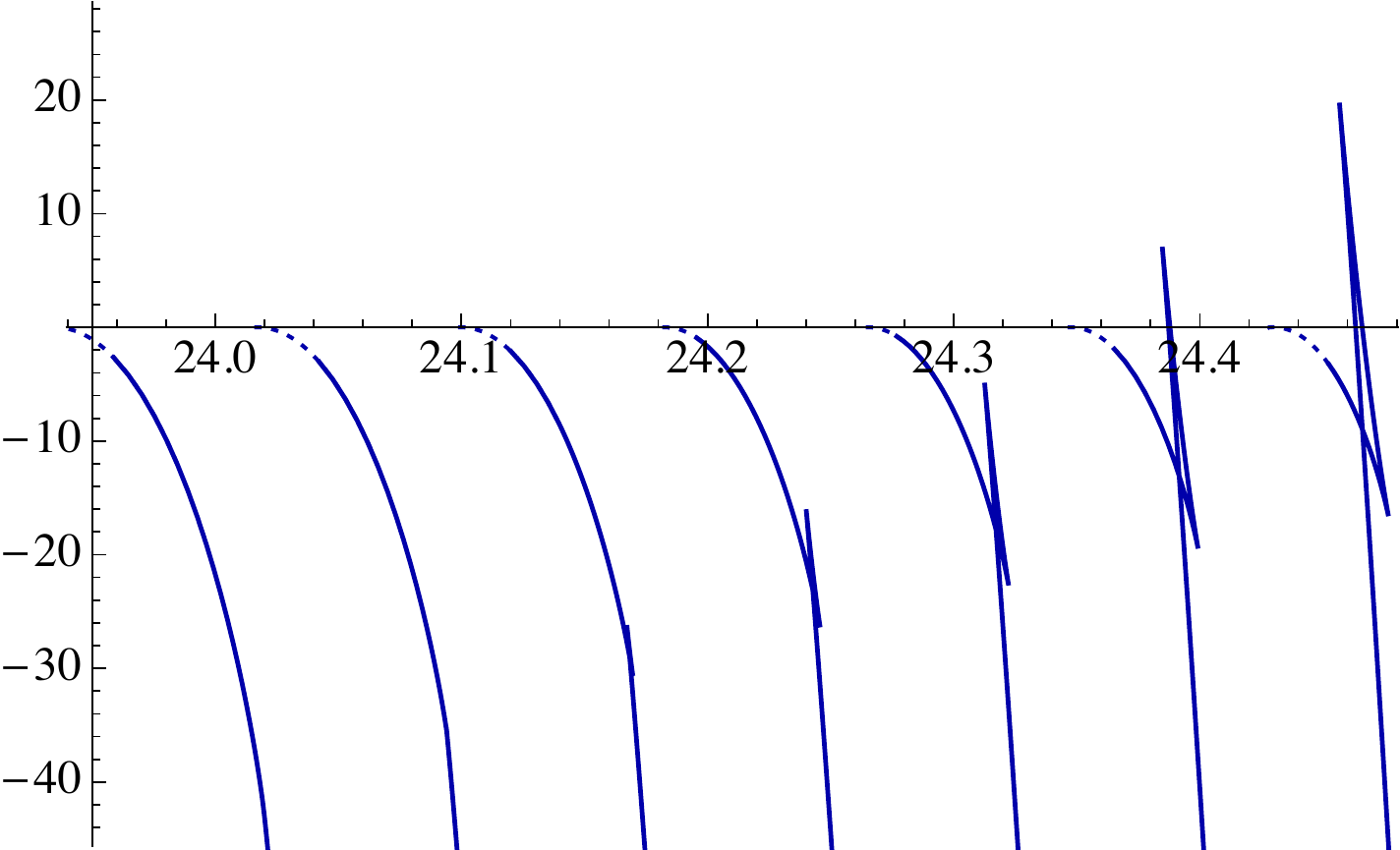} 
			\put(-290,180){$\overline F$}
			\put(-3,130){$\overline T$}
		\end{subfigure}\vspace{9mm}
		\begin{subfigure}{.92\textwidth}
			\includegraphics[width=\textwidth]{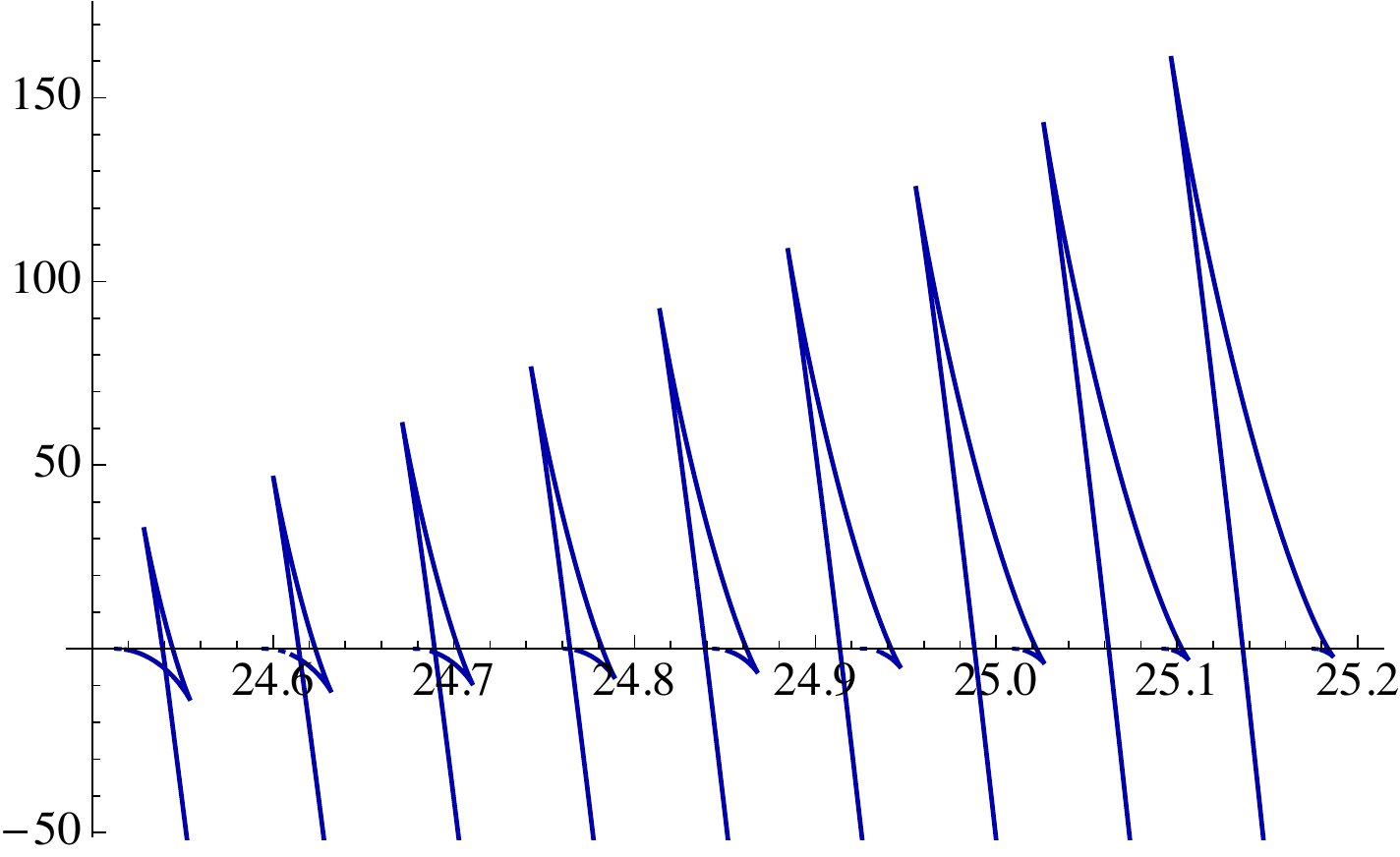} 
			\put(-290,180){$\overline F$}
			\put(-3,55){$\overline T$}
		\end{subfigure}\vspace{3mm}
		\caption{\small Free energy density  for different theories with $b_0$ varying from $b_0=0.6813 \lesssim b_0^\textrm{critical}$ to $b_0=0.6861\gtrsim b_0^\textrm{triple}$ from left to right and from top to bottom, with an approximate separation of $\Delta b_0 \approx 0.0003$ between adjacent curves. A triangle develops as we decrease $b_0$ (bottom panel). If $b_0$ is decreased further then the triangle disappears (top panel). The region close to the horizontal axis of the curve, shown with dashes, is the result of an extrapolation to zero entropy.}
		\label{fig:FreeEnergyCrossOver}
	\end{center}
\end{figure}
\begin{figure}[t!]
	\begin{center}
		\includegraphics[width=.75\textwidth]{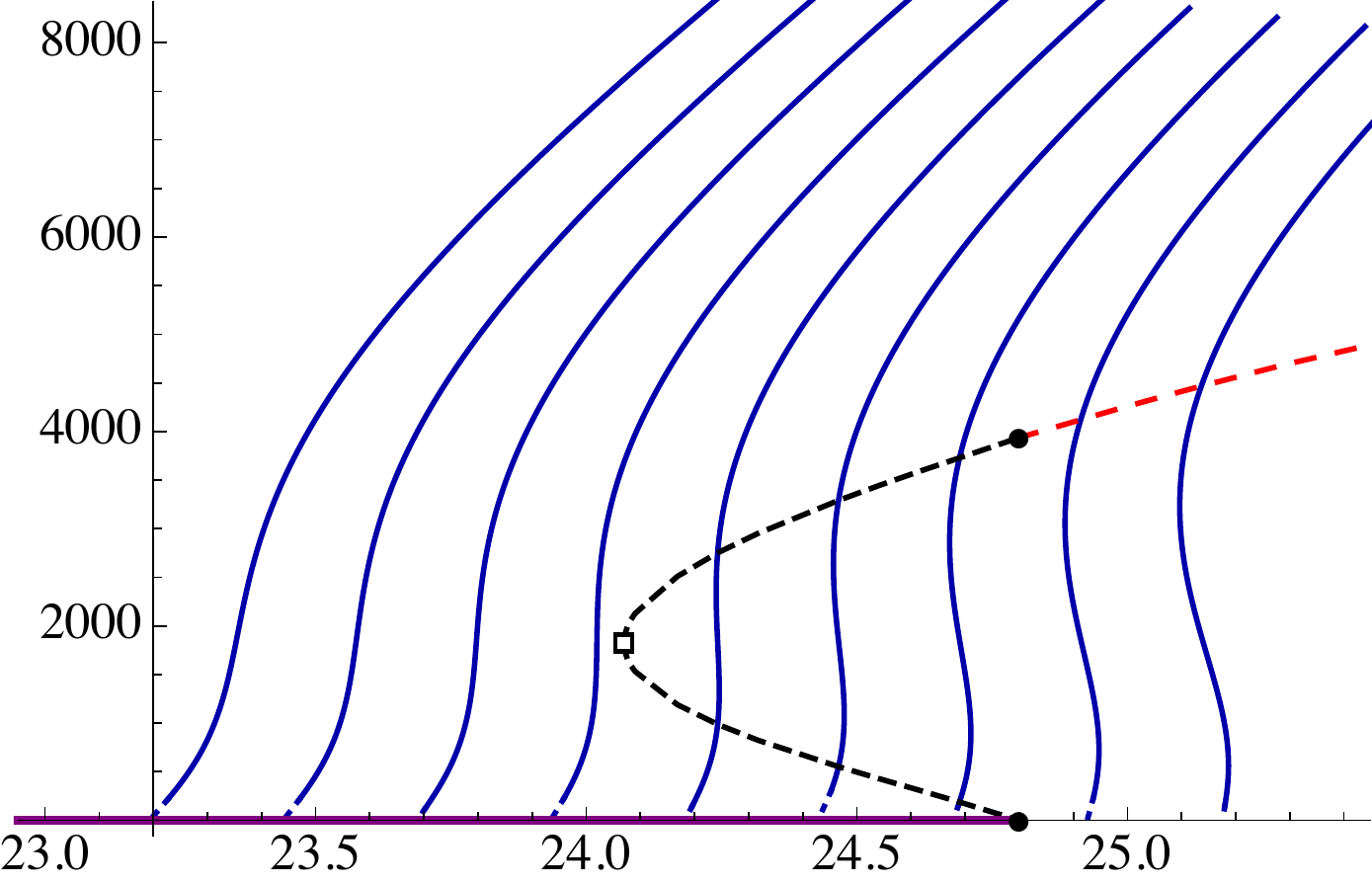} 
		\put(-220,160){\rotatebox{0}{$\overline S $}}
		\put(-5,20){$\overline T$}
		\caption{\small Entropy density for different theories with $b_0$ varying from $b_0=0.6783 \lesssim b_0^\textrm{critical}$ to \mbox{$b_0=0.6861\gtrsim b_0^\textrm{triple}$} from left to right, with an approximate separation of $\Delta b_0 \approx 0.001$ between adjacent curves. The dashed, red curve indicates first-order phase transitions between a black brane phase and a regular horizonless phase with a discontinuous jump in the entropy density. The dashed, black curve indicates first-order phase transitions between two black branes. The turning point of this line, indicated with a square, corresponds to the second-order phase transition at the critical point. The solid, purple line on the axis indicates first-order transitions between the zero-entropy limit of a black brane branch and a regular horizonless solution. The entropy density is continuous across these transitions. The phase transition corresponding to the triple point is indicated with two black dots. The region close to the horizontal axis of the blue curves, shown with dashes, is the result of an extrapolation to zero entropy.}
		\label{fig:EntropyCrossOver}
	\end{center}
\end{figure}
\begin{figure}[t!]
	\begin{center}
		\begin{subfigure}{.49\textwidth}
			\includegraphics[width=\textwidth]{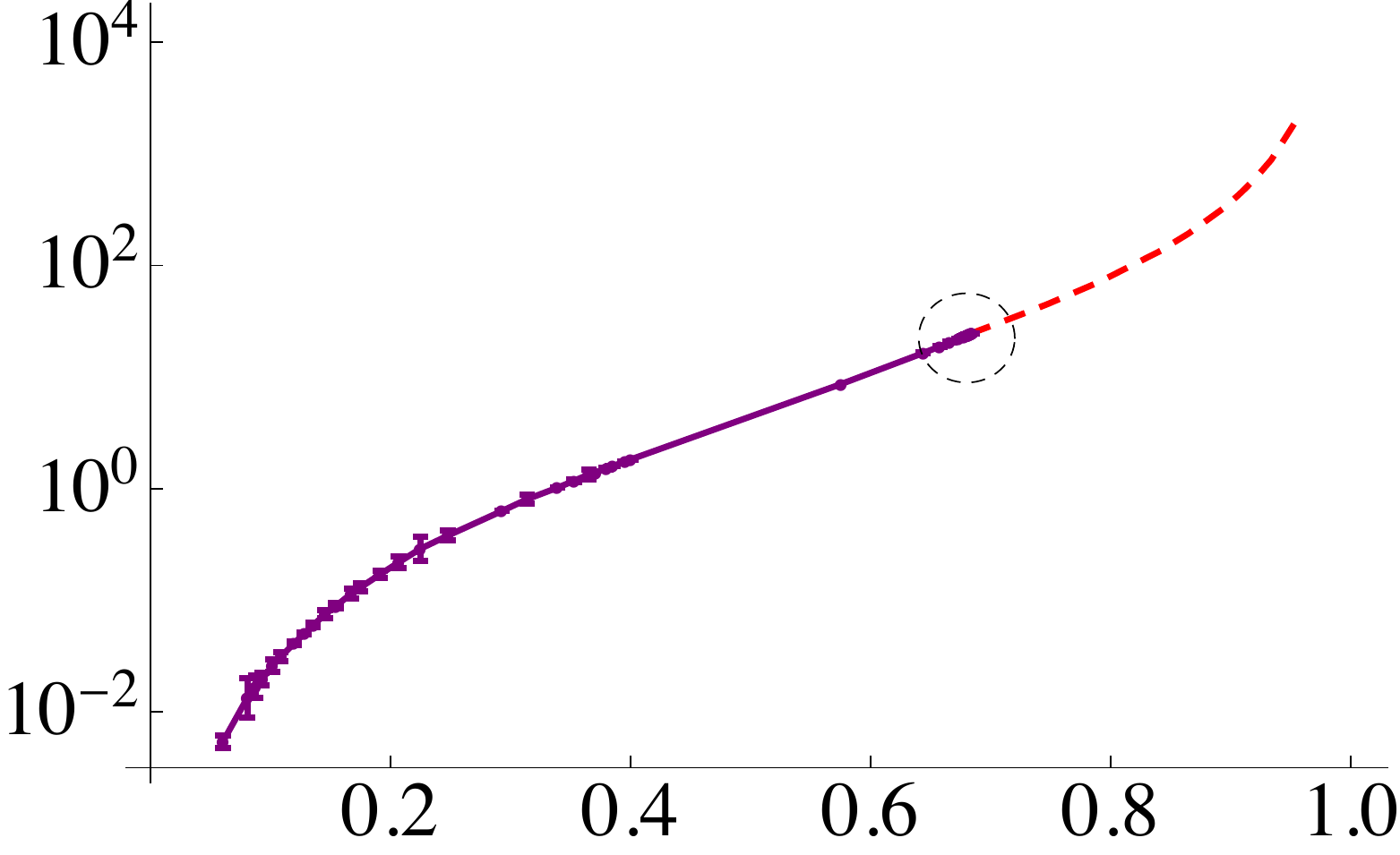} 
			\put(-140,100){\rotatebox{0}{$\overline T_c $}}
			\put(-10,20){$b_0$}
		\end{subfigure}\hfill
		\begin{subfigure}{.49\textwidth}
			\includegraphics[width=\textwidth]{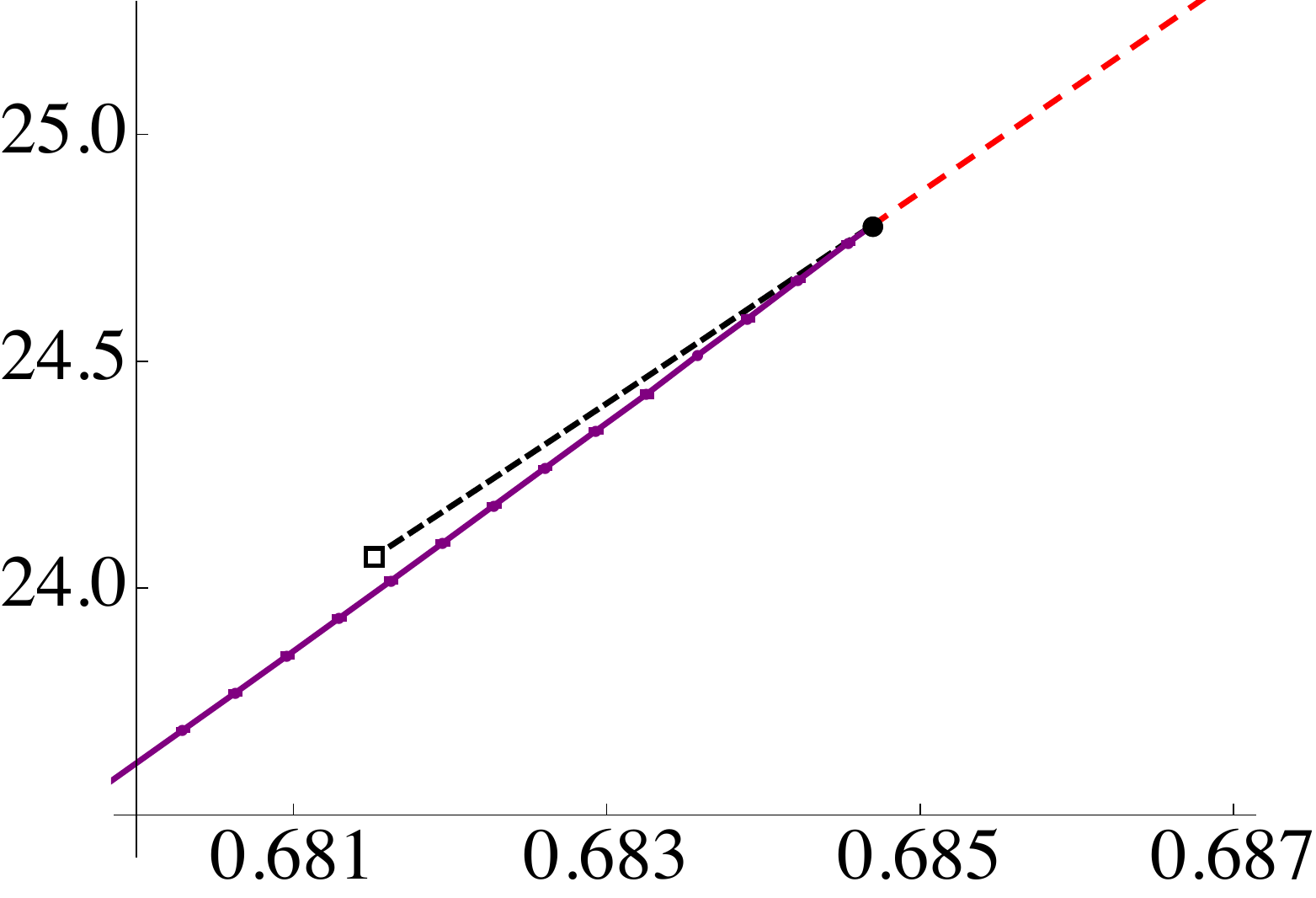} 
			\put(-140,100){\rotatebox{0}{$\overline T_c $}}
			\put(-10,20){$b_0$}
		\end{subfigure}
		\caption{\small Phase diagram for the entire family of gauge theories (left) and zoomed in version of the circle around $b_0\approx 0.68$. The dashed, red line represents  first-order transitions between a black brane and a regular horizonless solution with a discontinuity in the entropy density (Case A). The dashed, black line corresponds to phase transitions between two black brane solutions (Case B). The dotted, purple line indicates first-order transitions between the zero-entropy limit of a black brane branch and a regular horizonless solution. The entropy density is continuous across these transitions (Cases B and C). The purple curve requires an extrapolation to zero entropy. We have estimated the associated error bars, shown explicitly in the figure, by comparing a linear and a quadratic extrapolation.}
		\label{fig:PhaseDiagram}
	\end{center}
\end{figure}

\begin{description}
	\item[{\textbf{Case A:}} Phase transitions for \mbox{$b_0 \in (b_0^\textrm{triple},1]$}.]
	In this range the situation is similar to the typical Hawking--Page phase transition.
    The free energy crosses the horizontal axis at some critical temperature $T_c$. The preferred phase above $T_c$ is the black brane and below $T_c$ it is the horizonless, regular solution. Therefore at $T_c$ there is a phase transition between a gapped and an ungapped phase. At the critical temperature the entropy density jumps between some finite value and zero, signalling that the transition is first-order. In Figure~\ref{fig.TypeI_PT} we show the free energy and the entropy densities as a function of temperature for a representative case with  $b_0=0.7902$.
	\item[{\textbf{Case B:}} Phase transitions for  
	\mbox{$b_0 \in (b_0^\textrm{critical},b_0^\textrm{triple})$}.]
	In this small range the theories exhibit two phase transitions. 
	This is illustrated in Figure~\ref{fig.TypeII_PT}, where we show the free energy and the entropy of the black branes for the theory with $b_0=0.6835$. Decreasing the temperature from asymptotically high values we first find a first-order phase transition between two black brane geometries at which the entropy density changes discontinuously. Therefore this transition takes place between two ungapped phases and is indicated in Figure~\ref{fig.TypeII_PT} by the dashed, red line. Decreasing the temperature further along the low-temperature black-brane branch we find that the entropy and the free energy densities vanish at a finite value of the temperature. In this zero-entropy limit we recover the horizonless, singular solution described in Section~\ref{sec:zeroEntropyLimit}. Strictly speaking, since this solution is singular, its temperature is undetermined. However, the limit along the regular branch of black brane solutions results in a fixed, non-zero value of the temperature, indicated by the black dot in Figure~\ref{fig.TypeII_PT}. At this temperature we expect a transition between the supersymmetric,  horizonless, singular solution and the supersymmetric,  horizonless, regular solution with the same value of $b_0$. This transition is analogous to that in Case A  except for the fact that the entropy density is continuous across the transition. Nevertheless, the transition is still first-order because the solution changes discontinuously, as illustrated in Figure~\ref{fig.ZeroEntropyLimit}. In the gauge theory this would be reflected in a discontinuity in observable quantities such as $n$-point functions. Note that, strictly speaking, the supergravity description breaks down sufficiently close to the transition since the curvature diverges at that point, as shown in Figure~\ref{fig:Ricci}. 
	
	\item[{\textbf{Case C:}} Phase transitions for  \mbox{$b_0 \in (0, b_0^\textrm{critical})$}.]
	In this range of values the phase transition between black branes disappears. The qualitative behaviour of the free energy and entropy densities  is similar to that in Figure~\ref{fig.TypeIII_PT}, where we show the results for $b_0 = 0.5750$. For the theories in this range there is a unique phase transition between the ungapped black brane phase and the regular gapped solution. Again, this happens at zero entropy but finite temperature, as in Case B. 
\end{description}

The change from one behaviour to another happens smoothly as we vary $b_0$. Essentially, Case B is an intermediate scenario between Cases A and C. This can be seen by plotting the free energy and entropy densities for different values of the parameter in a small region that contains that of Case B entirely. The result is shown in Figs.~\ref{fig:FreeEnergyCrossOver} and \ref{fig:EntropyCrossOver}, respectively. The largest value of the parameter in the bottom panel of Figure~\ref{fig:FreeEnergyCrossOver} is $b_0\approx0.6861 \gtrsim b_0^\textrm{triple}$ and corresponds to the rightmost curve. As we decrease $b_0$, a small triangle, ending on the axis, starts to grow below the unstable branch. The lower side of this triangle is thermodynamically stable but not dominant. Decreasing further the parameter $b_0$, the lower side of the triangle ends up crossing the previous stable branch and becomes favoured. This explains the change from Case A to Case B. At this point the system goes from having one first-order transition to having two of them. Therefore this is a triple point at which three phases can coexist. 

As the size of the second stable branch increases, the size of the triangle decreases. This explains the change from Case B to Case C: by reducing $b_0$ the triangle eventually disappears. This can be seen in the top panel of Figure~\ref{fig:FreeEnergyCrossOver}, where the leftmost curve corresponds to $b_0\approx0.6813 \lesssim b_0^\textrm{critical}$.  The  point where the triangle disappears is a critical point, namely the end of a line of first-order transitions. At this point the transition becomes second-order, as can be seen in the plot of the entropy density of Figure~\ref{fig:EntropyCrossOver}. 

With all this information we can draw the phase diagram of the entire family in the $\left(b_0,T\right)$-plane, which is shown in Figure~\ref{fig:PhaseDiagram}. The region above the curves of critical temperatures in the left panel is dominated by black branes and corresponds to  ungapped phases in the dual gauge theories. Below these curves, the regular horizonless solutions are dominant and the preferred phases are gapped. The right panel zooms into the range of values that include Case B, where two transitions take place.  The triple point, where the three lines meet, is indicated with a black dot. The critical point, where the line of first-order phase transitions between black branes ends, is indicated with a square.  

\section{Limiting cases}

\subsection{Quasi-conformal thermodynamics}

As described in Chapter \ref{Chapter2_B8family}, there are two special values of the parameter with distinct IR dynamics. When $b_0=0$ the ground state RG flow, denoted by $\B_8^\infty$, ends on a fixed point, as indicated by the leftmost arrow in Figure~\ref{fig:triangle}. The properties of this particular theory at finite temperature were already discussed around equation.~\eqref{FreeECFT}. In this Section we comment on the imprints that this fixed point leaves on the thermodynamics of flows passing nearby. 

\begin{figure}[t]
	\begin{center} 
		\includegraphics[width=.65\textwidth]{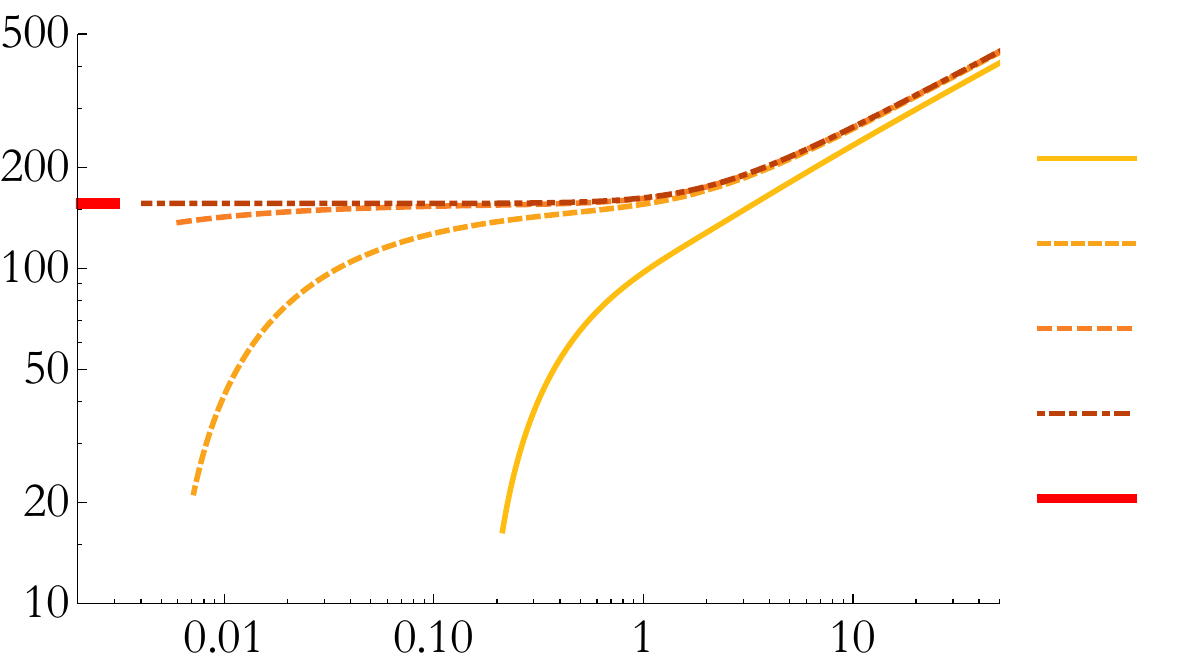} 
		\put(-220,130){$\overline S\  \overline{T}^{-2}$}
		\put(-30,5){$\overline T$}
		\put(0 ,95){$b_0 = 0.1914$}
		\put(0 ,79){$b_0 = 0.0604$}
		\put(0 ,62){$b_0 = 0.0191$}
		\put(0 ,45){$b_0 = 0 $ $(\B_8^\infty)$}
		\put(0, 30){OP | CFT}
		\caption{\small Dimensionless combination $S\,T^{-2}$ for different flows with $b_0\gtrsim0$. This quantity is constant in a CFT at finite temperature, as in the IR of $\Binf$.}
		\label{fig:quasi_conformal}
	\end{center}
\end{figure}

Conformal invariance dictates that when a three-dimensional CFT is heated up, the entropy grows as $S\propto T^2$. Flows with $b_0\gtrsim0$ approach the fixed point but eventually fail to reach it, so it is expected that the entropy should scale approximately as in a CFT for some range of temperatures. This behaviour is confirmed by the results plotted in Figure~\ref{fig:quasi_conformal}. For $\B_8^\infty$, the dimensionless quantity $S T^{-2}$ becomes nearly independent of $T$ for all temperatures below \eqref{Tconformal}. For black brane solutions with a small but non-zero value of $b_0$, a plateau develops before their zero-entropy limit is eventually reached at a finite temperature. The size and the steepness of the plateau, in which there is quasi-conformal behaviour, is controlled by the smallness of $b_0$. 
\begin{figure}[t!]
	\begin{center}
		\begin{subfigure}{.47\textwidth}
			\includegraphics[width=\textwidth]{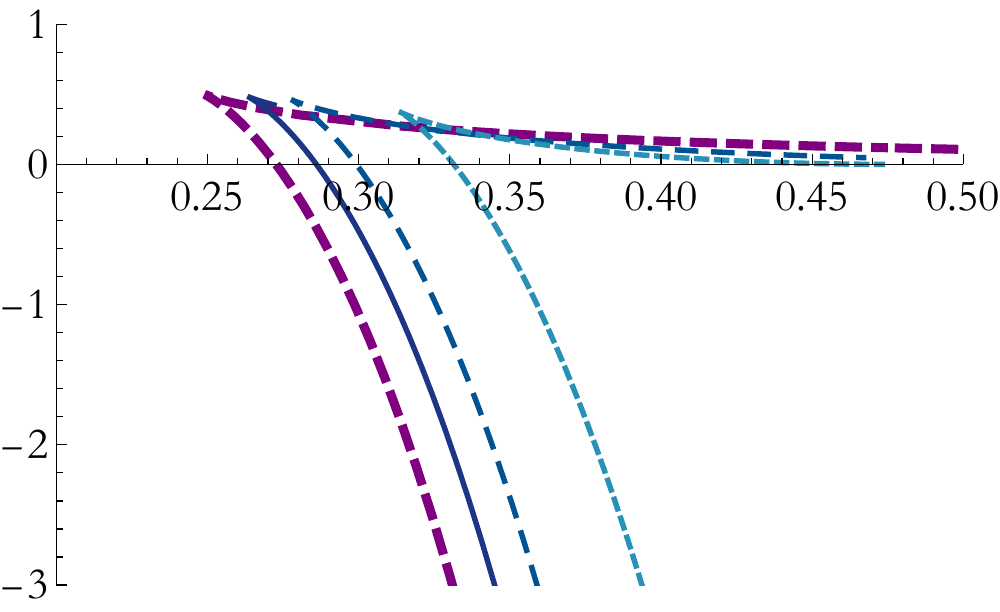} 
			\put(-145,100){$\overline F$}
			\put(-10,50){$\overline T$}
		\end{subfigure}\hfill
		\begin{subfigure}{.47\textwidth}
			\includegraphics[width=\textwidth]{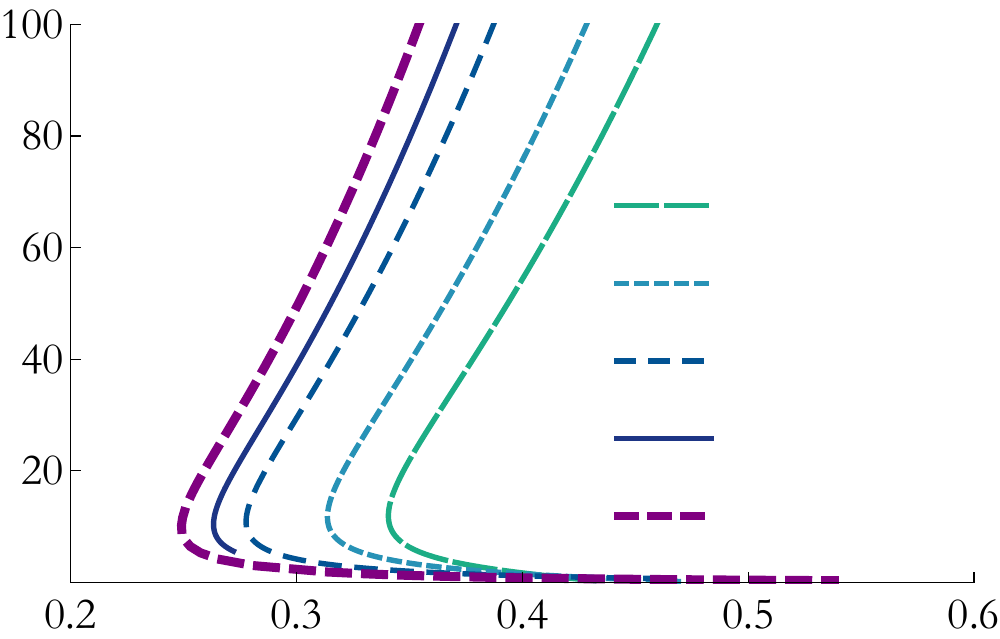} 
			\put(-140,100){$\overline S$}
			\put(-5,-10){$\overline T$}
			\put(-40,67){\footnotesize $b_0 = 0.7903$}
			\put(-40,55){\footnotesize $b_0 = 0.8400$}
			\put(-40,43){\footnotesize $b_0 = 0.9201$}
			\put(-40,31){\footnotesize $b_0 = 0.9581$}
			\put(-40,19){\footnotesize $b_0 = 1 $ $(\Bconf)$}
		\end{subfigure}
		\caption{\small Free energy (left) and entropy (right) for different $\B_8$ metrics with $b_0 \lesssim 1$, with the appropriate scaling of the charges, in dashed and dotted lines. In the limit $b_0\to1$, we recover those of $\Bconf$, shown in dashed purple line.} 
		\label{fig:QuasiConfining}
	\end{center}
\end{figure}
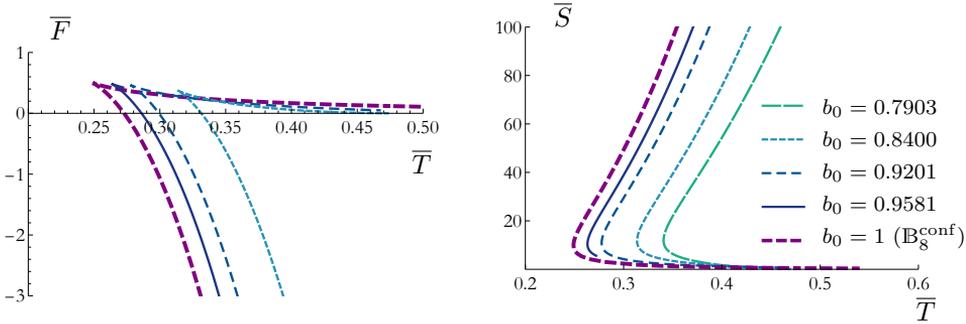

\subsection{Quasi-confining thermodynamics}

For $b_0=1$ the solution is $\Bconf$, at the right of Figure~\ref{fig:triangle}. Recall that this theory is not only gapped but truly confining.
In Chapter~\ref{Chapter2_B8family} we argue this was due to the vanishing of the CS level, $k=0$. Because of this, if we aim to understand the solution as a limit of the $\B_8$ family of metrics, we have to rescale the charge $Q_k$ --- related to the CS level by \eqref{eq:gauge_parameters} --- before taking $k\to0$ (see Section~\ref{sec:limiting} for further details). Moreover, the radial coordinate given by \eqref{ucoord} is no longer valid, so we define a new one through\footnote{In this radial coordinate, $\xi_H$ indicates the position of the horizon. It should not be confused with $\xi_h$, which is the value of $\BB_J$ at the horizon (see \eqref{Horizon_expansions})}
\begin{equation}
\dd r=-\frac{\rho_0}{\xi ^2 \sqrt{1-\xi ^4}}\ \dd\xi\,, \qquad\qquad \xi \in (0,\xi_H)\,,
\end{equation}
with $\rho_0$ a constant with dimensions of length. We factor out the charges from the metric functions and the fluxes to obtain dimensionless functions as
\begin{eqnarray}
e^f \,=\, \rho_0\,e^{\FF}\,, &\quad& e^g\,=\, \rho_0\,e^{\GG}\,,\nonumber\\[2mm]
b_J\,=\,\frac{Q_c}{4 q_c}+\frac{2q_c}{3 \rho _0}\  \mathcal{B}_J\,,&\quad& b_X\,=\,-\frac{Q_c}{4 q_c}-\frac{2q_c}{3 \rho _0}\  \mathcal{B}_X\,,\nonumber\\[2mm] a_J\,=\, -\frac{q_c}{2}-q_c \ \mathcal{A}_J\,,&\quad& h = \frac{128 q_c^2 }{9 \rho _0^6} \ \mathbf{h}\,.
\end{eqnarray}
The free parameters in the expansions both at the UV and at the horizon are essentially the same as in the rest of the $\mathds{B}_8$ family, since $\xi$ and $u$ coincide asymptotically. However, now we need to gauge fix $f_1=0$ to compare with the ground state solution. After the appropriate rescaling in order to compare the same dimensionless quantities, it can be seen that the free energies and entropies for different values of $b_0$ approach those of $\Bconf$ as $b_0\to1$. This is shown in Figure~\ref{fig:QuasiConfining}. In the case of $\Bconf$, as we remove the horizon, that is, in the limit $S \to 0$, the temperature diverges, as in the case of small black branes in global AdS spacetimes. This branch is however unstable: before reaching it, there is a first-order phase transition at a critical temperature 
$\overline{T}_c\approx 0.27$. In the dual gauge theory this corresponds to a genuine confinement/deconfinement first-order phase transition.   

\vspace{5mm}

\subsection{Regime of validity}
\label{sec:validity}

In this Section we determine the range of validity of the supergravity solutions we have studied. For the ground state, 
as well as for the low temperature solutions considered in Section~\ref{LowT},
there are two requisites, already discussed in Chapter \ref{Chapter2_B8family}, Section \ref{sec:validity_gs}. There we showed that the type IIA picture is trustable for energies in the gauge theory 
\begin{equation}\label{val1}
U \ll \lambda \left(1+\frac{\overline{M}^2}{2 N|k| }\right)(1-b_0^2)\,,
\end{equation}
above which the perturbative regime is adequate.\footnote{Recall that the three-dimensional gauge theories are superrenormalisable and asymptotically free.} On the other hand, the IR is regular in eleven dimensions. In order for the curvature to be small it is required that
\begin{equation}\label{val2}
\frac{\overline{M}^2}{2}+N|k| \gg 1\,.
\end{equation}

\vspace{1mm}

On top of this, we have seen for instance in Figure~\ref{fig:Ricci} that the curvature at the horizon grows unbounded as we remove it. This is important for phase transitions in Cases B and C, since they occur precisely in this limit. As we approach the critical temperature, the eleven-dimensional Ricci scalar in Planck units grows as
\begin{equation}
\ell_p^2 R\sim \left(\frac{\overline{M}^2}{2} +N|k|\right)^{-1/3}(\overline T-\overline T_c)^{-2/3}\,.
\end{equation}
Therefore, there is always an interval of temperatures
\begin{equation}
\overline T\in \left(\overline T_c,\overline T_c+\left(\frac{\overline{M}^2}{2} +N|k|\right)^{-1/2}\right)\,,
\end{equation}
where curvatures are large and higher-order curvature corrections must be considered. Nevertheless, this region is small provided \eqref{val2} is satisfied.

\vspace{1mm}

\section{Discussion}
\label{sec:conclusions}

In this Chapter we have studied the finite-temperature physics of the SYM--CSM theories theories introduced in Chapter~\ref{Chapter2_B8family}  by means of holography. We have seen that their gap is lost at some non-zero critical temperature, as we have shown by constructing the black brane geometries dual to the ungapped phase and demonstrating their dominance above $T_c$. The nature of the phase transition depends on the value of $b_0$ and there are three distinct regions. For large values of the parameter there is a single degapping first-order transition. At intermediate values, there is a small range in which two transitions take place: as the temperature is decreased from high values, there is a first-order transition between two different black branes, corresponding to two ungapped states, while at a lower critical temperature there is a degapping first-order transition at which the entropy is continuous. Finally, at even lower values of the parameter there is only a phase transition of this second type. This is summarised in the phase diagram of Figure~\ref{fig:PhaseDiagram}, where a triple point and a critical point are indicated by a black dot and by a square, respectively.

At the critical point the phase transition is of course second-order. An analogous  situation was found in \cite{Attems:2018gou}, where (in that case a bottom-up)  holographic model with a critical point was studied.  As expected on general grounds, the order parameter at the second-order phase transition can be constructed out of physical observables that jump discontinuously across the line of first-order phase transitions and whose susceptibilities diverge at the critical point. As in \cite{Attems:2018gou}, in our case this will be a combination of the energy density $E$ and the VEVs of (some) scalar operators dual to the scalar fields in the bulk. The specific combination could be determined via a calculation on the gravity side. In this description fluctuations of the metric mix with those of the scalar fields and the mode whose correlation length diverges at the critical point will be a linear combination of these fluctuations. 

\begin{figure}[t]
	\begin{center}
		\begin{subfigure}{.42\textwidth}
			\includegraphics[width=\textwidth]{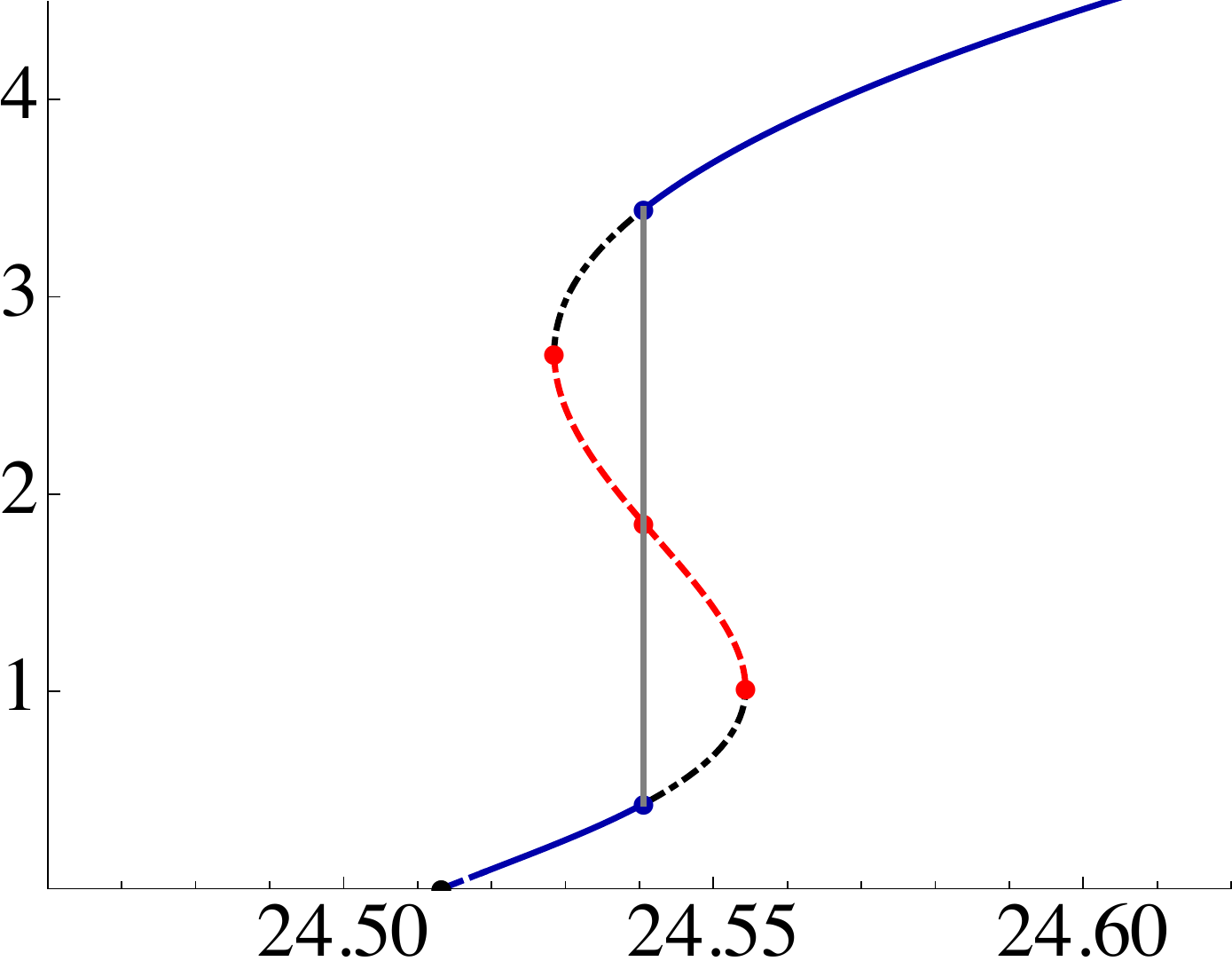} 
			\put(-140,130){$ 10^{-3}\ \overline S$}
			\put(-10,15){$\overline T$}
			\put(-90,70){$a$}			
			\put(-55,30){$b$}			
			\put(-78,48){$c$}			
			\put(-80,20){$e$}			
			\put(-75,90){$d$}	
		\end{subfigure}\hfill
		\begin{subfigure}{.55\textwidth}
			\includegraphics[width=\textwidth]{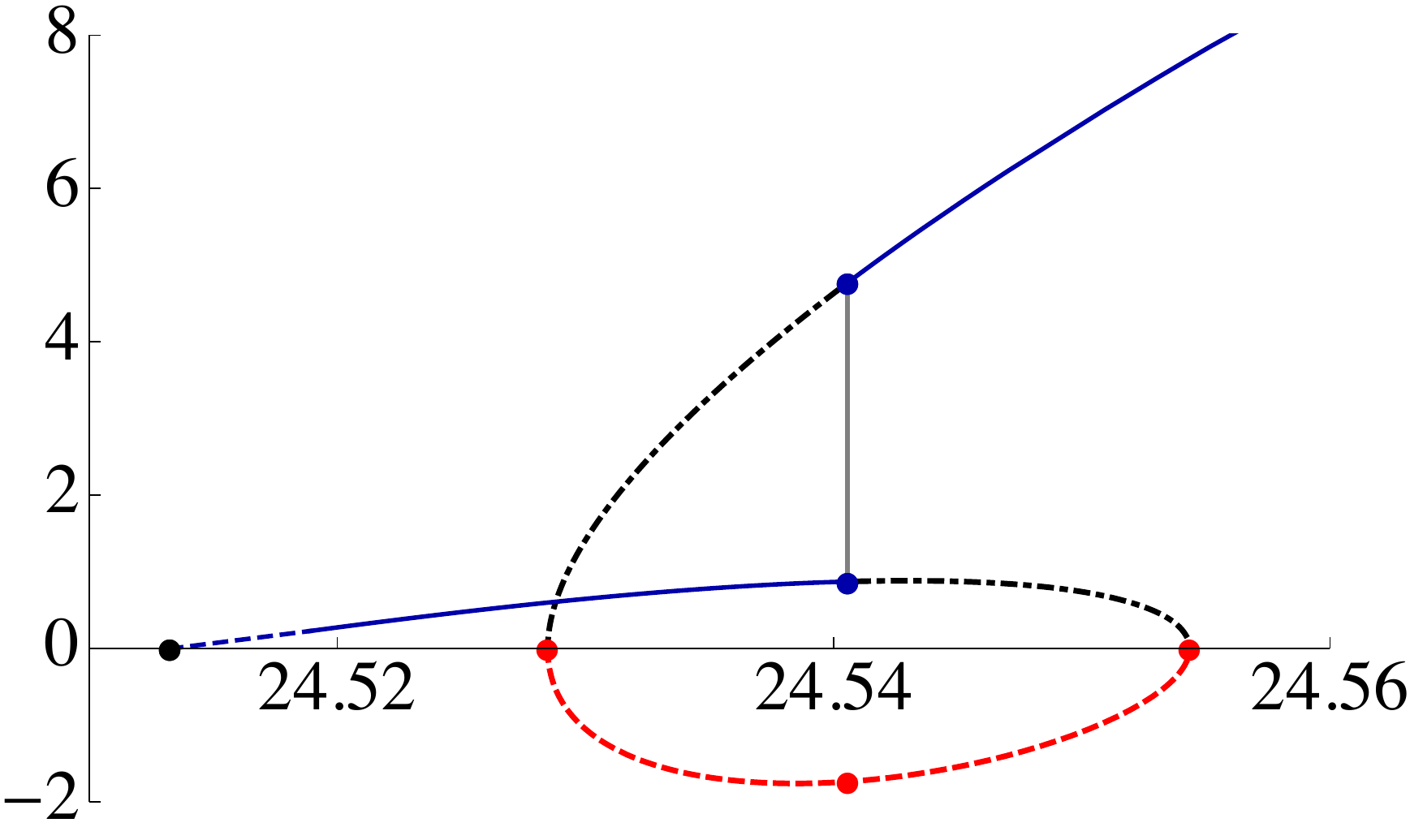} 
			\put(-185,120){$ 10^{3}\ c_s^2$}
			\put(-10,30){$\overline T$}
			\put(-125,20){$a$}			
			\put(-30,33){$b$}			
			\put(-81,0){$c$}			
			\put(-73,40){$e$}			
			\put(-83,80){$d$}
		\end{subfigure}
		\caption{\small Entropy density (left) and speed of sound squared (right) as a function of  temperature for black branes in the theory with $b_0=0.6835$ (Case B).}\label{fig.csplot}
	\end{center}
\end{figure}
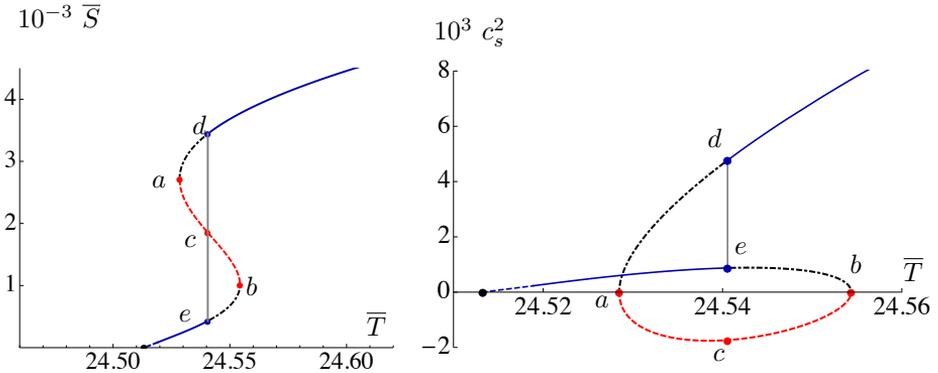

In Figure~\ref{fig.csplot}(left) we reproduce again the entropy density of Figure~\ref{fig.TypeII_PT} for Case B, which is the richest. A universal feature of a first-order phase transition is that the so-called ``spinodal region'' between points $a$ and $b$ is locally thermodynamically unstable, since the specific heat given in  the first equation in \eqref{cs} is negative, $c_v<0$. It follows from the second equation in \eqref{cs} that this region is also locally dynamically unstable, since it also has negative speed of sound squared. This is confirmed by the plot in Figure~\ref{fig.csplot}(right). 
We see that, as expected, the speed of sound squared crosses zero at the points $a$ and $b$ but, interestingly, also at the zero-entropy point where the transition to the confined phase takes place.  For detailed discussions of the spinodal instability and the associated dynamics in the holographic context see e.g.~\cite{Attems:2017ezz,Attems:2019yqn,Janik:2017ykj,Bellantuono:2019wbn,Bea:2020ees}.

As noted below Eq.~\eqref{below}, the size of the intermediate region where two phase transitions take place is very small compared to the range of the $b_0$ parameter. In other words, \mbox{$b_0^\textrm{triple}-b_0^\textrm{critical} =0.0032 \ll 1$}. Nevertheless, we have verified that the existence of this intermediate region is not a numerical artefact by integrating the equations with two different integrators and by varying the control parameters in each of them. It would be interesting to understand in detail how such a small region arises dynamically in our model. A dynamically-generated small parameter  was also encountered in \cite{Basu:2011yg} in the phase diagram of a holographic colour superconductor. 

\begin{figure}[t]
	\begin{center}
		\includegraphics[width=.50\textheight]{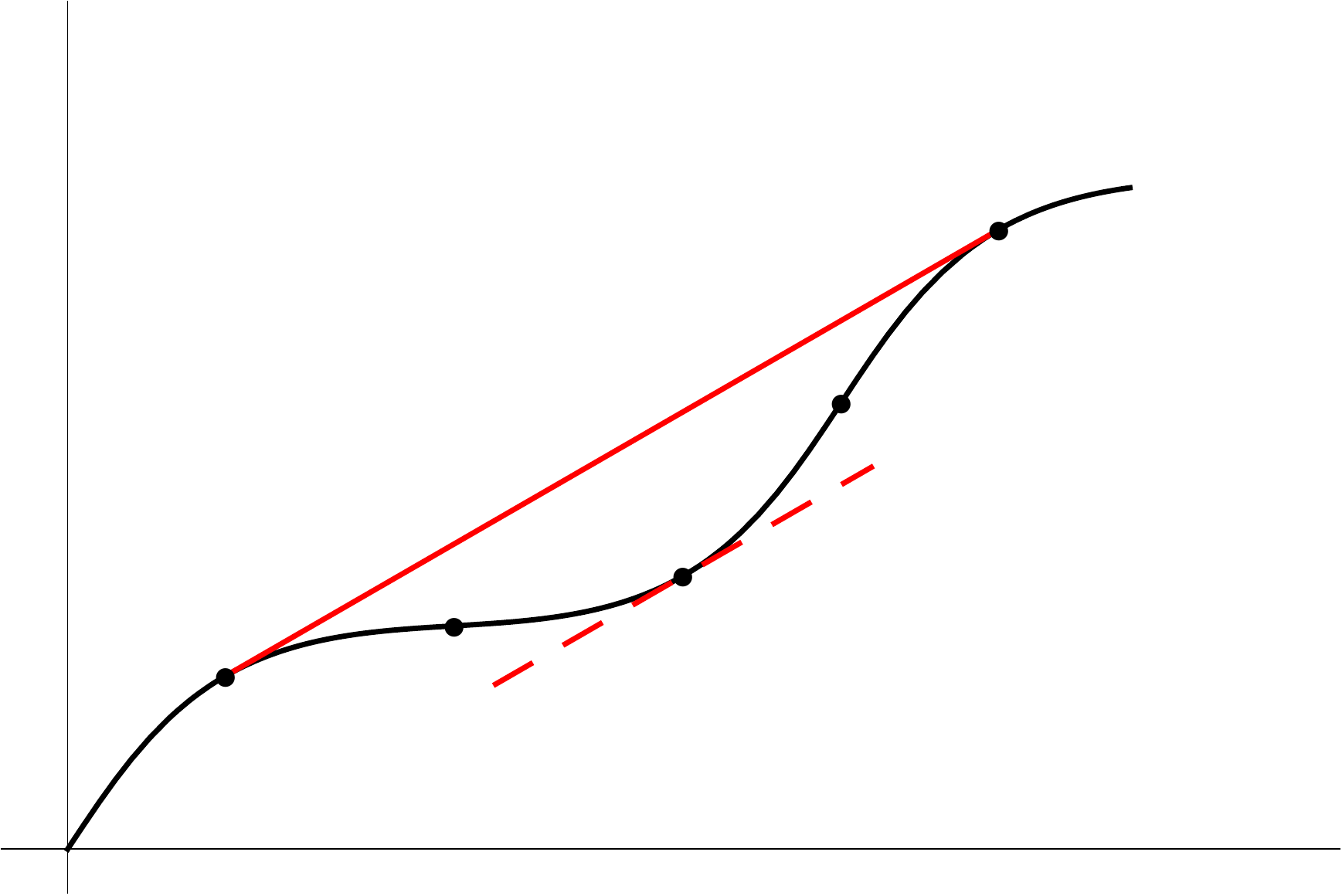} 
		\put(-270,170){$\overline{S}$}
		\put(-10,15){$\overline E$}
		\put(-185,43){$b$}			
		\put(-95,95){$a$}			
		\put(-130,55){$c$}			
		\put(-80,135){$d$}			
		\put(-225,35){$e$}			
		\caption{\small  Qualitative form of the entropy density in the microcanonical ensemble. The solid red segment corresponds to the average entropy density of phase-separated configurations, as explained in the text. The dashed red segment indicates that the slope of the tangent at point $c$ is the same as that of the solid red segment.}
		\label{entropyQ}
	\end{center}
\end{figure}
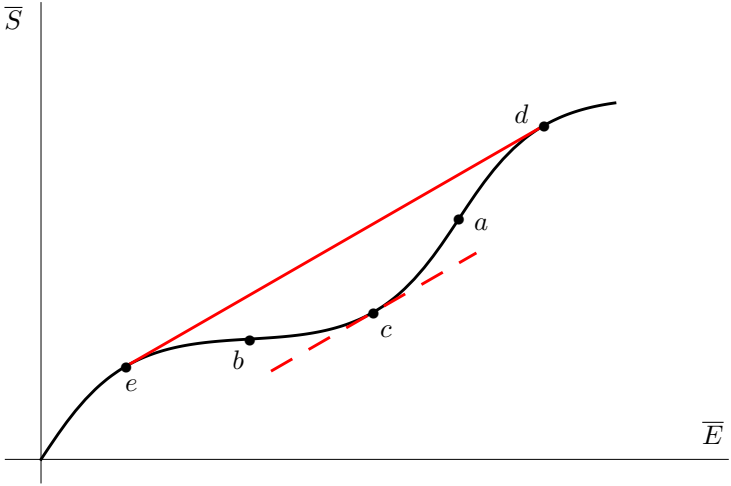

\begin{figure}[t]
	\begin{center}
		\begin{subfigure}{.47\textwidth}
			\includegraphics[width=\textwidth]{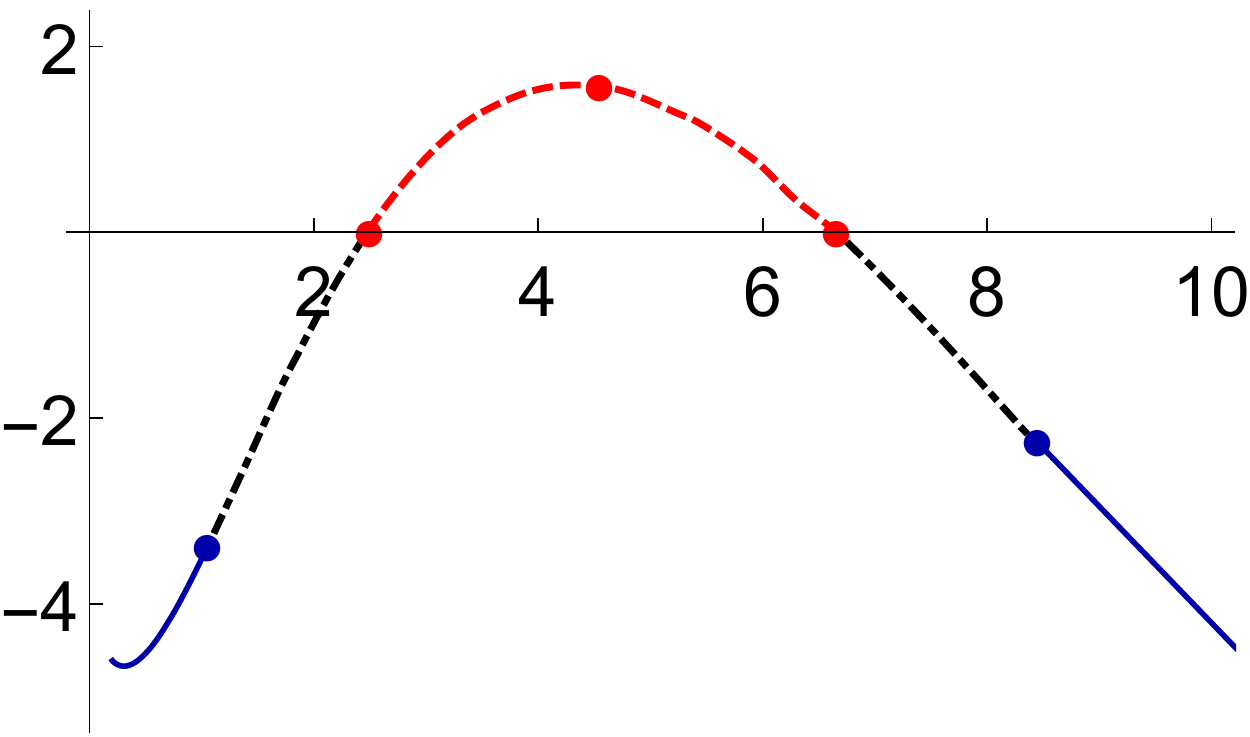} 
			\put(-122,70){$b$}			
			\put(-55,70){$a$}			
			\put(-90,90){$c$}			
			\put(-132,20){$e$}			
			\put(-35,25){$d$}												
			\put(-160,100){$10^9\ \overline{S}''(\overline{E})$}	
			\put(-30,70){$10^{-4}\overline{E}$}				
		\end{subfigure}\hfill
		\begin{subfigure}{.47\textwidth}
			\includegraphics[width=\textwidth]{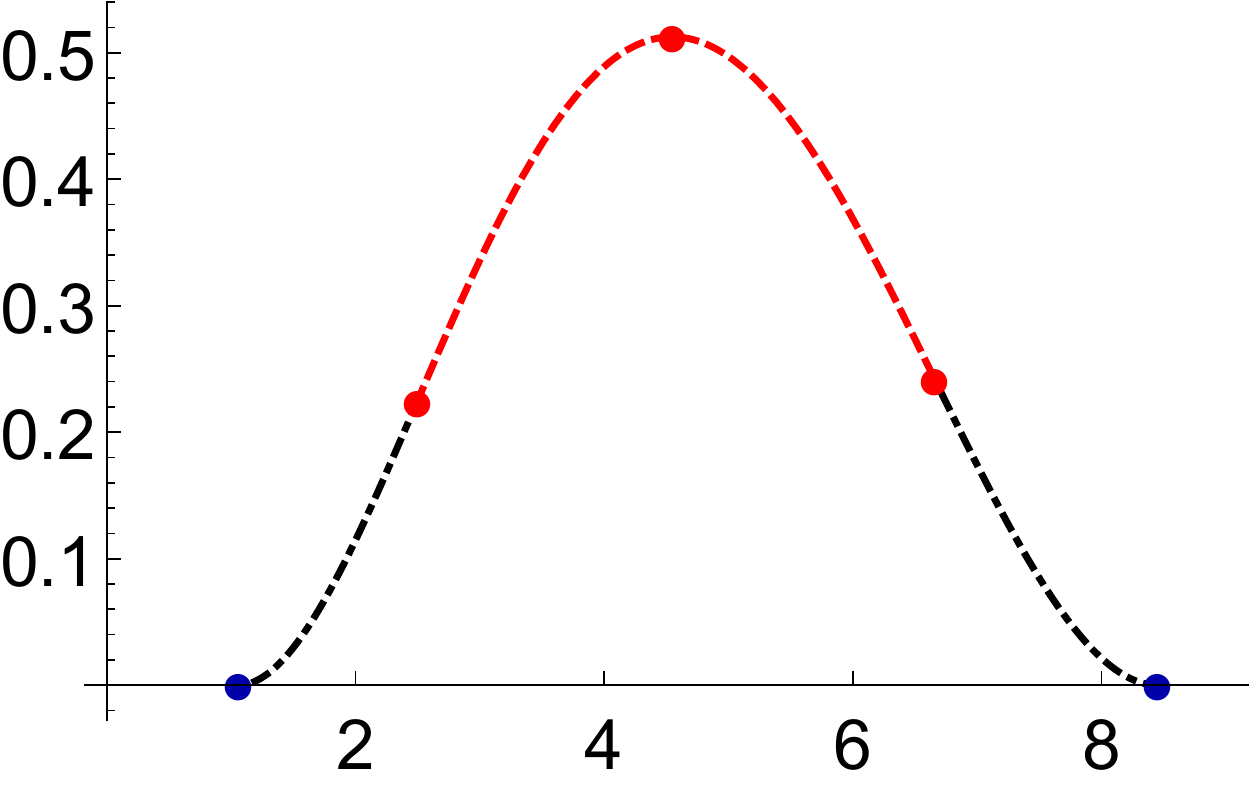} 
			\put(-117,52){$b$}			
			\put(-40,60){$a$}			
			\put(-78,105){$c$}			
			\put(-140,20){$e$}			
			\put(-10,20){$d$}												
			\put(-160,110){$\overline{S}_{\text{PS}}-\overline{S}$}	
			\put(-35,-10){$10^{-4}\ \overline{E}$}							
		\end{subfigure}
		\caption{\small (Left) Second derivative $S''(E)$ of the entropy density with respect to the energy density, showing the convexity/concavity properties discussed in the text. (Right) Difference between the average entropy density of the phase-separated configurations and the entropy density of the homogeneous solutions, showing that the former are preferred in the region between $a$ and $b$. As in Figure~\ref{fig.csplot} these plots correspond to the theory with $b_0=0.6835$ (Case B).}
		\label{microcanonical}
	\end{center}
\end{figure}
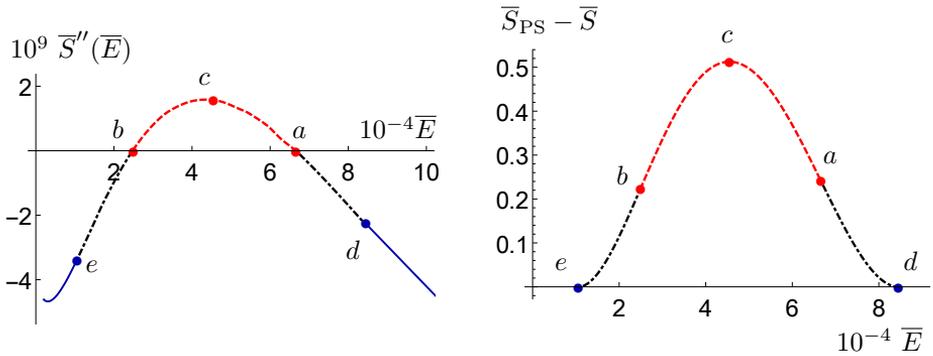

In this thesis we have focused on the behaviour of the system in the canonical ensemble, in which the control parameter is the temperature and the preferred state is the one that minimises the free energy. However, the behaviour in the microcanonical ensemble in the infinite-volume limit that we work in follows from general arguments (see \cite{Bea:2020ees} for a discussion at finite volume in the holographic context). Consider again Case B, since it is the richest. In the microcanonical ensemble the control parameter is the energy and the preferred state is the one that maximises the entropy. The qualitative form of the function $S(E)$ is shown in Figure~\ref{entropyQ}. 
The key features are as follows. $S$ is convex ($S''(E)>0$) in the region between $a$ and $b$. This indicates local thermodynamical instability, since the system can increase its total entropy by rising the energy slightly in part of its volume and lowering in another so as to keep the total energy fixed. In the regions $e$-$b$ and $a$-$d$ the entropy function is concave ($S''(E)<0$) but there are states with the same total energy and higher total entropy, namely phase-separated configurations in which the phases $e$ and $d$ coexist at the critical temperature. These states are characterised by the fractions $0\leq x, (1-x) \leq1$ of the total volume occupied by each phase, so their total entropy is of the form $S_e + (S_d-S_e) x$, as indicated by the red segment in Figure~\ref{entropyQ}. Therefore the regions $e$-$b$ and $a$-$d$ are locally but not globally thermodynamically stable. Finally, all states outside the region $e$-$d$ are globally stable. For our system, these qualitative features are difficult to appreciate directly on a plot of $S$ versus $E$ because the curve $S(E)$ is very close to a straight line. For this reason we show the convexity/concavity property (the second derivative) in Figure~\ref{microcanonical}(left) and the difference between the phase-separated configurations and the homogeneous solutions in Figure~\ref{microcanonical}(right). 

The black branes that we have found reach a finite temperature in the limit in which the horizon is removed and the entropy vanishes. As a result, the transition for low values of $b_0$ seems to take place at zero entropy but finite temperature. However, the geometries recovered in this limit, which were identified in Section~\ref{sec:zeroEntropyLimit}, are singular. This means that,  for temperatures slightly above the critical one,  the curvature is large and the supergravity approximation is unreliable in a small region near the horizon, as detailed in Section~\ref{sec:validity}. The singularity is a good one according to the classification of \cite{Gubser:2000nd}, since by construction it can be cloaked behind a horizon. It would be interesting to discern how string theory resolves the singularity and how the phase transition manifests itself in that picture.  

An analogous phase transition was observed in \cite{Dias:2017opt}, where black branes in a similar system were obtained. In that case the ground state is conformal in the UV and preserves $\mathcal{N}=2$, being a higher-supersymmetric version of the RG flow denoted $\BOP$ at the left of Figure~\ref{fig:triangle}. Our results can also be compared to those in \cite{Aharony:2007vg,Buchel:2018bzp}, where the Klebanov--Tseytlin and Klebanov--Strassler theories at finite temperature were studied. The phase transition shown to happen in those models is comparable to the one encountered here for  $b_0=1$. Indeed, when the CS-level vanishes, which happens for the flow called $\Bconf$ at the right of Figure~\ref{fig:triangle},\footnote{A black brane solution in this background, perturbative in the number of fractional branes, was constructed in \cite{Giecold:2009wj}.} the first-order phase transition is genuinely a confinement/deconfinement transition. Yet, the UV of our models, which are asymptotically free, is simpler than the infinite cascade of their four-dimensional cousins in \cite{Aharony:2007vg,Buchel:2018bzp}. It would also be interesting to compare to the phase structure of the $\mathcal{N}=1^*$ super Yang--Mills theory, the high temperature deconfined phase of which was studied in \cite{Bena:2018vtu}.

We have computed the different  Maxwell charges of our black brane solutions and verified that they do not change sign along the RG flows up to the horizon. In other words, the charges hidden behind the horizon have the same sign as those measured at infinity. This is expected \cite{Bena:2012ek,Bena:2013hr,Cohen-Maldonado:2015ssa,Cohen-Maldonado:2016cjh} since our solutions are smeared along the internal directions of the geometry, as in \cite{Buchel:2018bzp,Dias:2017opt}.
Some arguments (see e.g.~\cite{Hartnett:2015oda,Armas:2018rsy,Armas:2019asf} for recent analyses) suggest that a change of sign in the charges may be realised in solutions with localised horizons along the internal directions, whose existence may lend support to KKLT-type constructions of de Sitter vacua \cite{Kachru:2003aw}. We leave the investigation of these aspects for the future. 

\newpage
\thispagestyle{empty}


\newpage
\chapter{Holographic Complex Conformal Field Theories}
\label{Chapter5_HoloCCFTs} 

Fixed-point annihilation (FPA) is an interesting phenomenon in which two fixed points (FPs) of the renormalisation group (RG) flow merge and disappear as some parameter is varied. In the context of phase transitions and critical phenomena it is associated to a change from continuous to weak 
first-order transitions. Examples include the superconducting transition in the Abelian Higgs model \cite{Halperin:1973jh,Ihrig:2019kfv}, the related N\'eel-valence bond-solid transition in antiferromagnets \cite{Senthil:2004aza,Nahum:2015jya,Wang:2017txt,Serna:2018tct}, the ferromagnetic transition in the Potts model \cite{Nienhuis:1979mb,Nauenberg:1980nv,Gorbenko:2018dtm}, metal-Mott insulator transitions \cite{Herbut:2014lfa} and six-dimensional O$(N)$ models \cite{Fei:2014xta,Gracey:2018khg}. FPA has also been associated to the boundaries of the conformal window in gauge theories with flavours, both in $(2+1)$-dimensional quantum electrodynamics \cite{Appelquist:1988sr,Kubota:2001kk,Kaveh:2004qa,Herbut:2016ide} and in non-Abelian gauge theories in $3+1$ dimensions \cite{Gies:2005as,Pomoni:2008de,Kaplan:2009kr,Antipin:2012kc,Hansen:2017pwe}.  

In the previous Chapters of this thesis we have seen how a fixed point, dual to the Oogury--Park CFT, induces quasi-conformal dynamics in the SYM--CSM theories whose RG flows close to it. We wish to investigate whether a similar behaviour can take place in the context of FPA.

More generally, FPA has been proposed as a natural mechanism to produce ``walking behaviour'' in gauge theories \cite{Kaplan:2009kr,Gorbenko:2018ncu}. The idea is that, just after the merging, the critical points leave a footprint in the form of approximate scale invariance over a large range of scales. Typically the range of the walking region increases exponentially as parameters are tuned to the merging point, following  Miransky (or Berezinsky--Kosterlitz--Thouless) scaling \cite{Miransky:1984ef,Berezinsky:1970fr,Kosterlitz:1973xp}. This behaviour can be explained by continuing the theory to complex values of the couplings, so that the annihilation is understood as a migration of the FPs to the complex plane after the merger \cite{Kaplan:2009kr}. Their effect on the RG flow is noticeable as long as they remain close to the real axis. It has been recently conjectured  \cite{Gorbenko:2018ncu} that a non-unitary, complex conformal field theory (cCFT) exists at each of the two complex fixed points (cFPs), so that the properties of the theory in the walking region can be derived from perturbations of the cCFTs. Each cCFT has a complex spectrum of operators that is the conjugate of its companion's, implying that cFPs should always come in pairs.

Although FPA and cFPs are expected to exist generically, their study has been mostly limited to weakly coupled theories (see e.g.~\cite{Sieg:2016vap,Grabner:2017pgm,Pittelli:2019ceq,Benini:2019dfy} for recent examples), as their identification requires computing the beta functions for the different couplings in the theory, a task that  often can only be done via perturbation theory. In this Chapter we will construct a simple holographic model that realises FPA and cFPs, thus showing that these phenomena  can also occur at strong coupling. In addition, our analysis provides non-perturbative evidence that cFPs have the conjectured properties of cCFTs regarding the spectrum of local operators. 

\section{Fixed-point annihilation and complex CFTs}
\label{sec:FPAandcCFT}

Consider  a system with a dimensionless coupling $g$ whose $\beta$-function depends on an external parameter $\alpha$ in such a way that, for $\alpha\simeq \alpha_*$, 
\begin{equation}\label{eq:beta}
\beta\left(g\right)\,\simeq \,\left(\alpha-\alpha_*\right)-\left(g-g_*\right)^2.
\end{equation} 
We will see an explicit  example in Section~\ref{sec:cFPandRGflows}. 
If $\alpha>\alpha_*$, discarding higher-order terms, the $\beta$-function vanishes at two values  
\begin{equation}
g_{\pm}\,=\,g_*\pm\sqrt{\alpha-\alpha_*}\,.
\end{equation} 
To make sure that the theory is well defined in the far ultraviolet (UV)  we may imagine that $\beta$ has another zero at some $g_{\infty}<g_-$.
Decreasing the control parameter $\alpha$ the FPs $g_\pm$ approach each other until they merge at \mbox{$\alpha=\alpha_*$}. If we decrease $\alpha$ further, $\beta(g)$ loses these (real) zeroes and the theory ceases to have a (real) conformal phase in the infrared (IR). However, for \mbox{$|\alpha-\alpha_*|$} sufficiently small, $\beta(g)$ has cFPs close to the real axis at $g_\pm=g_*\pm i \sqrt{\alpha_*-\alpha}$. In this regime the theory  exhibits approximate scale invariance between UV and IR  scales $\mu_{\text{\tiny UV}}$ and $\mu_{\text{\tiny IR}}$ defined by the values of the coupling $g_{\text{\tiny UV}}\lesssim g_*\lesssim g_{\text{\tiny IR}}$. The ratio between these two scales becomes exponentially large as $\alpha$ approaches $\alpha_*$ and shows the characteristic Miransky scaling
\begin{equation}\label{eq:BKT}
\log\frac{\mu_{\text{\tiny UV}}}{\mu_{\text{\tiny IR}}}\,=\,\int_{g_{\text{\tiny IR}}}^{g_{\text{\tiny UV}}}\frac{\dd g}{\beta(g)}\simeq \frac{\pi}{\sqrt{\alpha_*-\alpha}}\,,
\end{equation}
where we assumed that $\left|g_{\text{\tiny IR,UV}}-g_*\right|\gg\sqrt{\alpha_*-\alpha}$. Thus if $|\alpha-\alpha_*|$ is small then the RG flow is slow in a large energy range, hence the term ``walking'' flow. 

The scenario conjectured in \cite{Gorbenko:2018dtm,Gorbenko:2018ncu} is that cFPs correspond to pairs of non-unitary cCFTs that control the walking flow, which passes precisely in between them. Each cCFT should have operators of complex dimensions in the spectrum that are not matched by other operators with complex conjugate dimension in the same theory. Instead, the missing operators of complex conjugate dimensions in one cCFT should be part of the the spectrum of the companion cCFT. Thus in this sense the two cCFTs are complex conjugate of one another. 

An important case is the operator associated to the coupling $g$ itself, whose complex dimensions at each cFP, 
\begin{equation}\label{eq:Delta}
\Delta_{\pm}\,=\,d+\beta^\prime(g_{\pm})\,\simeq\,d\mp 2i\,\sqrt{\alpha_*-\alpha}\,,
\end{equation}
with $d$ the spacetime dimension, 
are indeed complex conjugates of one another. Moreover, this operator is close to marginality when $\alpha \lesssim \alpha_*$, with the leading deviation being imaginary and small. Using this, the hierarchy \eqref{eq:BKT} can be rewritten as 
\be
\label{Miransky}
\log \frac{\mu_{\text{\tiny UV}}}{\mu_{\text{\tiny IR}}} \simeq 
\frac{2\pi}{|\operatorname{Im}{\Delta}|} \,.
\ee

\section{Holographic realisation}
\label{sec:holorealization}

Previous holographic realisations of walking behaviour with Miransky-like scaling in gravity duals \cite{Jarvinen:2011qe,Kaplan:2009kr,Jensen:2010ga,Iqbal:2010eh,Pomarol:2019aae} are based on flows where the mass of a scalar field on the gravity side violates the Breitenlohner--Freedman (BF) bound\footnote{See however \cite{Alanen:2010tg,Alanen:2011hh} for a holographic model whose $\beta$-function shares some properties with the lower one in our Figure~\ref{fig:beta}.}. In these cases there is a real FP that becomes dynamically unstable but no cFPs have been identified. We will present a different  construction in which both FPA and the resulting cFPs are explicitly realised. 

\begin{figure}[t]
	\begin{center}
		\includegraphics[width=0.60\textwidth]{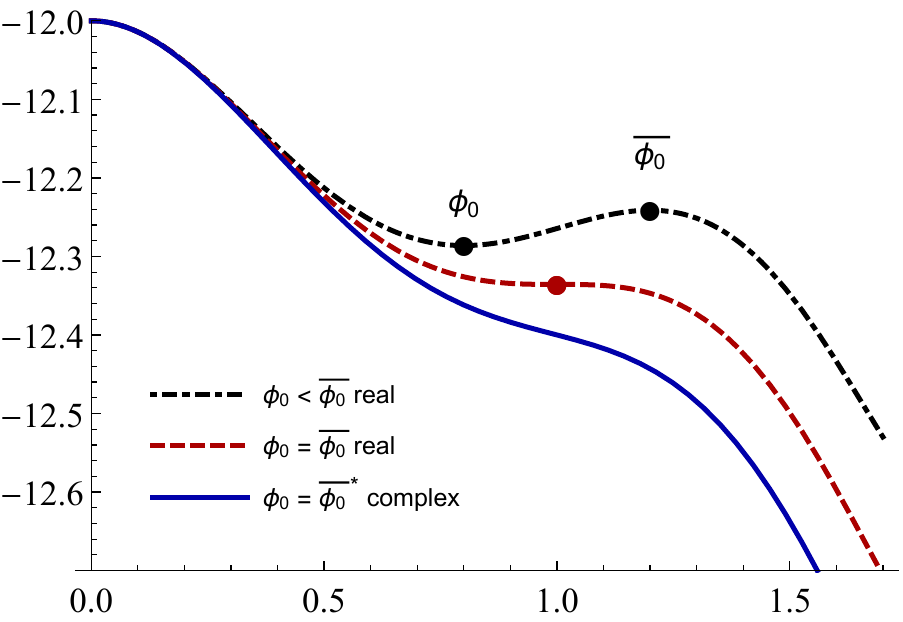} 
		\put(-200,145){ $L^2 V(\phi)$}
		\put(-7,0){$ \phi$}
		\caption{\small  Potential  of our model for the cases 
			$\{\phi_0=0.8, \overline{\phi}_0=1.2\}$ (top curve), 
			$\{\phi_0=\overline{\phi}_0=1\}$ (middle curve) and 
			$\{\phi_0= \overline{\phi}_0^*=1+0.2i\}$ (bottom curve).}
		\label{fig:pot}
	\end{center}
\end{figure}

Couplings of the gauge theory are holographically dual to fields on the gravity side. For simplicity we focus on a single coupling dual to a scalar field. The action on the gravity side is thus
\begin{equation}\label{eq:action}
S\,=\,\frac{1}{2\kappa^2}\int d^{d+1}x\sqrt{-g}\left(R-\frac12 (\partial\phi)^2-V(\phi)\right)\,.
\end{equation}
For each critical point $\phi_c$ of the potential with $V(\phi_c)<0$ there is an anti-de Sitter (AdS) solution with its corresponding $d$-dimensional CFT dual. We choose to write $V$ in terms of a (fake) superpotential $W$ through the usual relation
\begin{equation}\label{W}
V\,=\,\left(d-1\right)\left[2\left(d-1\right)\left(\frac{\dd W}{\dd\phi}\right)^2-d\,W^2\right] \,.
\end{equation} 
The only reason for this is to simplify the presentation. In particular, this choice implies nothing regarding the possible presence of supersymmetry in the system. Critical points of $W$ are also critical points of $V$ (but not vice versa). Since we wish to model three FPs we take a superpotential with derivative
\begin{equation}\label{dW}
\frac{\dd W}{\dd\phi}\,=\,\frac{W_0}{L}\,\phi\left(\phi-\phi_0\right)\left(\phi-\overline{\phi}_0\right).
\end{equation}
The resulting potential is shown in Figure~\ref{fig:pot}.
The UV FP dual to the AdS solution at $\phi=0$ is the analogue of the FP at \mbox{$g=g_{\infty}$} in Section~\ref{sec:FPAandcCFT}. The constants $\phi_0$ and $\overline{\phi}_0$ are parameters of the model analogous to $\alpha$. If both $\phi_0$ and $\overline{\phi}_0$ are real then there are two additional real FPs at \mbox{$\phi=\phi_0$} and $\phi=\overline{\phi}_0$, in analogy with  $g=g_\pm$ in Section~\ref{sec:FPAandcCFT}. 
When $\phi_0=\overline{\phi}_0$ these FPs merge into a single one. If 
$\phi_0$ and $\overline{\phi}_0$ become complex then they must be conjugate to one another since $W$ must be real for real $\phi$. In this case the potential looses two real critical points, giving a holographic realisation of FPA \footnote{The potential determined from \eqref{dW} may have additional FPs that will  play no role in our discussion.}.

At the UV FP the AdS radius is fixed by the integration constant $W(0)=1/L$, while the dimension $\Delta_{\text{\tiny UV}}$ of the operator $\mathbf{O}$ dual to $\phi$ is determined by $W_0= \delta_{\text{\tiny UV}}/[2(d-1)\phi_0\overline{\phi}_0]>0$. There are two possible choices depending on whether the flow is triggered by a source for $\mathbf{O}$, in which case $\delta_{\text{\tiny UV}}=  d- \Delta_{\text{\tiny UV}}$, or by a non-zero expectation value for $\mathbf{O}$, in which case $ \delta_{\text{\tiny UV}}= \Delta_{\text{\tiny UV}}$.

Expanding the superpotential around $\phi=\phi_0$ we get
\begin{equation}\label{expansion}
W(\phi) =\frac{1}{L_0}+\frac{\delta_0}{4(d-1)L_0}(\phi-\phi_0)^2+\cO(\phi-\phi_0)^3\,,
\end{equation}
where
\begin{equation}\label{eq:L0W}
L_0 =L \left[1+\frac{W_0}{6}\left(\phi_0^3\overline{\phi}_0-\frac{\phi_0^4}{2} \right) \right]^{-1}
\end{equation}
is the AdS radius at $\phi=\phi_0$ and 
\begin{equation} \label{eq:D0W}
\delta_0 = 2(d-1) W_0\frac{L_0}{L}\phi_0(\phi_0-\overline{\phi}_0)=\delta_{\text{\tiny UV}}\frac{L_0}{L}\frac{(\phi_0-\overline{\phi}_0)}{\overline{\phi}_0}\,.
\end{equation}
The expansion around $\overline{\phi}_0$ gives analogous results with the replacements 
\begin{equation}
\{ \phi_0, \overline{\phi}_0, L_0, \delta_0 \}\to \{ \overline{\phi}_0,\phi_0,  \overline{L}_0, \overline{\delta}_0 \}.
\end{equation}
Assuming \mbox{$0<\phi_0 \lesssim \overline{\phi}_0$} we have that 
\be
\delta_0<0 \,, \qquad 0 < \overline{\delta}_0 < \frac{d}{2}-1 \,.
\ee
In this case $\overline{\phi}_0$ corresponds to an UV FP deformed by a relevant scalar operator of dimension \mbox{$\overline{\Delta}_0=d-\overline{\delta}_0$}, whereas $\phi_0$ corresponds to an IR FP deformed by an irrelevant  scalar operator of dimension  
$\Delta_0=d-\delta_0$. When $\phi_0=\overline{\phi}_0$ the two points merge and the dual operator becomes marginal. 

\subsection{Complex FPs and RG flows}
\label{sec:cFPandRGflows}

\begin{figure}[t]
	\begin{center}
		\includegraphics[width=0.60\textwidth]{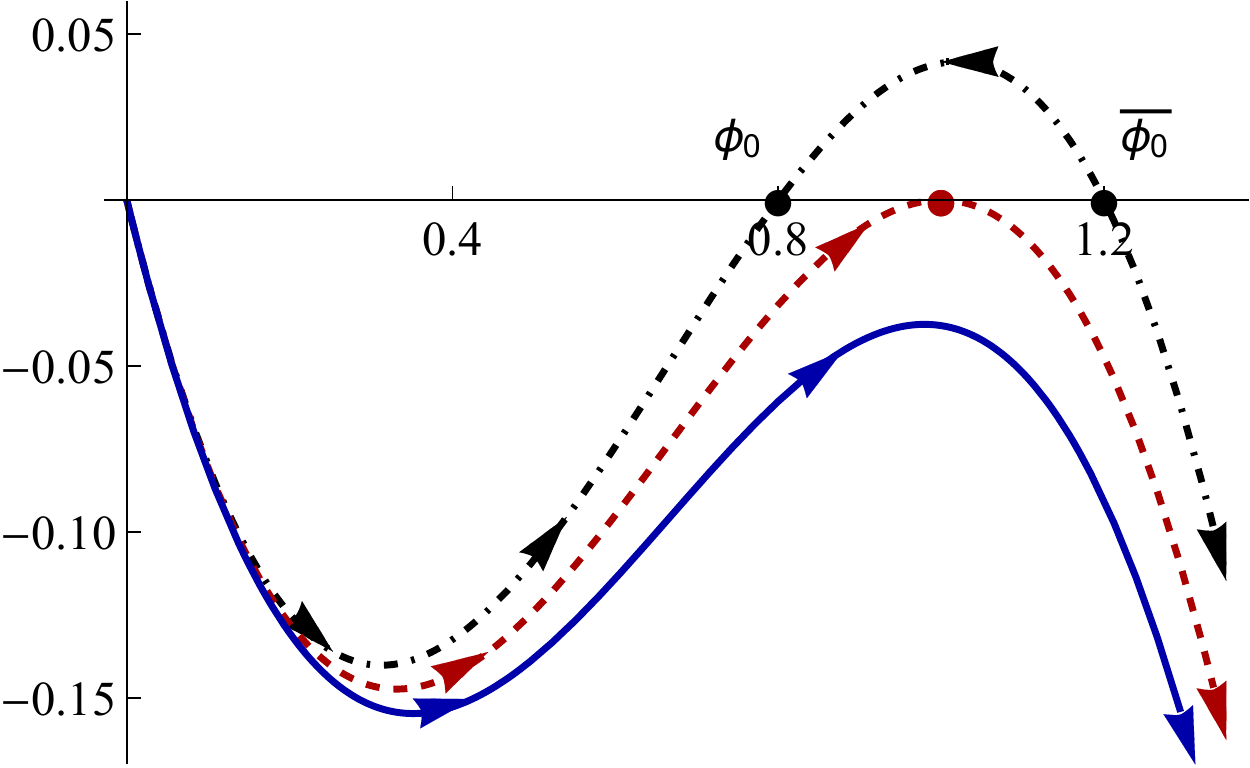} 
		\put(-200,135){$\beta(\phi)$}
		\put(10,95){$ \phi$}
		\caption{\small  $\beta$-functions associated to each of the three potentials of Figure~\ref{fig:pot}, as defined in \eqref{eq:betaphi}. The arrows indicate the direction of the RG flow from the UV to the IR. }
		\label{fig:beta}
	\end{center}
\end{figure}

In the so-called ``domain wall'' coordinates in which the metric takes the form 
\begin{equation}
\dd s_{d+1}^2\,=\,g_{MN}dx^M dx^N\,=\, e^{2A\left(\rho\right)}\dd x_{1,d-1}^2+\dd\rho^2
\end{equation}
the solution is determined by the  equations
\begin{equation}\label{BPS}
\frac{\dd A}{\dd\rho}\,=\, W\,,\qquad\qquad \frac{\dd \phi}{\dd\rho}\,=\,-2\left(d-1\right)\frac{\dd W}{\dd\phi}\,.
\end{equation}
In these coordinates the metric is foliated by copies of $d$-dimensional Minkowski space with scale factor $e^{A(\rho)}$, which is therefore interpreted as dual to the RG scale in the gauge theory. Similarly, the scalar field $\phi=\phi(\rho)$ is dual to a running coupling constant in a particular scheme, whose  $\beta$-function is therefore 
(see e.g.~\cite{Anselmi:2000fu})
\begin{equation}\label{eq:betaphi}
\beta\left(\phi\right)\,=\,\frac{\dd \phi}{\dd A}\,=\,-2\left(d-1\right)\frac{\dd\log W}{\dd\phi}\,.
\end{equation}
Close to any of the three real FPs $\phi_c=\{ 0,\phi_0,\overline{\phi}_0\}$ one finds the expected behaviour 
\be\label{UVRG}
\phi \simeq g_c \,e^{-(d-\Delta_c)\rho/L_c}
\,, \qquad A(\rho)\simeq \frac{\rho}{L_c}\sim \log \frac{\mu}{\Lambda_c},
\ee
and 
\be
\beta(\phi)\simeq -(d-\Delta_c)(\phi-\phi_c)+ \cO \left((\phi-\phi_c)^2\right)\,,
\ee
where $\mu$ is the RG scale, $\Lambda_c$ is the scale that triggers the flow away or into the FP, and $g_c$ is the corresponding (dimensionless) coupling at the FP. UV and IR FPs are approached for $\rho \to \infty$ and $\rho \to -\infty$, respectively. The $\beta$-functions for our model are shown in Figure~\ref{fig:beta}, where we see that they exhibit the  behaviour discussed in Section~\ref{sec:FPAandcCFT}.

Since the scalar field is dual to the coupling constant in the gauge theory, we propose that the holographic dual of the extension of this coupling to complex values consists of extending the scalar field on the gravity side to complex values as well. Since the scalar field couples to the metric, we also extend the metric components to complex values. We assume that, in this extension, the action \eqref{eq:action} is a holomorphic function of $g_{MN}$ and $\phi$. In other words, we do not introduce any explicit dependence on the complex conjugates of these fields in the action. Put yet another way, the equations of motion are obtained by varying the action with respect to $\phi$ and $g_{MN}$ as complex variables, as opposed to varying independently with respect to their real and imaginary parts.

With this extension, the FPs do not disappear at $\overline{\phi}_0=\phi_0$ but simply move to the complex-$\phi$ plane. The dimensions of the operators dual to the scalar field at each cFP \eqref{eq:D0W} are complex conjugate of one another, $\overline{\Delta}_0=\Delta_0^*$. In addition, there are formally AdS solutions with metrics 
$g_{MN}=\{ h_{MN},\overline{h}_{MN} \}$ whose complex radii \eqref{eq:L0W} are also complex conjugates, $\overline{L}_0=L_0^*$. Since we assume that the coordinates are real, this relates the two metrics by complex conjugation $\overline{h}_{MN}=h_{MN}^*$. Accordingly, any quantities that can be computed holographically at the cFPs purely in terms of geometric quantities will be related by complex conjugation. These include the expectation value of Wilson loops \cite{Maldacena:1998im,Rey:1998bq}, the entanglement entropy \cite{Ryu:2006bv},  and holographic $c$-functions \cite{Freedman:1999gp} that are related to central charges and anomaly coefficients.

\begin{figure}[t]
	\begin{center}
		\includegraphics[width=0.60\textwidth]{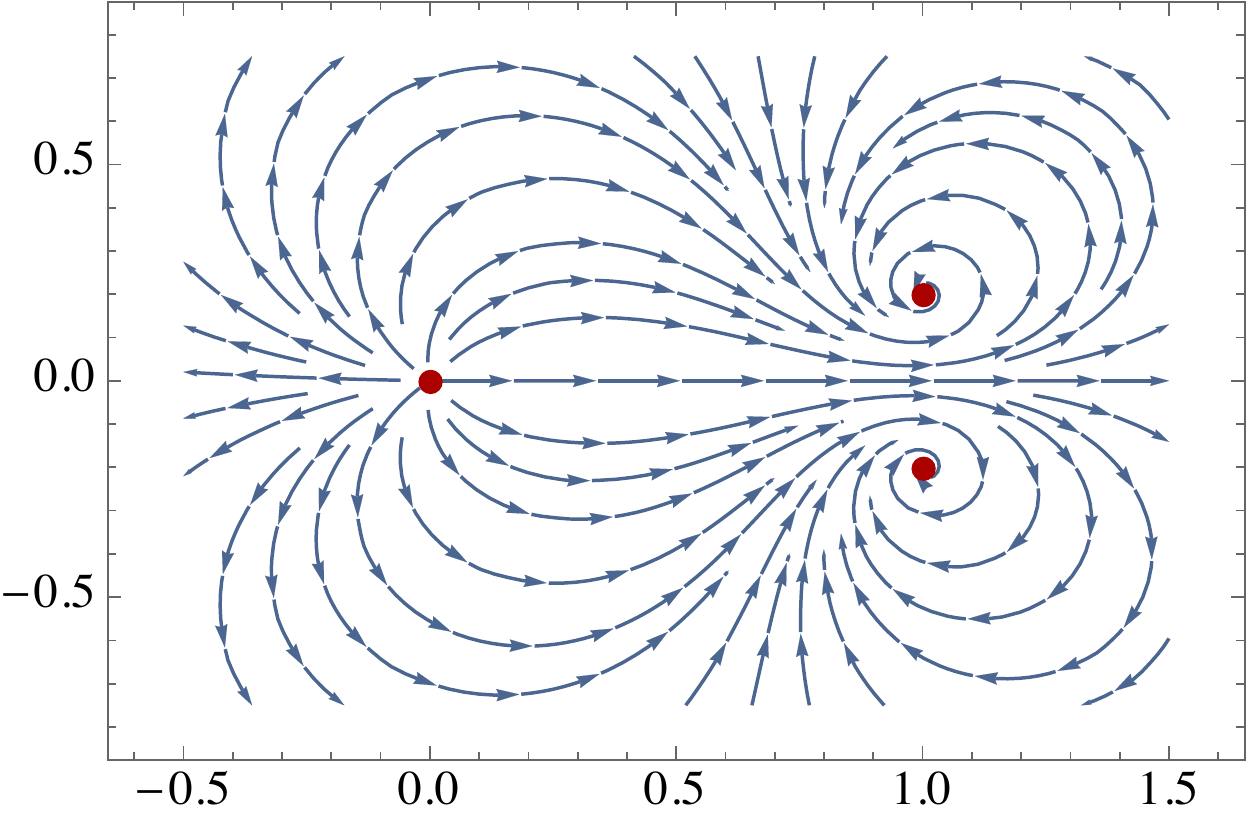} 
		\put(-200,140){$\operatorname{Im}\phi $}
		\put(0,20){$ \operatorname{Re}\phi$}
		\caption{\small  Examples of complex RG flows for $d=4$ displaying the characteristic spiralling behaviour around the cFPs. The flow in the UV is triggered by a source for a $\Delta_{\text{\tiny UV}}=3$ operator and the cFPs are located at $\phi_0=\overline{\phi}_0^*=1+0.2i$.}\label{fig:RGflow}
	\end{center}
\end{figure}

This result extends to complex RG flows between the UV FP at $\phi=0$ and the cFPs. Once we continue the scalar to complex values, the first-order equations \eqref{BPS} split into real and imaginary parts and the solutions describe the RG flow of the real and imaginary parts of the dual coupling. The scalar field still approaches  the cFPs as given by the first equation in \eqref{UVRG}. Since both $\Delta_c$ and $L_c$ are complex the coupling oscillates. In the particular example of Figure~\ref{fig:RGflow} the cFPs are IR FPs but this is not generic, i.e.~cFPs can also be UV FPs. The only purely real flow is the straight horizontal line that passes exactly in between the cFPs and should exhibit walking behaviour  We have collected some explicit formulas for the flows in Appendix \ref{ap:complexRGflows}.

\subsection{Holographic complex conformal field theories}
\label{sec:holocCFT}

The holographic cFPs show many of the properties expected for a pair of cCFTs. In addition to the operator dual to $\phi$, we will show that the spectrum of dimensions of local operators at one cFP is the complex conjugate of its companion's. Let $X^I$ denote the real components of a field in an arbitrary Lorentz representation. Then the action expanded to quadratic order around a cFP at $\phi_c=\{\phi_0,\overline{\phi}_0\}$, $h_c=\{ h,\overline{h}\}$ will be
\be
{\mathcal L}_c\simeq - \sqrt{-h_c}\left( \frac{1}{2}X^I K_{c\,IJ} X^J+\frac{1}{2}M^2_{c\,IJ} X^I X^J\right),
\ee
where we have separated a kinetic part determined by a differential operator $K_{c\,IJ}=K_{IJ}(h_c,\phi_c)=\{ K_{IJ},\overline{K}_{IJ}\}$ and a mass term $M^2_{c\,IJ}=M^2_{IJ}(h_c, \phi_c)=\{ M^2_{IJ},\overline{M}^2_{IJ}\}$. This gives the field equations
\be\label{eq:eom}
K_{c\,IJ} X_c^J+M^2_{c\,IJ} X_c^J=0.
\ee
The general solution near the AdS boundary  will be a superposition of exponentials of the form \footnote{For integer valued $\Delta_{c\,n}$ there can be additional powers of $\rho$ in the solutions, but the exponents do not change.}
\be\label{eq:XI}
X_c^I=\sum_n a_{c\,n}^I e^{-\Delta_{c\,n} \rho/L_c}+b_{c\,n}^Ie^{-(d-\Delta_{c\,n}) \rho/L_c}.
\ee
Introducing this in equation \eqref{eq:eom} one finds a homogeneous system of equations for the coefficients $a_{c\,n}^I$, $b_{c\,n}^I$ that has solutions when $\Delta_{c\,n}=\{ \Delta_n,\overline{\Delta}_n\}$ take the values corresponding to the conformal dimensions of the dual operators \mbox{$\mathbf{O}_{c\,n}=\{ \mathbf{O}_n,\overline{\mathbf{O}}_n\}$}. 
We can now use holomorphicity of the action to show that 
\be
\overline{K}=K(\overline{h},\overline{\phi}_0)=K(h^*,\phi_0^*)=  \left( K(h,\phi_0) \right)^*=(K)^*
\ee
and, similarly, that  $\overline{M}^2=(M^2)^*$, where we have suppressed the $IJ$-indices for simplicity.
This implies that the spectra of operators at the two cFPs are related by complex conjugation, as anticipated:
\be
\overline{\Delta}_n=\Delta_n^*.
\ee

\subsection{Walking behavior and Miransky scaling}
\label{sec:walking}

Our simple holographic model correctly describes the physics of walking and the associated Miransky scaling when the cFPs are close to the real axis. In particular, the $\beta$-function \eqref{eq:betaphi} reproduces  \eqref{eq:BKT}. We will illustrate this scaling further by heating up the real RG flow that passes exactly in between the cFPs in Figure~\ref{fig:RGflow}. On the gravity side this corresponds to constructing black hole solutions that start in the UV as deformations of a $d=4$ CFT by a $\Delta_{\text{\tiny UV}}=3$ operator with source $\Lambda$. We set $\phi_0= 1+i\epsilon$ and construct  solutions for several small values of $\epsilon$, following the procedure described in \cite{Gubser:2008ny}. Some explicit technical details can be found in Appendix \ref{ap:BH}.

\begin{figure}[t]
	\begin{center}
		\includegraphics[width=0.60\textwidth]{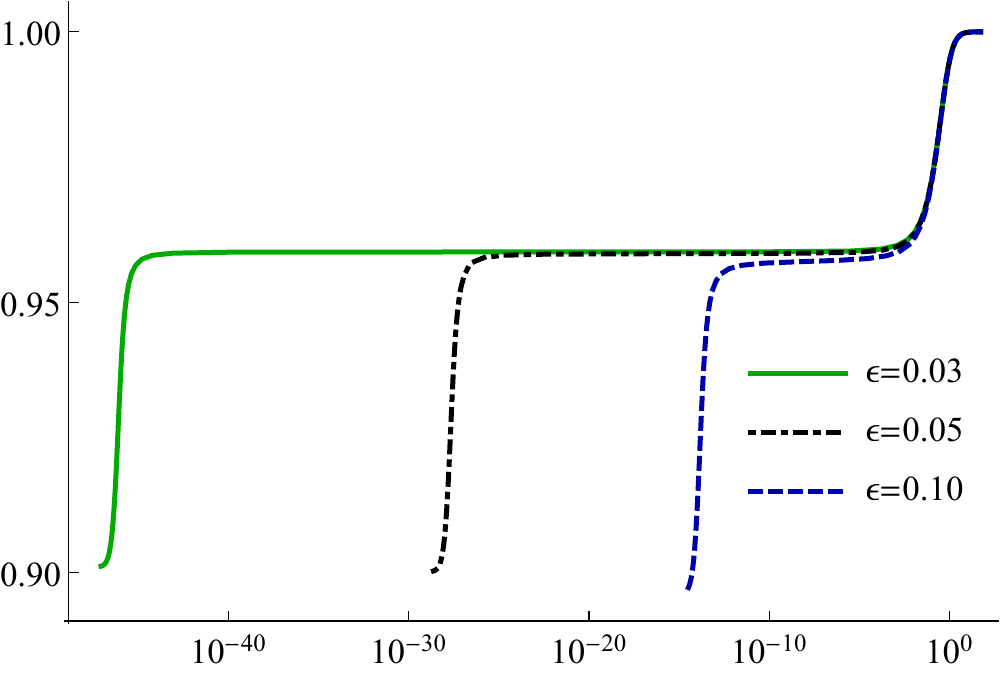} 
		\put(-210,145){$ S/S_{\text{\tiny UV}}$}
		\put(0,17){$ T/\Lambda$}
		\caption{\small Log-linear plots of the entropy density $S$ as a function of the temperature $T$, normalised to that of the UV CFT, for different values of $\phi_0= 1+ i\epsilon$.}\label{fig:entropy}
	\end{center}
\end{figure}

\begin{figure}[t]
	\begin{center}
		\includegraphics[width=0.60\textwidth]{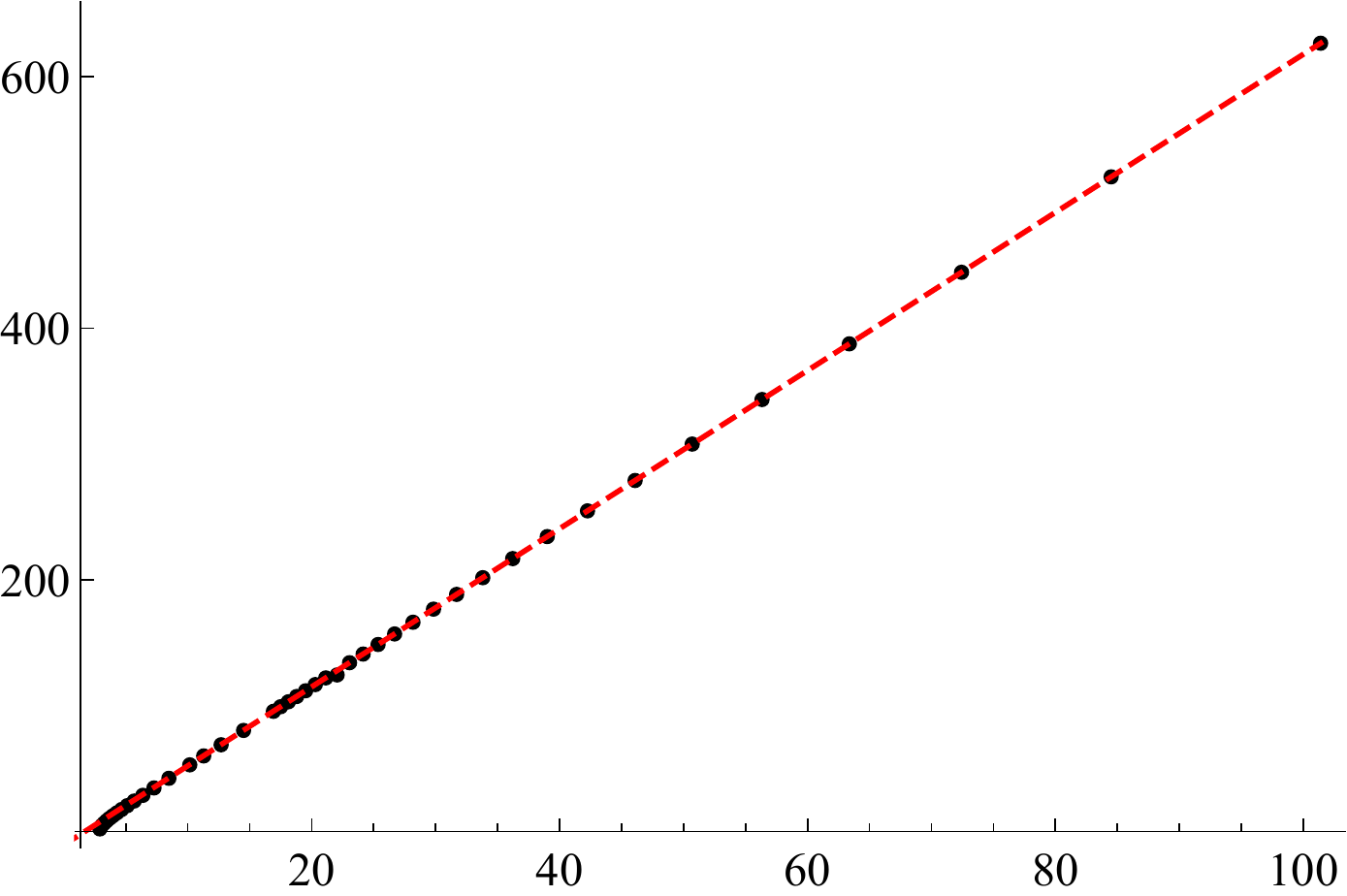} 
		\put(-185,130){$\log\frac{\mu_{\text{\tiny UV}}}{\mu_{\text{\tiny IR}}} $}
		\put(-35,18){$ |\operatorname{Im}\Delta_0|^{-1}$}
		\caption{\small  Size of the walking region as a function of the imaginary part of the dimension of the operator at the cFP. The black dots are the values computed in our model while the red dashed line  is a line with slope $2\pi$ that passes through the last black dot.}\label{fig:Miransky}
	\end{center}
\end{figure}

At very high temperature the thermodynamics is dominated by the physics of the UV CFT, hence $S_{\text{\tiny UV}} \propto T^3$. In a region with walking the entropy would show a similar temperature scaling, so $S/S_{\text{\tiny UV}}$ should be approximately constant. This plateaux can be clearly seen in the log-linear plot of  Figure~\ref{fig:entropy}. 
To quantify the size of the plateaux, we declare that the flow is in the walking region if the following derivative is smaller than a certain control parameter $\nu$
\begin{equation}
\frac{\dd \log(S/S_{UV})}{\dd \log (T/\Lambda)} < \nu\,. 
\end{equation}
For definiteness we take $\nu=\frac{1}{2}10^{-3}$. The condition is satisfied for values of the temperature in a bounded range $T_{\text{\tiny UV}}>T>T_{\text{\tiny IR}}$.  We identify these temperatures with the energy scales $\mu_{\text{\tiny UV}}$ and $\mu_{\text{\tiny IR}}$. We then vary 
$\epsilon$ and plot the ratio of these scales as a function of the imaginary part of the scaling dimension $\Delta_0$ at the cFP. The result is 
Figure~\ref{fig:Miransky}, which exhibits the expected scaling \eqref{Miransky}.

\section{Discussion}

The simple holographic model that we have presented captures the physics of FPA. Continuing the scalar field to complex values in such a way that the action remains holomorphic, it is also possible to describe cFPs. Then a straightforward extension of the rules of the gauge/gravity duality allows us to study not only the properties of conjectured cCFTs at strong coupling but also the complex RG flows between them. When the cFP are close to the real axis the real RG flow that passes exactly between them walks and displays the associated Miransky scaling behaviour 

The holomorphic gravitational action defined in this way is complex and non-Hermitian, so its meaning beyond the classical level is  unclear. It would be nice to relate our proposal to those in \cite{Witten:2010zr,Behtash:2015zha}, where it was argued that complex saddle points may give important contributions to the path integral in particular cases. 

Our work shows that, contrary to some belief in the community, the RG flow leading to FPA in large $N$ theories may be driven by a single-trace operator, the operator dual to the scalar field. Our model avoids the argument in Appendix D of \cite{Gorbenko:2018ncu} because the the  coupling constant associated to this operator is $\cO(N)$ instead of $\cO(1)$.

Our construction  is based on a bottom-up model. It would  be interesting to find top-down, string theory realisations of FPA and cFPs. While the scalar potentials obtained in consistent truncations of string and M-theory generically possess cFPs, an open question is whether their distance to the real axis is controlled by some parameter that can be freely varied. This issue is currently under investigation. 

\chapter{Transport in Strongly Coupled Quark Matter}
\label{Chapter6_Transport} 

\section{Introduction}

The first recorded observations of binary neutron star (NS) mergers, including both gravitational wave (GW) \cite{TheLIGOScientific:2017qsa,Abbott:2018exr} and electromagnetic (EM) \cite{GBM:2017lvd} signals, have opened intriguing new avenues for the study of strongly interacting matter at ultrahigh densities. While most of the attention in the field has so far been directed to the macroscopic properties of the stars and correspondingly to the Equation of State (EoS) of NS matter (see e.g.~\cite{Annala:2017llu,Rezzolla:2017aly,Capano:2019eae} and references therein), there exists ample motivation to inspect also transport properties of dense Quantum Chromodynamics (QCD), determined by viscosities and conductivities.
This is in particular due to the fact that 
while near the transition from nuclear to quark matter (QM) the EoSs of the two phases may largely resemble one another, the corresponding transport properties are expected to witness much more dramatic changes,
potentially enabling a direct detection of QM either in quiescent NSs or their binary mergers \cite{Schmitt:2017efp,Most:2018eaw}. In addition, understanding the relative magnitudes of different transport coefficients may turn out useful for the relativistic hydrodynamic simulations of NS mergers \cite{Baiotti:2016qnr,Alford:2017rxf}.

In the QM phase, expected to be found in the cores of massive NSs \cite{Annala:2019puf} and very likely created in NS mergers \cite{Most:2018eaw,Bauswein:2018bma,Chesler:2019osn,Ecker:2019xrw}, very few robust results exist for the QCD contribution to even the most central transport coefficients, including the bulk and shear viscosities and the electrical and heat conductivities. In fact, the only first-principles determination of these quantities in unpaired quark matter dates back to the early 1990s, amounting to a leading-order perturbative QCD (pQCD) calculation  \cite{Heiselberg:1993cr}. Owing to the strongly coupled nature of QM in the density regime relevant for NSs \cite{Kraemmer:2003gd,CasalderreySolana:2011us}, these results are, however, of limited predictive value. A nonpertubative analysis at strong coupling would be required for robust predictions, but the well-known problems that lattice Monte-Carlo simulations face at nonzero baryon densities presently prohibit the use of this standard tool (see e.g.~\cite{deForcrand:2010ys}).

Another complication in the determination of transport quantities in QM has to do with the fact that the physical phase of QCD realised at moderate densities is at present unknown \cite{Alford:2007xm}. While the general expectation is that some type of quark pairing is likely present all the way down to the deconfinement transition at low temperatures \cite{Rajagopal:2000wf}, it is currently unclear which particular phases are realised in Nature. Assuming that the physical moderate-density phase contains at least some nonzero fraction of unpaired quarks, it is, however, often considered a reasonable approximation to first inspect transport coefficients in the somewhat simpler case of unpaired QM \cite{Schmitt:2017efp}.

In this Chapter, we approach the most important transport properties of dense unpaired QM with the only first-principles machinery currently capable of describing strongly coupled quantum field theories at high baryon density: the gauge/gravity duality or, in short, holography. This correspondence provides a link between classical gravity and strongly coupled quantum field theories, relating their observables in a detailed manner (see e.g.~\cite{CasalderreySolana:2011us,Ramallo:2013bua,Brambilla:2014jmp} for reviews). In the context of heavy-ion physics, many important insights have been gained through the study of questions that are difficult to address with traditional field theory machinery, such as the details of equilibration dynamics (see e.g.~the recent review \cite{Berges:2020fwq}). In addition, the conjectured lower limit of the shear-viscosity-to-entropy ratio, $\eta/s\geq 1/(4\pi)$, has hinted towards universality in strongly coupled systems, which has had a profound effect on many subfields of theoretical physics \cite{Kovtun:2003wp}. 

In the context of NS physics, promising progress has  been achieved in applying holographic methods to 
determining the EoS of QCD matter
\cite{Hoyos:2016zke,Hoyos:2016cob,Ecker:2017fyh,Fadafa:2019euu,Ishii:2019gta}, but no attempts have been made to analyse transport in strongly coupled dense QM. Here, we shall take the first steps in this direction by utilising one of the most highly developed 
``bottom-up'' frameworks
designed to mimic a gravity dual of QCD at finite temperature and density \footnote{Note that at low densities, transport has been studied in the IHQCD and V-QCD frameworks in~\cite{Gursoy:2009kk,Gursoy:2010aa,Iatrakis:2014txa}.}, the Veneziano limit of the Improved Holographic QCD (IHQCD) model, V-QCD  \cite{Jarvinen:2011qe,Alho:2013hsa,Jokela:2018ers}. Interestingly, the results obtained for the most important transport coefficients of dense QM are in qualitative disagreement with known perturbative predictions, calling for caution in the application of the latter results to phenomenological studies of NSs.

For comparison and completeness, we shall contrast our V-QCD results not only to pQCD but also to one of the most widely studied top-down holographic model describing deconfined quark matter, \textit{i.e.}~the D3-D7 system in its unbackreacted limit \cite{Kruczenski:2003be,Kobayashi:2006sb}, implying we neglect the corrections from flavours This corresponds to the quenched approximation in field theory language \cite{Nunez:2010sf}. 
The latter provides reliable results for quantities that can be derived from the free energy \cite{Karch:2008uy}, which in the present case translates to the shear viscosity, obtainable via the relation $\eta/s=1/(4\pi)$. Other physical quantities are on the other hand expected to receive corrections from backreacted (unquenched) quark matter, which we shall indeed witness in our results. 

\section{Setup}

As explained above, we approach the description of dense strongly coupled matter via the string-theory inspired V-QCD model~\cite{Jarvinen:2011qe,Alho:2013hsa,Jokela:2018ers}. The way we have set up our analysis is, however, more general, and can accommodate other holographic models as well, such as the probe-brane limit of the D3-D7 system \cite{Kobayashi:2006sb,Itsios:2016ffv}. Both setups have been thoroughly applied to the study of the bulk properties of NS matter, and we refer the interested reader to references~\cite{Hoyos:2016zke,Annala:2017tqz,Jokela:2018ers,Chesler:2019osn,Ecker:2019xrw,Fadafa:2019euu}.

A significant difference between two holographic models lies in how the effects of the flavour sector are treated. V-QCD has these effects systematically built in, whereas the 
D3-D7 system treats the flavours as a probe
(note, however, that for massless quarks 
references \cite{Bigazzi:2009bk,Bigazzi:2011it,Bigazzi:2013jqa,Faedo:2016cih,Faedo:2017aoe} have gone beyond the probe approximation). Despite this difference, we can define the two models via the same gravitational action consisting of two terms $S_\mt{total} = S_\mt{g} + S_\mt{f}$, where
\begin{eqnarray}
S_\mt{g} &=& N_c^2 M_\text{Pl}^3\int \d^5 x \sqrt{-g} \left( R - \frac{1}{2} \partial_\rho \phi \, \partial^\rho\phi - \pot(\phi) \right) \label{Eq.ActionDefs} \\
S_\mt{f} & =& -N_f N_c M_\text{Pl}^3 \int \d^5 x \Z(\phi,\chi) \sqrt{-\det \left(\Gamma_{\mu\nu} \right) } \label{Eq.ActionWithTachyon}\ .
\end{eqnarray}
Here, $g$ is the determinant of the metric $g_{\mu\nu}$, $M_\text{Pl}$ denotes the rescaled five-dimensional Planck mass, $R$ is the Ricci scalar, and we have defined \be
\Gamma_{\mu\nu} =  g_{\mu\nu} + \k(\phi,\chi)  \partial_\mu \chi  \partial_\nu\chi + \W(\phi,\chi)  F_{\mu\nu}\ .
\ee
The potentials and couplings $V(\phi)$, $\Z(\phi,\chi)$, $\k(\phi,\chi)$, and $\W(\phi,\chi)$ will be different functions in D3-D7 and in V-QCD as we shall discuss below.

Of the two parts of the action, $S_\mt{g}$ is related to the glue sector of a gauge theory with rank $N_c$\footnote{In this Chapter we will use $N_c$ to denote the number of colours rather than $N$, in order to make the distinction from the number of flavours $N_f$ clear.}. The scalar field $\phi$ is identified with the dilaton, which according to the holographic dictionary is dual to the Yang--Mills running coupling constant. For V-QCD, following the IHQCD model \cite{Gursoy:2007cb,Gursoy:2007er}, the potential $V$ in~\eqref{Eq.ActionDefs} is chosen to reproduce the known physics of Yang--Mills theory upon comparison with perturbative results and lattice data~\cite{Gursoy:2009jd,Panero:2009tv,Jokela:2018ers}. In the D3-D7 model, $S_\mt{g}$ is on the other hand determined by the closed string sector of type IIB supergravity and $V=-12/L^2$. To fix units, we demand that the asymptotically AdS space has unit radius $L=1$.

At the same time, the physics of $N_f$ flavours of fundamental quarks is captured by the Dirac--Born--Infeld (DBI) action $S_\mt{f}$, where the tachyon field $\chi$ is dual to the chiral condensate $\bar qq$, whose boundary value is related to the masses of the quarks~\footnote{For V-QCD, the flavour setup is based on~\cite{Bigazzi:2005md,Casero:2007ae,Iatrakis:2010zf,Iatrakis:2010jb}.}. We set the quark masses to zero in V-QCD, but keep them non-vanishing in the D3-D7 model to achieve the breaking of conformal symmetry. In the D3-D7 model, a quark mass corresponds to the energy necessary to introduce an additional quark over the ground state, implying that it corresponds to a constituent quark mass, with a value of the order of $1/N_c$ times the baryon mass \cite{Hoyos:2016zke}.
The field strength $F_{\mu\nu}=\partial_\mu A_\nu-\partial_\nu A_\mu$ on the other hand provides the dynamics for the U(1)$_\mt{B}$ gauge field $A_\mu$ corresponding to the conserved baryonic charge of the dual field theory. 

\begin{figure*}[t!]
	\center
	\includegraphics[width=0.49\textwidth]{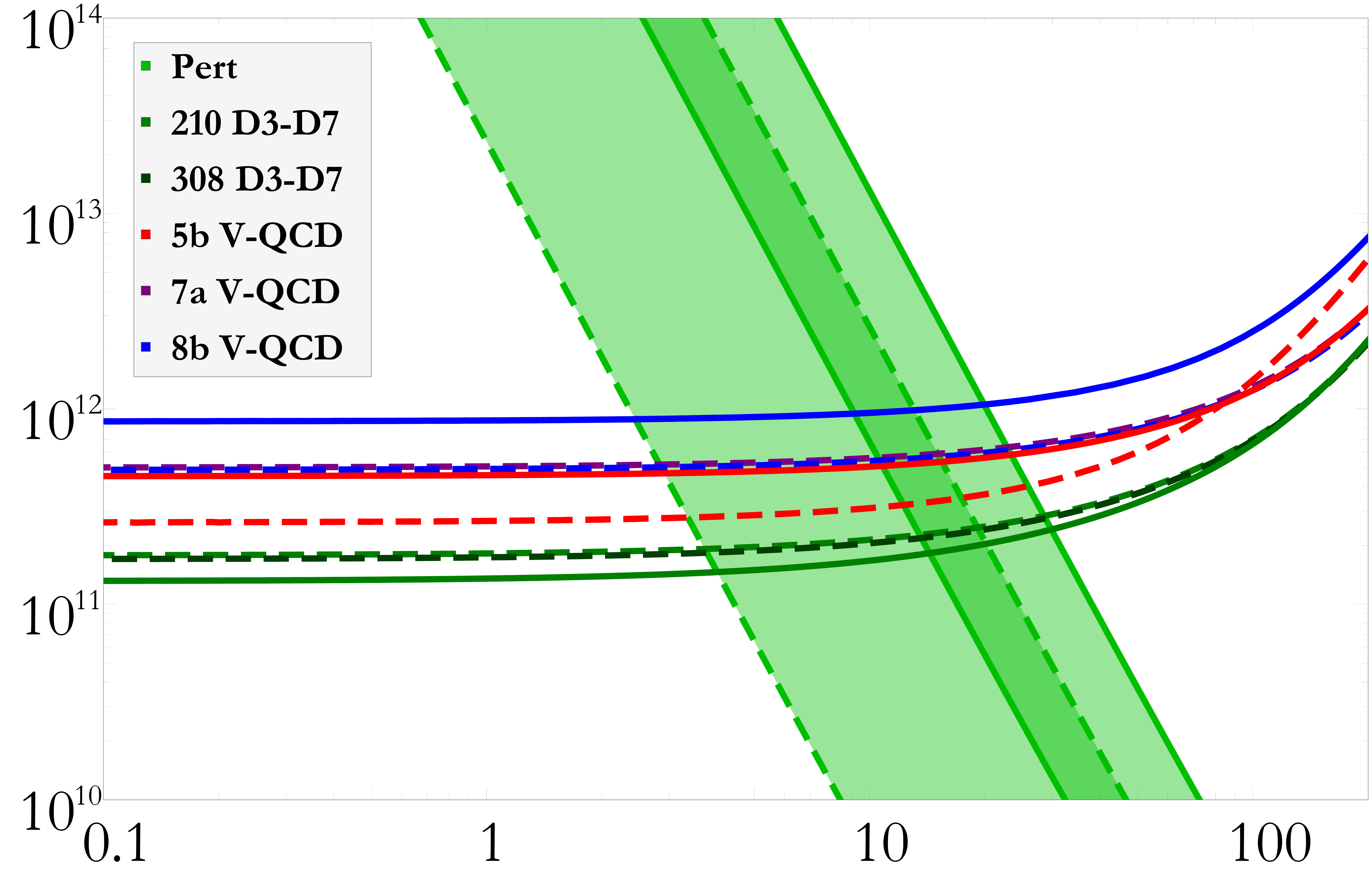}
	\put(-160,115){\footnotesize $\eta\, [\text{g}\, \text{cm}^{-1}\, \text{s}^{-1}]$}
	\put(-100,-5){\footnotesize $T\, [\text{MeV}]$}
	\hfill
	\includegraphics[width=0.49\textwidth]{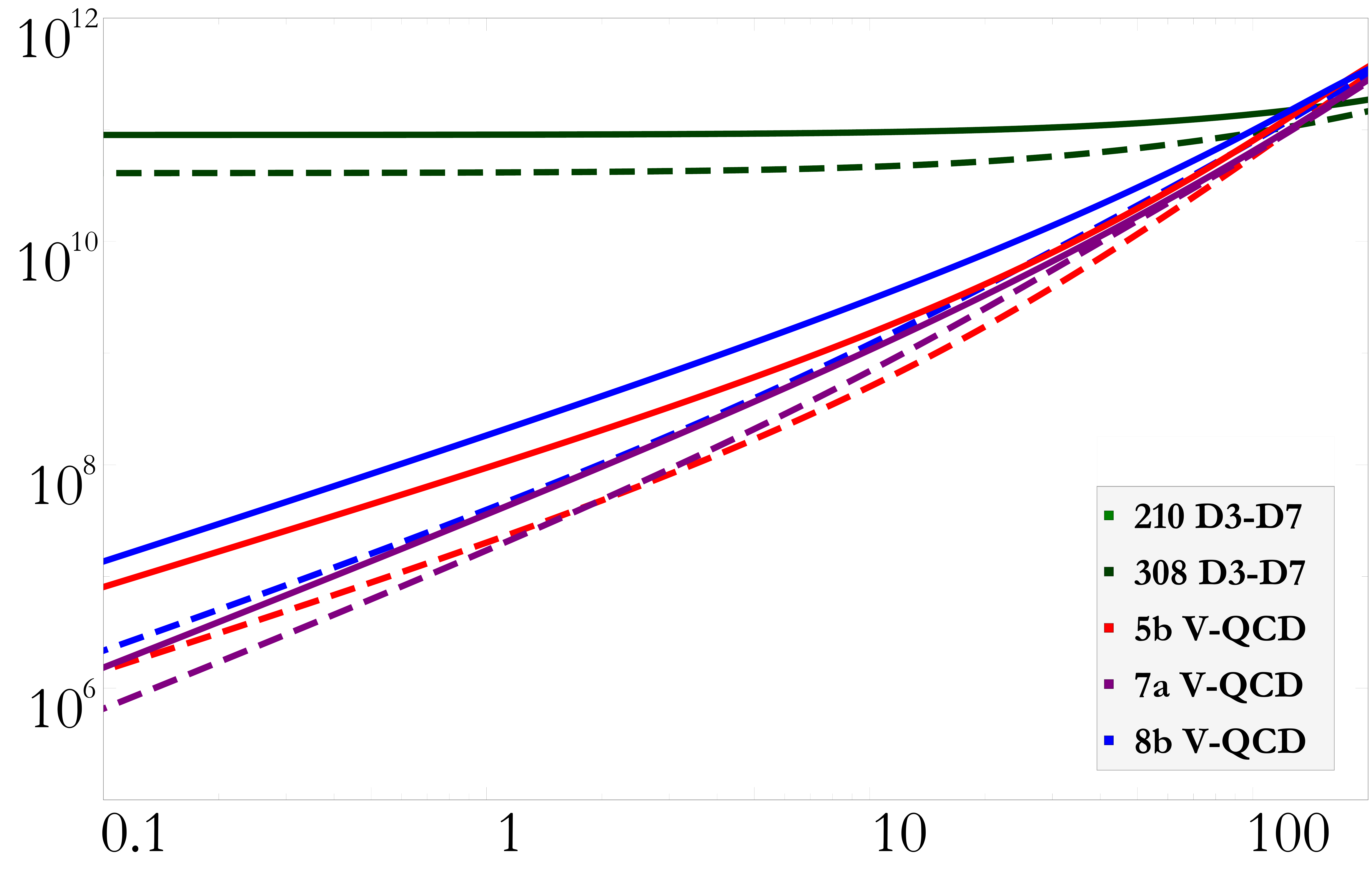}
	\put(-160,115){\footnotesize $\zeta\, [\text{g}\, \text{cm}^{-1}\, \text{s}^{-1}]$}
	\put(-100,-5){\footnotesize $T\, [\text{MeV}]$}
	\caption{Shear (left) and bulk (right) viscosities as functions of temperature for $\mu=450\,\operatorname{MeV}$ (dashed lines) and $\mu=600\,\operatorname{MeV}$ (solid lines). The filled bands on the left correspond to the pQCD results (upper band obtained for $\mu=600$ MeV, lower for 450 MeV), and they have been generated by varying the parameter $x$ inside $\alpha_s$ (see the main text for the definition) inside the interval $1/2\leq x\leq 2$.}
	\label{fig:visc}
\end{figure*}

In the V-QCD model, the functions $\Z$, $\k$, and $\W$ are fixed to reproduce the desired features of QCD (such as confinement and asymptotic freedom) both for weak \cite{Gursoy:2007cb,Jarvinen:2011qe} and strong \cite{Gursoy:2007er,Jarvinen:2011qe,Arean:2013tja,Arean:2016hcs,Jarvinen:2015ofa,Ishii:2019gta} Yang--Mills coupling. Moreover, the potentials were tuned to match with lattice data for the EoS at small chemical potentials \cite{Jokela:2018ers}\footnote{see Section \ref{app:vqcd} from Appendix~\ref{Appendix:Transport}, which includes references~\cite{Donos:2014cya,Gouteraux:2018wfe,Alho:2015zua,Alho:2020gwl,Gubser:2000nd,Borsanyi:2013bia,Borsanyi:2011sw,Jokela:2020piw,Brodsky:2010ur}.}. In this way, the model in effect extrapolates the lattice results to the regime relevant for NS cores. We also determine $M_\text{Pl}$ by using lattice data and set $N_f/N_c=1$, and furthermore employ the couplings of the fits \textbf{5b}, \textbf{7a}, and \textbf{8b} given in Appendix~A of~\cite{Jokela:2018ers}. 

For the D3-D7 model, supergravity implies the simple relations
\begin{equation}
M_\text{Pl}^3=\frac{1}{8\pi^2},\, \Z=\frac{\lym}{2\pi^2}\cos^3\chi,\, \W=\frac{2\pi}{\sqrt{\lym}},\,  \k=1,
\end{equation}
where $\lym=g_{\text{YM}}^2 N_c$ is the 't~Hooft coupling (with $g_{\text{YM}}$ the Yang--Mills coupling) of the dual field theory. Following \cite{Hoyos:2016zke}, we fix $\lym\simeq 10.74$ so that the pressure matches the Stefan--Boltzmann value at asymptotically large chemical potentials.

The holographic determination of the thermodynamic quantities and transport coefficients is briefly reviewed in Appendix~\ref{Appendix:Transport} (see Section~\ref{app:setup} including references \cite{Mateos:2006nu,Eling:2011ms}). To translate the resulting expressions into numerical results, we need to choose, among other things, the characteristic perturbative energy scale $\Lambda_\text{UV}$ for V-QCD. We use the values $\Lambda_\text{UV}=226.24,\, 210.76,\, \text{and}\, 156.68\, \operatorname{MeV}$ for the fits \textbf{5b}, \textbf{7a}, and \textbf{8b} in the V-QCD model, following the choices made in \cite{Jokela:2018ers,Chesler:2019osn}. The quark mass in the D3-D7 model is on the other hand given  two values --- $M_q=210.76\, \operatorname{MeV}$ for direct comparison with the V-QCD potential  \textbf{7a} \footnote{Recall that these parameters, $M_q$ for D3-D7 and $\Lambda_\text{UV}$ for V-QCD, both control how the conformal symmetry is broken. Their values must therefore be comparable, even though not necessarily exactly the same.}, and $M_q=308.55\, \operatorname{MeV}$ following the logic of~\cite{Hoyos:2016zke} --- which are used to (partially) probe the systematic uncertainties of this model. 

An important point to note is that in both of our models, we consistently work with three mass-degenerate quark flavours assuming beta equilibrium, where all quark flavours share the same chemical potential $\mu=\mu_\text{B}/3$. This implies the absence of electrons in the system, which would only change upon taking flavour-dependent masses into account.

\section{Viscosities}

The shear and bulk viscosities of dense QCD matter describe the resistance of the system to deformations. They become relevant in settings where NSs are either strongly deformed or their interiors taken out of thermal equilibrium, both of which occur in different stages of binary NS mergers \cite{Alford:2017rxf,Fujibayashi:2017puw}. In addition, viscosities play a role in determining the damping of unstable r-modes in rapidly rotating stars \cite{Andersson:1997xt,Friedman:1997uh,Andersson:2000mf,Arras:2002dw,Alford:2010gw,Haskell:2015iia}.

Viscosities appear in contributions to stress forces due to an inhomogeneous motion of the fluid. Letting $v_i$, $i=1,2,3$, be the components of the velocity of a fluid moving at low 
velocities, the resulting stress becomes
\begin{equation}
T_{ij}=-\eta\  \left(\partial_{i}v_{j}+\partial_{j}v_{i}\right) -\left(\zeta-\frac{2}{3}\eta \right)\delta_{ij}\partial_k v^k\ ,
\end{equation}
where $\eta$ and $\zeta$ are the shear and bulk viscosities, respectively.

The values obtained from the holographic models are plotted in Figure~\ref{fig:visc} (left) together with the pQCD result for unpaired quark matter \cite{Heiselberg:1993cr,Schmitt:2017efp},
\begin{equation}
\eta\approx 4.4\times 10^{-3} \frac{\mu^2 m_\text{D}^{2/3}}{\alpha_s^2T^{5/3}}\ .
\end{equation}
In evaluating this expression, we have used the one-loop Debye mass $m_\text{D}^2=2\alpha_s(N_f \mu^2+(2N_c+N_f)\pi^2 T^2/3)/\pi$ (see e.g.~\cite{Vuorinen:2003fs}) and the two-loop strong coupling $\alpha_s$, related to the QCD gauge coupling via $\alpha_s=g_{_\text{QCD}}^2/(4\pi)$. Following typical conventions in the literature (see e.g.~\cite{Kurkela:2009gj,Kurkela:2016was} and the discussion in the beginning of Section~7 in \cite{Ghiglieri:2020dpq}), the renormalisation scale is set to $\bar{\Lambda}=x\sqrt{(2\pi T)^2+(2\mu)^2}$, where $x$ is a parameter that parametrises the renormalisation scale dependence of the pQCD result and is varied between the values 1/2 and 2. Finally, the QCD scale $\Lambda_\text{QCD}$ is given the value 378 MeV, obtained by demanding that $\alpha_s(\bar{\Lambda}=2\, \textrm{GeV})=0.2994$ \cite{Amsler:2008zzb}.

We observe that at high temperatures the shear viscosity of the strongly coupled fluid is qualitatively larger than the perturbative result, while at low temperatures it approaches a constant with the crossing of the holographic and pQCD results taking place around $T\sim 5-50\,\operatorname{MeV}$. We also note that in agreement with our naive expectation, both holographic models give comparable results, as the D3-D7 calculation is not hampered by problems related to backreaction in this case.

\begin{figure*}[t!]
	\center
	\includegraphics[width=0.49\textwidth]{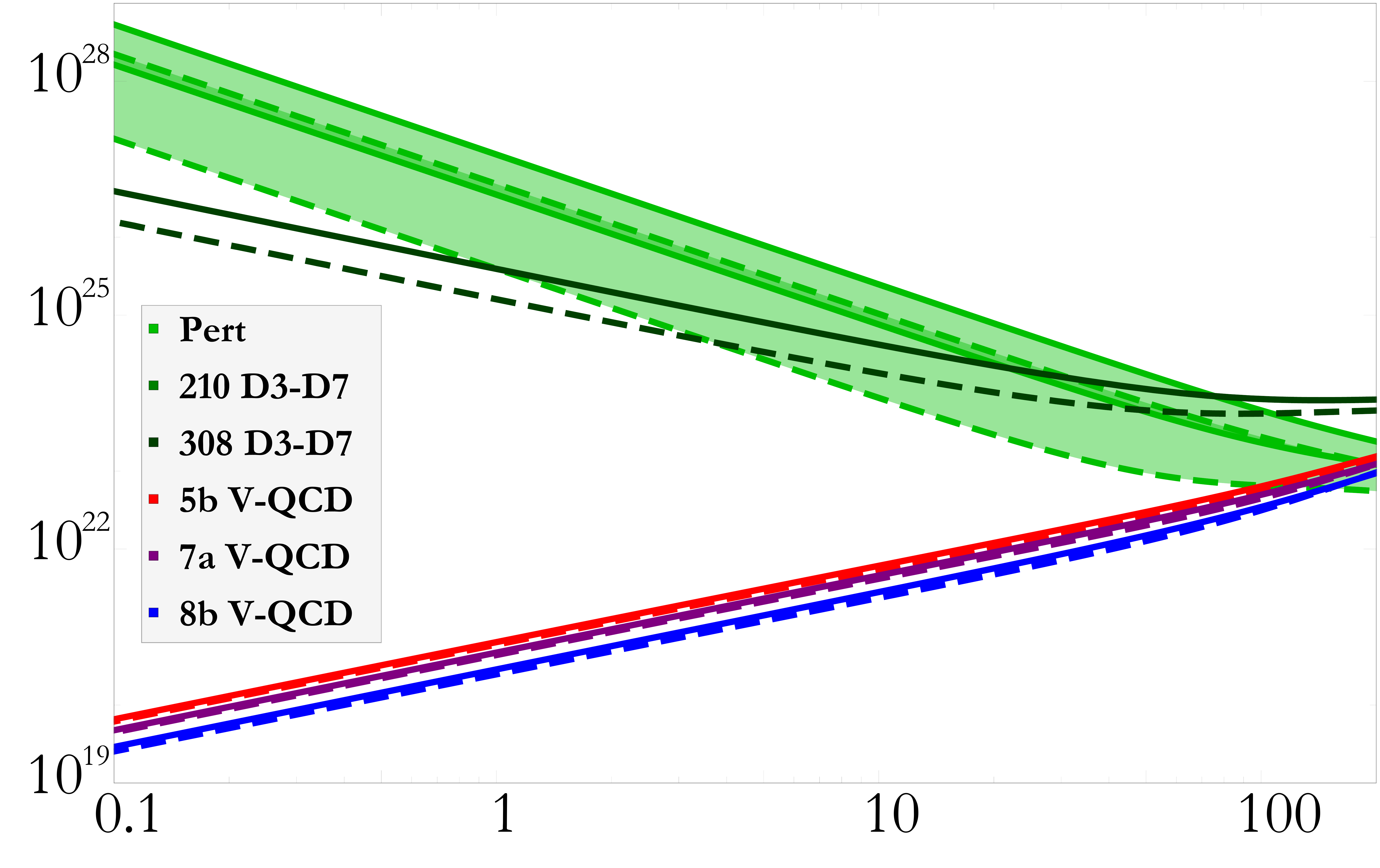}
		\put(-160,115){\footnotesize $\sigma\, [\text{s}^{-1}]$}
	\put(-100,-5){\footnotesize $T\, [\text{MeV}]$}
	\hfill 
	\includegraphics[width=0.49\textwidth]{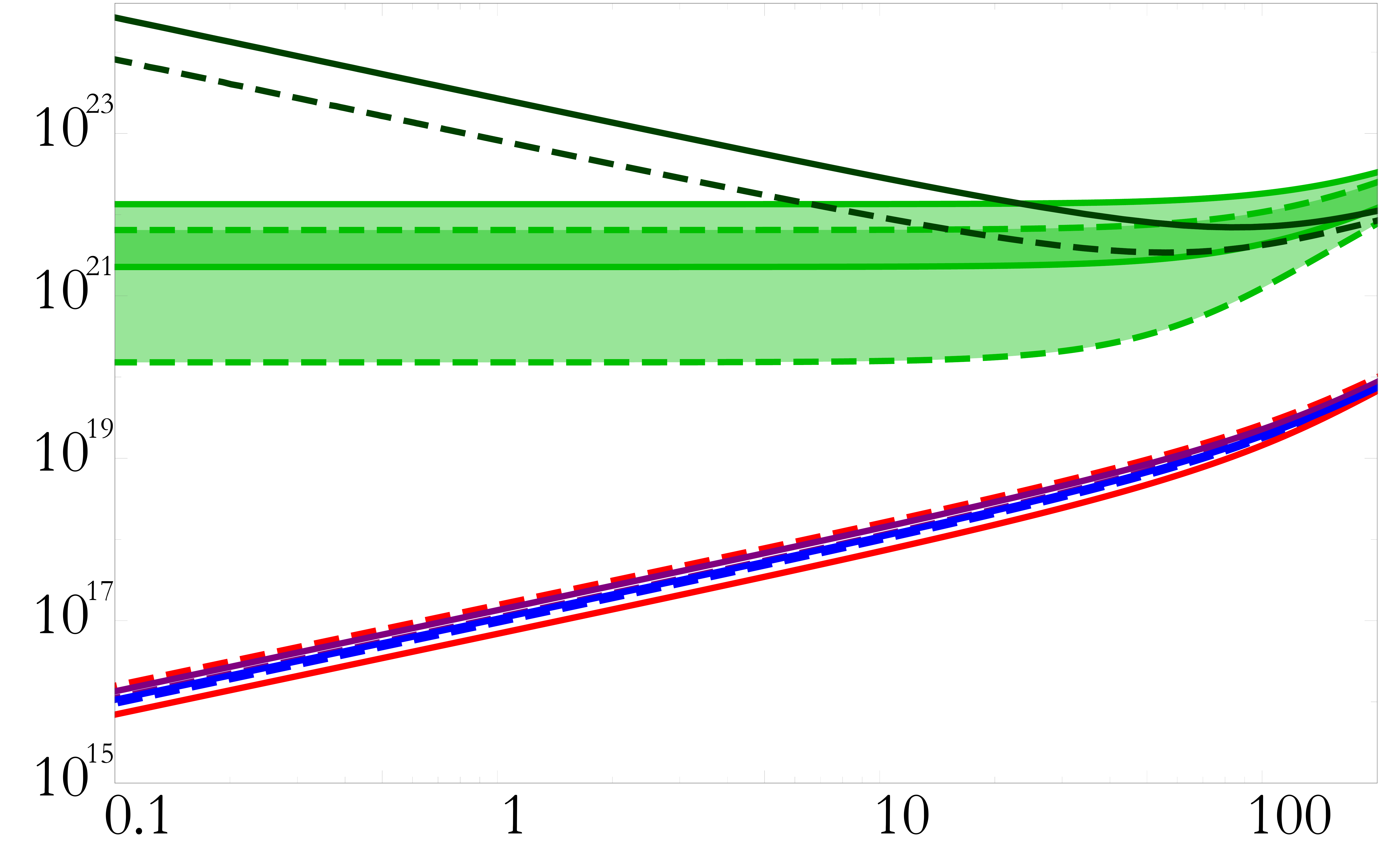}
		\put(-160,115){\footnotesize $\kappa\, [\text{erg}\, \text{cm}^{-1}\,\text{K}^{-1}\, \text{s}^{-1}]$}
	\put(-100,-5){\footnotesize $T\, [\text{MeV}]$}
	\caption{Electrical (left) and thermal (right) conductivities as functions of temperature for $\mu=450\,\operatorname{MeV}$ (dashed lines) and $\mu=600\,\operatorname{MeV}$ (solid). The filled bands again correspond to the pQCD results for $1/2\leq x\leq 2$, with the $\mu=600\,\operatorname{MeV}$ bands slightly above the 450 MeV ones.}
	\label{fig:cond}
\end{figure*}

In addition, we have determined the QCD contribution to the bulk viscosity $\zeta$, which  however represents only a subdominant contribution to r-mode damping \footnote{The dominant one originates from chemical equilibration through weak processes, $\zeta \sim 10^{24}-10^{29}\,\operatorname{g\, cm^{-1}s^{-1}}$ \cite{Alford:2010gw,Schmitt:2017efp}. A quantitative analysis of this effect is unfortunately outside the scope of our work.}. The corresponding results are shown in Figure~\ref{fig:visc} (right), where we observe that in V-QCD $\zeta$ is highly suppressed in comparison with $\eta$ at low temperatures, but reaches a value of around $10\%$ of the shear viscosity at high temperatures. The D3-D7 model on the other hand leads to a very flat curve in the range of temperatures studied, but the validity of this model is clearly questionable due to the flavour quenching. In leading-order pQCD, the bulk viscosity finally vanishes when the quark masses are negligible compared to the chemical potentials, and remains small even at high temperatures \cite{Arnold:2006fz} (note however that nonperturbative effects are expected to increase the value somewhat \cite{Kharzeev:2007wb,Wang:2011ur}), which explains the lack of a pQCD curve in this figure.

\section{Conductivities}

A newly formed NS undergoes a cooling process mainly through the emission of neutrinos from its interior. The neutrinos transport heat to the surface, where the energy is emitted as radiation. A closer inspection of this process shows that the thermal evolution of the star depends on several quantities, including the heat conductivity that determines the heat flux to the surface and the electrical conductivity that determines the magnitude of Joule heating through the decay of magnetic fields \cite{Yakovlev:2004iq,Page:2005fq,Vigano:2013lea}. In the postmerger phase of a NS binary merger, the electrical and thermal conductivities are furthermore relevant for equilibration and the evolution of magnetic fields \cite{Alford:2017rxf,Harutyunyan:2018mpe}. Finally, these quantities may in principle prove useful in distinguishing between different phases of QCD through the observation of thermal radiation.

In the strongly coupled theories that we study in the present work, matter resides in a state that can be described as a relativistic fluid. This implies that the electrical and thermal conductivities are not independent, but are determined by a single coefficient $\sigma$, defined by the constitutive relation of the current arising from a gradient of the chemical potential,
\begin{equation}
J_x=-\sigma\partial_x\left(\frac{\mu}{T}\right)\ .
\end{equation}
The electrical conductivity, defined as the ratio between the current and the electric field $E_x$, can be shown to take the form (see Section~\ref{app:cond} from Appendix~\ref{Appendix:Transport})
\begin{equation}
\sigma^{xx}=\frac{J^x}{E_x}=\frac{\varepsilon+p}{Ts}\sigma\ ,
\end{equation}
where $\varepsilon$, $p$, and $s$ are the energy density, pressure, and entropy density, respectively. Moreover, the thermal conductivity, defined as the ratio between the heat current $Q^x$ and the temperature gradient, becomes
\begin{equation}\label{eq:kappaxx}
\kappa^{xx}=\frac{Q^x}{-\partial_x T}=\frac{\mu}{T}\frac{\varepsilon+p}{\rho}\sigma =\frac{\mu\, s}{\rho} \,\sigma^{xx}\ ,
\end{equation}
where $\rho$ is the charge density. It should be stressed that these expressions hold only for a steady state, where the gradients of the temperature and chemical potential balance each other, so that the transport does not occur via convection.

With the above results established, we see that it suffices to compute the electrical conductivity in the holographic models. Our results for the two conductivities are displayed in Figure~\ref{fig:cond} together with the perturbative results for unpaired quark matter \cite{Heiselberg:1993cr,Schmitt:2017efp} 
\begin{equation}
\sigma^{xx}\approx 0.01\frac{\mu^2 m_\text{D}^{2/3}}{\alpha_s T^{5/3}}\ , \ \kappa^{xx}\approx 0.5 \frac{m_\text{D}^2}{\alpha_s^2} \ .
\end{equation}
Here, we continue to use the same values for $m_\text{D}$ and $\alpha_s$ as listed in the previous Section, and the electrical conductivity is given in units of $e^2/(\hbar c)$.

We observe that both in the perturbative and D3-D7 calculations the conductivities either decrease with temperature or are largely independent of it, while in the V-QCD model they are increasing functions of $T$. This qualitative discrepancy can be easily understood
as a suppression of momentum dissipation of charged particles coming from the assumptions of these models: in the pQCD case, the coupling is assumed  small by construction, whereas in the D3-D7 case, only the leading order effect in an $N_f/N_c$  expansion is retained \cite{Karch:2007pd}.
There are strong reasons to expect that properly including backreaction in this holographic model would bring the D3-D7 result into at least qualitative agreement with the V-QCD one (see e.g.~the discussion in \cite{Tarrio:2013tta}).

\section{Discussion}

A first-principles microscopic determination of the fundamental properties of dense QCD matter is a notoriously difficult problem, not least due to the strongly coupled nature of the system at phenomenologically relevant energies. In recent years, the gauge/gravity duality has shown considerable promise as a potential tool, as it allows approaching the problem from an angle complementary to traditional field theory methods; indeed, promising results have been obtained for many bulk thermodynamic quantities, leading to predictions for observables such as the NS mass-radius relation and the phase diagram of the theory \cite{Hoyos:2016zke,Jokela:2018ers,Chesler:2019osn}. 

In this Chapter, we have used the holographic machinery to tackle a more challenging class of physical quantities characterising the response of the medium to external perturbations. In particular, we have studied the behaviour of the transport coefficients most relevant for the physics of NSs and their mergers, i.e.~the shear and bulk viscosities and the thermal and electrical conductivities. All  these quantities have been evaluated in a highly developed bottom-up framework mimicking a gravity dual for QCD, V-QCD, and the corresponding results subsequently compared to those from the D3-D7 probe brane setup and perturbative QCD \cite{Heiselberg:1993cr}.

Our main results are depicted in Figs.~\ref{fig:visc} and \ref{fig:cond}. Inspecting these plots, an issue that stands out immediately is the stark contrast between the V-QCD and pQCD predictions for all quantities, for which both results are available: in V-QCD, transport coefficients typically increase with  temperature, while a qualitatively different behaviour is predicted by pQCD. As a result, in the $T=0$ limit V-QCD coefficients are comparatively strongly suppressed. In addition, while we observe qualitative agreement between the V-QCD and D3-D7 predictions for the shear viscosity, the same is not true for the other three quantities studied. These observations clearly call for physical interpretation.

As flavours are quenched in the D3-D7 model, the main dissipative effect affecting quark matter is due to drag by the thermal plasma \cite{Gubser:2006bz,Herzog:2006gh}. The increasing trend of the electrical conductivity at lower temperatures reflects the decreasing drag force, which is however expected to be cut off by radiation effects not captured by the D3-D7 analysis
\cite{Mikhailov:2003er,Chernicoff:2008sa,Karch:2007pd}. Furthermore, at very low temperatures, the quenched approximation is expected to break down altogether \cite{Bigazzi:2011it,Bigazzi:2013jqa}. The V-QCD model on the other hand does not suffer from these issues, as flavours are unquenched, and consequently the trends exhibited by  both conductivities can be expected to reflect the true behaviour of these quantities  in strongly coupled unpaired quark matter.

For the viscosities, the situation is somewhat different. In both holographic setups, the shear viscosity is proportional to a quantity derived from the free energy, which leads to a fair agreement between the two predictions. On the other hand, in the pQCD calculation flavour contributions give rise to a strong increase of the quantity in the $T\to 0$ limit, leading to a stark disagreement with the holographic results. In contrast, for the bulk viscosity, for which no pQCD result is available, we witness a marked sensitivity of the result to the pattern in which conformal invariance is broken (by a running coupling vs.~quark masses) in our holographic models, which essentially invalidates the prediction of the D3-D7 model for this quantity.

In summary, we have seen dramatic differences arise between the predictions of a strongly coupled unquenched theory and its perturbative or quenched approximations for the transport properties of dense quark matter. This observation clearly calls for significant caution in the application of the perturbative results in any phenomenological study within NS physics, and highlights the necessity of further developing the holographic approach to the problem.

\newpage
\thispagestyle{empty}

\vspace{0.3cm}


\chapter{Conclusions}


In this thesis we have studied a one-parameter family of quiver-like, super Yang--Mills Chern--Simons--matter gauge theories by means of their gravitational duals. These theories generically posses a mass gap but are not confining in the sense of a linear growth of the potential between quarks at large separations. Thus, they provide a counterexample to the expectation that holographic duals of gauge theories with a mass gap also exhibit confinement. 

For the two limiting values of the parameters the theories enjoy distinct infrared behaviour though: they flow either to a conformal field theory in one case or to a gapped and confining phase in the other. 
Additionally, nearby theories exhibit quasi-conformal dynamics and  quasi-confining dynamics respectively. These limiting behaviours showed up in many physical quantities such as the quark-antiquark potential and entanglement entropy measures in Chapter \ref{Chapter3_observables} or thermodynamical quantities in Chapter~\ref{Chapter4_thermo}.

Quasi-conformal dynamics is caused in our case by the presence of the Oogury--Park conformal field theory \cite{Ooguri:2008dk} near the Renormalisation Group flow of the theories, in the way it is depicted in Figure~\ref{fig:triangle}. The fact that this scale invariant theory possesses an exact moduli space and the quasi-conformal behaviour of nearby theories motivated our search for a light dilaton in their spectrum. The spectrum of spin-0 and spin-2 particles was computed and shown in Chapter~\ref{Chapter3_observables}. In this way we provided evidence on the theories being gapped, but no light dilaton was found. We argued that the reason behind its absence was that the flow departing from the fixed point was triggered by both sources and VEVs of comparable sizes.

Inspired by \cite{Klebanov:2007ws}, we also wanted to question whether entanglement entropy measures are sensitive to confining dynamics. In Chapter~\ref{Chapter3_observables} we realised that, despite being quite sensitive to the presence of a conformal field theory, they do not discriminate between confining and non-confining theories. 

Not only did we studied these family of theories at zero temperature, but also considered the finite temperature case by computing black brane solutions in the gravity side. This calculation was performed in Chapter \ref{Chapter4_thermo}, unveiling a rich phase diagram with first and second order phase transitions, with both a critical and a triple point. Finite temperature thermodynamic quantities approach those of the limiting conformal and confining cases smoothly. In particular, dimensionless combinations of the free energy, the entropy and the temperature flattened in theories flowing close to the fixed point. This was a nice manifestation of quasi-conformal dynamics being realised.

In Chapter~\ref{Chapter5_HoloCCFTs} we wanted to explore more deeply the effects that a conformal field theory near the Renormalisation Group flow of a theory could have. For that, we constructed the holographic duals of complex conformal field theories. In this case, the conformal field theories between which the theory flows are placed at complex valued operators couplings. We also investigated the scenario in which fixed point annihilation is realised in our holographic approach. In the holographic construction, the expected scaling is reproduced. Also, our contribution shows that, contrary to some belief in the community, the RG flow leading to fixed point annihilation in large $N$ theories may be driven by a single-trace operator, the operator dual to the scalar field

Finally, we studied transport coefficients in strongly coupled unpaired quark matter in Chapter \ref{Chapter6_Transport}, with the aim of providing a phenomenological application of the gauge/gravity correspondence. We argued that transport properties may witness much more dramatic changes in the presence of different phases inside neutron stars, which could potentially enable a direct detection of quark matter. Because of that, we computed the most central transport coefficients, mainly bulk and shear viscosities, as well as thermal and electrical conductivities in a top-down model (probe D7-branes in the D3-background) and in one of the most developed bottom-up frameworks constructed to describe Quantum Chromodynamics (the V-QCD model). Interestingly, the results obtained for the most important transport coefficients of dense quark matter were in qualitative disagreement with known perturbative predictions, calling for caution in the application of the latter results to phenomenological studies of neutron stars.

\chapter*{Resumen}
\addcontentsline{toc}{chapter}{Resumen}  

En esta tesis hemos utilizado la dualidad hologr\'afica para entender el r\'egimen no perturbativo de una familia uni-param\'etrica de teor\'ias con m\'ultiples escalas. T\'ipicamente, esta dualidad motivada por la teor\'ia de (super)cuerdas identifica teor\'ias gauge en espacio plano con teor\'ias de cuerdas en espacio curvo. Su utilidad radica en que relaciona el r\'egimen donde las teor\'ias cu\'anticas de campos son fuertemente acopladas con gravedad cl\'asica gobernada por las ecuaciones de Einstein. En la introducci\'on de esta tesis (Cap\'itulo~\ref{Chapter1_IntroductionThesis}), hemos hecho un breve repaso a los ingredientes esenciales que necesitamos de teor\'ia de cuerdas, a la vez que introducimos algunos resultados previos que son el punto de partida de nuestras investigaciones.

En el segundo Cap\'itulo de esta tesis recogemos expl\'icitamente todas las soluciones de supergravedad que describen la familia de teor\'ias cu\'anticas de campos en tres dimensiones que investigamos en la mayor parte de esta tesis. Gen\'ericamente, comparten la misma f\'isica a altas energ\'ias, dada por interacciones de Yang--Mills. Adicionalmente, describen sistemas con interacciones de Chern--Simons. El inter\'es de su estudio radica en la gran variedad de fenomenolog\'ia no perturbativa que se obtiene a bajas energ\'ias.

Dicha fenomenolog\'ia es muy rica. En particular, para un valor gen\'erico del par\'ametro que distingue las distintas teor\'ias, estas desarrollan un salto de masa (mass gap) en su espectro de estados. Ahora bien, las teor\'ias que se obtienen al tomar los dos valores l\'imites del par\'ametro que las distingue son especiales. En un caso, la teor\'ia fluye a un punto fijo en el infrarrojo, que corresponde a una teor\'ia de campos conforme. En el otro caso l\'imite, se obtiene una teor\'ia confinante, en la que el potencial entre una pareja de quark y antiquark crece linealmente con la distancia entre ellos cuando la separaci\'on es suficientemente grande.

Todos estos fen\'omenos, junto con el c\'alculo del espectro de estados con esp\'in 0 y esp\'in 2, se estudian en el Cap\'itulo~\ref{Chapter3_observables}. El hecho de que en este sistema el grupo de renormalizaci\'on pueda fluir arbitrariamente cerca del punto fijo mencionado anteriormente, motiv\'o la b\'usqueda de un estado escalar ligero en dicho espectro. No obstante, no se encontr\'o ninguno, debido a que los valores de la fuente y del valor de expectaci\'on en el vac\'io del operador que evitaba llegar al punto fijo eran del mismo orden.

Adem\'as, en ese Cap\'itulo se estudian diferentes medidas de entrelazamiento cu\'antico. Esta \'ultima investigaci\'on est\'a motivada por el hecho de que en la literatura se propusieran cantidades derivadas del entrelazamiento cu\'antico entre regiones del espaciotiempo como magnitudes que pudieran caracterizar teor\'ias confinantes. Nuestros c\'alculos muestran que, cuando estas cantidades se consideran en la familia de teor\'ias en la que nos hemos centrado, no son capaces de discriminar entre teor\'ias con confinamiento (en el sentido de un crecimiento lineal en el potencial entre quarks) y teor\'ias con un simple salto de masa y sin comportamiento confinante.

Pero en nuestro proyecto no hemos querido detenernos en el estudio de estas teor\'ias supersim\'etricas a temperatura cero. Por esa raz\'on, en el Cap\'itulo~\ref{Chapter4_thermo} hemos construido num\'ericamente soluciones de branas negras (en cierto sentido similares a agujeros negros en cuatro dimensiones, pero cuya superficie se extiende a lo largo de direcciones no compactas). Estas soluciones describen estados t\'ermicos de las diferentes teor\'ias. As\'i, hemos descubierto un diagrama de fases muy rico, dotado de transiciones de fase de primer y segundo orden. En particular, dicho diagrama contiene un punto cr\'itico y un punto triple.

Interesados por el efecto que una teor\'ia conforme de campos pudiera tener si es cercana al flujo del grupo de renormalizaci\'on de otra teor\'ia, en el Cap\'itulo~\ref{Chapter5_HoloCCFTs} nos adentramos en el estudio de teor\'ias conformes de campos complejas. Hemos construido el dual hologr\'afico a dicha propuesta y analizado algunas de sus propiedades en el caso en que el acoplo es fuerte.

Finalmente, en el \'ultimo Cap\'itulo de la tesis, hemos estudiado coeficientes de transporte en teor\'ias hologr\'aficas que pretenden describir fases densas de Cromodin\'amica Qu\'antica. Hemos visto que los resultados del c\'alculo hologr\'afico difieren de los que se obtienen perturbativamente. Estos estudios podr\'ian tener consecuencias fenomenol\'ogicas y encontrar su aplicaci\'on en las observaciones referentes a estrellas de neutrones.

\appendix

\chapter{Geometry of $\CP^3$}\label{ap:CP3}

In this Appendix we give all the details of the geometry of the complex projective plane $\CP^3$. In this thesis, it is seen as the coset $\rm{Sp}(2)/\rm{U}(2)$, consisting of a two-sphere $\rm{S}^2$ (described by the vielbeins $E^1$ and $E^2$) fibred over a four-sphere $\rm{S}^4$, whose metric is $\dd\Omega^2_4$. A convenient choice of coordinates goes as follows \cite{Conde:2011sw,Jokela:2012dw}. Let $\omega^i$ be a set of left-invariant one-forms on the three-sphere. The metric of the four-sphere with unit radius can be written as 
\begin{equation}
\dd\Omega_4^2\,=\,\frac{4}{\left(1+\xi^2\right)^2}\left[\dd \xi^2+\frac{\xi^2}{4}\omega^i\omega^i\right]\ ,
\end{equation}
where $\xi$ is a non-compact coordinate. Choosing $\theta$ and $\varphi$ as angles to parametrise the two-sphere, the non-trivial fibration is described by the vielbeins
\begin{equation}
\label{introduced}\begin{aligned}
E^1&=\dd \theta+\frac{\xi^2}{1+\xi^2}\left(\sin\varphi\,\omega^1-\cos\varphi\,\omega^2\right)\,, \\
E^2&=\sin\theta\left(\dd\varphi-\frac{\xi^2}{1+\xi^2}\omega^3\right)+\frac{\xi^2}{1+\xi^2}\cos\theta\left(\cos\varphi\,\omega^1+\sin\varphi\,\omega^2\right)\,.
\end{aligned}
\end{equation}
For our purposes, it is better to consider a rotated version of \eqref{introduced} in the four-sphere, namely
\begin{equation}\begin{aligned}
\mathcal{S}^1&=\frac{\xi}{1+\xi^2}\left[\sin\varphi\,\omega^1-\cos\varphi\,\omega^2\right]\,, \\
\mathcal{S}^2&=\frac{\xi}{1+\xi^2}\left[\sin\theta\,\omega^3-\cos\theta\left(\cos\varphi\,\omega^1+\sin\varphi\,\omega^2\right)\right] \,,\\
\mathcal{S}^3&=\frac{\xi}{1+\xi^2}\left[\cos\theta\,\omega^3+\sin\theta\left(\cos\varphi\,\omega^1+\sin\varphi\,\omega^2\right)\right] \,,\\
\mathcal{S}^4&=\frac{2}{1+\xi^2}\,\dd\xi \ .
\end{aligned}
\end{equation}
Even though ${\mathcal{S}}^i$ depend on the angles of the two-sphere, it holds that $\mathcal{S}^n\mathcal{S}^n=\dd \Omega_4^2$.
Using these vielbeins we can construct a set of left-invariant forms on the coset. This set contains the two-forms
\begin{equation}
X_2\,=\,E^1\wedge E^2  \, , \qquad  J_2\,=\,\mathcal{S}^1\wedge\mathcal{S}^2+\mathcal{S}^3\wedge\mathcal{S}^4\ ,
\end{equation}
as well as the three-forms 
\begin{equation}\begin{aligned}
X_3&=E^1\wedge\left(\mathcal{S}^1\wedge\mathcal{S}^3-\mathcal{S}^2\wedge\mathcal{S}^4\right)-E^2\wedge\left(\mathcal{S}^1\wedge\mathcal{S}^4+\mathcal{S}^2\wedge\mathcal{S}^3\right)
\,,\\
J_3&=-E^1\wedge\left(\mathcal{S}^1\wedge\mathcal{S}^4+\mathcal{S}^2\wedge\mathcal{S}^3\right)-E^2\wedge\left(\mathcal{S}^1\wedge\mathcal{S}^3-\mathcal{S}^2\wedge\mathcal{S}^4\right)\ .
\end{aligned}
\end{equation}
These are related by exterior differentiation as  
\begin{equation}
\dd X_2\,=\,\dd J_2\,=\,X_3 \, , \qquad  \dd J_3\,=\,2\left(X_2\wedge J_2+J_2\wedge J_2\right)\ .
\end{equation}

Additionally, we can construct higher forms by wedging these which will also be left-invariant. Then we have the two four-forms $X_2\wedge J_2$ and $J_2\wedge J_2$, appearing in the equation above, together with the volume form 
\begin{equation}
\omega_6 = - (E_1 \wedge E_2)\wedge (\mathcal{S}^1\wedge\mathcal{S}^2\wedge\mathcal{S}^3\wedge\mathcal{S}^4)\,.
\end{equation}
There are no left-invariant one- or five-forms. 

\chapter{Expansions of the metrics}\label{ap:expansions}

In this Appendix we are going to present different expansions of the metrics we have been discussing. In each Section of this Appendix we consider different situations.

\section{Finite temperature case UV expansions}\label{ap:UVexp}

Let us first present some details of the UV expansions in the finite temperature case. The undetermined parameters are those in equation~\eqref{UVexpansions}, while the remaining ones, shown here, are given in terms of them. In the following expressions we have already fixed $f_1=-1$ as explained in Chapter~\ref{Chapter4_thermo}. Although we only show a few terms, we indicate in the sums the order up to which we solved the equations --- for example, up to order $u^{22}$ in the case of $e^\FF$. 

For the metric function $e^\FF$ we expanded around $u=0 $ as
$$e^\FF = \frac{1}{u\sqrt{2}} \left[1+\sum_{n=1}^{22}  f_n \ u^n + \OO(u^{23})  \right]\,.$$
The first few coefficients read
\begin{equation}\begin{aligned}
f_2&=-\frac{5}{2}\,,\qquad\qquad\qquad f_3\,=\,-\frac{13}{2}\,,\\[2mm]
f_6&=\frac{1}{120 \left(b_0^2-1\right)}\Big[3 b_0^2 \left(52 \mathsf{b}_5-12436 f_4+224 f_5-33671\right)\\ &\qquad\qquad\qquad-\,96 \left(196 b_4-5 b_6\right) b_0  +4 \mathsf{b}_5-3 \left(364 f_4+224 f_5+11705\right)\Big]\,,\\[2mm]
f_7 &= \frac{1}{2520 \left(b_0^2-1\right)} 
\Big[4 \left(5321 b_0^2-141\right) \mathsf{b}_5- 21 \Big(b_0^2 \left(229456 f_4-3308 f_5+599069\right)\\ &\qquad\qquad\qquad  +592 \left(196 b_4-5 b_6\right) b_0+7344 f_4+3308 f_5+240387\Big)\Big]\,.
\end{aligned}
\end{equation}
Similarly, the function $e^\GG$ enjoys an expansion
$$e^\GG = \frac{1}{{2u}} \left[1+\sum_{n=1}^{22}  g_n \ u^n + \OO(u^{23})  \right]\,,$$
with the first coefficients being
\begin{equation}\begin{aligned}
g_1&=-2\,,\qquad g_2\,=\,-4\,,\qquad g_3\,=\,-8\,,\\[2mm] g_4 &= -2 f_4-\frac{205}{4}\,,\qquad g_5 \,=\,3 f_4+f_5+\frac{283}{4}\,,\\[2mm]
g_6&= -\frac{1}{15 \left(b_0^2-1\right)}\Big[b_0^2 \left(-36 \mathsf{b}_5+8160 f_4-72 f_5+22917\right)+21 \left(196 b_4-5 b_6\right) b_0\\ &
\qquad\qquad\qquad+\mathsf{b}_5+240 f_4+72 f_5+6861\Big]\,,\\[2mm]
g_7 &= \frac{1}{315 \left(b_0^2-1\right)}
\bigg[\left(4624 b_0^2-354\right) \mathsf{b}_5  \\ 
&\qquad\qquad\qquad -21 \Big(b_0^2 \left(47780 f_4-308 f_5+141577\right)
 \\
&\qquad\qquad\qquad\qquad+122 \left(196 b_4-5 b_6\right) b_0+1020 f_4+308 f_5+31419
\Big)\bigg]\,.\\[2mm]
\end{aligned}
\end{equation}

Analogously, for the function $e^\Lambda$ we have
$$e^\Lambda  = 1+\sum_{n=1}^{22}  \lambda_n \ u^n + \OO(u^{23}) \,, $$
where
\begin{equation}\begin{aligned}
\lambda_1 &=\, 0\,,\qquad \lambda_2\, =\, -4\,,\qquad \lambda_3 \,=\, -16\,,\qquad \lambda_4 \,=\, -48\,,\\[2mm] \lambda_5 &=\, \frac{6 }{5}f_4+2 f_5-\frac{71}{10}\,,\\[2mm]
\lambda_6 &= \frac{2}{3 \left(b_0^2-1\right)} \Big[b_0^2 \left(5 \mathsf{b}_5-1164 f_4+20 f_5-3102\right)+\left(15 b_6-588 b_4\right) b_0
\\&\qquad\qquad\qquad -4 \left(9 f_4+5 f_5+288\right)\Big]\,,\\[2mm]
\lambda_7 &= \frac{4}{105 \left(b_0^2-1\right)}
\Big[ b_0^2 \left(680 \mathsf{b}_5-151788 f_4+1740 f_5-412161\right)  \\
&\qquad\qquad\qquad -3 \left(10 \mathsf{b}_5+1404 f_4+580 f_5+46953\right)-390 \left(196 b_4-5 b_6\right) b_0\Big]\,.\\[2mm]
\end{aligned}
\end{equation}

The first non-trivial coefficient in the expansion of the blackening factor appears at order $u^5$ due to the D2-brane asymptotics imposed and is undetermined. The expansion can be written as
$$\mathsf{b} = 1 +  \sum_{n=5}^{23}  \ \mathsf{b}_n \ u^n + \OO(u^{24}) \,, $$
with the parameters 
\begin{eqnarray}
\mathsf{b}_6&=& \frac{20 \mathsf{b}_5}{3}\,,\qquad \mathsf{b}_7 \,=\, \frac{240 \mathsf{b}_5}{7}\,,\qquad\mathsf{b}_8 \,=\, 160 \mathsf{b}_5\,,\qquad \mathsf{b}_9 \,=\, \frac{6400 \mathsf{b}_5}{9}\,,\nonumber\\ [2mm]
\mathsf{b}_{10} &=& \frac{1}{20} \mathsf{b}_5 \left(4 f_4-20 f_5+60513\right)\,,\nonumber\\ [2mm]
\mathsf{b}_{11} &=& -\frac{2 \mathsf{b}_5}{33 \left(b_0^2-1\right)}\Big[b_0^2 \left(25 \mathsf{b}_5-5916 f_4+220 f_5-223857\right)  \nonumber\\ 
&&\qquad\qquad\qquad + \left(75 b_6-2940 b_4\right) b_0- 84 f_4-220 f_5+202587\Big]\,.\nonumber\\
\end{eqnarray}

Likewise, the leading order in the expansion of the warp factor is $u^5$ so that the D2-brane asymptotics is maintained
$$\mathbf{h} = \sum_{n=5}^{27}  \ h_n \ u^n + \OO(u^{28}) \,. $$
The first few coefficients are
\begin{eqnarray}
h_5 &=& -\frac{16}{15} \left(b_0^2-1\right)\,,\quad h_6 \,=\, -\frac{64}{9} \left(2 b_0^2-1\right)\,,\quad h_7 \,=\, -\frac{64}{315} \left(581 b_0^2-201\right) \,, \nonumber\\ [2mm]
h_8&=& -\frac{512}{45} \left(69 b_0^2-19\right)\,,\qquad\qquad h_9 \,=\, -\frac{1024}{945} \left(4263 b_0^2-1003\right)\,,\nonumber\\ [2mm]
h_{10} &=& \frac{4}{1575} \, \Big[7 b_0^2 \left(252 f_4+180 f_5-1387925\right)\nonumber\\ 
&&\qquad\qquad+\,784 b_4 b_0-5 \left(84 f_4+252 f_5-401183\right)\Big]\,, \nonumber\\ [2mm]
h_{11} &=& \frac{32}{10395}\,  \Big[21 b_0^2 \left(85 \mathsf{b}_5-15904 f_4+1000 f_5-1896225\right) \nonumber\\ 
&& \qquad\qquad-\,21 \left(8092 b_4-225 b_6\right) b_0 -4 \left(3213 f_4+3465 f_5-1794298\right)\Big]\,.\nonumber\\
\end{eqnarray}

\newpage

Finally, the fluxes are expanded as
\begin{equation}
\mathcal{B}_J = \sum_{n=0}^{23} \ \mathcal{B}_{J,n} u^n + \OO (u^{24})\nonumber,\qquad 
\mathcal{B}_X= \sum_{n=0}^{23} \ \mathcal{B}_{X,n} u^n + \OO (u^{24})\nonumber, \end{equation}
\begin{equation}
\mathcal{A}_J = \sum_{n=0}^{22} \ \mathcal{A}_{J,n} u^n +  \OO(u^{23})\nonumber\,.
\end{equation}
For convenience, we keep the undetermined parameters in the same function, namely $\BB_J$. The parameter $b_0$ labelling the representative of the family is the leading order of both $\BB_J$ and $\BB_X$. The following coefficients in the expansion of $\BB_J$ are
\begin{eqnarray}
\mathcal{B}_{J,1} &=& 4 b_0\,,\qquad \mathcal{B}_{J,2} \,=\, 8 b_0\,,\qquad \mathcal{B}_{J,3} \,=\, -16 b_0\,,\nonumber \\ [2mm]\mathcal{B}_{J,5} &=& \frac{8}{5} b_0 \left(8 f_4+45\right)+\frac{34 b_4}{5}\,,\nonumber \\ [2mm]
\mathcal{B}_{J,7} &=& -\frac{4}{525 \left(b_0^2-1\right)}\Big[ b_0^3 \left(108 b_5+21108 f_4-244 f_5+246963\right)  \nonumber\\ 
&& \qquad\qquad\qquad -\,b_0 \left(178 b_5+4308 f_4-244 f_5+187407\right)\nonumber \\[1mm]&&\qquad\qquad\qquad + \,42 \left(139 b_4-20 b_6\right) b_0^2+126 \left(19 b_4+5 b_6\right)\Big]\,,\nonumber\\
\end{eqnarray}
while for $\BB_X$ we have
\begin{eqnarray}
\mathcal{B}_{X,1} &=& 4 b_0\,,\qquad\quad \mathcal{B}_{X,2} \,=\, 12 b_0\,,\nonumber\\ [2mm] \mathcal{B}_{X,3}&=& 32 b_0\,,\qquad\quad\mathcal{B}_{X,4} \,=\, -64 b_0-\frac{b_4}{2}\,,\nonumber\\ [2mm]
\mathcal{B}_{X,5} &=& \frac{1}{5} \left(-b_0 \left(56 f_4+1755\right)-6 b_4\right)\,,\nonumber\\ [2mm] \mathcal{B}_{X,6} &=& -16 b_0 \left(8 f_4+93\right)-42 b_4+b_6\,,\nonumber\\ [2mm]
\mathcal{B}_{X,7} &=& \frac{2}{525 \left(b_0^2-1\right)} \Big[4 b_0^3 \left(72 b_5-87918 f_4-46 f_5-511593\right)  \nonumber\\ 
&& \qquad\qquad\qquad +\,4 b_0 \left(173 b_5+29118 f_4+46 f_5+303147\right) \nonumber\\[2mm]&&\qquad\qquad\qquad -\,21 \left(7332 b_4-185 b_6\right) b_0^2+38724 b_4-945 b_6\Big]\,.\nonumber\\ [2mm]
\end{eqnarray}

\newpage
The remaining flux, $\AAA_J$, has coefficients 

\begin{eqnarray}
\mathcal{A}_{J,0}&=&\frac{b_0}{6}\,,\qquad\quad \mathcal{A}_{J,1} \,=\, \frac{4 b_0}{3}\,,\qquad\quad \mathcal{A}_{J,2} \,=\, 8 b_0\,,\\ \mathcal{A}_{J,3} &=& \frac{1}{6} \left(-32 b_0-b_4\right)\,,\qquad \quad\mathcal{A}_{J,4} \,=\,b_0 \left(-2 f_4-\frac{167}{12}\right)-\frac{2 b_4}{3}\,,\nonumber\\ [2mm]
\mathcal{A}_{J,5} &=& -\frac{16}{15} \left[b_0 \left(8 f_4+45\right)+3 b_4\right]\,, \nonumber\\ [2mm]
\mathcal{A}_{J,6} &=& \frac{1}{225 \left(b_0^2-1\right)}\Big[b_0^3 \left(4 b_5-3 \left(5732 f_4+44 f_5+22325\right)\right)\nonumber\\ 
&& \qquad\qquad\qquad +\,3 b_0 \left(12 b_5+2532 f_4+44 f_5+10981\right)\nonumber\\&&\qquad\qquad\qquad-\,24 \left(336 b_4-5 b_6\right) b_0^2+3360 b_4\Big] \,,\nonumber\\ [2mm]
\mathcal{A}_{J,7} &=& -\frac{4}{17325 \left(b_0^2-1\right)} \Big[b_0^3 \left(7594 b_5+1537164 f_4+39748 f_5+5893635\right) \nonumber\\
&& \qquad\qquad\qquad  - \,b_0 \left(12004 b_5+478764 f_4+39748 f_5+2141607\right) 
\nonumber\\[1mm] 
&& \qquad\qquad\qquad +\,84 \left(9204 b_4-145 b_6\right) b_0^2-210 \left(1212 b_4+5 b_6\right)\Big]\,.\nonumber\\ [2mm]
\end{eqnarray}

Now that we have given the UV expansions, lets move to the expansions about the regular horizon.
\vfill

\newpage
\section{Finite temperature case near the horizon} \label{ap:horizonexp}

In this Appendix we consider the expansion around the horizon, see~\eqref{Horizon_expansions}. Although we have solved the equations to sixth order, we only show the first term for the different functions. They read
\begin{eqnarray}
f^h_1 &=& \frac{1}{18 h_h \mathsf{b}_h f_h^3 g_h^4 u_h^4}\Big[\lambda _h^2 \left(9 h_h \left(f_h^4+g_h^4\right)+\left(-6 \alpha _h+\xi _h+\chi
	_h\right){}^2\right)\nonumber\\ &&\qquad -\,2 g_h^2 \left(9 h_h g_h^2 \left(g_h^2-3 f_h^2\right)+\left(\xi
	_h-\chi _h\right){}^2\right)\Big]\,,\nonumber\\[3mm]
g^h_1&=& \frac{g_h^2 \lambda _h^2 \left(9 h_h f_h^4+2 \left(6 \alpha _h+\xi
	_h\right){}^2\right)+f_h^4 \left(9 h_h \left(f_h^4+g_h^4\right)-\left(\xi _h-\chi
	_h\right){}^2\right)}{9 h_h \mathsf{b}_h f_h^8 g_h u_h^4}\,,\nonumber\\[3mm]
\lambda_1^h&=& \frac{\lambda _h}{9
	h_h \mathsf{b}_h f_h^8 g_h^4 u_h^4} \Big[f_h^4 \Big(18 h_h g_h^4 \lambda _h^2-2 g_h^2 \left(\xi _h-\chi
_h\right){}^2 \nonumber \\ && \qquad +\,\lambda _h^2 \left(-6 \alpha _h+\xi _h+\chi _h\right){}^2\Big)+9 h_h
f_h^8 \lambda _h^2+2 g_h^4 \lambda _h^2 \left(6 \alpha _h+\xi _h\right){}^2\Big]\,,\nonumber\\[3mm]
h_1^h&=&-\frac{\lambda _h^2}{81 h_h \mathsf{b}_h f_h^8 g_h^4 u_h^4} \Big[18 h_h f_h^4 \left(9 h_h g_h^4+\left(-6 \alpha _h+\xi _h+\chi
_h\right){}^2\right)+81 h_h^2 f_h^8\nonumber \\ && \qquad  +\,36 h_h g_h^4 \left(6 \alpha _h+\xi
_h\right){}^2 + \left(2 \chi _h \left(6 \alpha _h+\xi _h\right)-12 \alpha _h \xi _h+\xi
_h^2-3\right){}^2\Big]\,,\nonumber\\[3mm]
\mathsf{b}_1^h&=& \frac{9 h_h f_h^4 \left(-\mathsf{b}_h g_h^2 u_h^3-1\right)-54 h_h f_h^2 g_h^2+9 h_h
	g_h^4+\left(\xi _h-\chi _h\right){}^2}{9 h_h f_h^4 g_h^2 u_h^4}\,,\nonumber\\[3mm]
\BB_{J,1}^h &=&
\frac{1}{9 h_h \mathsf{b}_h f_h^4 g_h^4 u_h^4}\bigg[9 h_h f_h^4 \left(2 g_h^2 \left(\xi _h-\chi _h\right)+\lambda _h^2 \left(-6 \alpha
_h+\xi _h+\chi _h\right)\right) \nonumber \\ &&  \qquad +\,\lambda _h^2 \Big(\left(6 \alpha _h-\xi _h-\chi _h\right) \left(12 \alpha _h \left(\xi
_h-\chi _h\right)-\xi _h \left(\xi _h+2 \chi _h\right)\right)\nonumber \\ && \qquad\qquad +18 h_h g_h^4 \left(6 \alpha _h+\xi
_h\right) - 3 \left(\xi _h+\chi
_h\right)\Big)+18 \alpha _h \lambda _h^2\bigg]\,,\nonumber\\[3mm]
\BB_{X,1}^h &=&
\frac{1}{9 h_h \mathsf{b}_h f_h^8 u_h^4}\Big[18 h_h f_h^4 \left(2 g_h^2 \left(\chi _h-\xi _h\right)+\lambda _h^2 \left(-6
\alpha _h+\xi _h+\chi _h\right)\right)\nonumber \\ && \qquad-\,2 \lambda _h^2 \left(6 \alpha _h+\xi
_h\right) \left(12 \alpha _h \left(\xi _h-\chi _h\right)-\xi _h \left(\xi _h+2 \chi
_h\right)+3\right)\Big]\,,
\nonumber\\[3mm]
\AAA_{J,1}^h&=& -\frac{1}{27 h_h \mathsf{b}_h f_h^4 g_h^2 u_h^4} \Big[9 h_h f_h^4\left(-6 \alpha _h+\xi _h+\chi _h\right)-18 h_h g_h^4 \left(6 \alpha
_h+\xi _h\right)\nonumber \\ && \qquad\ +\,\left(\xi _h-\chi _h\right) \left(-12 \alpha _h \left(\xi _h-\chi
_h\right)+2 \xi _h \chi _h+\xi _h^2-3\right)\Big]\,.\nonumber\\[3mm]
\end{eqnarray}

\section{Supersymmetric solutions at the UV} \label{ap:UVexpSUSY}

In this Appendix we present the UV of the regular supersymmetric solutions in the $u$ \eqref{ucoord} coordinate. This choice of coordinate, in contrast with the $y$ coordinate introduced in \eqref{eq:coordy} Chapter \ref{Chapter2_B8family}, simplifies the UV expansion, which turns out to be crucial for different computations such as the renormalisation of the entanglement entropy of disks. In fact, the UV expansion of our system can be directly read off from the finite temperature ones in Section \ref{ap:UVexp} by setting the parameter that renders the temperature $b_5$ to zero, which leads to
\begin{equation}\label{eq:solutionUV}
\begin{aligned}
e^{2\FF} &= \frac{1}{2 u^2} \left[ 1-2u-4u^2-6u^3+\parentsq{\frac{77}{4}+2f_4}u^4+\parentsq{\frac{65}{2}-2f_4+2f_5}u^5+\cdots \right] \\[2mm]
e^{2\GG} &= \frac{1}{4u^2} \left[1-4u-4u^2-\parentsq{-\frac{109}{2}-4f_4}u^4 + \parentsq{\frac{821}{2}+14f_4+2f_5}u^5+\cdots\right] \\[2mm]
e^{\Lambda} &= 1 - 4u^2 -16 u^3 -48u^4+\parentsq{-\frac{71}{10}+\frac{6}{5}f_4+2f_5}u^5+\cdots \\[2mm]
\BB_J &= b_0\left[ 1+4u+8u^2-16u^3+\frac{b_4}{b_0} u^4 +\cdots \right] \\[2mm]
\BB_X&= b_0\left[1+4 u+12 u^2+32 u^3- \left(64 + \frac{b_4}{2 b_0} \right) u^4+\cdots\right] \\[2mm]
\textbf{h}&= \frac{16}{15}  (1-b_0^2)\ u^5 \left[1+\frac{20(1-2b_0^2)}{3(1-b_0^2)}u+\frac{4(581b_0^2-201)}{21(-1+b_0^2)}u^2+\cdots\right] \ .
\end{aligned}
\end{equation}
Because these are the expansions for the supersymmetric case, some of the undetermined parameters we found in the finite temperature case are going to be fixed now by supersymmetry. Mathematically, we see that this is the case when we impose that \eqref{eq:solutionUV} satisfies the first order BPS equations from \eqref{BPSsystem} and \eqref{BPSsystem_fluxes}. For example, $f_5$ is fixed in terms of $f_4$,
\begin{equation}
f_5 = -3f_4 - \frac{411}{4} .
\end{equation}
Similarly, the parameters $b_6$, $b_9$ and $f_{10}$ are now fixed in terms of $f_4$, $b_0$, and $b_4$:
\begin{equation}\begin{aligned}
b_6 &= \frac{2}{5} \, \left(\, b_0\,  \left(\, 200 \, f_4\, +\, 709\, \right)\, +\, 98 b_4\, \right)\,, \\
b_9 &= \frac{1}{140} \left(\, 28 b_4 \, \left(\, 104 f_4\, +\, 26441 \, \right)\, +\, b_0\,  \left(\, 3008 f_4^2\, +\, 1202736 f_4\, -\, 1641593\, \right)\right)\,,  \\
f_{10} &= - \frac{1}{1120}\, \left( \, 6472 f_4^2\, +\, 1028728 f_4\, +\, 33459213\, \right) \,.
\end{aligned}
\end{equation}
So in the end there are only three UV parameters in the supersymmetric case, namely $b_0$, $b_4$, and $f_4$. On the one hand, $b_0$ is the parameter distinguishing the different solutions appearing in Figure~\ref{fig:triangle}. On the other hand, once $b_0$ is fixed, $b_4$ and $f_4$ are fixed by requirement of regularity at the IR.

Recall that the supersymmetric case was constructed in a different radial coordinate $y$, given by \eqref{eq:coordy}, in Chapter \ref{Chapter2_B8family}. There, we found a parameter $y_0$, which stood for the value of $y$ at the regular end-of-space. This parameter $y_0$ is in a one-to-one (numerical) correspondence with $b_0$ (see Figure \ref{fig.numparam}). Knowing that, we can give explicit expressions for $f_4$ and $b_4$ in terms of $y_0$ in the supersymmetric case, namely
\footnote{As we mentioned in Chapter \ref{Chapter2_B8family}, the parameter $\beta_4$ was called $b_4$ in \cite{Faedo:2017fbv}.}
\begin{equation}
f_4 = \frac{1}{24}\left(- 423 - (w_0^\pm)^4\right)\,,\qquad b_4 = \frac{ \ b_0 \ \beta_4 }{2} \ ,
\end{equation}
where the superscript in $w^\pm_0$ consistently refers to either $\B_8^+$ (when $b_0\in(0,2/5)$) or $\B_8^-$ (when $b_0\in(2/5,1)$) and their expressions in each case are given by
\begin{eqnarray}\label{eq:expressionsw_0}
w_0^+(y_0) &=& \frac{\Gamma\left[\frac{1}{4}\right]^2}{\sqrt{8\pi}} + 2 \left({1-y_0^2}\right)^{\frac{1}{4}}-y_0 \, _2F_1\left(\frac{1}{2},\frac{3}{4};\frac{3}{2}; y_0^2\right)\,, \nonumber\\
w_0^-(y_0) &=& \sqrt{8 \pi }\cdot \frac{\Gamma
	\left(\frac{5}{4}\right)}{\Gamma \left(\frac{3}{4}\right)}+2\left(y_0^2-1\right)^{\frac{1}{4}} -\frac{2}{\sqrt{y_0}} \,_2F_1\left(\frac{1}{4},\frac{3}{4};\frac{5}{4};\frac{1}{y_0^2}\right).\qquad 
\end{eqnarray}

\section{Supersymmetric solutions at the regular IR}

Apart from the UV expansion, let us also give the IR expansion of the metric in the supersymmetric case. After a straightforward calculation, the first few orders of the IR expansions for $b_0\in (0,1)$ read:\footnote{Note that these expansions are not valid in the limiting cases $b_0=0,\, 1$.}
\begin{eqnarray}\label{eq:IRexpansionmetric}
e^{2\FF} &=& \frac{f_s^2}{u_s^2} \, (u_s-u)\  +\  \left(\frac{3}{u_s^4}+f_s^2 \lambda _s\right)\,  (u_s-u)^2  \nonumber\\[0mm]
&& +\  \frac{
	3 f_s^4 \lambda _s^2 u_s^8 \ +\  10 f_s^2 \lambda _s u_s^4\ + \ 8 f_s^2  u_s \ +6}{3f_s^2u_s^6}\, (u_s-u)^3 
+ \OO(u_s-u)^4 \nonumber\\[3mm]
e^{2\GG} &=&\frac{1}{u_s^4} \,  (u_s-u)^2 \ +\ \frac{2}{u_s^6} \left(u_s-\frac{1}{f_s^2}\right) \, (u_s-u)^{3} + \OO(u-u_s)^4\ \nonumber\\[3mm]
e^{\Lambda} &=& \frac{1}{u_s^2}\, (u_s-u) \ +\  \lambda_s\,  (u_s-u)^2  \nonumber\\[0mm] 
&&  +\ \frac{\ 3 f_s^4 \lambda _s^2 u_s^8 \ + \ 4 f_s^2 u_s \left(\lambda _s u_s^3 \ - 1\right) \ + \ 9}{3 f_s^4 u_s^6} \, (u-u_s)^3 + \OO(u_s-u)^4 \nonumber\\[3mm]\nonumber
\end{eqnarray}
\begin{eqnarray}
\BB_J &=& 1\ -\ \frac{u_s-u}{f_s^2u_s^2} \ +\  \frac{2-f_s^2u_s}{f_s^4u_s^4}\, (u_s-u)^2 \nonumber\\[0mm]
&& -\ \frac{2 f_s^2 u_s \left(\lambda _s u_s^3-31\right)+15 f_s^4 u_s^2+72}{15 f_s^6 u_s^6}\ (u_s-u)^3\ +\ \OO(u_s-u)^4 \nonumber\\[3mm]\nonumber
\BB_X &=& 1-\frac{3}{f_s^4u_s^4}\, (u_s-u)^2  + \frac{2 \left(f_s^2 u_s \left(\lambda _s u_s^3-4\right)+9\right)}{f_s^6 u_s^6}\,  (u_s-u)^3 \nonumber\\&&+\  \OO(u_s-u)^4 \nonumber\\[3mm]
\mathbf{h} &=& h_{\text{IR}}\cdot\frac{1}{u_s-u} - \left(h_{\text{IR}} \lambda _s u_s^2+\frac{7}{3 f_s^8} \right)\nonumber \\[0mm]
&& + \ \frac{1}{9 f_s^{10} u_s^4}\Big[28 u_s^2 \left(f_s^2 u_s \left(\lambda _s u_s^3-1\right)+4\right) \nonumber \\[0mm]
&&\qquad\qquad\qquad  -3 h_{\text{IR}} f_s^6 \left(4 f_s^2 u_s
	\left(\lambda _s u_s^3-1\right)+9\right)\Big] \, (u_s-u)  \nonumber\\[0mm]
&&  + \  \OO(u_s-u)^2
\ .
\end{eqnarray}

Note that in \eqref{eq:IRexpansionmetric} we found four IR parameters which are not fixed by regularity in the IR expansion. For $u_s$, $f_s$ and $\lambda_s$ it is easy to find expressions as a function of the parameter $y_0$ from Chapter \ref{Chapter2_B8family}:
\begin{eqnarray}
f_s^2 &=& \frac{ (1+y_0)^{\frac{3}{4}}}{\left(\pm (1-y_0)\right)^{\frac{1}{4}}}\cdot \frac{w_0^\pm}{2}\,, \qquad
\lambda_s = \frac{1}{u_s^3} \ \pm \ \frac{2y_0^2+2y_0-4}{w_0^\pm\, u_s^4\,\left(\pm(1-y_0^2)\right)^{\frac{3}{4}}}\,, \nonumber\\ [2mm]
\end{eqnarray}
\begin{eqnarray}\label{eq:expression_us}
u_s &=& \frac{6}{w_0^\pm}\Big[-2^{3/4} \left(y_0\pm 1\right)^{\frac{3}{4}} \, _2F_1\left(\frac{1}{4},\frac{3}{4};\frac{7}{4};\frac{1\pm y_0}{2} \right)+3\frac{
	\left(1+y_0\right){\frac{3}{4}}}{\left(\pm(1-y_0\right))^{\frac{1}{4}}}\nonumber\\
&&\qquad\qquad\qquad+6 \sqrt{\pi }\, \sigma_\pm \, \frac{\Gamma \left(\frac{3}{4}\right)}{\Gamma \left(\frac{1}{4}\right)}\Big]^{-1}\, \ .
\end{eqnarray}
Again, the sign depends on whether the theory belongs to $\B_8^+$ or $\B_8^-$. Also $\sigma_+ = 1$ and $\sigma_- = 0$ and
\begin{equation}
h_{\text{IR}} = \frac{64}{(w_0^\pm)^3}u_s^2\ \mathcal{H}_{\text{IR}} \ .
\end{equation}
Here $\mathcal{H}_{\text{IR}}$ is the parameter appearing in \eqref{eq:fluxesExpansionsSUSYcoordyIR}.

\newpage
\thispagestyle{empty}

\chapter{Four dimensional reduction and Holographic renormalisation}
\label{app:4Deffectivetheory}
To study aspects like spectra (Chapter \ref{Chapter3_observables}, Section \ref{sec:spectrum}) or thermodynamic properties (Chapter \ref{Chapter4_thermo}) it is convenient to work with a four-dimensional consistent truncation. This truncation was presented in \cite{Faedo:2017fbv}. In order to reduce from ten to four dimensions, we take the following metric
\begin{equation}\label{different_ansatz}
\dd s_{\rm st}^2\,=\,e^{\Phi/2}\left(e^{-2U-4V}\dd s_4^2+e^{2U}\,\frac14\left[\left(E^1\right)^2+\left(E^2\right)^2\right]+e^{2V}\,\frac12\dd\Omega_4^2\right)\,.
\end{equation}
By comparing \eqref{different_ansatz} with \eqref{10Dansatz} it is possible to translate into the variables used in the bulk of the thesis. 
The precise relations are
\begin{eqnarray}\label{eq:relation10Dto4D}
e^{\Phi}&=&h^{1/4}\,e^{\Lambda}\,,\nonumber\\[2mm]
e^{2U}&=&4\,h^{3/8}\, e^{2g-\Lambda/2}\,,\nonumber\\[2mm]
e^{2V}&=&2\,h^{3/8}\, e^{2f-\Lambda/2}\,,\nonumber\\[2mm]
e^{2A}&=&16\,h^{1/2}\, e^{4f+2g-2\Lambda}\,.
\end{eqnarray}
The forms read exactly as in equation~\eqref{eqfluxesansatzFT}. Assuming that all the functions depend just on the coordinates in $\dd s_4^2$, the resulting equations of motion can be obtained from the action 
\begin{eqnarray}\label{reduced_Action}
S_4&=&\frac{1}{2\kappa_4^2}\,\int\,\left[R*1+\LL_{\mbox{\footnotesize kin}}-\mathcal{V}*1\right]\nonumber\\[2mm]
&=&\frac{1}{2\kappa_4^2}\,\int\,\Big[(R-\mathcal{V})*1-\frac12\left(\dd\Phi\right)^2-4\left(\dd U\right)^2-12\left(\dd V\right)^2-8\dd U\cdot\dd V\nonumber\\[2mm]
&&\qquad-4e^{-4V-\Phi}\left(\dd b_J\right)^2-8e^{-4U-\Phi}\left(\dd b_X\right)^2-32e^{-2U-4V+\Phi/2}\left(\dd a_J\right)^2\Big]\,,\nonumber
\\[2mm]
\end{eqnarray}
with the potential
\begin{eqnarray}\label{potential}
\mathcal{V}&=&128\,e^{- 6 U -12 V-\Phi/2} \Big[Q_c + 4 a_J \left(b_J + b_X\right) +Q_kb_J\left(b_J-2b_X\right) \nonumber\\[-1mm]
&& \qquad\qquad\qquad\qquad\qquad\qquad\qquad\qquad\qquad\qquad\qquad + 2 q_c \left(b_X - b_J\right)\Big]^2  \nonumber\\[-2mm]
&+& 32 \left(b_J + b_X\right)^2 e^{-4 U - 8 V - \Phi} \nonumber\\[2mm]
&+& 64 \left[2 a_J +Q_k\left(b_J-b_X\right) -  q_c\right]^2 e^{-6 U - 8 V + \Phi/2} \nonumber\\[2mm]
&+& 32 \left(2 a_J -Q_kb_J + q_c\right)^2 e^{-2 U - 12 V + \Phi/2} +4Q_k^2 e^{-2 U - 8 V + 3\Phi/2 } \nonumber\\[2mm]
&+&8Q_k^2 e^{-6 U - 4 V + 3\Phi/2}- 24 e^{-2 U - 6 V} -  8 e^{-4 U - 4 V} + 2 e^{-8 V} \,. 
\end{eqnarray}
Thus, the action contains six scalars: $U$ and $V$ from the metric, $b_X$, $b_J$ and $a_J$ from the fluxes (RR and NS forms) plus the dilaton $\Phi$.

The potential can be recovered from the superpotential  
\begin{eqnarray}\label{eq:superpotential}
\mathcal{W}&=&e^{-4 V} + 2 e^{-2U -2 V} + Q_k\,e^{-3 U - 2 V + 3\Phi/4} - Q_k\,e^{-U - 4 V + 3\Phi/4} \\[2mm]
&-& 4 e^{-3 U - 6 V - \Phi/4} \left[Q_c + 4 a_J \left( b_J + b_X\right) +Q_k b_J \left(b_J-2b_X\right)+2q_c \left(b_X - b_J\right) \right]\nonumber
\end{eqnarray}
through the usual relation
\begin{equation}
\mathcal{V}\,=\,4\,\mathbf{G}^{ij}\partial_i\mathcal{W}\partial_j\mathcal{W}-6\,\mathcal{W}^2\,,
\end{equation}
where $\mathbf{G}$ is the sigma model metric defined implicitly via \begin{equation}
\LL_{\mbox{\footnotesize kin}} = -\,\mathbf{G}_{ij} \dd\phi^i\wedge*\dd \phi^j 
\end{equation}. We now perform standard holographic renormalisation on this action. As a first step, we cut off the radial coordinate at some $\mu_{\mbox{\tiny UV}}$. The action diverges in the limit $\mu_{\mbox{\tiny UV}}\to \infty$, so we have to find the counterterms $S_{\mbox{\footnotesize ct}}$ that regularise it. If these are correctly chosen, adding them to the action and removing the regulator gives a finite quantity. Additionally, we have to include the Gibbons--Hawking term \cite{Emparan:1999}. The regularised action is thus
\begin{eqnarray}\label{Sren}
S_{\mbox{\footnotesize reg}}=\frac{1}{2\kappa_4^2}\,\int_{\MM}\,\Big[R*1+\LL_{\mbox{\footnotesize kin}}-\mathcal{V}*1\Big] +S_{\mbox{\tiny GH}} + S_{\mbox{\footnotesize ct}},
\end{eqnarray}
where $\MM$ is the whole four dimensional spacetime, $S_{\mbox{\footnotesize ct}}$ is the counterterm piece and $S_{\mbox{\tiny GH}}$ is the Gibbons--Hawking term
\begin{eqnarray}
S_{\mbox{\tiny GH}}&=&\frac{1}{\kappa_4^2}\,\int_{\partial \MM}\,K*1,
\end{eqnarray}
being $K$ the extrinsic curvature induced on the boundary $\partial \MM$.

Introducing a horizon in the gravitational description and thus considering the gauge theory at finite temperature is an infrared deformation. On the other hand, it is known that the superpotential renormalises the supersymmetric solution. If the black brane enjoys the same UV asymptotic behaviour as the ground state, the UV divergences will be cancelled out in the same way. Therefore, the counterterm we consider is
\begin{eqnarray}
S_{\mbox{\footnotesize ct}}&=&-\frac{2}{\kappa_4^2}\,\int_{\partial \MM}\, \mathcal{W}*1,
\end{eqnarray}
where $\mathcal{W}$ is the superpotential, Eq.~\eqref{eq:superpotential}. The renormalised on-shell action used to extract the thermodynamic quantities is obtained through the usual manipulations as 
\begin{eqnarray}\label{eq:Full_Sren}
S_{\mbox{\footnotesize ren}}&=&\lim_{\mu_{\mbox{\tiny UV}}\to\infty}S_{\mbox{\footnotesize reg}}\ =\nonumber\\[2mm]
&&\, - \frac{\beta V_2}{2\kappa_4^2}\,\lim_{\mu_{\mbox{\tiny UV}}\to\infty}\left[\sqrt{\gamma}\ (2{K^t}_t - 2K + 4\mathcal{W})\big|_{\mu_{\mbox{\tiny UV}}} -2\sqrt{\gamma} {K^t}_t\big|_{H}\right]\nonumber\\
\end{eqnarray}
where the last term is to be evaluated at the horizon. Here $\gamma$ is the determinant of the boundary metric,  $\beta$ is the period of the Euclidean time and $V_{2}$ is the volume of $\mathds{R}^2$. From this renormalised action we can compute the energy-momentum tensor of the dual gauge theory by varying with respect to the induced metric, evaluated at the boundary
\begin{equation}\label{EMtensor}
{T^i}_j = - \frac{1}{ \kappa_4^2} \lim_{\mu_{\mbox{\tiny UV}}\to\infty}\left\{\sqrt{\gamma}\ ({K^i}_j - {\delta^i}_j (K-2\mathcal{W}))\big|_{\mu_{\mbox{\tiny UV}}}\right\} = \mbox{diag}(-E,P,P)\,.
\end{equation}

There is a final observation which is useful in the numerical computations. Given a four-dimensional metric of the form
\begin{eqnarray}\label{domain_wall_ansatz}
\dd s_4^2&=&-g_{tt}\ \dd t^2+g_{xx}\ \dd x_1^2 +g_{xx}\ \dd x_2^2+g_{rr}\ \dd r^2\,,
\end{eqnarray}
it can be seen that the action is invariant under
\begin{equation}
g_{tt}\mapsto \mu^2 g_{tt}\qquad \qquad g_{xx}\mapsto \mu^{-1} g_{xx}\,.
\end{equation}
Using Noether's theorem there must be a conserved current associated to this symmetry, which is
\begin{equation}
j^\mu =\frac{g_{tt}g_{xx}'-g_{tt}'g_{xx}}{\sqrt{g_{tt}g_{rr}}}\,\delta^\mu_r\,,
\end{equation}
satisfying
\begin{equation}\label{conserved_current}
\partial_\mu j^\mu = \partial_r \left(\frac{g_{tt}g_{xx}'-g_{tt}'g_{xx}}{\sqrt{g_{tt}g_{rr}}}\right) = 0,\qquad \Rightarrow\qquad \frac{g_{tt}g_{xx}'-g_{tt}'g_{xx}}{\sqrt{g_{tt}g_{rr}}}=\mbox{ constant}\,.
\end{equation}
This equation relates a combination of UV parameters with horizon data as in Eq.~\eqref{b5_conserved}. This can be employed either as a check of the numerical coefficients obtained while solving the equations or, using it as an input, to reduce the number of unknown constants in the shooting problem.  

\newpage
\thispagestyle{empty}

\chapter{Computation of entanglement entropies}\label{ap:compEntanglementEntropies}
\section{Computation of the entanglement entropy of the strip}\label{ap:strip}

In this appendix we give more details on the computation of the entanglement entropy of the strip. Let us first discuss the connected configuration, whose embedding \eqref{eq:embedding_strip1} leads to expression \eqref{eq:EE_strip} which we collect here for ease of reference
\begin{equation}
S_\cup(l) = \frac{V_6 L_y}{4 G_{10}} \int_{-\frac{l}{2}}^{\frac{l}{2}}\dd \sigma^1\  \Xi^{\frac{1}{2}} (1+h\ \dot r ^2)^{\frac{1}{2}}\ ,
\end{equation}
where  $\Xi = h^2 e^{8f+4g-4\Phi}$. There is a conserved quantity in this integral, which can be used to find a simple expression for the embedding:
\begin{equation}\label{eq:FO_embedding}
\dot r = \pm \ h^{-\frac{1}{2}} \sqrt{\frac{\Xi\ }{\Xi_*}-1}\ ,
\end{equation}
where $\Xi_* = \Xi(r_*)$ and the dot indicates differentiation with respect to $\sigma^1$. This allows us to write \eqref{eq:EE_strip} as 
\begin{equation}\label{eq:Scupdiv}
S_\cup = 2\frac{V_6\ L_y}{4G_{10}} \int_{r_*}^\infty \frac{\Xi\ h^{\frac{1}{2}}}{\sqrt{\Xi-\Xi_*}}\dd r \ .
\end{equation}
As alluded to in Section~\ref{sec:strip}, this quantity is UV divergent. Due homogeneity, (\ref{eq:Scupdiv}) possesses the same divergence as the $\sqcup$ configuration \eqref{eq:EE_disconnected}. Then, the difference between them, defined in \eqref{eq:EE_strip_reg} as $\Delta S$, can be computed by performing the integral
\begin{equation}\label{eq:formula_EE_reg}
\begin{aligned}
\Delta S &= S_\cup -S_\sqcup =\frac{V_6\ L_y}{4G_{10}}\left[2\int_{r_*}^\infty \left[\frac{\Xi^{\frac{1}{2}}}{\sqrt{\Xi-\Xi_*}}-1\right] 
\Xi^{\frac{1}{2}}h^{\frac{1}{2}}\, \dd r \, -\, 2\int_{r_s}^{r_*}\,\Xi^{\frac{1}{2}}h^{\frac{1}{2}}\, \dd r\ \right].
\end{aligned}
\end{equation}
Interestingly, \eqref{eq:FO_embedding} also allows us to write the width of the strip as
\begin{equation}\label{eq:widthStrip}\begin{aligned}
l &= \int_{-\frac{l}{2}}^{\frac{l}{2}}\ \dd x^1\ = \ 2\int_{r_*}^\infty\frac{\Xi_*^{\frac{1}{2}}}{\sqrt{\Xi-\Xi_*}} h^{\frac{1}{2}}\ \dd r\ ,
\end{aligned}
\end{equation}
in such a way that scanning the parameter space of the turning point of the embedding $r_*\in (r_s,\infty)$, we find the corresponding values of the entanglement entropy and the strip width by simple integration of \eqref{eq:formula_EE_reg} and \eqref{eq:widthStrip}, respectively. 

Again, for non-vanishing CS level, changing to the coordinate $u$ \eqref{ucoord} and performing the rescalings \eqref{eq:dimlessfunctions}, the charges can be factored out. This allows us to redefine $\Delta S$ in such a way that it does not depend on the charges (or the rank of the gauge groups):
\begin{equation}
\label{eq:dimensionlessEEstrip}
\Delta \overline S \ =\  \frac{4G_{10}}{|Q_k|(4 q_c^2 + 3Q_c|Q_k|)}  \times \frac{\Delta S}{L_yV_6} \ =\  \frac{2^8 \pi^4}{9 \lambda}\cdot \frac{N}{|k|\left(\bar{M}^2+2|k|N\right)}
\times \frac{\Delta S}{L_yV_6} \ .
\end{equation}     
We can also define a dimensionless strip width \eqref{eq:widthStrip}, namely
\begin{equation}
\overline l\ =\  \frac{|Q_k|^2}{(4q_c^2+3Q_c|Q_k|)^{\frac{1}{2}}}\cdot l\ = \  \frac{\lambda}{6 \pi  N}\frac{|k|^2}{(\bar M + 2|k| N)^\frac{1}{2}}\cdot l\ .
\end{equation}
Similar expressions can be written in order to factor charges out when CS level is vanishing.

\section{Computation of the entanglement entropy of the disk}\label{ap:disk}

Let us now discuss the computation of the entanglement entropy of disks. As in Appendix \ref{ap:strip}, we will be discussing the case when $|Q_k|\neq0$ (\textit{i.e.}~the whole family of $\B_8$ excluding $\Bconf$). An analogous analysis can be performed in the case of vanishing CS level. Because the procedure is conceptually identical, we will not discuss it here, the main difference being that the coordinate $u$ is not well defined when $|Q_k|=0$ and a distinct radial coordinate has to be used in that case.

First of all, it is useful to change to $r$ as the integration variable in \eqref{eq:EE_disk}. Doing so, the entanglement entropy of the disk reads
\begin{equation}\label{eq:EE1Ddisk}
S_{\text{disk}} =
\frac{V_6}{4 G_{10}} \ 2\pi \int_{r_*}^{\Lambda|Q_k|}\ \dd r \ (\rho'^2+h )^\frac{1}{2} \ \rho \ \Xi^{\frac{1}{2}} \ .
\end{equation}
Note that, because \eqref{eq:EE1Ddisk} is UV divergent, we have explicitly introduced the cut-off $\Lambda$, which after the regularisation will be taken to infinity. The embedding is now given by the function $\rho(r)$, which satisfies the second order differential equation coming from \eqref{eq:EulerLagrange}
\begin{equation}\label{eq:EOMembeddingDisk}
\frac{\dd}{\dd r}\left[\frac{\rho'\ \rho\  \Xi^{\frac{1}{2}}}{\sqrt{\rho'^2 + h}}\right] \ - \ \Xi^{\frac{1}{2}}\ \sqrt{\rho'^2 + h}\ =\ 0\ .
\end{equation}
As we mentioned in Section~\ref{sec:disk}, there are three types of solutions to \eqref{eq:EOMembeddingDisk} whose boundary conditions will be discussed below. To simplify notation, it is convenient to define the rescaled quantities
\begin{eqnarray}
\rhot &=& \frac{|Q_k|^2}{(4q_c^2+3Q_c |Q_k|)^{\frac{1}{2}}} \ \rho \nonumber\\
\overline{S}_{\text{disk}} &=& \frac{4G_{10}\ |Q_k|}{(4q_c^2 + 3Q_c|Q_k|)^{\frac{3}{2}}}\times \frac{{S}_{\text{disk}}}{V_6} \ =\  \frac{2^7\pi^3\ |k|}{3^3\ (\bar M + 2|k| N)^\frac{3}{2}} \times \frac{{S}_{\text{disk}}}{V_6} \ .
\end{eqnarray}
Let us first explain how to solve the embedding equation \eqref{eq:EOMembeddingDisk} and then explain the regularisation of \eqref{eq:EE1Ddisk}. We use the $u$ coordinate as defined in \eqref{ucoord}. 
We can first consider the UV expansion of the metric functions (see Section~\ref{ap:UVexpSUSY} in Appendix~\ref{ap:expansions}) and solve \eqref{eq:EOMembeddingDisk} perturbatively about the UV. Our solutions eventually deviate from the D2-brane metric and hence from the expansion in \cite{vanNiekerk:2011yi}, in particular by logarithmic terms. We obtain:
\begin{equation}\label{eq:UVrhot}
\begin{aligned}
\rhot &= c_0 - \frac{8(1-b_0^2)}{45 c_0}u^3 +\frac{16(-2+7b_0^2)}{45c_0}u^4 + \parentsq{c_5 -\frac{64(-1+21b_0^2)}{315c_0}\log u}u^5 \ +  \\
&+ \parentsq{\frac{20 c_5}{3}-\frac{8
		\left(1-b_0^2\right){}^2}{2025 c_0^3}-\frac{128 \left(861 b_0^2-181\right)}{2835
		c_0}  -\frac{256 \left(21 b_0^2-1\right) }{189 c_0} \log u } u^6+\cdots.
\end{aligned}
\end{equation}

Note there are two undetermined parameters, $c_0$ and $c_5$. Taking \eqref{eq:radiusat infty} into consideration, we realise that $c_0$ is the (rescaled) value of the radius of the disk, $\overline R = c_0$. This UV expansion is valid for the three cases represented in Figure~\ref{fig:disk_conf}, the difference between them being determined by the other boundary condition, as explained in Section~\ref{sec:disk}, elaborated upon here:
\begin{itemize}
	\item The embedding from Figure~\ref{fig:disk_conf}(a) has to satisfy the boundary condition \eqref{eq:disk_BC1}, which in the $u$ coordinate reads
	\begin{equation}
	u(0) = u_* \qquad , \qquad \dot u(0) = 0 .
	\end{equation}
	To this end, we solve $\rhot(u)$ about $u=u_*$ by using a series expansion
	\begin{equation}\label{eq:bounday_condition_disk1}
	\rhot(u)\ =\ (u-u_*)^{\frac{1}{2}} \sum_{k=0}^{\infty} B_k (u-u_*)^k\ .
	\end{equation}
	For a given theory of the $\B_8$ family (\textit{i.e.} for a given $b_0$), all the $B_k$ coefficients are determined in terms of $u_*$. For each choice of $u_*$ we use a shooting technique in order to determine the corresponding value of the UV parameters $c_0$ and $c_5$. More precisely, we impose the expansion \eqref{eq:bounday_condition_disk1} near $u_*$ and integrate up to some value $\epsilon_{\text{UV}}$ where we can trust the UV expansion \eqref{eq:UVrhot} within our numerical precision. Then, using a Newton-Raphson routine, we fixed the values of $c_0$ and $c_5$ which render $\rhot(u)$ continuous and differentiable at $\epsilon_{\text{UV}}$. This procedure should then be repeated for each $u_*$.
	
	Since $c_0$ is essentially the dimensionless radius $\overline R$, scanning values for $u_*\in (0,u_s)$ we get all the embeddings for the RT surfaces associated to disks of radius $\overline R\in (0,\overline R_c)$.
	\item Similarly, embeddings represented in Figure~\ref{fig:disk_conf}(b) satisfy the analogous boundary condition \eqref{eq:disk_BC1}, which after changing to the $u$ coordinate gives
	\begin{equation}
	\lim_{\rhot\to \rhot_*} u(\rhot)= u_s \ .
	\end{equation}
	We then need to solve the equation for $\rhot(u)$ about $u_s$, which leads to a series expansion of the form
	\begin{equation}
	\rhot(u) = \rhot_* + \sum_{k=1}^\infty C_k (u-u_s)^k \ .
	\end{equation}
	For a given theory, the only free parameter in this expansion is $\rhot_*$. Consequently, after imposing this expansion in the equation for $\rhot(u)$ near $u=u_s$, we can find the corresponding values of $c_0$ (\textit{i.e.} $\overline{R}$) and $c_5$ imposed by a shooting procedure analogous to the aforementioned one. In this case, scanning over all the values for $\overline \rho_*\in (0,\infty)$ leads to the corresponding embeddings of RT surfaces of disks with radius $\overline{R}\in \left(\overline{R}_c,\infty\right)$. Note that this type of embedding is not realised when  $b_0=0$, which is the case of the theory which flows to an IR fixed point.
	\item Finally, there is one further type of embedding we are interested in, pictorially represented in Figure~\ref{fig:disk_conf}(c). The boundary condition in this case is \eqref{eq:embeddding_teo_disks}, which in the $u$ coordinate is given by
	\begin{equation}
	u(\rhot_*) = u_*\qquad , \qquad \dot u(\rhot_*) = 0
	\end{equation}
	with $u_*\neq u_s$ and $\rhot_*\neq 0$. In this case, for the solution about $u=u_*$ we get
	\begin{equation}
	\rhot(u) = \rhot_* + \sum_{k=1}^\infty\ D_k^{\pm} \ (u-u_*)^{\frac{k}{2}} \ .
	\end{equation}
	As the superscript in $D_k^\pm$ suggests, there are two different series, depending on a choice of sign that has to be made while solving the first coefficient. The rest of the coefficients are fixed in each case as functions of $\rhot_*$ and $u_*$. Each choice is giving a distinct branch of the embedding corresponding to Figure~\ref{fig:disk_conf}(c). For each of the two branches, the shooting method gives a different value of $c_0$, corresponding to the two radius of the two disks to which this embedding is attached. We refer to them as $\overline{R}_1$ and $\overline{R}_2$. Scanning the parameters space $u_*\in (0,u_s)$ and $\rhot_* \in (0,\infty)$ we efficiently get the embeddings corresponding to all possible values of $\overline R_1$ and $\overline R_2$.
\end{itemize}

Now that we understand the different embeddings we encounter, we can turn to the issue of regulating the action functional. Knowing the UV expansion of all the functions which are involved in the computation, it is possible to study the divergence structure of the entanglement entropy in this particular problem. First, let us write the integral \eqref{eq:EE1Ddisk} in the $u$ coordinate:
\begin{equation}\label{eq:actionDisk}
\overline S_{\text{disk}} =  \int_{u_*}^{\Lambda^{-1}} \dd u\parent{-\frac{1}{u^2}\ \overline \Xi^{\frac{1}{2}}\ \rhot\ (\mathbf{h}+u^4\rhot'^2)^\frac{1}{2} } \equiv \int_{u_*}^{\Lambda^{-1}} \dd u\  L_D \ ,
\end{equation}
where $u_* = |Q_k| \ r_*^{-1}$  is the value of the radial coordinate at the turning point (or $u_*=u_s$ if the embedding reaches the bottom of the geometry) and $\overline{\Xi}$ is the dimensionless version of $\Xi$, namely 
\begin{equation}
\overline \Xi \ =\ \frac{4q_c^2 + 3Q_c |Q_k|}{|Q_k|^6} \ \mathbf{h}^2 e^{8\FF+4\GG-4\Phi}\ .
\end{equation}
Also, last equality in \eqref{eq:actionDisk} defines $L_D$. Replacing the functions by their UV expression and performing the integral we get
\begin{equation}
\begin{aligned} \label{eq:actionUVexpandsion}
\overline S_{\text{disk}} =& \ \int_{u_*}^{\Lambda^{-1}} \dd u\parent{-\frac{1}{u^2}\ \overline \Xi^{\frac{1}{2}}\ \rhot\ (\mathbf{h}+u^4\rhot'^2)^\frac{1}{2} } \\[2mm]
=&\  c_0\, \Big[\, \frac{1}{30} \left(1-b_0^2\right) \Lambda ^2 \ - \ \frac{4}{45} \left(4
b_0^2+1\right) \Lambda\  - \  \frac{4}{63} \left(21 b_0^2-1\right)
\log \Lambda \ + \ S_0  \\[2mm]
&\ +\ \frac{2 \left(14 b_0^2 \left(720
	c_0^2-1\right)+7 b_0^4-480 c_0^2+7\right)}{4725 c_0^2 
} \frac{1}{\Lambda}\ +\ \OO(\Lambda^{-2})\ \Big].
\end{aligned}
\end{equation}

A few important remarks are in order:
\begin{itemize}
	\item The leading divergence in \eqref{eq:actionUVexpandsion} is of the same order as the one found in \cite{vanNiekerk:2011yi}, namely $\Lambda^2$.
	\item Because we deviate from D2-brane asymptotics eventually \eqref{eq:solutionUV}, there are new terms appearing. Nevertheless, the counterterms are readily available
	\begin{equation}\label{eq:counterterm}
	\overline  S_{\text{disk}}^{\text{ct}} (\Lambda^{-1})\ =\ c_0 \, \Big[\,  \frac{1}{30} \left(1-b_0^2\right) \Lambda ^2\ -\ \frac{4}{45} \left(4
	b_0^2+1\right) \Lambda \ -\ \frac{4}{63} \left(21 b_0^2-1\right)
	\log \Lambda \, \Big].
	\end{equation}
	\item In \eqref{eq:actionUVexpandsion} we see that the finite contribution $S_0$ is just the integration constant and we can set it to zero (or absorb it in \eqref{eq:counterterm}). One can show that this corresponds to picking a particular renormalisation scheme on the field theory \cite{Hoyos:2016cob,Ecker:2017fyh}.
	\item Because there is a global $c_0$ multiplying \eqref{eq:counterterm}, this tells us that at the end of the day $S_{\text{disk}}^{\text{ct}}$ is proportional to the perimeter of the disk, thus fulfilling and area law.
	\item Note also that the counterterms do not depend on the precise embedding. The counterterms depend on $c_0$, which is essentially the radius of the disk, but not on $c_5$, since its value is determined after solving the equation of the embedding and therefore depend on the IR data.
	\item Finally, note that the counterterms not only are independent of $c_5$, but also all the subleading parameters $f_4$ and $b_4$ appearing in \eqref{eq:solutionUV}.
	This means that the counterterms are indeed completely independent of the subleading parameters of the UV expansions, which are associated to VEVs and determined by imposing some IR condition.
\end{itemize}

Taking into account that we can express the counterterms in \eqref{eq:counterterm} as
\begin{equation}
\label{eq:regDiskFromLagrangian}
\begin{aligned}
\overline  S_{\text{disk}}^{\text{ct}} (\Lambda^{-1}) = \int_{u_*}^{\Lambda^{-1}}\dd u \ &L_D^{\text{ct}}  \ +\  \overline  S_{\text{disk}}^{\text{ct}} (u_*)\\
\mbox{where }\qquad &L_D^{\text{ct}}  \equiv \frac{\left(b_0^2-1\right) c_0}{15 u^3}+\frac{4 \left(4 b_0^2+1\right) c_0}{45
	u^2}+\frac{4 \left(21 b_0^2-1\right) c_0}{63 u}\ ,
\end{aligned}\end{equation}
the regularised entanglement entropy of the disk can be written in a way that makes the numerical computations easier. This yields
\begin{equation}\label{eq:formulaEE}
\overline  S_{\text{disk}}^{\text{reg}} = \lim_{\Lambda\to\infty} \left[\overline S_{\text{disk}} - \overline S_{\text{disk}}^{\text{ct}}(\Lambda^{-1})\right] = \int_{u_*}^{0} \dd u \parent{L_D-L_D^{\text{ct}}} -S_{\text{disk}}^{\text{ct}}(u_*)\ .
\end{equation}
Finally, we use \eqref{eq:formulaEE} in order to define a version of $\ffunc_{\text{disk}}$ where the charges has been factored out
\begin{equation}
\label{eq:Ffunction_disk_dimless}
\overline\ffunc_{\text{disk}}(\overline R) =\overline R\cdot \frac{\dd \overline S^{\text{reg}}_{\text{disk}}}{\dd \overline R} - \overline S^{\text{reg}}_{\text{disk}} \ .
\end{equation}

\chapter{Numerical output}
\label{ap:solutions}
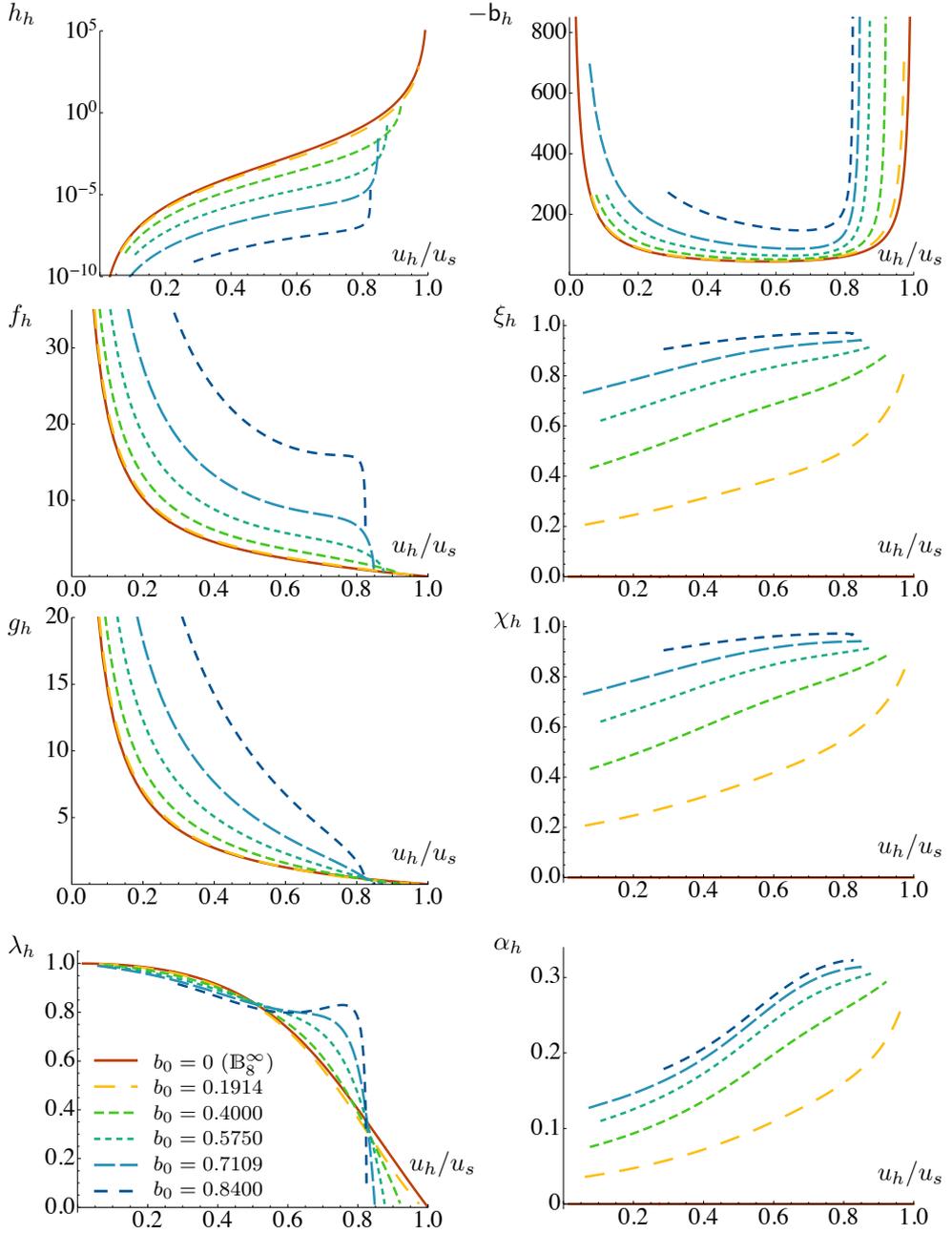
\begin{figure}[t!]
	\begin{center}
		\begin{subfigure}{0.45\textwidth}
			\includegraphics[width=\textwidth]{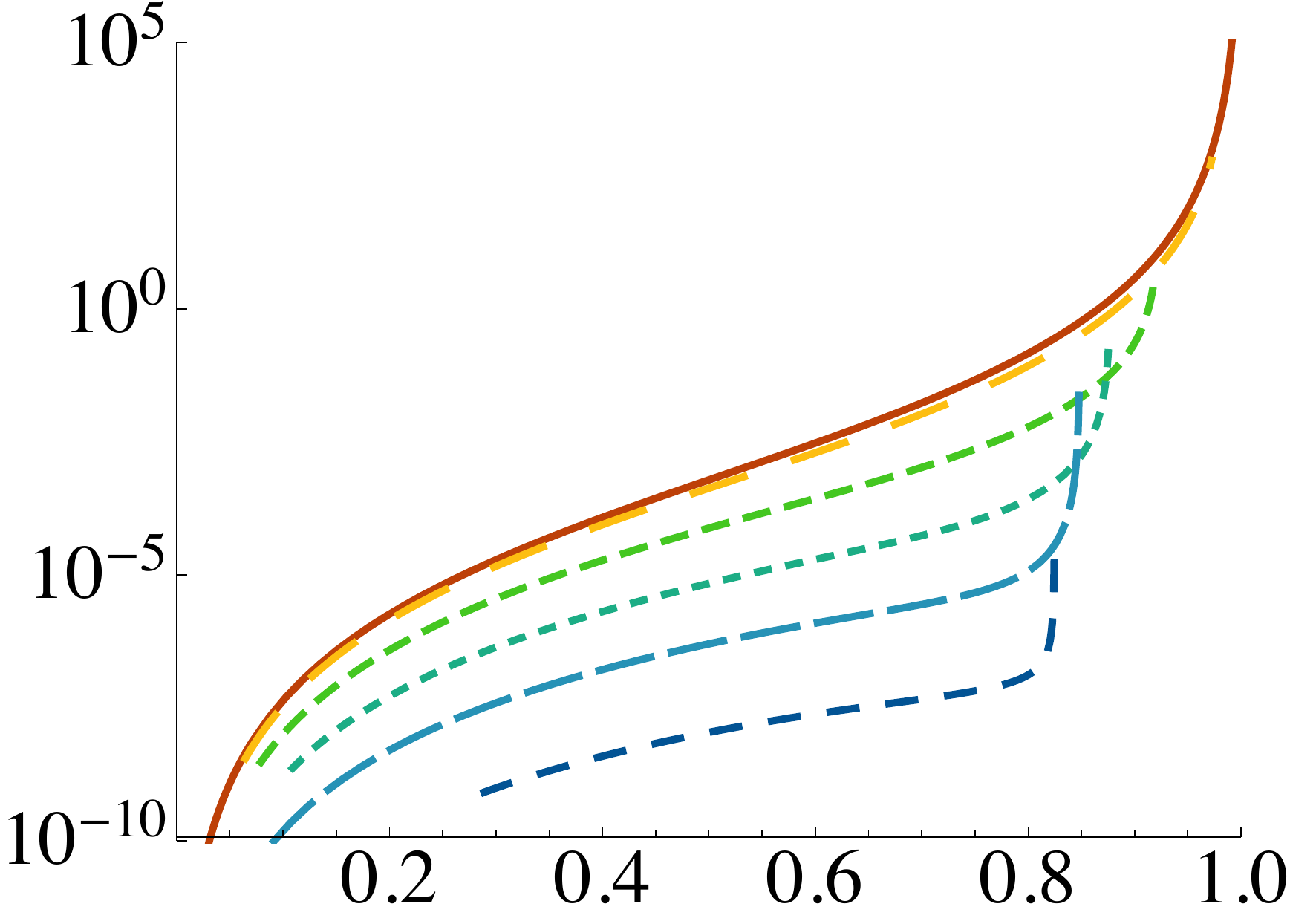} 
			\put(-170,110){$h_h$}
			\put(-20,16){$u_h/u_s$}
		\end{subfigure}\hfill
		\begin{subfigure}{.45\textwidth}
			\includegraphics[width=\textwidth]{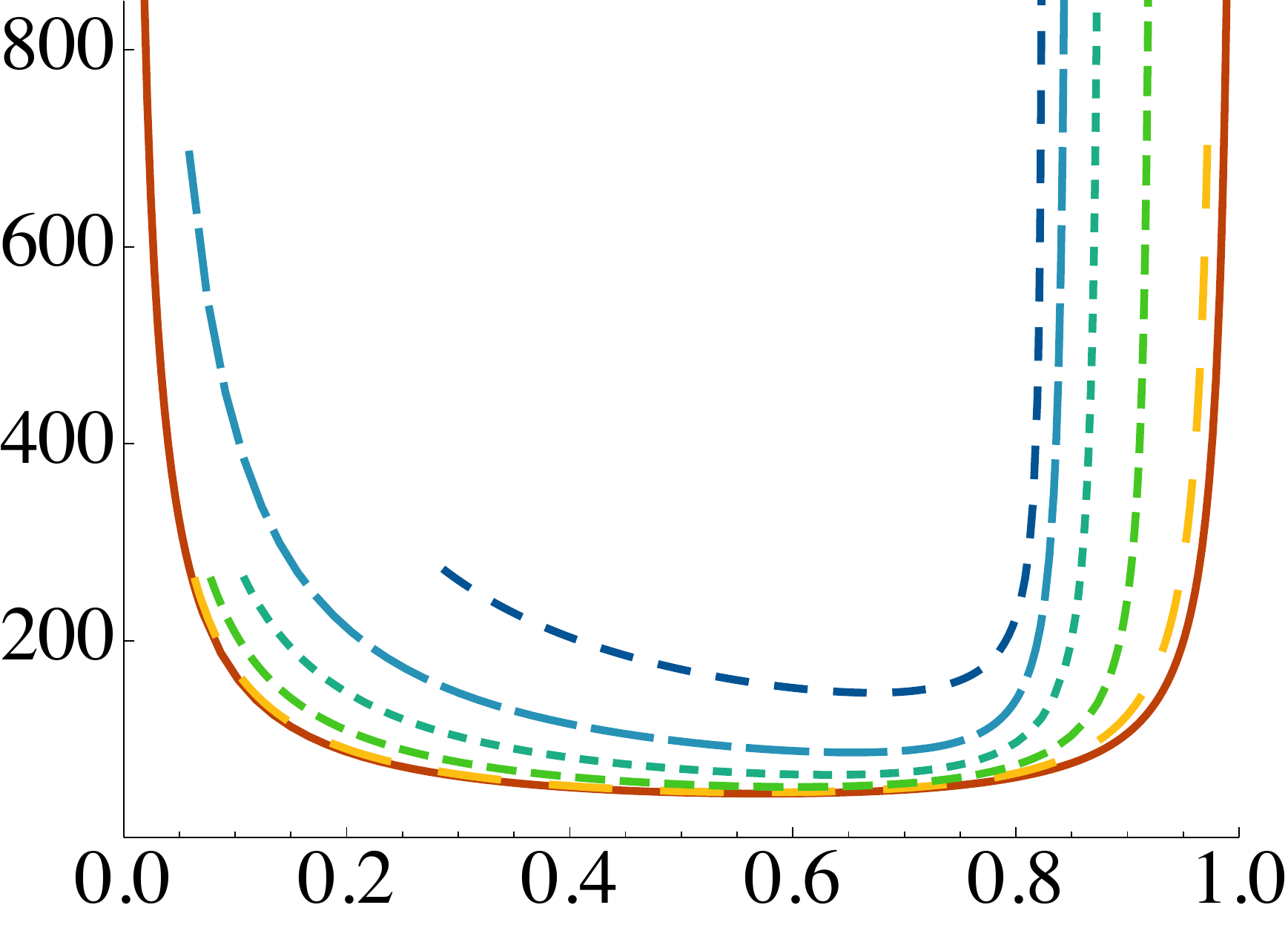} 
			\put(-180,110){$-\textsf{b}_h$}
			\put(-20,16){$u_h/u_s$}
		\end{subfigure}
		\begin{subfigure}{0.45\textwidth}
			\includegraphics[width=\textwidth]{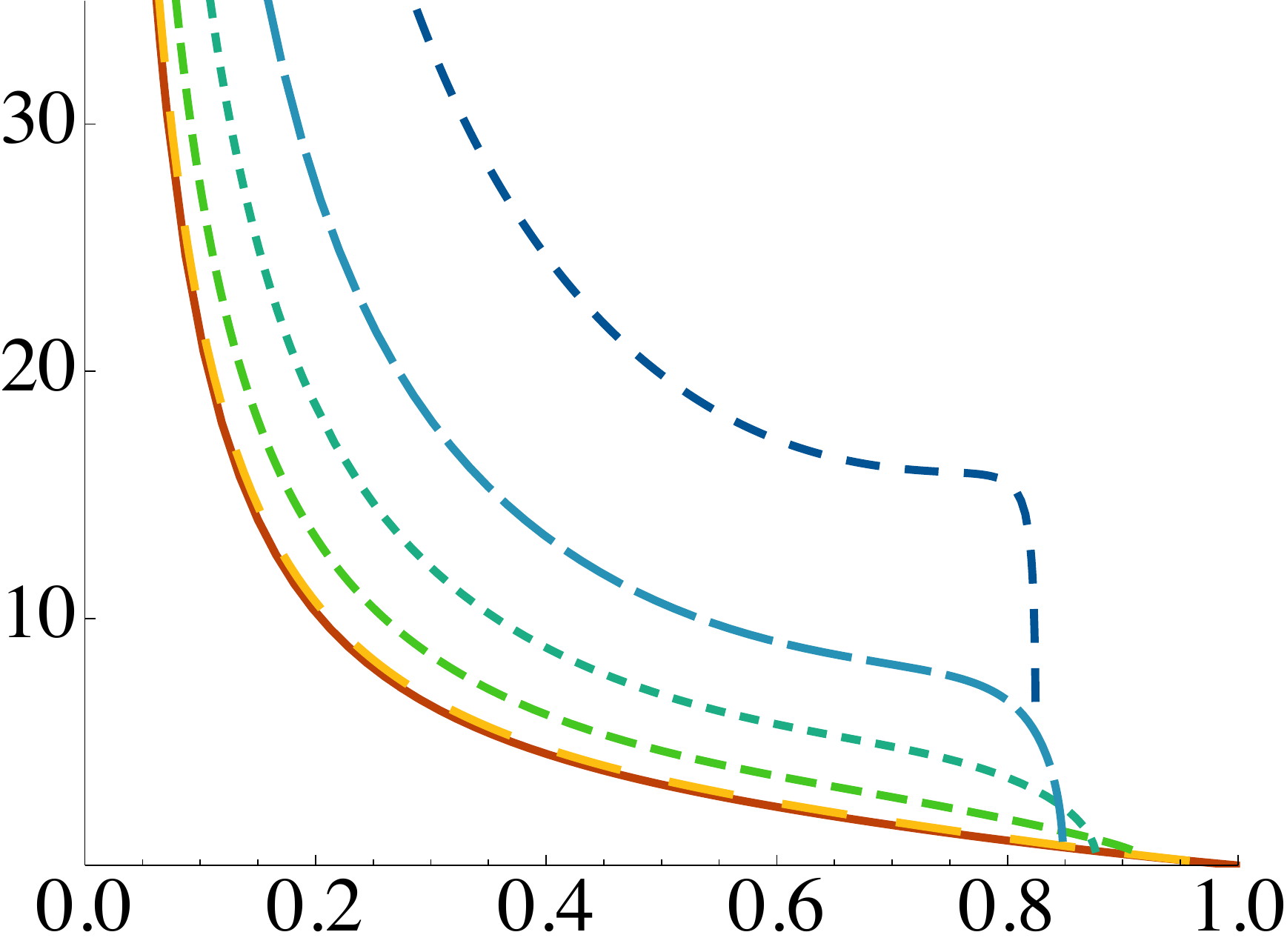} 
			\put(-170,110){$f_h$}
			\put(-20,20){$u_h/u_s$}
		\end{subfigure}\hfill
		\begin{subfigure}{0.45\textwidth}
			\includegraphics[width=\textwidth]{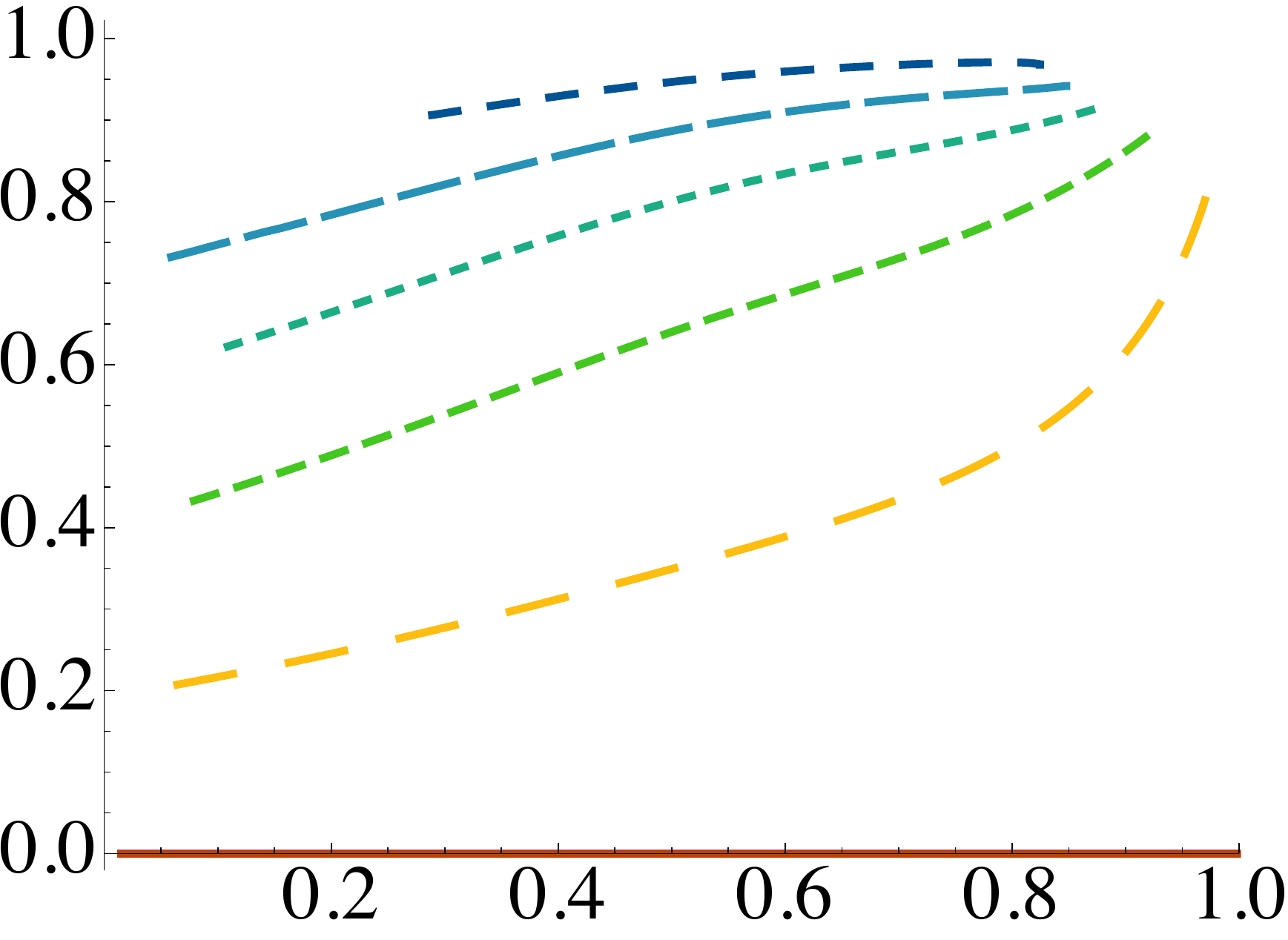} 
			\put(-170,110){$\xi_h$}
			\put(-20,20){$u_h/u_s$}
		\end{subfigure} \vspace{3mm}
		\begin{subfigure}{.45\textwidth}
			\includegraphics[width=\textwidth]{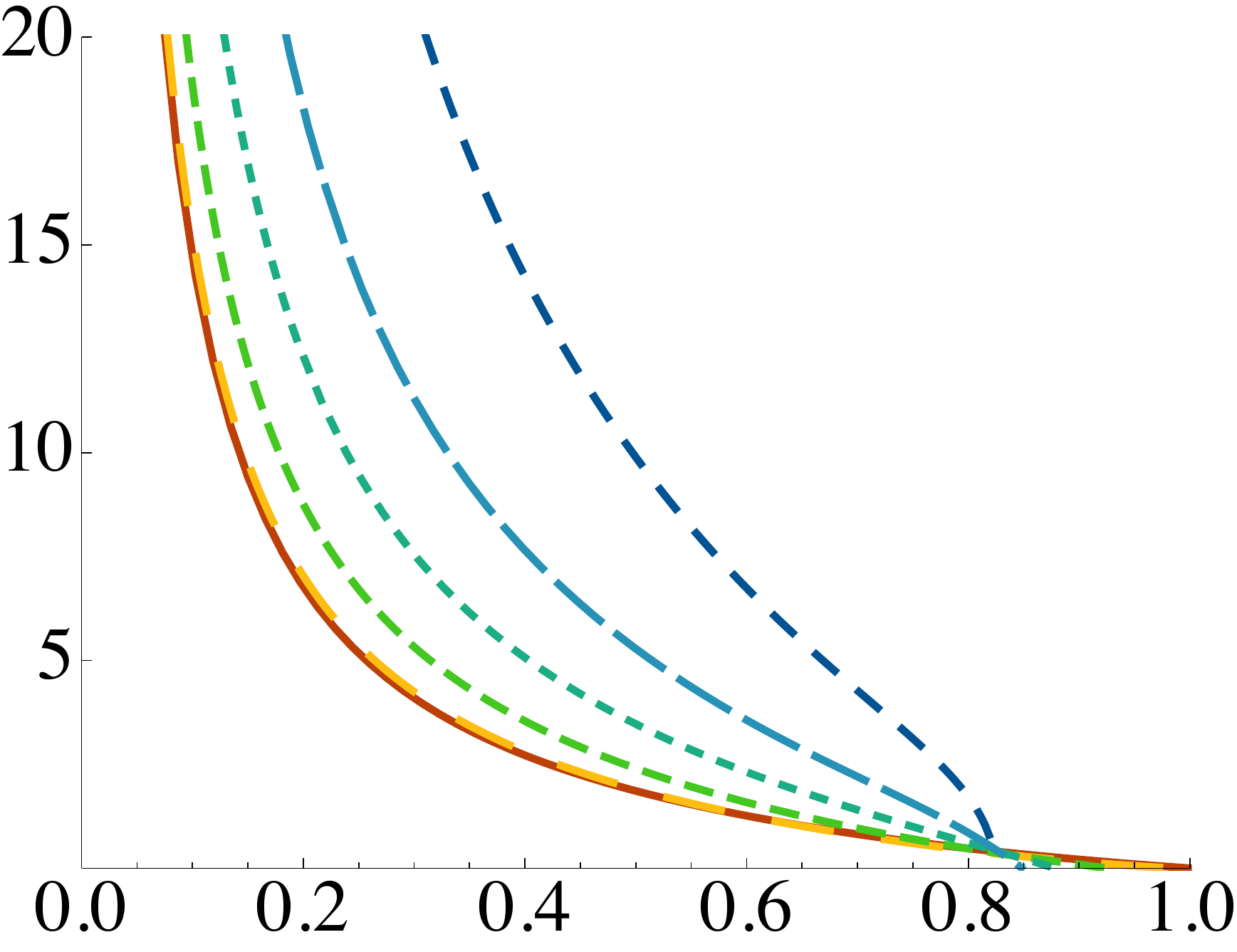} 
			\put(-170,110){$g_h$}
			\put(-20,20){$u_h/u_s$}
		\end{subfigure}\hfill
		\begin{subfigure}{.45\textwidth}
			\includegraphics[width=\textwidth]{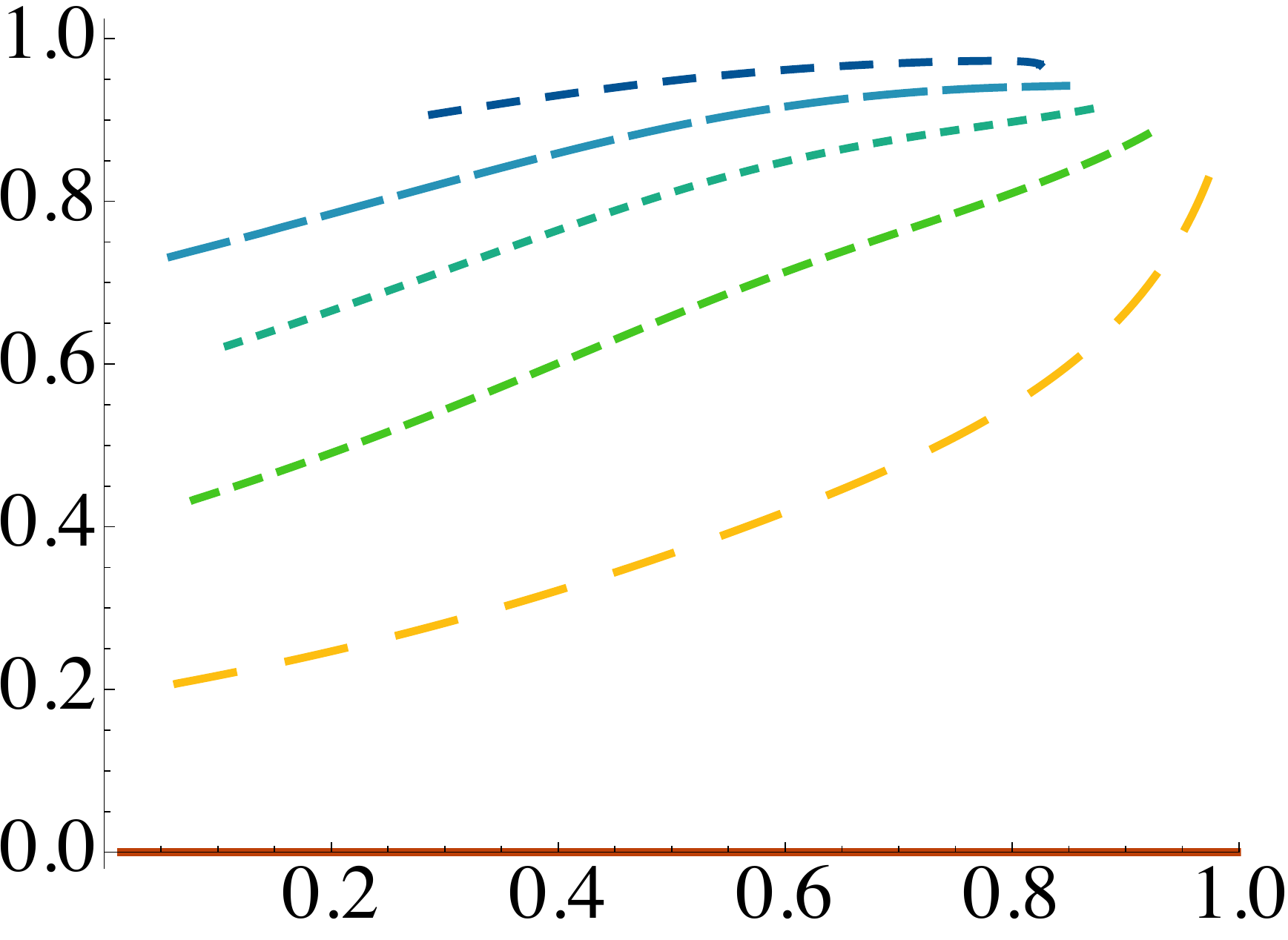} 
			\put(-170,110){$\chi_h$}
			\put(-20,20){$u_h/u_s$}
		\end{subfigure}
		\begin{subfigure}{.45\textwidth}
			\includegraphics[width=\textwidth]{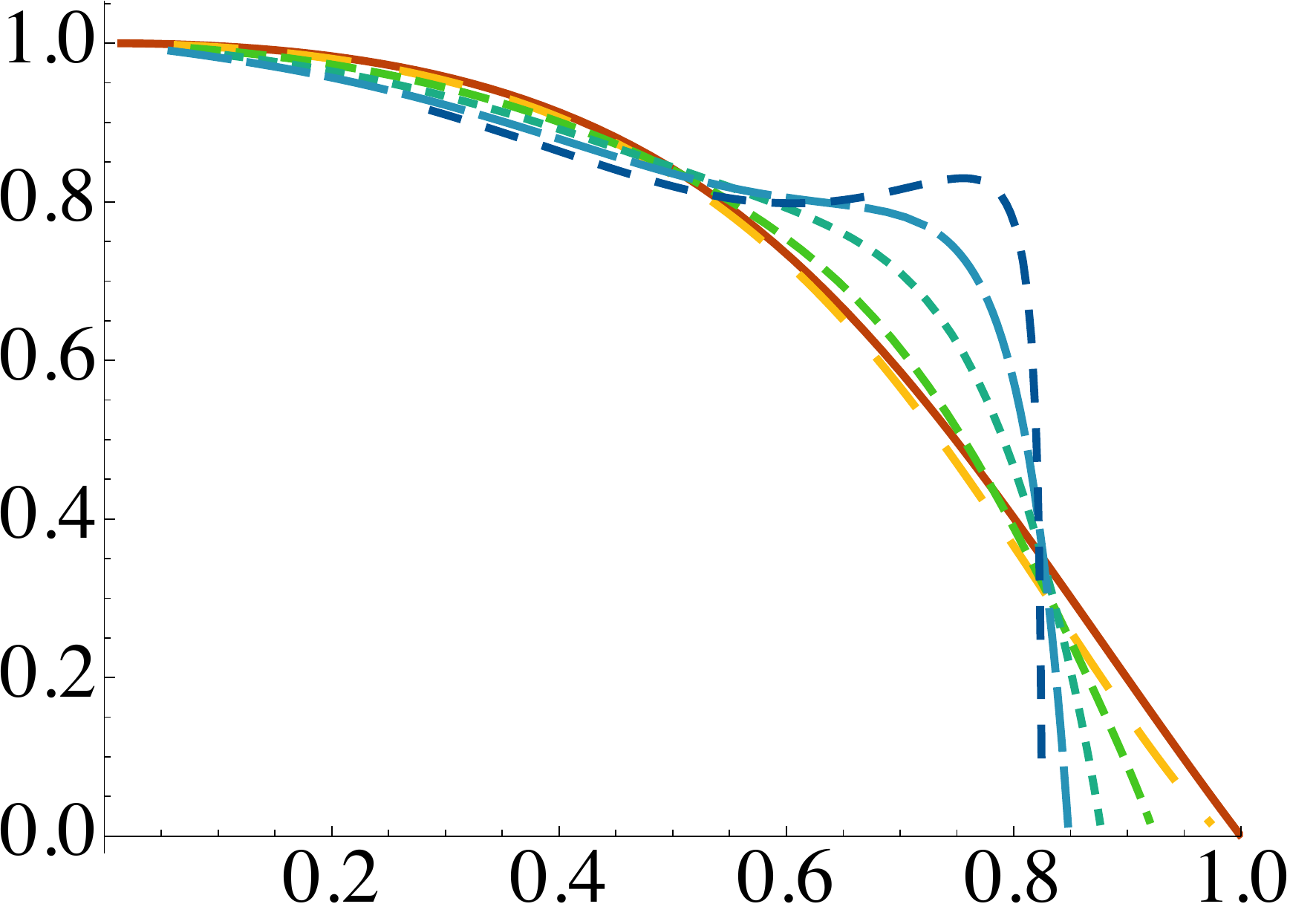} 
			\put(-144,10){\includegraphics[scale=.59]{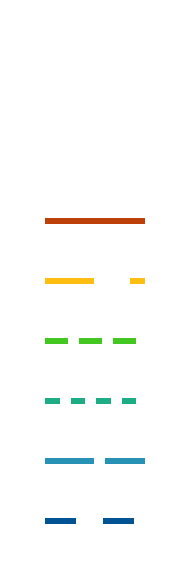} }
			\put(-170,110){$\lambda_h$}
			\put(-13,25){$u_h/u_s$}
			\put(-113,65){\footnotesize$b_0 = 0 $ $(\Binf)\qquad$}
			\put(-113,55){\footnotesize$b_0 = 0.1914$}
			\put(-113,45){\footnotesize$b_0 = 0.4000$}
			\put(-113,35){\footnotesize$b_0 = 0.5750$}
			\put(-113,25){\footnotesize$b_0 = 0.7109$}
			\put(-113,15){\footnotesize$b_0 = 0.8400$}
		\end{subfigure}\hfill
		\begin{subfigure}{.45\textwidth}
			\includegraphics[width=\textwidth]{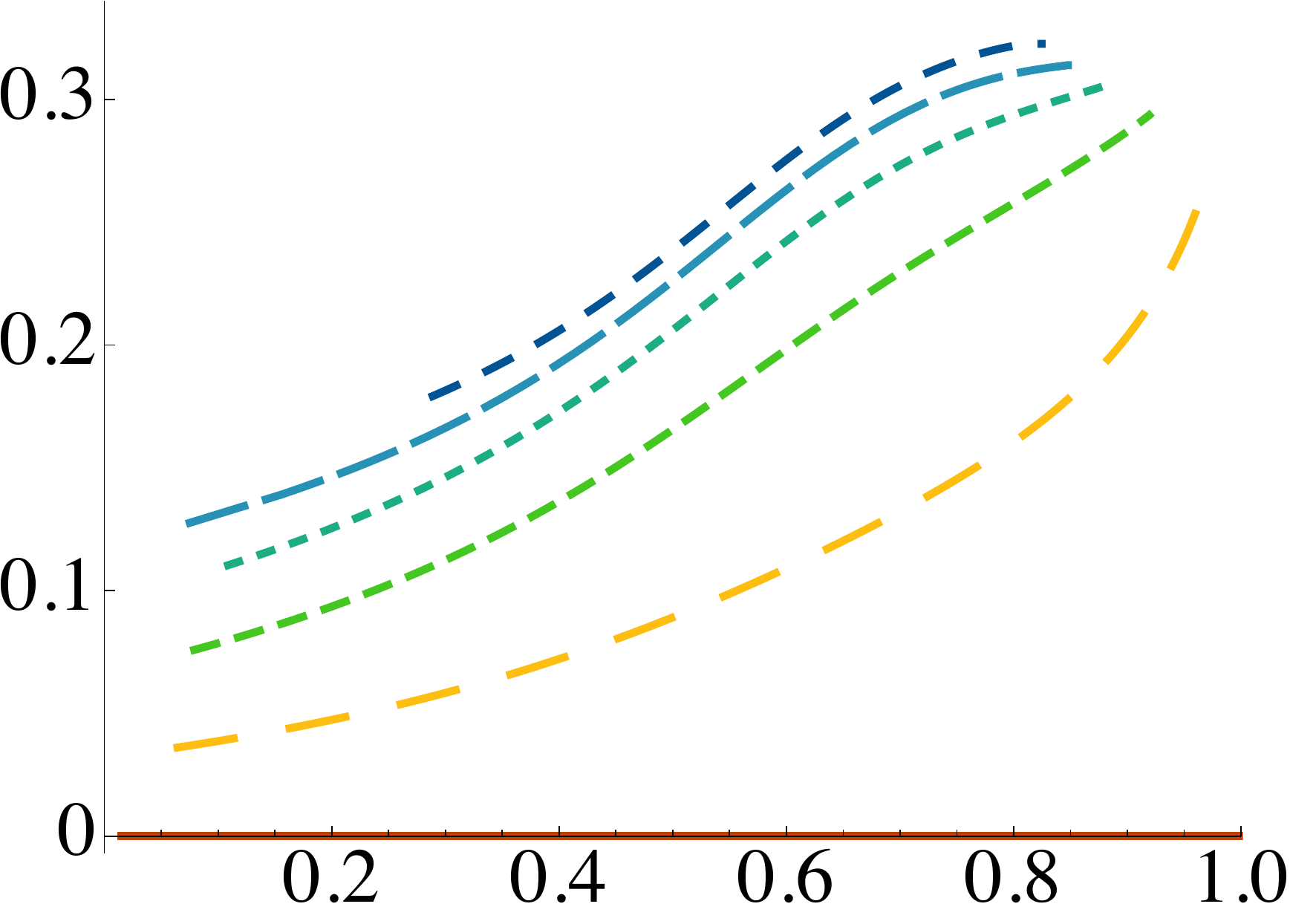} 
			\put(-170,110){$\alpha_h$}
			\put(-20,20){$u_h/u_s$}
		\end{subfigure}
		\caption{\small Value of the metric functions and the fluxes at the horizon as a function of the position of the horizon normalised to $u_s$ for the metrics showed in the legend (see main text).}\label{fig.FunctionsBXBJAJ_Horizon}
	\end{center}
\end{figure}

In this Appendix we show some of the data obtained in our numerical computation, which can be found in Figure~\ref{fig.FunctionsBXBJAJ_Horizon}. 
Let us highlight the main features of these plots, which gather the value of the metric functions and the fluxes for some representative values of $b_0$, as a function of the position of the horizon normalised to the point where the regular solution ends. 

First of all, recall that we already argued that there was a maximum $u_N<u_s$ above which we cannot place the horizon. This can be clearly seen in the plots corresponding to $e^\FF$, $e^\GG$ and $e^\Lambda$ (\textit{i.e.} $f_h$, $g_h$ and $\lambda_h$ respectively), since the values at the horizon go to zero at some point $u_N/u_s<1$. Moreover, in these plots we see that as $b_0$ increases, the ratio between $u_N$ and $u_s$ decreases. For completeness, the value of the fluxes at the horizon has also been shown. 

It is often the case that, when ``heating up'' a geometry, the value of the scalar fields at the horizon is similar to those in the zero-temperature solution at the same value of the radial coordinate (this statement is meaningful if the radial gauge is completely fixed, and in the same way, in both geometries). In our case this does not happen, as can be seen, for example, in the plot of $e^\Lambda$: whilst for the supersymmetric solutions (both regular and irregular) the function $e^\Lambda$ is monotonically decreasing, its horizon values develop a minimum and a maximum, resulting in a manifest difference in the dilaton behaviour.

In Figure~\ref{fig.FunctionsBXBJAJ_Horizon} we have also plotted the value of the warp factor at the horizon and the value of $\mathsf{b}_h$, which is the derivative of the blackening factor at the horizon. Notice that both magnitudes diverge as $u_h$ approach $u_N$: it is the contribution of both divergences (one in the denominator and the other in the numerator) that gives a finite temperature in the zero-entropy limit.

\chapter{Complex RG flows}
\label{ap:complexRGflows}

The $\beta$-function \eqref{eq:betaphi} corresponding to \eqref{dW} has the desired properties to describe both walking and the physics of cCFTs. Expanding to quadratic order in $g\equiv\phi-\phi_0$ around the fixed point at $\phi=\phi_0$ we find
\begin{equation}\label{betaW}\begin{aligned}
\beta=&-2\left(d-1\right)W_0\phi_0\left(\phi_0-\overline{\phi}_0\right)\frac{L_0}{L}\,g\\[2mm]&-2\left(d-1\right)W_0\phi_0\left(2\phi_0-\overline{\phi}_0\right)\frac{L_0}{L}\,g^2\,+\,\cO(g^3)\,,
\end{aligned}
\end{equation}
where $L_0$ is given in \eqref{eq:L0W}. To this quadratic order, in addition to the fixed point at $g=0$, there is another one at
\begin{equation}
\overline{g}\,\equiv\,-\frac{\phi_0\left(\phi_0-\overline{\phi}_0\right)}{2\phi_0-\overline{\phi}_0}\,.
\end{equation}
When the fixed point at $\phi=\phi_0$ is close to the real axis, i.e.~if 
$\phi_0=1+i \,\epsilon$ with $\epsilon\ll1$, then this second fixed point is its complex conjugate since
\be
\overline{\phi}_0 = \phi_0 + \overline{g} = 1-i\epsilon + \cO(\epsilon^2)
\ee
Under these circumstances  the model is expected to exhibit walking behaviour when the flow passes exactly between the two complex fixed points. In this regime the $\beta$-function is obtained by evaluating \eqref{betaW} at $g_{\text{\tiny W}}+\overline{g}/2$ with $g_{\text{\tiny W}}$ real. By rescaling the coupling $g_{\text{\tiny W}}$, it can be seen that this $\beta$-function is of the form \eqref{eq:beta} with
\begin{equation}
\alpha-\alpha_*\,=\,-\frac{576\left(d-1\right)^2W_0^2}{\left(12+W_0\right)^2}\,\epsilon^2\,=\,-\frac14|\operatorname{Im}{\Delta_0}|^2\,,
\end{equation}
recovering the results of Chapter \ref{Chapter5_HoloCCFTs}.

It is also straightforward to study the complex flows of this model by assuming that $\phi$ is a complex field so that the action is a holomorphic function. Both the field and the complex extrema of the potential have now both real and imaginary parts
\begin{equation}
\phi\,=\,\phi_{\tiny{\rm R}}+i\,\phi_{\tiny{\rm I}}\,,\qquad\qquad\qquad\phi_0\,=\,\phi_0^{\tiny{\rm R}}+i\,\phi_0^{\tiny{\rm I}}\,.
\end{equation}
In terms of these the  first-order equations (14) take the form
\begin{equation}\label{complexflows}
\begin{array}{rcl}
\phi_{\tiny{\rm R}}'&=&\frac{1}{L|\phi_0|^2}\left[\left(3\phi_{\tiny{\rm I}}^2+2\phi_0^{\tiny{\rm R}}\phi_{\tiny{\rm R}}-|\phi_0|^2-\phi_{\tiny{\rm R}}^2\right)\phi_{\tiny{\rm R}}-2\phi_{\tiny{\rm I}}^2\phi_0^{\tiny{\rm R}}\right]\,,\\[4mm]
\phi_{\tiny{\rm I}}'&=&\frac{1}{L|\phi_0|^2}\left(\phi_{\tiny{\rm I}}^2-|\phi_0|^2-3\phi_{\tiny{\rm R}}^2+4\phi_0^{\tiny{\rm R}}\phi_{\tiny{\rm R}}\right)\phi_{\tiny{\rm I}}\,,
\end{array}
\end{equation}
where we fixed $W_0$ so that the scalar is dual to an operator of dimension $\Delta_\text{\tiny UV}=3$ at the UV fixed point and moreover we are working in a four-dimensional gauge theory.
At the linear level around the UV fixed point, the system is solved by
\begin{equation}
\delta\phi_{\tiny{\rm R}}\,=\,v_{\tiny{\rm R}}\,e^{-\frac{\rho}{L}}\,,\qquad\qquad\delta\phi_{\tiny{\rm I}}\,=\,v_{\tiny{\rm I}}\,e^{-\frac{\rho}{L}}\,,
\end{equation}
showing that both the real and imaginary parts correspond indeed to sources for a $\Delta_\text{\tiny UV}=3$ operator. On the other hand, around $\phi=\phi_0$ the linear solution is
\begin{equation}
\begin{array}{rcl}
\delta\phi_{\tiny{\rm R}}&=&c_1\,e^{\frac{2(\phi_0^{\tiny{\rm I}})^2}{|\phi_0|^2}\frac{\rho}{L}} \cos\left(\frac{2\phi_0^{\tiny{\rm R}}\phi_0^{\tiny{\rm I}}}{|\phi_0|^2}\frac{\rho}{L}\right) +c_2\,e^{\frac{2(\phi_0^{\tiny{\rm I}})^2}{|\phi_0|^2}\frac{\rho}{L}} \sin\left(\frac{2\phi_0^{\tiny{\rm R}}\phi_0^{\tiny{\rm I}}}{|\phi_0|^2}\frac{\rho}{L}\right)  \,,\\[2mm]
\delta\phi_{\tiny{\rm I}}&=&c_2\,e^{\frac{2(\phi_0^{\tiny{\rm I}})^2}{|\phi_0|^2}\frac{\rho}{L}} \cos\left(\frac{2\phi_0^{\tiny{\rm R}}\phi_0^{\tiny{\rm I}}}{|\phi_0|^2}\frac{\rho}{L}\right) -c_1\,e^{\frac{2(\phi_0^{\tiny{\rm I}})^2}{|\phi_0|^2}\frac{\rho}{L}} \sin\left(\frac{2\phi_0^{\tiny{\rm R}}\phi_0^{\tiny{\rm I}}}{|\phi_0|^2}\frac{\rho}{L}\right) \,,
\end{array}
\end{equation}
for real constants $c_1$ and $c_2$. Notice that the modulus is exponentially decreasing as $\rho\to-\infty$, while the phase has an infinite number of oscillations. Examples of full non-linear flows are depicted in Figure~\ref{fig:RGflow}. The result is similar to the cartoon in Figure 11 of \cite{Gorbenko:2018dtm}.

\chapter{Black hole solutions in Einstein-dilaton gravity}
\label{ap:BH}

In order to construct the black hole solutions in Chapter~\ref{Chapter5_HoloCCFTs} we followed the procedure described in \cite{Gubser:2008ny}, which we now review for any dimensionality. This is applicable to any gravitational action of the form \eqref{eq:action} with arbitrary potential admitting a UV fixed point. 

We are interested in solutions which are asymptotically AdS$_{d+1}$, so we choose the following ansatz for the metric
\begin{equation}\label{eq:Ansatz}
\dd s^2_{d+1} = e^{2A} \parent{- h\  \dd t^2 + \dd \vec{x}^2} + e^{2B} \ \frac{\dd r^2}{h}\,,
\end{equation}
where $\dd\vec{x}^2 = \dd x_1^2+\cdots + \dd x_{d-1}^2$ and we are assuming that all the functions depend only on the radial coordinate. Note that, since we are including the $e^{2B}$ factor, the radial gauge is not fixed yet. It is possible to use this freedom to consider the scalar $\phi$ as the radial coordinate, so that
\begin{equation}
\label{eq:ansatzPhi}
\dd s^2_{d+1} = e^{2A} \parent{- h\  \dd t^2 + \dd \vec{x}^2} + e^{2B} \ \frac{\dd \phi^2}{h}\,.
\end{equation}
The price to pay is a dynamical equation determining $B$. With this choice it is implicitly assumed that $\phi$ is either monotonically increasing or decreasing. Then, from \eqref{eq:action} and \eqref{eq:ansatzPhi} we obtain the following equations of motion
\begin{equation}\label{eq:EOM}
\begin{aligned}
0&=h'' + \parent{d A' - B'}h'\,,\\[2mm]
0&=A'' -A'B' + \frac{1}{2(d-1)}\,,\\[2mm]
0&=d\ A'-B' +\frac{h'}{h} -\frac{e^{2B}}{h}V'\,,\\[2mm]
0&=2(d-1)\ A'h'+h\ \parent{\ 2d (d-1)\ A'^2 -1}+2e^{2B} V\,,
\end{aligned}
\end{equation}
where primes denote differentiation with respect to $\phi$. It turns out to be useful to define the function $G(\phi) = A'(\phi)$, in such a way that the solution can be given in terms of the following integrals involving $G$
\begin{equation}
\label{eq:solution}\begin{aligned}
A(\phi)&= A_0+\int_{\phi_0}^{\phi} \dd\tilde{\phi} \ G(\tilde{\phi})\,,\\
B(\phi)&= B_0+\int_{\phi_0}^{\phi} \dd\tilde{\phi}\  \frac{1}{G(\tilde{\phi})} \parent{ {G'(\tilde{\phi})+\frac{1}{2(d-1)}}}\,,\\
h(\phi)&= h_0 + h_1\int_{\phi_0}^{\phi}\ \dd\tilde{\phi}\ e^{-d A(\tilde{\phi}) + B(\tilde{\phi})}\,.
\end{aligned}
\end{equation}
Moreover, there is a relation between the potential and this new function 
\begin{equation}\label{eq:potential}
V(\phi) = \frac{h}{2} e^{-2B}\ \parent{1-2d(d-1)\  G^2 -2(d-1) \ G\ \frac{h'}{h}}\,.
\end{equation}
By differentiating some combinations of \eqref{eq:solution} and \eqref{eq:potential} it can be seen that $G$ satisfies the following master equation 
\begin{equation}\label{ap:eq:equationOfG}
\begin{aligned}
&\frac{G'(\phi)}{\displaystyle G(\phi)+ \frac{V(\phi)}{(d-1)\ V'(\phi)}} \, =\\[2mm]
&\qquad =\, \frac{\dd}{\dd \phi}\left[\frac{G'(\phi)}{G(\phi)} + \frac{1}{2(d-1)G(\phi)}-d\  G(\phi) - \frac{G'(\phi)}{\displaystyle G(\phi)+ \frac{V(\phi)}{(d-1)\ V'(\phi)}}\right]\,.\\[2mm]
\end{aligned}
\end{equation} 
As usual, the boundary conditions at the horizon (placed at $\phi=\phi_H$) are such that $h$ has a simple zero, whereas the remaining functions are finite. In order to see what conditions should be imposed on $G$ at the horizon, we evaluate the last two equations in \eqref{eq:EOM} to obtain 
\begin{equation}
V(\phi_H)\,=\,-(d-1)\  e^{-2B(\phi_H)}\ G(\phi_H)h'(\phi_H)\,,\qquad V'(\phi_H) \,=\, e^{-2B(\phi_H)}h'(\phi_H)\,.
\end{equation}
Consequently, $G+\frac{V}{(d-1)V'}$ must vanish at the horizon. Taking this into account, \eqref{ap:eq:equationOfG} can be solved perturbatively near the horizon
\begin{equation}
\label{eq:GnearH}
G(\phi)\,=\, -\frac{1}{d-1}\frac{V(\phi)}{V'(\phi)}+ \frac{1}{2(d-1)}\parentsq{\frac{V(\phi_H)V''(\phi_H)}{V'(\phi_H)^2}-1}(\phi-\phi_H) + \OO(\phi-\phi_H)^2\,.
\end{equation}
This series can be extended to any order without finding further integration constants. The numerical strategy to solve \eqref{ap:eq:equationOfG} is as follows. We choose a value $\phi_H$ for the scalar and evaluate \eqref{eq:GnearH} near the horizon. Then this value is introduced as a seed in a numerical integrator such as Mathematica's \verb|NDSolve|, which can be used to solve \eqref{ap:eq:equationOfG} numerically up to a value close to the UV fixed point, corresponding to AdS.

The temperature $T$ and entropy $S$ associated to these black holes are computed as usual from the Euclidean time period and the horizon area, respectively. Indeed, they can be written as simple integrals involving $G$
\begin{equation}
\label{eq:SandT}
\begin{aligned}
T&= \frac{d\ \phi_H^{1/(\Delta-d)}}{4\pi L}  \frac{V(\phi_H)}{V(0)}\exp \left\{ \int_0^{\phi_H}\dd \phi \left[G(\phi)-\frac{1}{(\Delta-d)\phi}+\frac{1}{2(d-1)G(\phi)}\right]\right\}\,,\\[2mm]
S&= \frac{2\pi}{2\kappa_{d+1}^2} \phi_H^{\frac{d-1}{\Delta-d}} \exp\left\{ (d-1)\int_0^{\phi_H} \dd\phi \ \left[G(\phi) - \frac{1}{(\Delta-d)\phi}\right]  \right\}\,,
\end{aligned}
\end{equation}
where the appropriate values for the integration constants appearing in \eqref{eq:solution} were chosen so that the same AdS asymptotics is recovered for the different $\phi_H$ that we consider (see \cite{Gubser:2008ny} for details). 

\newpage
\thispagestyle{empty}

\chapter{Technicalities of the transport computations}\label{Appendix:Transport}

In this Appendix, we list some formulae that may be useful in the reproduction of the calculations presented in the main text. In Section~\ref{app:setup}, we review some intermediate results that are used in the evaluation of the transport coefficients,
in Section~\ref{app:vqcd} we detail the functions and parameters of the V-QCD model,
and in Section~\ref{app:cond} we recall how the thermal and electrical conductivities are defined for a relativistic fluid.

\section{Technical details of the holographic calculation}\label{app:setup}

For the metric, the Ansatz corresponding to a homogeneous, rotationally-invariant background reads
\be\label{Eq.Ansatz}
\d s^2  =  g_{tt}(r) \d t^2 + g_{xx}(r) \d \vec x ^2 + g_{rr}(r) \d r^2\ ,
\ee
where $r$ is the holographic radial coordinate. In addition, we use the fact that the gauge potential and scalar fields are only functions of the holographic radial coordinate, $A_t=A_t(r),\;\phi=\phi(r),\;\chi=\chi(r)$. Owing to the fact that we are interested in the deconfined phase of the field theory, there is a black hole of radius $r_\text{H}$ in the interior, with $r_\text{H}$ determined by the condition $g_{tt}(r_\text{H})=0$. This is utilised e.g.~when we write down closed formulas for the transport coefficients in terms of the potentials and fields evaluated at $r=r_\text{H}$, with subscript $H$ referring to this evaluation.

We also recall generic expressions for the temperature $T$, entropy density $s$, and charge density $\rho$:
\be
4\pi T = \left|\frac{d}{dr}\sqrt{-\frac{g_{tt}(r)}{g_{rr}(r)}}\right|_{r=r_\text{H}},\nonumber
\ee
\be
 s=4\pi N_c^2 M_\text{Pl}^3(g_{xx}^\mt{H})^{3/2} ,
 \nonumber
 \ee
 \be \frac{4\pi\rho}{s}=-\frac{N_f}{N_c}\frac{ \Z_\mt{H}\W_\mt{H}^2 F_{rt}^\mt{H}}{\sqrt{1-\W_\mt{H}^2 (F_{rt}^\mt{H})^2}}\ .
\ee
In contrast, the quark chemical potential $\mu\equiv \mu_\text{B}/N_f$ is determined by the boundary value of the gauge potential, $\mu=A_t|_{\textrm{bdry}}$, 
given the regularity condition $A_t(r_\text{H})=0$. In the D3-D7 system, in the canonical gauge for the radial coordinate $r$, the quark mass is determined from the asymptotic expansion of the scalar \cite{Mateos:2006nu}:
$M_q/T=\lym^{1/2}/2\,\lim_{r\to\infty}r\chi(r)/r_\text{H}$.

The thermodynamic energy density $\varepsilon$ and the pressure $p$ can finally be derived from the thermodynamic relations $\varepsilon+p=Ts+\mu \rho$ and $\partial_T p=s$. Alternatively, $p$ may also be computed by evaluating the action of the holographic model in the on-shell limit. To solve the metric and thermodynamics in the V-QCD setup, we used the Mathematica package available at~\cite{VQCDThermo}.

In many holographic models, the shear viscosity saturates the  Kovtun--Son--Starinets (KSS) bound $\eta/s=1/(4\pi)$ \cite{Kovtun:2003wp}. This is in particular the case for both of our holographic models V-QCD and D3-D7, which allows us to evaluate the quantity with ease. Similarly, to compute the bulk viscosity we use the Eling-Oz formula \cite{Eling:2011ms}
\begin{equation}
\begin{split}
\frac{\zeta}{\eta}= \left(s\frac{\partial \phi_H}{\partial s}+\rho \frac{\partial \phi_H}{\partial\rho}\right)^2+\frac{N_f}{N_c}c_H\left(s\frac{\partial \chi_H}{\partial s}+\rho \frac{\partial \chi_H}{\partial\rho}\right)^2\ ,  
\end{split}
\end{equation}
where $c_H=\k_\mt{H}\Z_\mt{H}/\sqrt{1-\W_\mt{H}^2(F_{rt}^\mt{H})^2}$.
The conductivities can finally be computed by extending the methods of \cite{Donos:2014cya,Gouteraux:2018wfe} to a generic DBI action. 
The result obtained in this fashion reads \footnote{One can show that Eq.~(\ref{Eq.DCconductivity}) agrees with the original calculation for the D3-D7 model \cite{Karch:2007pd}.}
\begin{equation}
\label{Eq.DCconductivity}
\sigma^{xx} = N_f N_c M_\text{Pl}^3\frac{s\,T}{\varepsilon+p} \frac{(g_{xx}^\mt{H})^{1/2} \Z_\mt{H}\W_\mt{H}^2}{\sqrt{1-\W_\mt{H}^2(F_{rt}^\mt{H})^2}} \ .
\end{equation}
It should be noted that this expression corresponds to the conductivity computed under the steady state condition \eqref{eq:notdep}.

\section{Definitions for the V-QCD model}\label{app:vqcd}

In order to define the V-QCD model precisely, we need to specify the numerical values of the model parameters and the various functions of $\phi$ and $\chi$ in the gravitational action.
In the glue sector, the function $V(\phi)$ in Eq. (1) of the letter is conveniently expressed by using the field $\lambda = e^{\sqrt{3/8}\,\phi}$ which is identified as the 't~Hooft coupling:
\begin{equation}
V(\lambda) = -12\left[1 + V_1 \lambda + V_2 \frac{\lambda^{2}}{1+\lambda/\lambda_0} +V_\mathrm{IR} e^{-\lambda_0/\lambda}(\lambda/\lambda_0)^{4/3}\sqrt{\log(1+\lambda/\lambda_0)}\right] \ .
\end{equation}
Here the asymptotics at strong coupling have been chosen such that the model is confining and has  asymptotically linear trajectories for the squared masses of radially excited glueballs~\cite{Gursoy:2007cb,Gursoy:2007er}.
The parameters $V_1$ and $V_2$ are fixed by requiring agreement with the running of the  Yang--Mills coupling (see discussion below), which leads to 
\begin{equation}
V_1 = \frac{11}{27\pi^2} \ , \qquad V_2 = \frac{4619}{46656 \pi ^4} \ ,
\end{equation}
so that in particular asymptotic freedom is implemented.
The strong coupling parameters $\lambda_0$ and $V_\text{IR}$ are determined by comparing to lattice data~\cite{Panero:2009tv} for the thermodynamics of pure Yang--Mills at large $N_c$~\cite{Gursoy:2009jd,Alho:2015zua,Jokela:2018ers,Alho:2020gwl}:
\begin{equation}
\lambda_0 = 8\pi^2/3 \ , \qquad V_\mathrm{IR} = 2.05 \ .
\end{equation}

We then discuss the potentials of the flavour sector in Eq.~(2) of the main text. We first write $\Z(\lambda,\chi)= V_{f0}(\lambda)e^{-\chi^2}$ and employ the Ansatz
\begin{align}
V_{f0}(\lambda) &= W_0 + W_1 \lambda +\frac{W_2 \lambda^2}{1+\lambda/\lambda_0} + W_\mathrm{IR} e^{-\lambda_0/\lambda}(\lambda/\lambda_0)^{2},  & \\
\frac{1}{\W(\lambda)} &=  w_0\left[1 + \frac{w_1 \lambda/\lambda_0}{1+\lambda/\lambda_0} + 
\bar w_0 
e^{-\hat\lambda_0/\lambda}\frac{(\lambda/\hat\lambda_0)^{4/3}}{\log(1+\lambda/\hat\lambda_0)}\right] \  .
\end{align}
Notice that we do not need the function $\kappa(\lambda)$, because it is the kinetic coupling of the tachyon field $\chi$, and for V-QCD we only consider chirally symmetric configurations at zero quark mass so that $\chi=0$. The strong coupling asymptotics of $V_{f0}$ and $\W$ is chosen such that the asymptotics of the spectra of radially excited mesons are linear~\cite{Jarvinen:2011qe,Arean:2013tja}, solutions are regular at zero~\cite{Jarvinen:2011qe,Arean:2013tja} and at finite $\theta$-angle~\cite{Arean:2016hcs} (in particular the geometry ends in a ``good'' kind of singularity in the classification of~\cite{Gubser:2000nd}), 
and the phase diagram at finite chemical potential includes a confining phase~\cite{Ishii:2019gta}. The parameters $W_1$ and $W_2$ are determined by requiring agreement with the perturbative running of the 't~Hooft coupling at two loops for QCD in the Veneziano limit of large $N_c$ and $N_f$ with $N_f/N_c=1$, which gives
\begin{equation}
W_1 = \frac{8+3\, W_0}{9 \pi ^2} \ , \qquad W_2 = \frac{6488+999\, W_0}{15552 \pi ^4} \ .
\end{equation}
The remaining parameters are then fitted to QCD thermodynamics from lattice with 2+1 dynamical quarks as follows. The function $V_{f0}$ controls the flavour contributions to the thermodynamics at zero chemical potential. Therefore we fit~\cite{Jokela:2018ers} the parameters $W_0$ and $W_\mathrm{IR}$ as well as the Planck mass $M_\text{Pl}$ to the interaction measure $(\epsilon-3p)/T^4$~\cite{Borsanyi:2013bia}. Being a coupling of the field strength tensor, the function $\W$ controls the thermodynamics at finite $\mu$. Therefore we fit the parameters $\hat \lambda_0$, $w_0$, $w_1$, and $\bar w_0$ to the lattice QCD data for the baryon number susceptibility $\chi_2 = d^2p/d\mu^2|_{\mu=0}$ with 2+1 dynamical flavours~\cite{Borsanyi:2011sw}. The results for the three fits  \textbf{5b}, \textbf{7a}, and \textbf{8b} used in this letter can be found in Table~2 of~\cite{Jokela:2020piw}.

Finally, the energy scale of the model is a parameter which, as in real QCD, does not appear in the action but is a property of the solutions. It is convenient to specify the scale by studying the asymptotics of the solution in the weak coupling regime. For a different approach for comparison to the QCD scale and running coupling in a holographic framework see~\cite{Brodsky:2010ur}.

We choose a gauge where $g_{tt}(r)g_{rr}(r)=g_{xx}(r)^2$ and 
choose the coordinate value where the coupling vanishes to be $r=0$. The Einstein equations then imply that
\begin{align}
g_{xx}(r) &= \frac{\ell^2}{r^2}\left[1+\frac{8}{9 \log r \Lambda_\text{UV}} +\mathcal{O}(\log r\Lambda_\text{UV})^{-2}\right]  & \\[2mm]
\lambda(r) &= e^{\sqrt{3/8}\, \phi(r)} = -\frac{8\pi^2}{3 \log r \Lambda_\text{UV}}   -\frac{28 \pi ^2}{27}  \frac{\log(-\log r\Lambda_\text{UV} )}{(\log r \Lambda_\text{UV})^2} +   \mathcal{O}(\log r\Lambda_\text{UV})^{-3} \ ,
\end{align}
where $\ell = 1/\sqrt{1-W_0/12}$ is the asymptotic AdS radius. These expansions define the energy scale $\Lambda_\text{UV}$. Moreover, the renormalisation scale in field theory is identified as 
$\sqrt{g_{xx}(r)}$~\cite{Gursoy:2007cb}, which was used to map these expansions to the perturbative renormalisation group flow of QCD in order to determine the coefficients $V_{1}$, $V_2$, $W_1$, and $W_2$ above. Comparing our results to the lattice data for the interaction measure, we find that $\Lambda_\text{UV}=226.24,\, 210.76,\, \text{and}\, 156.68\, \operatorname{MeV}$ for the fits \textbf{5b}, \textbf{7a}, and \textbf{8b}, respectively.

\section{Conductivities in a relativistic fluid}\label{app:cond}

In the  background metric $G_{\mu\nu}$ and in the presence of a background gauge field $A_\mu$, the constitutive relations for the energy-momentum tensor and charge current of a relativistic fluid read
\be
T^{\mu\nu}=(\varepsilon+p) u^\mu u^\nu+p\, G^{\mu\nu}+\tau^{\mu\nu}\ ,\ J^\mu=\rho u^\mu+\nu^\mu\, ,
\ee
where $u^\mu$ is the fluid velocity ($u^\mu u_\mu=-1$), $p$ the pressure, $\varepsilon$ the energy density, and $\rho$ the charge density. The thermodynamic potentials depend on the temperature $T$ and chemical potential $\mu$ that will be the dynamical variables together with the velocity. 

In the above energy-momentum tensor, the terms $\tau^{\mu\nu}$ and $\nu^\mu$ contain derivatives of the fields. We work in the Landau frame, where
\be
u_ \mu \tau^{\mu\nu}=0\ , \ u_\mu\nu^\mu=0\ .
\ee
We note that in the absence of parity violation, the most general derivative terms in the current compatible with the second law of thermodynamics read, to first order in derivatives,
\be
\nu^\mu=\sigma \left( E^\mu-T P^{\mu\nu}\nabla_\nu \left( \frac{\mu}{T}\right)\right)\ .
\ee
Here, $P^{\mu\nu}$ is the projector transverse to the velocity, and  $E^\mu=F^{\mu\nu}u^\nu$ is the electric field. The derivative terms of the energy-momentum tensor will not be relevant in the following.

The dynamics of the fluid is determined by the conservation equations
\be
\nabla_\mu T^{\mu\nu}=F^{\nu\lambda} J_\lambda\ , \ \nabla_\mu J^\mu=0 \ .
\ee
In the absence of sources $G_{\mu\nu}=\eta_{\mu\nu}$, $A_\mu=0$, the energy and charge densities and the pressure are constant, and the fluid is at rest.

We now turn on small homogeneous time-dependent perturbations $h_{\mu\nu}$, $a_\mu$:
\be
G_{\mu\nu}=\eta_{\mu\nu}+ h_{\mu\nu}(t),\ \ A_\mu= a_\mu(t)\, ,
\ee
where $\eta_{\mu\nu}$ is the Minkowski metric. These perturbations will induce a small change in the hydrodynamic variables $T$, $\mu$, and $u^i$ that can be found by solving the hydrodynamic equations to linear order in the sources.
The value of the energy-momentum tensor and the current in the presence of the external sources is then obtained by inserting the solutions for the hydrodynamic variables back in the constitutive relations and expanding to linear order. For the calculation of the conductivities, we can set $h_{00}=h_{ij}=a_0=0$.

The explicit dependence of the currents on the sources turns out to read
\be\label{eq:hydrosol}
\begin{split}
	& J_i  = -\frac{\rho^2}{\varepsilon+p} a_i-\rho h_{0i}-\sigma \partial_t a_i \\
	& T^0_{\ i}  =-\rho a_i-(\varepsilon+p) h_{0i}\, ,
\end{split}
\ee
while a constant electric field and temperature gradient $\zeta_i=-\partial_i T/T$ correspond to sources linear in time, i.e.
\be
a_i=-t(E_i-\mu \zeta_i)\ , \ h_{0i}=-t \zeta_i \ .
\ee
Introducing these expressions in Eq.~\eqref{eq:hydrosol}, we readily obtain
\be
\begin{split}
	&J_i= \frac{\rho}{\varepsilon+p}t\left((\varepsilon+p-\mu \rho)\zeta_i +\rho E_i\right) +\sigma (E_i-\mu\zeta_i) \\
	&T^0_{\ i}=t\left[\rho E_i+(\varepsilon+p-\mu\rho)\zeta_i\right] \ .
\end{split}
\ee

One can impose the condition that the fluid remains at rest, $T^0_i=0$ (no convection), by imposing the following relation between the electric field and the gradient of temperature
\be\label{eq:notdep}
\rho\, E_i+T s\, \zeta_i=0\, ,
\ee
where we have used the thermodynamic relation $\varepsilon+p-\mu\rho=T s$. Physically, the forces induced by the electric and temperature gradients compensate each other, so all transport will occur through diffusion.

The charge and heat currents finally become
\be
\begin{split}
	&J_i= \sigma E_i-\mu\sigma \zeta_i \\
	&Q_i=T^0_{\ i}-\mu J^i=-\mu\sigma  E_i+\mu^2\sigma \zeta_i\, .
\end{split}
\ee
Using Eq.~\eqref{eq:notdep} to solve for $\zeta_i$ in terms of $E_i$, or vice versa, we then obtain for the currents
\be
\begin{split}
	&J_i= \sigma\frac{\varepsilon+p}{Ts} E_i \\ 
	&Q_i=T^0_{\ i}-\mu J^i=\mu\sigma\frac{\varepsilon+p}{\rho}\zeta_i\ ,
\end{split}
\ee
so that the electrical and  thermal conductivities read
\be
\sigma^{ij}= \sigma\frac{\varepsilon+p}{Ts}  \delta^{ij} \ , \ \kappa^{ij}=\frac{\mu}{T}\sigma\frac{\varepsilon+p}{\rho}\delta^{ij} \ .
\ee

\backmatter

\newpage
\thispagestyle{empty}
\backgroundsetup{
	scale=1.05,
	angle=0,
	opacity=1,  
	contents={\includegraphics[]{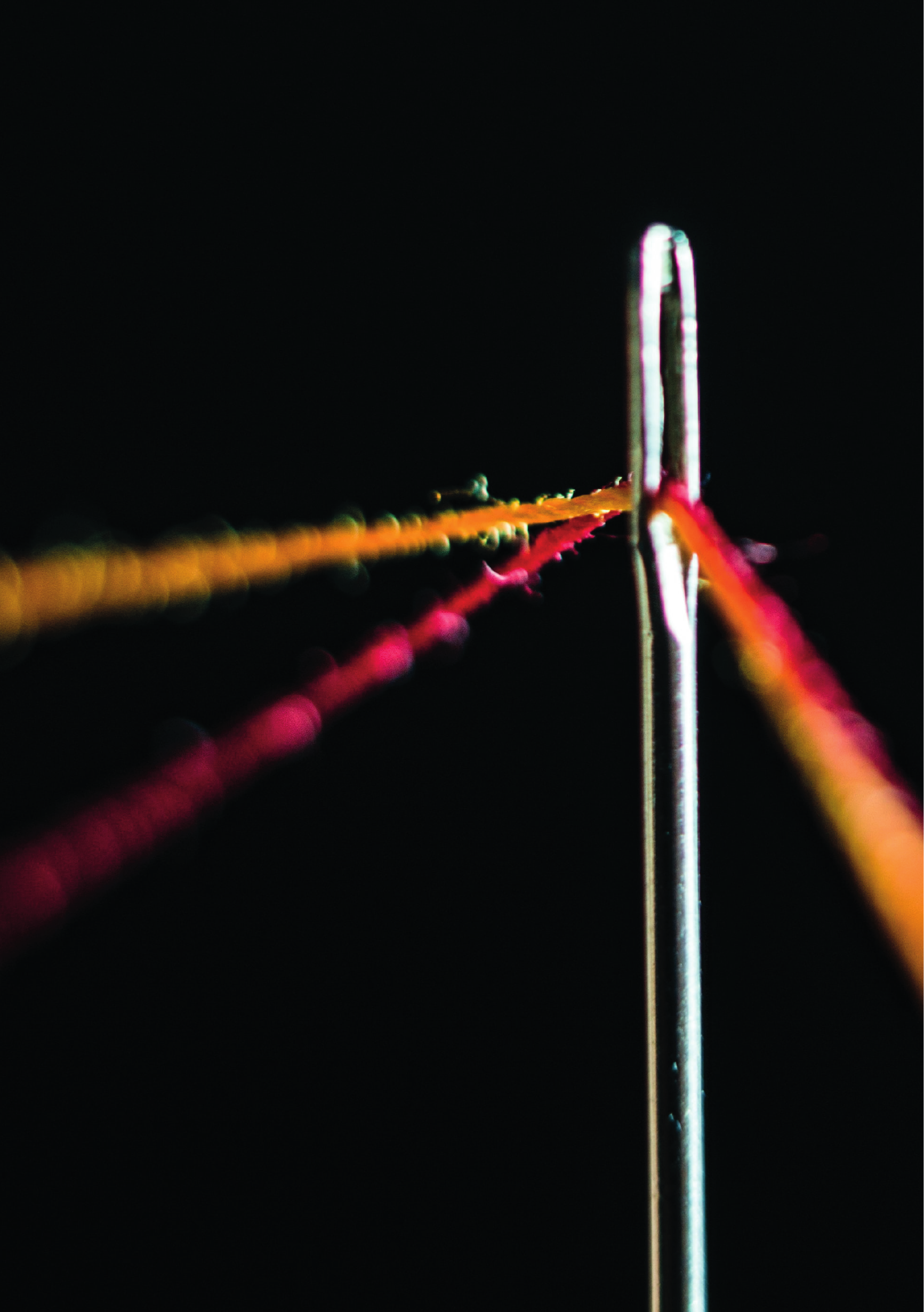}}
}
$\mbox{ }$
\newpage

\end{document}